\documentclass[1p,sort&compress]{elsarticle}
\usepackage{amsmath}
\usepackage{amsfonts}
\usepackage{amssymb}
\usepackage{mathtools}
\usepackage[latin1]{inputenc}
\usepackage{slashed}
\usepackage{bm}
\usepackage{graphicx}
\usepackage{tensor}
\usepackage{setspace}

\usepackage[normalem]{ulem}

\usepackage[usenames,dvipsnames]{color}

\usepackage[caption=false]{subfig}

\numberwithin{equation}{section}

\newcommand{\rv}[1]{\vec{#1}}
                   
\newcommand{\abs}[1]{\lvert #1 \rvert}                   
\newcommand{\bv}[1]{\mathbf{#1}}
\newcommand{\mean}[1]{\langle #1 \rangle} 
\newcommand{\smean}[1]{\left\langle #1 \right\rangle} 
\newcommand{\cc}{{\rm c}}

\newcommand\epigraph[3]{
\vspace{1em}\hfill{}\begin{minipage}{#1}{\begin{spacing}{0.9}
\small\noindent\textit{#2}\end{spacing}
\vspace{1em}
\hfill{}{#3}}\vspace{2em}
\end{minipage}}

\journal{Physics Reports}

\begin{document}

\begin{frontmatter}
\title{Spacetime algebra \\as a powerful tool for electromagnetism}
\author[river,cems]{Justin Dressel}
\author[cems,itsrg]{Konstantin Y. Bliokh}
\author[cems,umich]{Franco Nori}
\address[river]{Department of Electrical and Computer Engineering, \\University of California, Riverside, CA 92521, USA}
\address[cems]{Center for Emergent Matter Science (CEMS), RIKEN, Wako-shi, Saitama, 351-0198, Japan}
\address[itsrg]{Interdisciplinary Theoretical Science Research Group (iTHES), RIKEN, Wako-shi, Saitama, 351-0198, Japan}
\address[umich]{Physics Department, University of Michigan, Ann Arbor, MI 48109-1040, USA}

\begin{abstract}
  We present a comprehensive introduction to \emph{spacetime algebra} that emphasizes its practicality and power as a tool for the study of electromagnetism. We carefully develop this natural (Clifford) algebra of the Minkowski spacetime geometry, with a particular focus on its intrinsic (and often overlooked) \emph{complex structure}. Notably, the scalar imaginary that appears throughout the electromagnetic theory properly corresponds to the unit 4-volume of spacetime itself, and thus has \emph{physical} meaning. The electric and magnetic fields are combined into a single complex and frame-independent bivector field, which generalizes the \emph{Riemann-Silberstein complex vector} that has recently resurfaced in studies of the single photon wavefunction. The complex structure of spacetime also underpins the emergence of electromagnetic waves, circular polarizations, the normal variables for canonical quantization, the distinction between electric and magnetic charge, complex spinor representations of Lorentz transformations, and the \emph{dual} (electric-magnetic field exchange) \emph{symmetry} that produces helicity conservation in vacuum fields.  This latter symmetry manifests as an arbitrary global phase of the complex field, motivating the use of a complex vector potential, along with an associated transverse and gauge-invariant bivector potential, as well as complex (bivector and scalar) Hertz potentials.  Our detailed treatment aims to encourage the use of spacetime algebra as a readily available and mature extension to existing vector calculus and tensor methods that can \emph{greatly simplify} the analysis of fundamentally relativistic objects like the electromagnetic field. 
\end{abstract}

\end{frontmatter}


\tableofcontents
\listoftables
\listoffigures


\vspace{1cm}

\epigraph{4.5in}{If we include the biological sciences, as well as the physical sciences, Maxwell's paper was second only to Darwin's ``Origin of Species''.  But the importance of Maxwell's equations was not obvious to his contemporaries.  Physicists found it hard to understand because the equations were complicated.  Mathematicians found it hard to understand because Maxwell used physical language to explain it.}{Freeman J. Dyson \cite{Dyson1999}}

\epigraph{4.5in}{This shows that the mathematical language has more to commend it than being the only language which we can speak; it shows that it is, in a very real sense, the correct language. \ldots~The miracle of the appropriateness of the language of mathematics for the formulation of the laws of physics is a wonderful gift which we neither understand nor deserve.}{Eugene P. Wigner \cite{Wigner1960}}

\section{Introduction}



\epigraph{4.5in}{For the rest of my life I will reflect on what light is.}{Albert Einstein \cite{Perkowitz1999}}


\subsection{Motivation}


At least two distinct threads of recent research have converged upon an interesting conclusion: the electromagnetic field has a more natural representation as a \emph{complex} vector field, which then behaves in precisely the same way as a relativistic quantum field for a single particle (i.e., a \emph{wavefunction}).  It is our aim in this report to clarify why this conclusion is warranted from careful considerations of the (often overlooked) intrinsic complex structure implied by the \emph{geometry of spacetime}.  We also wish to draw specific parallels between these threads of research, and unify them into a more comprehensive picture.  We find that the coordinate-free, geometrically motivated, and manifestly reference-frame-independent \emph{spacetime algebra} of Hestenes \cite{Hestenes1966,Hestenes1987,Hestenes2003,Hestenes2003b,Hestenes2005,Doran2007} is particularly well suited for this task, so we provide a thorough introduction to this formalism in order to encourage its use and continued development.

The first thread of preceding research concerns the development of a \emph{first-quantized approach to the single photon}.  Modern quantum optical experiments routinely consider situations involving only a single photon or pair of photons, so there has been substantial motivation to reduce the full quantized field formalism to a simpler `\emph{photon wave function}' approach that more closely parallels the Schr\"odinger or Dirac equations for a single electron \cite{Laporte1931,Oppenheimer1931,Mignani1974,Archibald1955,Good1957,Giannetto1985,Fushchich1987,Berry1990,Bialynicki-Birula1994,Bialynicki-Birula1996,Sipe1995,Esposito1998,Stepanovsky1998,Kobe1999,Gersten2001,Smith2006,Smith2007,Dragoman2007,Tamburini2008,Thide2011,
Barnett2014,Thide2014}. Such a first-quantization approach is equivalent to classical electromagnetism: it represents Maxwell's equations in the form of a relativistic wave equation for massless spin-1 particles.  The difficulty with this pursuit has stemmed from the fact that the photon, being a manifestly relativistic and massless spin-1 boson, is \emph{nonlocalizable}. More precisely, the position of a single photon is ill-defined \cite{Pryce1948,Newton1949,Wightman1962,Skagerstam1992,Keller2005,Bacry1988}, and photons have no localizable number density \cite{Bialynicki-Birula1996,Sipe1995,Smith2007,Keller2005}, which makes it impossible to find a position-resolved probability amplitude in the same way as for electrons\footnote{The absence of a probability density for photons follows from the Weinberg--Witten theorem \cite{Weinberg1980} that forbids conserved four-vector currents for relativistic massless fields with spins higher than 1/2 \cite{Stepanovsky1998,Smith2007}.}. Nevertheless, the \emph{energy} density of a single photon in a particular inertial frame \emph{is} localizable, and closely corresponds to what is actually measured by single photon counters.  

Following this line of reasoning, a suitable `photon wave function' can be derived from general considerations as an energy-density amplitude in a particular inertial reference frame \cite{Bialynicki-Birula1994,Bialynicki-Birula1996,Gersten2001,Smith2006}, which directly produces a complex form of the usual electric and magnetic fields $\rv{E}$ and $\rv{B}$ in that frame
\begin{align}\label{eq:RiemannSilberstein}
  \rv{F} = [\rv{E}/\cc + \rv{B}i]/\sqrt{\mu_0}.
\end{align}
Furthermore, the quantum mechanical derivation that produces \eqref{eq:RiemannSilberstein} also produces Maxwell's vacuum equations in an intriguing form that resembles the Dirac equation for massless spin-1 particles \cite{Laporte1931,Oppenheimer1931,Mignani1974,Archibald1955,Good1957,Giannetto1985,Fushchich1987,Berry1990,Bialynicki-Birula1994,Bialynicki-Birula1996,Sipe1995,Esposito1998,Stepanovsky1998,Kobe1999,Gersten2001,Smith2006,Smith2007,Dragoman2007,Tamburini2008,Thide2011,
Barnett2014,Thide2014} (shown here in both the momentum-operator notation and the equivalent coordinate representation) 
\begin{subequations}\label{eq:MaxwellRiemann}
\begin{align}
  i\hbar\, \partial_t \rv{F} &= \cc|p|\,\hat{\chi}\,\rv{F}~, & i\hbar\, \partial_t \rv{F} &= \cc \hbar\, \rv{\nabla}\times\rv{F}~, \\
 \label{eq:MaxwellRiemann2}
 \hat{\rv{p}}\cdot\rv{F} &= 0~, &~ -i\hbar\,\rv{\nabla}\cdot\rv{F} &= 0.
\end{align}
\end{subequations}
Here $\cc|p|$ is the energy of a single photon, and $\hat{\chi} = \hat{\rv{s}}\cdot\hat{\rv{p}}/|p|$ is the \emph{helicity} operator that projects the spin-1 matrix operator $\hat{\rv{s}}$ onto the direction of the momentum operator $\hat{\rv{p}} = -i\hbar\rv{\nabla}$, which produces the curl $(\hat{\rv{s}}\cdot\rv{\nabla})\rv{F} = i\rv{\nabla}\times\rv{F}$. The second equation \eqref{eq:MaxwellRiemann2} appears as an auxiliary transversality condition satisfied by radiation fields far from matter \cite{Cohen1997,Keller2005}. 

Treating the complex vector \eqref{eq:RiemannSilberstein} as a single-particle wave function in the first-quantized quantum mechanical sense produces consistent results, and suggests that this vector should be a more natural representation for classical electromagnetic fields as well.\footnote{The factor of $\hbar$ cancels in Maxwell's equations \eqref{eq:MaxwellRiemann} because the photon is massless. This explains the absence of this obviously quantum factor in classical electromagnetism.} Indeed, the practical classical applications of \eqref{eq:RiemannSilberstein} have been emphasized and reviewed by Bialynicki-Birula \cite{Bialynicki-Birula1996,Bialynicki-Birula2013}, who noted that this complex vector was also historically considered by Riemann \cite{Weber1901} and Silberstein \cite{Silberstein1907a,Silberstein1907b} in some of the earliest investigations of electromagnetism.  Remarkably, performing the standard second-quantization procedure using the complex Riemann--Silberstein wave function \eqref{eq:RiemannSilberstein} identically reproduces the correctly quantized electromagnetic field \cite{Smith2007,Tamburini2008}.  Moreover, if the number of photons in \eqref{eq:MaxwellRiemann} is increased to two using the standard bosonic symmetrization procedure, the resulting equations of motion precisely reproduce the Wolf equations from classical coherence theory \cite{Saleh2005,Smith2007} in a consistent way.  

The explicit appearance of the momentum, spin, and helicity operators in \eqref{eq:MaxwellRiemann} connects the complex field vector \eqref{eq:RiemannSilberstein} to the second thread of research into the \emph{local momentum, angular momentum, and helicity properties of structured optical fields} (such as optical vortices, complex interference fields, and near fields) \cite{Nye1974,Allen1992,Enk1994,Allen1999,Andrews2008,Torres2011,Andrews2013,O'Neil2002,Garces2003,Curtis2003,
Zhao2007,Adachi2007,Roy2014,Bliokh2014,Arlt2003,Leach2006,Yeganeh2013,Terborg2013,Dennis2008,
Berry2009,Bekshaev2011,Kocsis2011,Dennis2013,Bliokh2013,Bliokh2013a,Barnett2013,
Huard1978,Huard1979,Bliokh2012,Bliokh2013b,Bekshaev2014,
Tang2010,Hendry2010,Bliokh2011,Tang2011,Hendry2012,Schaferling2012,Canaguier2013,Cameron2012c,Tkachenko2014,Bliokh2013c}.  
The development of nano-optics employing structured light and local light-matter interactions has prompted the careful consideration of nontrivial local dynamical properties of light, even though such properties are usually considered to be unobservable in orthodox quantum-mechanical and field-theory approaches. 

Indeed, it was traditionally believed that \emph{spin and orbital angular momenta} of light are not meaningful separately, but only make sense when combined into the total angular momentum \cite{Akhiezer1965,Berestetskii1982,Cohen1997}. However, when small probe particles or atoms locally interact with optical fields that carry spin and orbital angular momenta, they clearly acquire separately observable intrinsic and extrinsic angular momenta that are proportional to the local expectation values of the spin $\hat{\rv{s}}$ and orbital $\hat{\rv{r}}\times\hat{\rv{p}}$ angular momentum operators for photons, respectively. \cite{Enk1994,O'Neil2002,Garces2003,Curtis2003,Zhao2007,Adachi2007,Roy2014,Bliokh2013b,Bliokh2014}. (Akin to this, the spin and orbital angular-momentum contributions of gluon fields in QCD are currently considered as measurable in spite of the field-theory restrictions \cite{Leader2014}.) Similarly, the local \emph{momentum density} of the electromagnetic field (as well as any local current satisfying the continuity equation) is not uniquely defined in field theory, so only the integral momentum value has been considered to be measurable. Nevertheless, when a small probe particle is placed in an optical field it acquires momentum proportional to the canonical momentum density of the field, i.e., the local expectation value of $\hat{\rv{p}}$ \cite{Berry2009,Bliokh2013a,Bliokh2013c}. Remarkably, this same canonical momentum density of the optical field (and \emph{not} the Poynting vector, see \cite{Berry2009,Bliokh2013,Bliokh2013b,Bliokh2013c}) is also recovered using other methods to measure the local field momentum \cite{Arlt2003,Leach2006,Yeganeh2013,Terborg2013,Huard1978,Huard1979,Barnett2013}, including the so-called \emph{quantum weak measurements} \cite{Kocsis2011,Wiseman2007,Traversa2013,Bliokh2013a,Dressel2014b}. Such local weak measurements of the momentum densities in light fields (i.e., analogues of the probability current for photons in the above `photon wave function' approach) simultaneously corroborate predictions of the Madelung hydrodynamic approach to the quantum theory \cite{Madelung1926,Madelung1927}, the Bohmian causal model \cite{Bohm1952a,Bohm1952b}, and the \emph{relativistic energy-momentum tensor in field theory} \cite{Hiley2012,Bliokh2013,Bliokh2013a,Bliokh2013b}.  In the quantum case, these local densities correspond to the associated classical mean (background) field, so can be observed only by averaging many measurements of the individual quantum particles \cite{Dressel2014b}.

Most importantly for our study, the reconsideration of the momentum and angular momentum properties of light has prompted discussion of the \emph{helicity} density of electromagnetic fields.  Recently it was shown that this helicity density naturally appears in local light-matter interaction with chiral particles \cite{Tang2010,Hendry2010,Bliokh2011,Tang2011,Andrews2012,Hendry2012,Schaferling2012,Bliokh2013c}. However, while in the quantum operator formalism the helicity $\hat{\chi}$ is an intuitive combination of the momentum and spin operators, the \emph{field-theory} picture of the electromagnetic helicity is not so straightforward.

Namely, electromagnetic helicity \cite{Candlin1965,Calkin1965,Oconnell1965,Zwanziger1968,Deser1976,Afanasiev1996,Trueba1996,
Cameron2012b,Bliokh2013,Cameron2012,Fernandez2012,Fernandez2013} is a curious physical electromagnetic-field property that is only conserved during vacuum propagation. Unlike the momentum and angular momentum, which are related to the Poincar\'{e} spacetime symmetries according to Noethers theorem \cite{Noether1918}, the helicity conservation corresponds to the more abstract \emph{dual symmetry} that involves the exchange of electric and magnetic fields \cite{Calkin1965,Zwanziger1968,Deser1976,Cameron2012b,Bliokh2013,Cameron2012,Fernandez2012,Fernandez2013} (which was first discussed by Heaviside and Larmor \cite{Heaviside1892,Larmor1897}, and has been generalized to field and string theories beyond standard electromagnetism \cite{Born1934,Schrodinger1935,Ferrara1977,Gaillard1981,Sen1993,Schwarz1994,Pasti1995b,Gibbons1995,Hull1995,Deser1998,Cremmer1998,Figueroa-O'Farrill,Aschieri2008,Aschieri2014}). Due to this intimate relation with the field-exchange symmetry, the definition of the helicity must involve the electric and magnetic fields and their two associated vector-potentials on equal footing.  Furthermore, the consideration of the dual symmetry, which is inherent to the vacuum Maxwell's equations \eqref{eq:MaxwellRiemann}, recently reignited discussion regarding the proper dual-symmetric description for all vacuum-field properties, including the canonical momentum and angular-momentum densities \cite{Berry2009,Barnett2010,Cameron2012,Bliokh2013}. These quantities are dual \emph{asymmetric} in the traditional electromagnetic field theory \cite{Soper1976,Bliokh2013}.  Notably, the dual electric-magnetic symmetry become particularly simple and natural when the electric and magnetic fields and potentials are combined into the complex Riemann--Silberstein form \eqref{eq:RiemannSilberstein} \cite{Gaillard1981,Bliokh2013}. Indeed, the continuous dual symmetry of the vacuum electromagnetic field takes the form of a simple $U(1)$ global phase-rotation 
\begin{align}
  \rv{F} \mapsto \rv{F}\,\exp(i\theta).
\end{align}
It is this continuous `gauge symmetry of the photon wave function' that produces the conservation of optical helicity as a consequence of Noether's theorem. Moreover, the whole Lagrangian electromagnetic field theory in vacuum, including its fundamental local Noether currents (energy-momentum, and angular momentum), becomes more natural and self-consistent with a properly complex and dual-symmetric Lagrangian \cite{Bliokh2013}. This provides further evidence that the complex form of \eqref{eq:RiemannSilberstein} is a more fundamental representation for the electromagnetic field.  

In this report we clarify why the complex Riemann--Silberstein representation \eqref{eq:RiemannSilberstein} has fundamental significance for the electromagnetic field.  To accomplish this goal, we carefully construct the natural (Clifford) algebra of spacetime \cite{Hestenes1966} and detail its intrinsic complex structure.  We then derive the entirety of the traditional electromagnetic theory as an inevitable consequence of this \emph{spacetime algebra}, paying close attention to the role of the complex structure at each stage.  The complex vector \eqref{eq:RiemannSilberstein} will appear naturally as a relative 3-space expansion of a \emph{bivector field}, which is a proper geometric (and thus frame-independent) object in spacetime.  Moreover, this bivector field will have a crucial difference from \eqref{eq:RiemannSilberstein}: the scalar imaginary $i$ will be replaced with an algebraic pseudoscalar $I$ (also satisfying $I^2 = -1$) that is intrinsically meaningful to and required by the geometric structure of spacetime.  This replacement makes the complex form of \eqref{eq:RiemannSilberstein} \emph{reference-frame-independent}, which is \emph{impossible} when the mathematical representation is restricted to the usual scalar imaginary $i$.  We find this replacement of the scalar $i$ with $I$ to be systematic throughout the electromagnetic theory, where it appears in electromagnetic waves, elliptical polarization, the normal variables for canonical field mode quantization, and the dual-symmetric internal rotations of the vacuum field.  Evidently, the intrinsic complex structure inherent to the geometry of spacetime has deep and perhaps under-appreciated consequences for even our classical field theories.

This report is organized as follows.  In Section \ref{sec:insights} we briefly summarize some of the most interesting insights that we will uncover from spacetime algebra in the remainder of the report.  In Section \ref{sec:history} we present a brief history of the formalism for the electromagnetic theory to give context for how the spacetime algebra approach relates.  In Section \ref{sec:algebra} we give a comprehensive introduction to spacetime algebra, with a special focus on the emergent complex structure.  In Section \ref{sec:calculus} we briefly extend this algebra to a spacetime manifold on which fields and calculus can be defined.  In Section \ref{sec:maxwell} we derive Maxwell's equations in vacuum from a \emph{single}, simple, and inevitable equation.  In Section \ref{sec:potentials} we consider three different potential representations, which are all naturally complex due to the dual symmetry of the vacuum field.  In Section \ref{sec:source} we add a source to Maxwell's equation and recover the theory of electric and magnetic charges, along with the appropriately symmetrized Lorentz force and its associated energy-momentum and angular momentum tensors.  In Section \ref{sec:lagrangian} we revisit the electromagnetic field theory. We start with the field Lagrangian and modify it to preserve dual symmetry in two related ways, derive the associated Noether currents, including helicity, and remark upon the breaking of dual symmetry that is inherent to the gauge mechanism.  We conclude in Section \ref{sec:conclusion}.


\subsection{Insights from the spacetime algebra approach}\label{sec:insights}


To orient the reader, we summarize a few of the interesting insights here that will arise from the spacetime algebra approach.  These lists indicate what the reader can expect to understand from a more careful reading of the longer text.  Note that more precise definitions of the quantities mentioned here are contained in the main text.

\vspace{3mm}

The Clifford product for spacetime algebra is initially motivated by the following basic benefits of its resulting \emph{algebraic structure}:
\begin{itemize}
  \item The primary reference-frame-independent objects of special relativity (i.e., scalars, 4-vectors, bivectors / anti-symmetric rank-2 tensors, pseudo-4-vectors, and pseudoscalars) are unified as distinct grades of a \uline{single object known as a multivector}.  An element of grade $k$ \emph{geometrically} corresponds to an oriented surface of dimension $k$ (e.g., 4-vectors are line segments, while bivectors are oriented plane segments).
  \item The associative Clifford product combines the dot and wedge vector-products into a 
\uline{single and often invertible product} (e.g., for a 4-vector $a^{-1} = a/a^2$).
  \item The symmetric part of the Clifford product is the \uline{dot product} (Minkowski metric): $a \cdot b = (ab + ba)/2$.  Thus, the square of a vector is its scalar pseudonorm $a^2 = \epsilon_a|a|^2$ with signature $\epsilon_a = \pm 1$.
  \item The antisymmetric part of the Clifford product is the \uline{wedge product} $a\wedge b = (ab - ba)/2$ (familiar from differential forms), which generalizes the 3-vector cross product $(\times)$.
  \item The sign ambiguity of the \uline{spacetime metric signature} (dot product) is fixed to $(+,-,-,-)$ from general considerations that embed spacetime into a larger sequence of (physically meaningful) nested Clifford subalgebras.
\end{itemize}

From this structure we obtain a variety of useful and enlightening \emph{mathematical unifications}:
\begin{itemize}
  \item The \uline{Dirac matrices} $\gamma_\mu$ (usually associated with quantum mechanics) appear as a matrix representation of an orthonormal basis for the 4-vectors in spacetime algebra, emphasizing that they are not intrinsically quantum mechanical in nature.  The matrix product in the matrix representation simulates the Clifford product.  One does \emph{not} need this matrix representation, however, and can work directly with $\gamma_\mu$ as purely algebraic 4-vector elements, which dramatically simplifies calculations in practice.
  \item The \uline{Dirac differential operator} $\nabla = \sum_\mu \gamma^\mu \partial_\mu$ appears as the proper vector derivative on spacetime, with \emph{no} matrix representation required.
  \item The usual \uline{3-vectors} from nonrelativistic electromagnetism are \uline{spacetime bivectors} that depend on a particular choice of inertial frame, specified by a timelike unit vector $\gamma_0$.  A unit 3-vector $\rv{\sigma}_i = \gamma_i\gamma_0 = \gamma_i\wedge\gamma_0$ that is experienced as a spatial axis by an inertial observer thus has the geometric meaning of a \emph{plane-segment} that is obtained by dragging the spatial unit 4-vector $\gamma_i$ along the chosen proper-time axis $\gamma_0$.
  \item The \uline{Pauli matrices} are a matrix representation of an orthonormal basis of relative 3-vectors $\rv{\sigma}_i$, emphasizing that they are also not intrinsically quantum mechanical in nature.  The matrix product in the matrix representation again simulates the Clifford product.
  \item Factoring out a particular timelike unit vector $\gamma_0$ from a 4-vector $v = \sum_\mu v^\mu\gamma_\mu = (v_0 + \rv{v})\gamma_0$ produces a \uline{paravector} $(v_0 + \rv{v})$ as a sum of a relative scalar $v_0$ and relative vector $\rv{v} = \sum_i v_i \rv{\sigma}_i$.  Thus relative paravectors and proper 4-vectors are \emph{dual} representations under right multiplication by $\gamma_0$.
  \item The usual \uline{3-gradient vector derivative} $\rv{\nabla} = \gamma_0\wedge\nabla = \sum_i \rv{\sigma}_i\partial_i $ is the spatial part of the Dirac operator in a particular inertial frame specified by $\gamma_0$.
  \item The \uline{d'Alembertian} is the square of the Dirac operator $\nabla^2 = \partial_0^2 - \rv{\nabla}^2$.
  \item \uline{Directed integration} can be defined using a Riemann summation on a spacetime manifold, where the measure $d^k\!x$ becomes an oriented $k$-dimensional geometric surface that is represented within the algebra. This integration reproduces and generalizes the standard results from vector analysis, complex analysis (including Cauchy's integral theorems), and differential geometry.
\end{itemize}

All scalar components of the objects in spacetime algebra are \emph{purely real numbers}.  Nevertheless, an intrinsic \emph{complex structure} emerges within the algebra due to the geometry of spacetime itself:
\begin{itemize}
  \item The \uline{``imaginary unit''} naturally appears as the pseudoscalar (unit 4-volume) $I = \gamma_0\gamma_1\gamma_2\gamma_3 = \rv{\sigma}_1\rv{\sigma}_2\rv{\sigma}_3$, such that $I^2 = -1$.  This pseudoscalar plays the role of the scalar imaginary $i$ throughout the electromagnetic theory, without the need for any additional \emph{ad hoc} introduction of a complex scalar field.
  \item The \uline{Hodge-star duality operation} from differential forms is simply the right multiplication of any element by $I^{-1} = -I$.  This operation transforms a geometric surface of dimension $k$ into its orthogonal complement of dimension $4-k$ (e.g., the Hodge dual of a 4-vector $v$ is its orthogonal 3-volume $vI^{-1}$, or pseudo-4-vector).
  \item The \uline{quaternion algebra of Hamilton} $(1,\bv{i},\bv{j},\bv{k})$ that satisfies the defining relations, $\bv{i}^2 = \bv{j}^2 = \bv{k}^2 = \bv{i}\bv{j}\bv{k} = -1$ also appears as the (left-handed) set of spacelike planes (bivectors) in a relative inertial frame: $\bv{i} = \rv{\sigma}_1 I^{-1}$, $\bv{j} = -\rv{\sigma}_2 I^{-1}$, $\bv{k} = \rv{\sigma}_3 I^{-1}$.  The bivectors $\rv{\sigma}_i I^{-1}$ directly describe the planes in which spatial rotations can occur, which explains why quaternions are particularly useful for describing spatial rotations.
\end{itemize}

In addition to the complex structure, \emph{Lie group} and \emph{spinor} structures also emerge:
\begin{itemize}
  \item The bivector basis forms the \uline{Lie algebra of the Lorentz group} under the Lie (commutator) bracket relation $[\bv{F},\bv{G}] = (\bv{F}\bv{G} - \bv{G}\bv{F})/2$.  Thus, bivectors directly generate \uline{Lorentz transformations} when exponentiated.
  \item The \uline{Lorentz group generators} are the 6 bivectors $\bv{S}_i = \rv{\sigma}_i I^{-1}$ and $\bv{K}_i = \rv{\sigma}_i$.  Thus, writing the 15 brackets of the Lorentz group $[\bv{S}_i,\bv{S}_j] = \epsilon_{ijk}\bv{S}_k$, $[\bv{S}_i,\bv{K}_j] = \epsilon_{ijk}\bv{K}_k$ and $[\bv{K}_i,\bv{K}_j] = -\epsilon_{ijk}\bv{S}_k $ in terms of a particular reference frame produces simple variations of the three fundamental commutation relations $[\rv{\sigma}_i,\rv{\sigma}_j] = \epsilon_{ijk}\rv{\sigma}_k I$ that are usually associated with quantum mechanical spin (using a Pauli matrix representation).  Indeed, the bivectors $\rv{\sigma}_i I^{-1}$ are spacelike planes that generate spatial rotations when exponentiated (just as in quantum mechanics), while $\rv{\sigma}_i$ are the timelike planes, and thus generate Lorentz boosts when exponentiated.
  \item The usual \uline{3-vector cross product} is the Hodge-dual of the bivector Lie bracket: $\bv{F}\times\bv{G} = [\bv{F},\bv{G}]I^{-1}$.  Thus, the set of commutation relations $[\rv{\sigma}_i,\rv{\sigma}_j] = \epsilon_{ijk}\rv{\sigma}_k I$ between the 3-vectors $\rv{\sigma}_i$ are simply another way of expressing the usual cross-product relations $\rv{\sigma}_i \times \rv{\sigma}_j = \epsilon_{ijk}\rv{\sigma}_k$.
  \item General \uline{spinors} $\psi$ appear as the (closed) even-graded subalgebra of spacetime algebra (i.e., scalars $\alpha$, bivectors $\bv{F}$, and pseudoscalars $\alpha I$). Lorentz and other group transformations acquire a simplified form as double-sided products with these spinors (e.g., exponentiated generators like $\psi = \exp(-\alpha \rv{\sigma}_3 I - \phi I) = [\cos\alpha - \sin\alpha\,\rv{\sigma}_3 I][\cos\phi - \sin\phi I]$), just as is familiar from unitary transformations in quantum mechanics and spatial rotations expressed using quaternions.
\end{itemize}

All these mathematical benefits produce a considerable amount of \emph{physical} insight about the \emph{electromagnetic theory} in a straightforward way:
\begin{itemize}
  \item The \uline{electromagnetic field} is a bivector field $\bv{F}$ with the same components $F^{\mu\nu}$ as the usual antisymmetric tensor.  This tensor is the corresponding multilinear function $\underbar{F}(v,w) = v\cdot\bv{F}\cdot w = \sum_{\mu\nu} v_\mu F^{\mu\nu} w_\nu$ that contracts its two 4-vector arguments ($v$ and $w$) with the bivector $\bv{F}$. 
  \item The electromagnetic field is an \uline{irreducibly complex} object with an intrinsic phase $\bv{F} = \bv{f}\exp(\varphi I)$.  This phase necessarily involves the intrinsic pseudoscalar (unit 4-volume) $I$ of spacetime, and is intimately related to the appearance of \uline{electromagnetic waves} and \uline{circular polarizations}, with no need for any \emph{ad hoc} addition of a complex scalar field.
  \item The continuous \uline{dual (electric-magnetic exchange) symmetry} of the vacuum field is a U(1) internal gauge-symmetry (phase rotation) of the field: $\bv{F} \mapsto \bv{F}\exp(\theta I)$.  This symmetry produces the conservation of helicity via Noether's theorem.
  \item Bivectors decompose in a relative inertial frame into a complex pair of 3-vectors $\bv{F} = \rv{E} + \rv{B}I$, which clarifies the origin of the \uline{Riemann--Silberstein vector}.  Both real and imaginary parts in a particular frame are thus physically meaningful.  Moreover, the geometric properties of $I$ keep $\bv{F}$ reference-frame-independent, even though the decomposition into relative fields $\rv{E} = (\bv{F}\cdot\gamma_0)\gamma_0$ and $\rv{B}I = (\bv{F}\wedge\gamma_0)\gamma_0$ still implicitly depends upon the chosen proper-time axis $\gamma_0$ in the relative 3-vectors $\rv{\sigma}_i = \gamma_i\gamma_0$ that form the basis for $\rv{E}$ and $\rv{B}$.
  \item All of \uline{Maxwell's equations in vacuum} reduce to a single equation: $\nabla \bv{F} = 0$.
  \item All of \uline{Maxwell's equations with sources} (both electric and magnetic) also reduce to a single equation: $\nabla \bv{F} = j$, where $j = j_e + j_m I$ is a complex representation of both types of source, $j_e = (c\rho_e + \rv{J}_e)\gamma_0$ and $j_m = (c\rho_m + \rv{J}_m)\gamma_0$.
  \item The scalar \uline{Lorentz invariants} of the electromagnetic field are the invariant parts of its square: $\bv{F}^2 = (|\rv{E}|^2 - |\rv{B}|^2) + 2(\rv{E}\cdot\rv{B})I$.
  \item A \uline{circularly polarized plane wave} is intrinsically complex with the simple exponential form $\bv{F}(x) = (sk)\exp[\pm(k\cdot x)I]$, with spacelike unit vector $s=\rv{E}_0\gamma_0$, coordinates $x = (\cc t + \rv{x})\gamma_0$, and null wavevector $k=(\omega/\cc +\rv{k})\gamma_0$ such that $k^2 = |\omega/\cc|^2 - |\rv{k}|^2 = 0$ is the usual dispersion relation.  The sign of $I$ corresponds to the invariant handedness (i.e., \emph{helicity}) of the wave.  Expanding the exponential in the relative frame $\gamma_0$ yields $\bv{F} = \rv{E} + \rv{B}I$ with the relative fields  $\rv{E} = \rv{E}_0\cos(\rv{k}\cdot\rv{x} - \omega t/\cc) \pm \rv{\kappa}\times\rv{E}_0\sin(\rv{k}\cdot\rv{x}-\omega t/\cc)$ and $\rv{B} = \rv{E}_0\sin(\rv{k}\cdot\rv{x}-\omega t/\cc) \mp \rv{\kappa}\times\rv{E}_0\cos(\rv{k}\cdot\rv{x} - \omega t/\cc)$, with unit vector $\rv{\kappa}=\rv{k}/|\rv{k}|$ indicating the propagation direction.  
  \item The \uline{vector-potential} representation is $\bv{F} = \nabla z = (\nabla\wedge a_e) + (\nabla\wedge a_m)I$, where $z = a_e + a_m I$ is a complex representation of both electric $a_e = (\phi_e + \rv{A})\gamma_0$ and magnetic $a_m = (\phi_m + \rv{C})\gamma_0$ 4-vector potentials, each satisfying the Lorenz-FitzGerald gauge conditions $\nabla\cdot a_e = \nabla\cdot a_m = 0$.  Maxwell's equation then has the simple wave equation form $\nabla^2 z = j$ with the correspondingly complex source current $j = j_e + j_m I$.
  \item A \uline{transverse and gauge-invariant bivector potential} representation can be defined in a particular inertial frame $\bv{Z} = z_\perp \gamma_0 = \rv{A}_\perp - \rv{C}_\perp I$, where $\bv{F} = (\nabla \bv{Z})\gamma_0$.  This complex potential appears in the definition of the conserved optical helicity.
  \item The complex scalar \uline{Hertz potential} $\Phi = \Phi_e + \Phi_m I$ for vacuum fields appears as $\bv{Z} = (\rv{\nabla}\times\rv{\Pi})I \Phi$ with a chosen unit direction vector $\rv{\Pi}$, and satisfies $\nabla^2 \Phi = 0$.
  \item The proper \uline{Lorentz force} is $d(mw)/d\tau = \bv{F}\cdot(qw) = \mean{\bv{F}(qw)}_1$, where $w$ is a proper 4-velocity of a particle with charge $q$ and mass $m$.
  \item Making the charge $q$ complex with a dual-symmetry phase rotation $q \mapsto qe^{\theta I} = q_e + q_m I$ produces an equivalent \uline{magnetic monopole} description using the corresponding phase-rotated field $\bv{F}\mapsto \bv{F}e^{\theta I}$.  The proper Lorentz force that includes these magnetic monopoles is still $d(mw)/d\tau = \mean{\bv{F}(qw)}_1$.  Choosing electric sources is an arbitrary convention that essentially fixes this duality gauge freedom and \emph{breaks the dual symmetry} of the fields.
  \item The \uline{symmetric (Belinfante) energy-momentum tensor} is a bilinear function of the electromagnetic field and a chosen proper-time direction $\gamma_0$ that has a simple quadratic form: $\overline{T}_{\text{sym}}(\gamma_0) = \bv{F}\gamma_0\tilde{\bv{F}}/2 = (\rv{E} + \rv{B}I)(\rv{E}-\rv{B}I)\gamma_0/2 = (\varepsilon + \rv{P})\gamma_0$.  This tensor recovers the usual energy density $\varepsilon = (|\rv{E}|^2 + |\rv{B}|^2)/2$ and Poynting vector $\rv{P} = \rv{E}\times\rv{B}$ in the relative frame $\gamma_0$, and has the same invariant mathematical form as the symmetric energy-momentum current for the Dirac theory of electrons.
  \item The corresponding (Belinfante) \uline{angular momentum tensor} is a wedge product $\overline{M}_{\text{sym}}(\gamma_0) = x \wedge \overline{T}_{\text{sym}}(\gamma_0) = [\varepsilon\rv{x} - (\cc t)\rv{P}] + \rv{x}\times\rv{P}I^{-1}$ of a radial coordinate $x$ with the energy-momentum tensor.  The resulting angular momentum is a \emph{bivector} that includes both boost and spatial-rotation angular momentum, which are distinguished by the extra factor of $I$ in the spatial-rotational part (exactly as $\rv{\sigma}_i I^{-1}$ indicates a plane of rotation, while $\rv{\sigma}_i$ indicates a boost plane).
  \item The \uline{dual-symmetric Lagrangian density} for the vacuum electromagnetic field is the scalar $\mathcal{L}_{\text{dual}}(x) = \mean{(\nabla z)(\nabla z^*)}_0/2$, which is a simple kinetic energy term for the complex vector potential field $z = a_e + a_m I$.  This expands to the sum of independent terms for the electric and magnetic vector potentials: $\mathcal{L}_{\text{dual}} = \mean{(\nabla a_e)^2}_0/2 + \mean{(\nabla a_m)^2}_0/2$.  Each term is identical in form to the traditional electromagnetic Lagrangian that only includes the electric part $a_e$.
  \item The \uline{conserved Noether currents} of the dual-symmetric Lagrangian produce the proper conserved dual-symmetric \uline{canonical energy-momentum tensor}, \uline{orbital and spin angular momentum tensors}, and the \uline{helicity pseudovector}.
  \item In the presence of only \emph{electric} sources $j_e$, the \uline{dual symmetry of the Lagrangian is broken}.  The symmetry-breaking of this neutral vector boson doublet is entirely analogous to the boson doublet that is broken in the electroweak theory by the Brout-Englert-Higgs-Guralnik-Hagan-Kibble (BEHGHK, or Higgs) mechanism.  According to this analogy, the second independent magnetic vector potential $a_m$ that appears in the simple dual-symmetric electromagnetic Lagrangian may become related to the neutral \uline{$Z_0$ boson} when additional interaction terms of the Standard Model are included.
\end{itemize}


\section{A brief history of electromagnetic formalisms}\label{sec:history}

\epigraph{4.5in}{In Science, it is when we take some interest in the great discoverers and their lives that it becomes endurable, and only when we begin to trace the development of ideas that it becomes fascinating.}{James Clerk Maxwell \cite{Scully2007}}


The electromagnetic theory has a meandering mathematical history, largely due to the fact that the appropriate mathematics was being developed in parallel with the physical principles.  As a result of this confusing evolution, the earlier and more general algebraic methods have been rediscovered and applied to electromagnetism only more recently. We summarize the highlights of this history in Table~\ref{tab:history} for reference.

The original nonrelativistic unification of the physical concepts by Maxwell \cite{Maxwell1861,Maxwell1861b,Maxwell1862,Maxwell1862b,Maxwell1865} was described as a set of 20 coupled differential equations of independent \emph{scalar} variables.  To conceptually simplify these equations down to 2 primary coupled equations that were easier to understand and solve, Maxwell later adopted and heavily advocated \cite{Maxwell1881} the algebra of \emph{quaternions}, which was developed by Hamilton \cite{Hamilton1853,Hamilton1901} as a generalization of the algebra of complex numbers to include three independent ``imaginary'' axes that can describe rotations in three-dimensional space.  The quaternionic formulation of electromagnetism was heavily attacked by Heaviside \cite{Heaviside1892,Heaviside1893}, who reformulated Maxwell's treatment into 4 coupled equations using an algebra of \emph{3-vectors} (with the familiar dot and cross products), which was developed by Gibbs \cite{Wilson1901} as a simple subset of the algebraic work of Grassmann \cite{Grassmann1844} and its extensions by Clifford \cite{Clifford1878}.  This 3-vector reformulation of Heaviside has since remained the most widely known and used formulation of electromagnetism in practice, with the electric field $\rv{E}$ as a polar 3-vector and the magnetic field $\rv{B}$ as an axial 3-vector.

The relativistic treatment of electromagnetism has subsequently followed a rather different mathematical path, however, and now deviates substantially from these earlier foundations.  Lorentz \cite{Lorentz1899,Lorentz1904,Lorentz1904b} first applied his eponymous transformations to moving electromagnetic bodies using the Heaviside formulation.  His initial results were corrected and symmetrized by Poincar\'e \cite{Poincare1905,Poincare1906}, who suggested that the Lorentz transformation could be considered as a geometric rotation of a \emph{4-vector} that had an \emph{imaginary} time coordinate $(\cc t)i$ scaled by a constant $\cc$ (corresponding to the speed of light in vacuum).  This was the first unification of space and time into a single 4-vector entity of \emph{spacetime}.  The need for this invariant constant $\cc$ was independently noticed by Einstein \cite{Einstein1905}, and elevated to a postulate for his celebrated special theory of relativity that removed the need for a background aether.  The 4-vector formalism suggested by Poincar\'e was fully developed by Minkowski \cite{Minkowski1907}, but Einstein himself argued against this construction \cite{Einstein1908} until his later work on gravitation made it necessary \cite{Einstein1912}.

\begin{table}
  \centering
  \begin{tabular}{ l  l  r }
    \hline
\noalign{\vskip 3mm} 
    \multicolumn{3}{c}{\textbf{History of Electromagnetic Formalisms}} \\
\noalign{\vskip 3mm} 
    \hline 
\noalign{\vskip 2mm}  
    \textbf{Year} & \textbf{Nonrelativistic} & \textbf{Relativistic} \\
\noalign{\vskip 2mm} 
    \hline 
    \\
    1844 & \multicolumn{2}{c}{\textit{Exterior Algebra (Grassman)}} \\
    \\
    1853 & \textit{Quaternions (Hamilton)} & \\
    \\
    1861 & Scalar Components (Maxwell) & \\
    \\
    1878 & \multicolumn{2}{c}{\textit{Geometric Algebra (Clifford)}} \\
    \\
    1881 & Quaternions (Maxwell) & \\
    \\
    1892 & \textbf{3-Vectors} (Gibbs \& Heaviside) & \\
    \\
    1899 & \multicolumn{2}{c}{\textit{Differential Forms (Cartan)}} \\
    \\
    1901 & Complex 3-Vectors (Riemann) & \\
    \\
    1905 & & \textit{4-Vectors with Imaginary Time (Poincar\'e)} \\
    \\
    1907 & Complex 3-Vectors (Silberstein) & \\
    \\
    1908 & & \textit{4-Vectors (Minkowski)} \\
    \\
    1910 & & Complex 6-Vectors (Sommerfeld) \\
    \\
    1911 & & Exterior Algebra (Wilson \& Lewis) \\
    \\
    1916 & & \textbf{Tensor Scalar Components} (Einstein) \\
    \\
    1918 & & Differential Forms (Weyl) \\
    \\
    \hline
    \\
    1966 & \multicolumn{2}{c}{\textbf{Spacetime Algebra} (Hestenes)} \\
    \\
    \hline 
 \end{tabular}
 \caption[History of electromagnetic formalisms]{A brief history of the development of mathematical formalisms for representing the electromagnetic theory, showing the purely mathematical developments in italic font and their use in electromagnetism unitalicized.  The bolded formalisms are the two most commonly used today: the nonrelativistic 3-Vectors of Gibbs preferred by Heaviside, and the relativistic tensor scalar components preferred by Einstein.  Spacetime algebra was introduced by Hestenes in 1966 following Clifford's 1878 generalization of Grassman's original 1844 exterior algebra that directly describes oriented geometric surfaces. Importantly, this spacetime algebra contains and generalizes \emph{all} the other formalisms in a simple and powerful way. }
 \label{tab:history}
\end{table}

The 4-vector formalism of Poincar\'e and Minkowski was meanwhile developed by Sommerfeld \cite{Sommerfeld1910,Sommerfeld1910b}, who emphasized that the electromagnetic field was not a 4-vector or combination of 4-vectors, but instead was a different type of object entirely that he called a ``\emph{6-vector}''.  This 6-vector became intrinsically \emph{complex} due to the imaginary time $(\cc t)i$; specifically, it had both electric and magnetic components that differed by a factor of $i$ and transformed into one another upon Lorentz boosts.  Riemann \cite{Weber1901} and Silberstein \cite{Silberstein1907a,Silberstein1907b} independently noted this intrinsic complexity while working in the Heaviside formalism, which prompted them to write the total nonrelativistic field as the single complex vector
\eqref{eq:RiemannSilberstein} in agreement with the relativistic 6-vector construction of Sommerfeld.  

Historically, however, the further development of the spacetime formalism of Poincar\'e led in a different direction.  Minkowski \cite{Minkowski1908} dropped the explicit scalar imaginary $i$ attached to the time coordinate in favor of a different definition of the 4-vector dot product that produced the needed factor of $-1$ directly. This change in 4-vector notation removed the \emph{ad hoc} scalar imaginary, but also effectively discouraged the continued development of the 6-vector of Sommerfeld (and the Riemann-Silberstein vector) by making the complex structure of spacetime implicit.  In the absence of an explicit complex structure to distinguish the electric and magnetic field components, these components were reassembled into a \emph{rank-2 antisymmetric tensor} (i.e., a multilinear antisymmetric function taking two vector arguments), whose characteristic components could be arranged into a 4x4 antisymmetric matrix
\begin{align}\label{eq:tensormatrix}
  [F^{\mu\nu}] &= \begin{bmatrix}0 & -E_x/\cc & -E_y/\cc & -E_z/\cc \\ E_x/\cc & 0 & -B_z & B_y \\ E_y/\cc & B_z & 0 & -B_x \\ E_z/\cc & -B_y & B_x & 0\end{bmatrix}.
\end{align}
Although the same components are preserved, this tensor formulation is more difficult to directly relate to the 3-vector formalism.

Of particular concern for practical applications, the familiar 3-vector cross product of Gibbs could no longer be used with the updated Minkowski 4-vector formalism, and its component representation with tensors like \eqref{eq:tensormatrix} was less conceptually clear.  Without this cross product, laboratory practitioners had reduced physical intuition about the formalism, which hampered derivations and slowed the adoption of the relativistic formulation.  In fact, an algebraic solution to this problem in the form of the \emph{wedge product} had already been derived by Grassmann \cite{Grassmann1844} and Clifford \cite{Clifford1878} (as we shall see shortly), and had even been adopted by Cartan's theory of differential forms \cite{Cartan1899}.  However, this solution remained obscure to the physics community at the time (though it was partly rediscovered by Wilson and Lewis \cite{Wilson1911}).  As such, the readily available methods for practical calculations with 4-vectors and tensors were either 
\begin{enumerate}[(a)]
  \item to convert the equations back into the nonrelativistic 3-vector notation and abandon manifest Lorentz covariance, or 
  \item to work in \emph{component-notation} like $F^{\mu\nu}$ above (and akin to the original papers by Maxwell), effectively obscuring the implicit algebraic structure of the relativistic theory.
\end{enumerate}

The choice between these two practical alternatives has essentially created a cultural divide in the development and understanding of electromagnetism: the approach using nonrelativistic 3-vectors is still dominant in application-oriented fields like optics where physical intuition is required (e.g., \cite{Jackson1999}), while the component notation is dominant in theoretical high energy physics and gravitational communities that cannot afford to hide the structural consequences of relativity (though Cartan's differential forms do occasionally make an appearance, e.g.,  \cite{Goeckeler1989}). To make the component notation less cumbersome, Einstein developed an index summation convention for general relativity \cite{Einstein1916} (where repeated component indices are summed\footnote{Note that we will not use this convention in this report for clarity.}: $v\cdot w = v^\mu w_\mu$), which has now been widely adopted by most practitioners who use component-based manipulations of relativistic quantities.  Indeed, the formal manipulations of components using index-notation has essentially become synonymous with tensor analysis.

An elegant solution to this growing divide in formalisms was proposed a half-century later by Hestenes \cite{Hestenes1966}, who carefully revisited and further developed the geometric and algebraic work of Grassmann \cite{Grassmann1844} and Clifford \cite{Clifford1878} to construct a full \emph{spacetime algebra}.  In modern mathematical terms, this spacetime algebra is the \emph{orthogonal Clifford algebra} \cite{Crumeyrolle1990} that is uniquely constructed from the spacetime metric (dot product) proposed by Minkowski.  This modern formalism has the great benefit of preserving and embracing all the existing algebraic approaches in use (including the complex numbers, the quaternions, and the 3-vectors of Gibbs), but it also extends them in a natural way to allow simple manipulations of relativistically invariant objects.  For example, using spacetime algebra the electromagnetic field can be written in a relativistically invariant way (as we shall soon see) as
\begin{align}\label{eq:embivector}
  \bv{F} &= [\rv{E}/\cc + \rv{B} I]/\sqrt{\mu_0},
\end{align}
which has the same complex form as the Riemann--Silberstein vector \eqref{eq:RiemannSilberstein}.  Importantly, however, the algebraic element $I$ satisfying $I^2 = -1$ is \emph{not} the scalar imaginary $i$ added in an \emph{ad hoc} way as in \eqref{eq:RiemannSilberstein}, but is instead an intrinsic \emph{geometric} object of spacetime itself (i.e., the unit 4-volume) that emerges automatically alongside the 3-vector formalism of Gibbs.  The geometric significance of the factor $I$ makes the expression \eqref{eq:embivector} a proper geometric object that is invariant under reference-frame changes, unlike \eqref{eq:RiemannSilberstein}.  Moreover, the scalar components of \eqref{eq:embivector} are precisely equivalent to the tensor components \eqref{eq:tensormatrix} even though $\bv{F}$ is not understood as an antisymmetric tensor.  It is a geometric object in spacetime called a \emph{bivector}, which is a modern refinement of the 6-vector concept introduced by Sommerfeld.  The manifestly geometric significance of the electromagnetic field made manifest in this approach reaffirms and augments the topological fiber bundle foundations of modern gauge-field theories that was observed by Yang and Mills \cite{Yang1954,Yang1980,Yang2014,Yang2014b}, Chern and Simons \cite{Chern1974}, and many others in theoretical high energy physics \cite{Aharonov1959,THooft1972,Tonomura1986,Witten1988,Weinberg1995,Nakahara2003,Wu2006}.

The spacetime algebra of Hestenes has been heavily developed by a relatively small community \cite{Hestenes2003,Hestenes2003b,Hestenes2005,Hestenes1968,Hestenes1968b,Hestenes1971,Sobczyk1989,Lasenby1998,Ablamowicz2004,DeSabbata2007,VanEnk2013}, but has recently been growing in popularity as a useful tool, and has been cross-pollinating with other modern mathematical investigations of Clifford algebras \cite{Crumeyrolle1990,Colombo2004,Delanghe2012}.  Indeed, spacetime algebra has been mentioned in a growing number of recent textbooks about algebraic approaches to physics and mathematics based on the work of Grassmann, Clifford, and Hestenes more generally \cite{Hestenes1987,Doran2007,Baylis1996,Hestenes1999,Corrochano2001,Baylis2002,Perwass2010,Dorst2011,Macdonald2011,Macdonald2012}.  Nevertheless, many of these treatments have under-emphasized certain features of spacetime that will be important for our discussion (such as its intrinsic complex structure) so we include our own introduction to this formalism that is tuned for applications in electromagnetism in what follows, assuming no prior background.


\section{Spacetime algebra}\label{sec:algebra}

\epigraph{4.5in}{Mathematics is taken for granted in the physics curriculum---a body of immutable truths to be assimilated and applied. The profound influence of mathematics on our conceptions of the physical world is never analyzed. The possibility that mathematical tools used today were invented to solve problems in the past and might not be well suited for current problems is never considered.}{David Hestenes \cite{Hestenes2003}}


Physically speaking, \emph{spacetime algebra} \cite{Hestenes1966,Hestenes1987,Hestenes2003,Hestenes2003b} is a complete and natural algebraic language for compactly describing physical quantities that satisfy the postulates of special relativity.  Mathematically speaking, it is the largest associative algebra that can be constructed with the vector space of spacetime equipped with the Minkowski metric.  It is an orthogonal Clifford algebra \cite{Crumeyrolle1990}, which is a powerful tool that enables manifestly frame-independent and coordinate-free manipulations of geometrically significant objects.  Unlike the dot and cross products used in standard vector analysis, the Clifford product between vectors is often \emph{invertible} and not constrained to three dimensions.  Unlike the component manipulations used in tensor analysis, spacetime algebra permits compact and component-free derivations that make the intrinsic geometric significance of physical quantities transparent.  

When spacetime algebra is augmented with calculus then it subsumes many disparate mathematical techniques into a single comprehensive formalism, including (multi)linear algebra, vector analysis, complex analysis, quaternion analysis, tensor analysis, spinor analysis, group theory, and differential forms \cite{Hestenes1987,Hestenes1968,Hestenes1968b,Hestenes1971}.  Moreover, for those who are unfamiliar with any of these mathematical techniques, spacetime algebra provides an encompassing framework that encourages seamless transitions from familiar techniques to unfamiliar ones as the need arises.  As such, spacetime algebra is also a useful tool for pedagogy \cite{Hestenes2003,Hestenes2003b}. 

During our overview we make an effort to illustrate how spacetime algebra contains and generalizes all the standard techniques for working with electromagnetism. Hence, one can appreciate spacetime algebra not as an obscure mathematical curiosity, but rather as a principled, practical, and powerful \emph{extension} to the traditional methods of analysis.  As such, all prior experience with electromagnetism is applicable to the spacetime algebra approach, making the extension readily accessible and primed for immediate use.


\subsection{Spacetime}\label{sec:spacetime}


Recall that in special relativity \cite{Einstein1905} one postulates that the scalar time $t$ and vector spatial coordinates $\rv{x}$ in a particular inertial reference frame always construct an invariant interval $(\cc t)^2 - |\rv{x}|^2$ that does not depend on the reference frame, where $\cc$ is the speed of light in vacuum.  An elegant way of encoding this physical postulate is to combine the scaled time and spatial components into a proper 4-vector $x = (\cc t,x_1,x_2,x_3)$ with a squared length equal to the invariant interval \cite{Poincare1905,Poincare1906,Minkowski1907}.  The proper notion of length is defined using the relativistic scalar (dot) product between two 4-vectors, which can be understood as a symmetric bilinear function $\eta(a,b)$ that takes two vector arguments $a$ and $b$ and returns their scalar shared length.  This dot product is known as the \emph{Minkowski metric} \cite{Minkowski1908}.  In terms of components $a = (a_0,a_1,a_2,a_3)$ and $b = (b_0,b_1,b_2,b_3)$ in a particular inertial frame, this metric has the form $a\cdot b \equiv b\cdot a \equiv \eta(a,b) = a_0 b_0 - a_1 b_1 - a_2 b_2 - a_3 b_3$.  The proper squared length of a 4-vector $x$ is therefore $x \cdot x$.  Since there is one positive sign and three negative signs in this dot product, we say that it has a mixed \emph{signature} $(+,-,-,-)$  and denote the vector space of 4-vectors as $\mathcal{M}_{1,3}$ to make this signature explicit\footnote{Note that there is a sign-ambiguity in the spacetime interval, so one can seemingly choose either $(+,-,-,-)$ or $(-,+,+,+)$ for the metric signature, the latter being the initial choice of Poincar\'e.  However, this choice is not completely arbitrary from an algebraic standpoint: it produces geometrically distinct spacetime algebras \cite{Crumeyrolle1990}.  We choose the signature here that will produce the spacetime algebra that correctly contains the relative Euclidean 3-space as a proper subalgebra, which we detail in Section~\ref{sec:paulidirac}.}.

All physical vector quantities in special relativity are postulated to be 4-vectors in $\mathcal{M}_{1,3}$ that satisfy the Minkowski metric.  These vectors are geometric quantities that do not depend on the choice of reference frame, so they shall be called \emph{proper} relativistic objects in what follows.  Importantly, the Minkowski metric is qualitatively different from the standard Euclidean dot product since it does not produce a positive length.  Indeed, the mixture of positive and negative signs can make the length of a vector positive, negative, or zero, which produces three qualitatively different classes of vectors.  We call these classes of vectors \emph{timelike}, \emph{spacelike}, and \emph{lightlike}, respectively.  As such, we can write the length, or \emph{pseudonorm}, of a 4-vector $a$ as $a\cdot a = \epsilon_a |a|^2$ in terms of a positive magnitude $|a|^2$ and a \emph{signature} $\epsilon_a = \pm 1$ that is $+1$ for timelike vectors and $-1$ for spacelike vectors.  For lightlike vectors the magnitude $|a|^2$ vanishes; thus, unlike for Euclidean vector spaces, a zero magnitude vector need not be the zero vector.

The Minkowski metric $\eta$ and the vector space $\mathcal{M}_{1,3}$ are sufficient for describing proper vector quantities in special relativity, such as time-space $(\cc t,\rv{x})$ and energy-momentum $(\mathcal{E}/\cc,\rv{p})$.  Since the laws of physics do not depend on the choice of reference frame, we expect that all physical quantities should be similarly represented by proper geometric objects like vectors.  However, the electromagnetic field presents us with a conundrum: the polar 3-vector $\rv{E}$ of the electric field and axial 3-vector $\rv{B}$ of the magnetic field in a particular reference frame do not combine into proper 4-vectors.  Relativistic angular momentum suffers a similar dilemma: the polar 3-vector $\rv{N} = (\cc t)\rv{p}-(\mathcal{E}/\cc)\rv{x}$ of the boost angular momentum (also known as the dynamic mass moment) and the axial 3-vector $\rv{L} = \rv{x}\times\rv{p}$ of the orbital angular momentum in a particular reference frame do not combine into proper 4-vectors.  To resolve these dilemmas, the components of these vectors are typically assembled into the components $F^{\mu\nu}$ and $M^{\mu\nu}$ of rank-2 antisymmetric \emph{tensors}, as we noted in \eqref{eq:tensormatrix} for the electromagnetic field \cite{Soper1976,Landau1975}. This solution, while formally correct at the component level, is conceptually opaque.  Why does a single rank-2 tensor $F^{\mu\nu}$ decompose into two 3-vector quantities $\rv{E}$ and $\rv{B}$ in a relative frame?  How does a rank-2 tensor, which is mathematically defined as a multilinear \emph{function} with two vector arguments, conceptually correspond to a physical quantity like the electromagnetic field or angular momentum?  Is there some deeper significance to the mathematical space in which the tensors $F^{\mu\nu}$ and $M^{\mu\nu}$ reside?  Do the proper tensor descriptions have any geometric significance in Minkowski space?  Evidently, the vector space $\mathcal{M}_{1,3}$ does not contain the complete physical picture implied by special relativity, since it must be augmented by quantities like $F^{\mu\nu}$ and $M^{\mu\nu}$.


\subsection{Spacetime product}


To obtain the complete picture of special relativity in a systematic and principled way, we make a critical observation: any physical manipulation of vector quantities uses not only addition, but also \emph{vector multiplication}.  Indeed, standard treatments of electromagnetism involving relative 3-vectors use both the symmetric dot product and the antisymmetric vector cross product to properly discuss the physical implications of the theory.  The vector space $\mathcal{M}_{1,3}$ only specifies the relativistic version of the dot product in the form of the Minkowski metric.  Without introducing the proper relativistic notion of the cross product the physical picture of spacetime is incomplete.

Mathematically, the introduction of a product on a vector space creates an \emph{algebra}.  Hence, we seek to construct the appropriate algebra for spacetime from the vector space $\mathcal{M}_{1,3}$ by introducing a suitable vector product.  We expect this vector product to be generally \emph{noncommutative}, since the familiar cross product is also noncommutative\footnote{Note that algebraic noncommutativity has nothing \emph{a priori} to do with the noncommutativity in quantum mechanics.}.  We also expect the vector product to enlarge the mathematical space in order to properly accommodate quantities like the electromagnetic field tensor $F^{\mu\nu}$.

To accomplish these goals, we define the appropriate spacetime product to satisfy the following four properties for any vectors $a,b,c \in \mathcal{M}_{1,3}$:
\begin{subequations}
\begin{align}
  a(bc) &= (ab)c \qquad\qquad & \mbox{(Associativity)} \\
  a(b+c) &= ab + ac & \mbox{(Left Distributivity)} \\
  (b+c)a &= ba + ca & \mbox{(Right Distributivity)} \\
  \label{eq:contraction}
  a^2 &= \eta(a,a) = \epsilon_a |a|^2 & \mbox{(Contraction)}
\end{align}
\end{subequations}
Note that we omit any special product symbol for brevity.  The contraction property \eqref{eq:contraction} distinguishes the resulting spacetime algebra as an \emph{orthogonal Clifford algebra} \cite{Crumeyrolle1990} that is generated by the metric $\eta$ and the vector space $\mathcal{M}_{1,3}$.  This Clifford algebra is the largest associative algebra that can be constructed solely from spacetime, so it will contain all other potentially relevant algebras as subalgebras.  Indeed, this nesting of algebras will be quite useful for practical calculations, as we shall see.

Decomposing the resulting associative vector product into symmetric and antisymmetric parts produces the proper spacetime generalizations to the 3-vector dot and cross products that we were seeking \cite{Clifford1878}:
\begin{equation}\label{eq:product}
  ab = a\cdot b + a\wedge b.
\end{equation}
The symmetric part of the product,
\begin{align}
  \label{eq:dot}
a\cdot b &\equiv \frac{1}{2}(ab + ba) = b \cdot a = \eta(a,b),
\end{align}
is precisely the scalar (dot) product inherited from the spacetime structure of $\mathcal{M}_{1,3}$.  The last equivalence follows from the contraction relation $(a + b)^2 = \eta(a+b,a+b)$ demanded by property \eqref{eq:contraction}.  

\begin{figure}[t]
  \begin{center}
    \includegraphics[width=0.7\columnwidth]{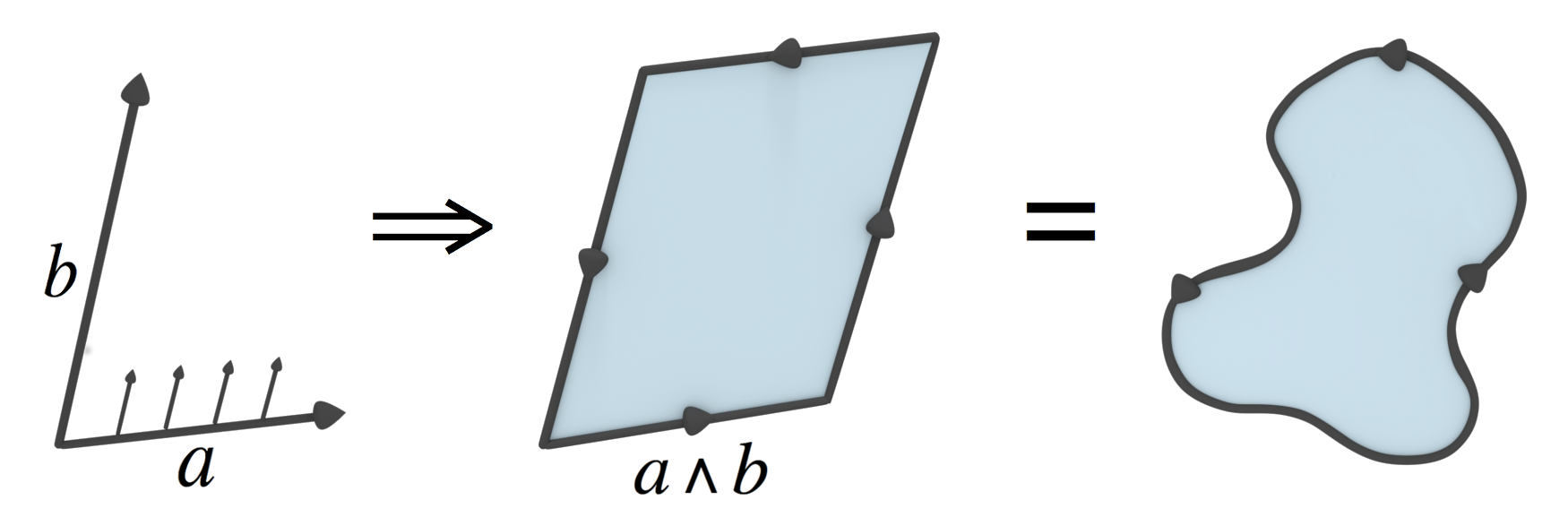}
  \end{center}
  \caption[Wedge product]{The wedge product $a \wedge b$ between spacelike vectors $a$ and $b$ produces an oriented plane segment known as a bivector.  Conceptually, the vector $a$ slides along the vector $b$, sweeping out a parallelogram with area $|a\wedge b|$.  The orientation of this area follows the right-hand rule, and may be visualized as a circulation around the boundary of the plane segment.  Importantly, a bivector is characterized entirely by its magnitude and orientation, so the plane segment resulting from a wedge product may be deformed into any shape that preserves these two properties.  Conversely, a bivector (of definite signature) may be factored into a wedge product between any two vectors that produce the same area and orientation $a\wedge b = c\wedge d$; if the two chosen factors are also orthogonal (i.e., $c\cdot d = 0$), then the bivector is a simple product of the orthogonal factors $c\wedge d = cd$.}
  \label{fig:bivector}
\end{figure}

The antisymmetric part of the product,
\begin{align}
  \label{eq:wedge}
  a\wedge b &\equiv \frac{1}{2}(ab - ba) = - b \wedge a,
\end{align}
is called the \emph{wedge product} and is the proper generalization of the vector cross product to relativistic 4-vectors\footnote{This is precisely Grassman's exterior wedge product \cite{Grassmann1844}, adopted by Cartan when defining differential forms \cite{Cartan1899}.}.  It produces a qualitatively new type of object called a \emph{bivector} that does not exist \emph{a priori} in $\mathcal{M}_{1,3}$, as anticipated.  A bivector $(a\wedge b)$ produced from spacelike vectors $a$ and $b$ has the geometric meaning of a \emph{plane segment} with magnitude equal to the area of the parallelogram bounded by $a$ and $b$, and a surface orientation (handedness) determined by the right-hand rule; this construction is illustrated in Figure~\ref{fig:bivector}.  Hence, the wedge product generalizes the cross product by directly producing an oriented plane segment, rather than a vector normal to that surface.  This generalization is important in spacetime since there is no unique normal vector to a plane in four dimensions.  We will see in Sections~\ref{sec:components}, \ref{sec:relative}, and \ref{sec:maxwell} that the electromagnetic field is properly expressed as precisely such a bivector.

The vector product \eqref{eq:product} combines the nonassociative dot and wedge products into a single \emph{associative} product.  The result of the product thus decomposes into the sum of distinct scalar and bivector parts, which should be understood as analogous to expressing a complex number as a sum of distinct real and imaginary parts.  Just as with the study of complex numbers, it will be advantageous to consider these distinct parts as composing a unified whole, rather than separating them prematurely.  We will explore this similarity more thoroughly in Section~\ref{sec:complex}.  

A significant benefit of combining both the dot and wedge products into a single associative product in this fashion is that an \emph{inverse} may then be defined 
\begin{equation}\label{eq:inverse}
  a^{-1} = \frac{a}{a^2},
\end{equation}
provided $a$ is not lightlike (i.e., $a^2 \neq 0$).  Note that $a^2 = \epsilon_a |a|^2$ is a scalar by property \eqref{eq:contraction}, so it trivially follows that $a^{-1}a = a a^{-1} = 1$.  Importantly, neither the dot product nor the wedge product alone may be inverted; only their combination as the sum \eqref{eq:product} retains enough information to define an inverse.


\subsection{Multivectors}\label{sec:multivectors}


By iteratively appending all objects generated by the wedge product \eqref{eq:wedge} to the initial vector space $\mathcal{M}_{1,3}$, we construct the full \emph{spacetime algebra} $\mathcal{C}_{1,3}$.  This notation indicates that the spacetime algebra is a \emph{Clifford} algebra generated from the metric signature $(+,-,-,-)$.  Importantly, all components in this Clifford algebra are purely \emph{real}---we will not need any \emph{ad hoc} addition of the complex scalar field in what follows.

\begin{figure}[t]
  \begin{center}
    \includegraphics[width=\columnwidth]{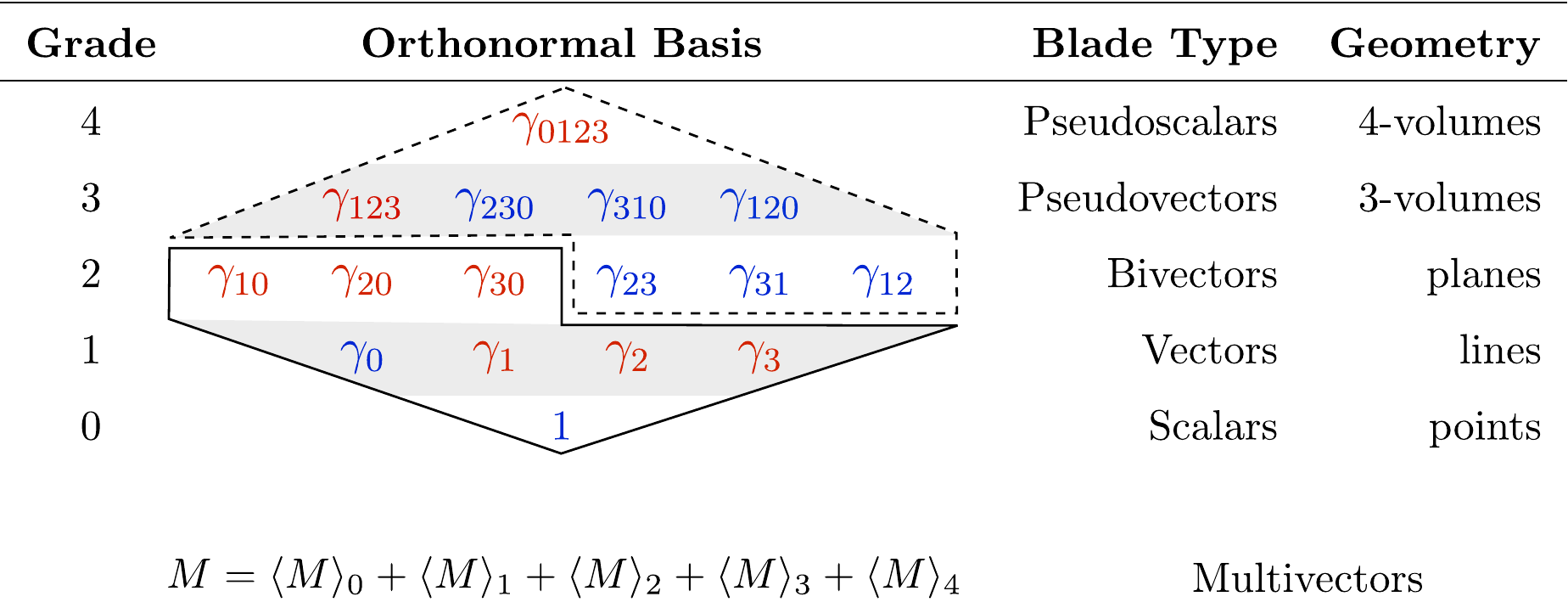}
  \end{center}
  \caption[Graded basis for spacetime algebra]{Graded basis for the spacetime algebra $\mathcal{C}_{1,3}$.  Each multivector $M\in\mathcal{C}_{1,3}$ decomposes into a sum of distinct and independent grades $k=0,1,2,3,4$, which can be extracted as grade-projections $\mean{M}_k$.  The oriented basis elements of grade-1 $\{\gamma_\mu\}_{\mu=0}^3$ are an orthonormal basis ($\gamma_\mu \cdot \gamma_\nu = \eta_{\mu\nu}$) for the Minkowski 4-vectors $\mathcal{M}_{1,3}$.  An oriented basis element of grade-$k$, such as $\gamma_{\mu\nu} \equiv \gamma_\mu\gamma_\nu = \gamma_\mu\wedge\gamma_\nu = - \gamma_\nu \wedge \gamma_\mu$ (with $\mu\neq\nu$), is constructed as a product of $k$ of these orthonormal 4-vectors.  Interchanging indices permutes the wedge products, which only changes the sign of the basis element; hence, only the independent basis elements of each grade are shown.  The color coding indicates the signature of each basis element (see Section~\ref{sec:reversion}), with blue being $+1$ and red being $-1$.  The boxes and shading indicate useful dualities of the algebra: the solid and dashed boxes are (Hodge) dual under right multiplication of the pseudoscalar $I = \gamma_{0123}$ (see Section~\ref{sec:complex}), while within each box the shaded region is dual to the unshaded region under right multiplication of the timelike basis vector $\gamma_0$ (see Section~\ref{sec:relative}).  These dualities are further detailed in Figures~\ref{fig:dual1} and \ref{fig:dual2}.}
  \label{fig:grades}
\end{figure}

The repeated wedge products produce 5 linearly independent subspaces of the total algebra, known as \emph{grades}, which are illustrated in Figure \ref{fig:grades}.  Each grade is a distinct type of ``directed number'' \cite{Ablamowicz2004}.  The real scalars (pure numbers) $\alpha\in \mathbb{R}$ are grade $0$, while the 4-vectors (line segments) $a\in\mathcal{M}_{1,3}$ are grade $1$.  The \emph{bivectors} (plane segments) $(a\wedge b)$ appearing in \eqref{eq:wedge} are grade $2$.  Successive wedge products will also produce \emph{trivectors} (\emph{pseudovectors}, 3-volume segments) $(a\wedge b\wedge c)$ of grade $3$ and \emph{quadvectors} (\emph{pseudoscalars}, 4-volume segments) $(a\wedge b\wedge c\wedge d)$ of grade $4$, which completes the algebra.  We will refer to the elements of a grade-$k$ subspace as $k$-\emph{blades} in what follows to disambiguate them from the grade-1 vectors.  

For concreteness, we systematically generate a complete graded basis for $\mathcal{C}_{1,3}$ as all independent products of the vectors $\{\gamma_\mu\}_{\mu=0}^3$ in a chosen basis of $\mathcal{M}_{1,3}$.  The rich structure of the resulting graded basis is also detailed in Figure \ref{fig:grades}.  We choose the starting vector basis to be orthonormal $\gamma_\mu\cdot\gamma_\nu = \eta_{\mu\nu}$ in the sense of the Minkowski metric, so $\gamma_0^2 = 1$ and $\gamma_j^2 = -1$ for $j=1,2,3$.  The choice of notation for the basis is motivated by a deep connection to the Dirac $\gamma$-matrices that we will clarify in Section \ref{sec:paulidirac}.

The basis of 0-blades is the real number $1$.  The basis of 1-blades is the chosen set of four orthonormal vectors $\gamma_\mu$ themselves.  The $(16-4)=12$ possible products of these vectors that produce bivectors (i.e., with $\mu\neq\nu$) $\gamma_\mu\gamma_\nu = \gamma_\mu \wedge \gamma_\nu \equiv \gamma_{\mu\nu} = -\gamma_{\nu\mu}$ produce only $(16-4)/2=6$ independent bivectors $\gamma_{10},\gamma_{20},\gamma_{30},\gamma_{12},\gamma_{23},\gamma_{31}$ due to the antisymmetry of the wedge product; these independent elements form the oriented basis of 2-blades.  Similarly, products of orthogonal vectors with these bivectors $\gamma_\mu\gamma_\nu\gamma_\delta = \gamma_\mu \wedge \gamma_\nu \wedge \gamma_\delta \equiv \gamma_{\mu\nu\delta}$ produce only $4$ independent trivectors $\gamma_{123},\gamma_{120},\gamma_{230},\gamma_{310}$ that form the oriented basis of 3-blades.  Finally, the product of all four orthogonal vectors produces only a single independent element $\gamma_0\gamma_1\gamma_2\gamma_3 = \gamma_0\wedge\gamma_1\wedge\gamma_2\wedge\gamma_3 \equiv \gamma_{0123}$ that serves as the basis for the 4-blades. Hence, there are $2^4 = 16$ independent basis elements for the spacetime algebra, partitioned into $5$ grades of ``4 choose $k$,'' $4!/k!(4-k)!$, independent basis $k$-blades.  

The 4-blade $\gamma_{0123}$ geometrically signifies the same unit 4-volume in any basis, and will be particularly important in what follows, so we give it a special notation 
\begin{align}
  I \equiv \gamma_{0123}.
\end{align}
This algebraic element is the \emph{pseudoscalar} for spacetime.  That is, its scalar multiples $\alpha I$ act as scalars that flip sign under an inversion of the handedness of spacetime (i.e., flipping the orientation of any one basis direction).  The notation of $I$ is motivated by the fact that $I^2 = \gamma_0^2 \gamma_1^2 \gamma_2^2 \gamma_3^2 = (1)(-1)(-1)(-1) = -1$, so it will perform a similar function to the scalar imaginary $i$.  Indeed, the element $I$ provides spacetime with an intrinsic \emph{complex structure} without any \emph{ad hoc} introduction of the scalar complex numbers\footnote{It is worth emphasizing that there is no unique square root of $-1$.  There are many algebraic elements in $\mathcal{C}_{1,3}$ that square to $-1$, each with distinct geometric significance.  The scalar imaginary $i$ of standard complex analysis does not specify any of this additional structure; hence, it is often enlightening to determine whether a particular $\sqrt{-1}$ is implied by a generic $i$ that appears in a traditional physics expression.  We will show in what follows that $I$ is indeed the proper physical meaning of $i$ throughout the electromagnetic theory.}.  We will detail this complex structure in Section \ref{sec:complex}.

\begin{table}
  \centering
  \begin{tabular}{l l}
    \hline
\noalign{\vskip 2mm} 
    \textbf{Notation} & \textbf{Description} \\ 
\noalign{\vskip 2mm} 
\hline
    \\
    $\alpha,\,\beta,\,\omega$ & scalars (real numbers) \\
    $a,b,v,w$ & 4-vectors \\
    $\bv{F},\,\bv{G}$ & bivectors \\
    $v I$ & pseudovectors \\
    $\alpha I$ & pseudoscalars \\
    $M$ & multivector \\
    $\zeta = \alpha + \beta I$ & complex scalar \\
    $z = v + w I$ & complex vector \\
    $\nabla$ & 4-gradient, Dirac operator \\
    $\underbar{T}(b),\overline{T}(b)$ & tensor, adjoint tensor \\
    \\
    \hline 
    \\
    $\gamma_\mu$  & orthonormal basis (Lorentz frame) for 4-vectors \\
    $\gamma^\mu = \gamma_\mu^{-1}$  & reciprocal basis for 4-vectors \\
    $\rv{E},\,\rv{B}$ & relative 3-vectors \\
    $\rv{\sigma}_i = \gamma_i\gamma_0$ & orthonormal basis for 3-vectors \\
    $\rv{\nabla} = \gamma_0\wedge\nabla$ & relative 3-gradient \\
    $\partial_0 = \gamma_0\cdot\nabla$ & relative time derivative \\
    $F^{\mu\nu},\,v^\mu$ & scalar components ($\mu=0,1,2,3$) \\
    $F^{ij},\,v^k$ & spatial scalar components ($i,j,k=1,2,3$) \\
    $\eta_{\mu\nu}$ & Minkowski metric tensor \\
    $\delta_{\mu\nu}$ & Euclidean metric tensor / Kronecker delta \\
$\epsilon_{ijk},\epsilon_{\mu\nu\delta\kappa}$ & anti-symmetric Levi-Civita symbols \\
    \\
    \hline
  \end{tabular}
  \caption[Notational distinctions]{Notational distinctions. Proper (frame-independent) quantities (top) have the different grades distinguished by font type, boldface, and capitalization for simplicity, while relative (frame-dependent) quantities (bottom) conform to traditional vector or index annotations for clarity.}
  \label{tab:notation}
\end{table}

Any \emph{multivector} $M\in\mathcal{C}_{1,3}$ in the spacetime algebra may be written as a sum of independent $k$-blades
\begin{equation}\label{eq:grades}
  M = \alpha + v + \bv{F} + \mathfrak{T} + \beta I
\end{equation}
where the Greek letters $\alpha,\beta$ are real numbers, the lowercase Roman letter $v$ is a vector with $4$ real components, the boldface Roman letter $\bv{F}$ is a bivector with $6$ real components, and the Fraktur letter $\mathfrak{T}$ is a trivector with $4$ real components.  Each of these independent grades are distinct and proper geometric objects.  We shall see in Section~\ref{sec:complex} that in practice we can dramatically simplify the description of multivectors by exploiting dualities of the algebra; in particular, we will be able to dispense with trivectors (and their elaborate notation) altogether by rewriting them as \emph{pseudovectors} $\mathfrak{T} = v I$. To keep these distinctions conceptually clear, we shall make an effort to maintain useful notational conventions throughout this work that are summarized in Table~\ref{tab:notation} for reference. 

Each $k$-blade in a multivector can be extracted by a suitable grade-projection, denoted as $\mean{M}_k$.  For example, the bivector projection of \eqref{eq:grades} is $\mean{M}_2 = \bv{F}$.  Notably, the scalar projection satisfies the cyclic property 
\begin{align}\label{eq:cyclicscalar}
  \mean{MN}_0 = \mean{NM}_0
\end{align}
for any two multivectors $M$ and $N$, so is an algebraic \emph{trace} operation.  In fact, this trace is entirely equivalent to the matrix trace operation, which can be verified by considering a (Dirac) matrix representation of the spacetime algebra $\mathcal{C}_{1,3}$ that simulates the noncommutative vector product using the standard matrix product.  We will comment more on such a matrix representation in Section \ref{sec:paulidirac}, but we shall not require it in what follows.

\begin{table}
  \centering
  \begin{tabular}{c  l  l}
    \hline
\noalign{\vskip 2mm} 
    \textbf{Grade} & \textbf{Blade Type} & \textbf{Physical Examples} \\ 
\noalign{\vskip 2mm} 
\hline
\noalign{\vskip 2mm} 
    0 & Scalars & charge $q$ \\
      &        & mass $m$ \\
      &        & proper time $\tau$ \\
\noalign{\vskip 2mm} 
    1 & Vectors & coordinates $x$ \\
      &        & proper velocity $w = dx/d\tau$ \\
      &        & energy-momentum $p$ \\
      &        & force $dp/d\tau$ \\
      &        & electric potential $a_e$ \\
      &        & electric current $j_e$ \\
\noalign{\vskip 2mm} 
      2 & Bivectors & electromagnetic field $\bv{F} = \rv{E} + \rv{B}I$ \\
      &        & angular momentum $\bv{M} = x\wedge p = -\rv{N} + \rv{L}I^{-1}$ \\
      &        & torque $d\bv{M}/d\tau$ \\
      &        & vorticity $\nabla\wedge w$ \\
\noalign{\vskip 2mm} 
    3 & Pseudovectors & magnetic potential $a_m I$ \\
      &        & magnetic pseudocurrent $j_m I$ \\
      &        & helicity pseudocurrent $X I$ \\
\noalign{\vskip 2mm} 
    4 & Pseudoscalars & magnetic charge $q_m I$ \\ 
      &               & phase $\varphi I$ \\
\noalign{\vskip 2mm} 
    \hline
  \end{tabular}
  \caption[Physical quantities by grade]{Examples of common physical quantities by grade.  All these physical quantities appear in any proper description of relativistic physics, and fit naturally within the graded structure of spacetime algebra.}
  \label{tab:physical}
\end{table}

All proper physical quantities in spacetime must either correspond to pure $k$-blades, or multivectors that combine $k$-blades of differing grades.  These are the only frame-independent objects permitted by the constraints of special relativity.  Hence, the expansion \eqref{eq:grades} of a general multivector indicates the full mathematical arena in which relativistic physics must occur.  The algebraic structure allows us to manipulate each of these proper objects with equal ease and finesse.  

The added complication of multivectors $M\in\mathcal{C}_{1,3}$ over the original 4-vectors $v\in\mathcal{M}_{1,3}$ may seem excessive and intimidating at first glance; however, as we have been suggesting, the added structure contained in a full multivector is in fact necessary for a proper description of spacetime quantities.  Indeed, as we will see in more detail in Sections~\ref{sec:components} and \ref{sec:relative}, each distinct blade of a general multivector is in fact a familiar type of object that appears in the standard treatments of relativistic physics.  We summarize several examples in Table~\ref{tab:physical} for reference.  The multivector construction unifies all these quantities into a comprehensive whole in a principled way.  Moreover, the Clifford product provides them with a wealth of additional structure, which we will exploit to make manipulations of multivectors simple in practice.


\subsubsection{Bivectors: products with vectors}\label{sec:bivectorvector}


We will see in the following sections that several dualities of spacetime algebra allow a complete understanding in terms of scalars, vectors, and bivectors.  Out of these three fundamental objects bivectors are the least familiar, so we shall make an effort to clarify their properties as we progress through this introduction.  

We now consider how bivectors and vectors relate to one another in more detail, defining two useful operations (\emph{contraction} and \emph{inflation}) in the process.  It is worth emphasizing that all operations in spacetime algebra are derived from the fundamental vector product and sum, so are introduced for calculational convenience, conceptual clarity, and for contact with existing formalisms.  For reference, we list all of the auxiliary operations that will be introduced in this report in Table~\ref{tab:operations}.

\begin{table}
  \centering
  \begin{tabular}{l l}
    \hline
\noalign{\vskip 2mm} 
    \textbf{Description} & \textbf{Notation \& Definition} \\ 
\noalign{\vskip 2mm} 
\hline
\noalign{\vskip 2mm} 
    Clifford product & $ab$ \\
    \\
    vector dot product & $a\cdot b = (ab + ba)/2 = b \cdot a$ \\
    vector wedge product & $a\wedge b = (ab - ba)/2 = - b \wedge a$ \\
    \\
    projection to grade-$k$ & $\mean{M}_k$ \\
    algebraic trace & $\mean{M}_0$ \\
    \\
    Hodge duality & $M\mapsto M I^{-1}$ \\
    relative frame duality & $M\mapsto M\gamma_0$ \\
    \\
    bivector contraction  & $a\cdot\bv{F} = (a\bv{F}-\bv{F}a)/2 = -\bv{F}\cdot a$ \\
    bivector inflation & $a\wedge\bv{F} = (a\bv{F}+\bv{F}a)/2 = \bv{F}\wedge a$\\
    bivector dot product & $\bv{F}\cdot\bv{G} = (\bv{F}\bv{G} + \bv{G}\bv{F})/2 = \bv{G}\cdot\bv{F}$ \\
    commutator bracket & $[\bv{F},\bv{G}] = (\bv{F}\bv{G} - \bv{G}\bv{F})/2 = -[\bv{G},\bv{F}]$ \\
    bivector cross product & $\bv{F}\times\bv{G} = [\bv{F},\bv{G}]I^{-1} = -\bv{G}\times\bv{F}$ \\
    \\
    reversion (transpose) & $(ab)^\sim = ba$ \\
                          & $(MN)^\sim = \widetilde{N}\widetilde{M}$ \\
    \\
    positive magnitude & $|M|^2 = |\mean{\widetilde{M}M}_0|^2 + |\mean{\widetilde{M}M}_4I^{-1}|^2$ \\
    signature & $\epsilon_M = \mean{\widetilde{M}M}_0 / |M|^2$ \\
    \\
    complex conjugation & $(\alpha + \beta I)^* = (\alpha - \beta I)$ \\
                        & $(v + wI)^* = (v - wI)$ \\
                        & $[\bv{f}\,e^{\varphi I}]^* = \bv{f}\,e^{-\varphi I}$ \\
                        \\
    relative reversion & $M^\dagger = \gamma_0\widetilde{M}\gamma_0$ \\
                       & $(\rv{E}+\rv{B}I)^\dagger = \rv{E} - \rv{B}I$ \\
\noalign{\vskip 2mm} 
    \hline
  \end{tabular}
  \caption[Algebraic operations]{Algebraic operations.  From the fundamental Clifford product we define a variety of convenient operations that are useful for calculations.}
  \label{tab:operations}
\end{table}

\begin{figure}[t]
  \begin{center}
    \includegraphics[width=\columnwidth]{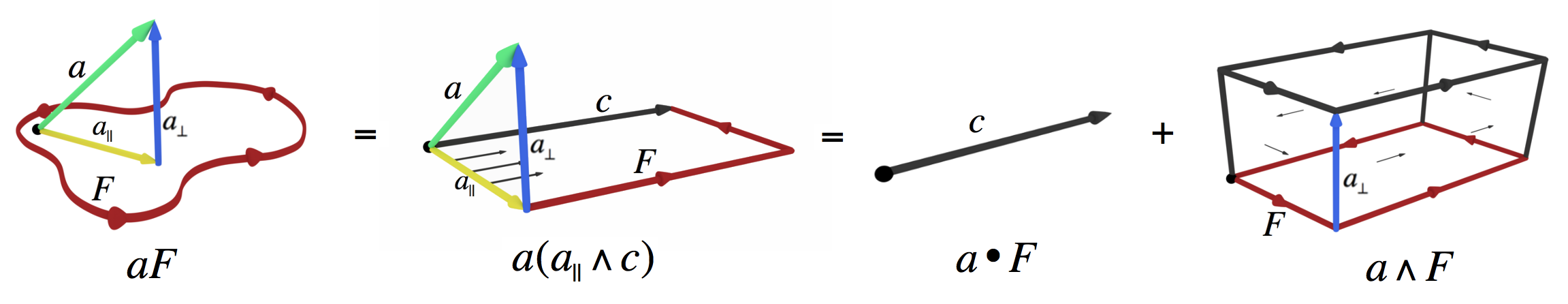}
  \end{center}
  \caption[Vector-bivector product]{The product $a \bv{F}$ between a vector $a$ (green) and bivector $\bv{F}$ (red) decomposes into a vector $a\cdot \bv{F}$ and a trivector $a\wedge \bv{F}$ part.  A spacelike $\bv{F}$ is an oriented plane segment that may be deformed into a parallelogram $\bv{F} = a_\parallel \wedge c$ that shares a side with the part of the spacelike vector $a = a_\parallel + a_\perp$ that is parallel to $\bv{F}$ (yellow).  The contraction extracts the orthogonal vector $c = a\cdot \bv{F}$ (black), while the inflation drags the bivector $\bv{F}$ along the perpendicular vector $a_\perp$ (blue) to produce an oriented trivector volume $a_\perp \wedge \bv{F}$. }
  \label{fig:bivectorproduct}
\end{figure}

A product between a vector $a$ and a bivector $\bv{F}$ can be split into two terms (vector and trivector) in an analogous way to the product of vectors in \eqref{eq:product} 
\begin{equation}\label{eq:productvecbi}
  a \bv{F} = a \cdot \bv{F} + a \wedge \bv{F}.
\end{equation}
The vector and trivector parts are constructed respectively as
\begin{align}\label{eq:dotwedgebivector}
  a\cdot \bv{F} &\equiv \mean{a\bv{F}}_1 = \frac{1}{2}(a\bv{F} - \bv{F}a) = -\bv{F} \cdot a, \\
  a\wedge \bv{F} &\equiv \mean{a\bv{F}}_3 = \frac{1}{2}(a\bv{F} + \bv{F}a) = \bv{F} \wedge a.
\end{align}
Notably, the grade-lowering dot product, or \emph{contraction}, is antisymmetric while the grade-raising wedge product, or \emph{inflation}, is symmetric, which is opposite to the vector product case \eqref{eq:product}.  The use of the dot product to lower the grade and the wedge product to raise the grade has become conventional \cite{Hestenes1987}.

The geometric meaning of these products can be ascertained for spacelike $\bv{F}$ by decomposing the vector $a = a_\perp + a_\parallel$ into parts perpendicular ($a_\perp$) and parallel ($a_\parallel$) to the plane of $\bv{F}$.  After factoring $\bv{F}$ into constituent orthogonal vectors (i.e., $b\cdot c = 0$), $\bv{F} = b\wedge c = bc$, we find
\begin{align}
  a \bv{F} &= abc = \frac{1}{2}a_\parallel(bc - cb) + a_\perp \wedge b \wedge c, \\
  \bv{F} a &= bca = \frac{1}{2}(bc - cb)a_\parallel + a_\perp \wedge b \wedge c, 
\end{align}
so the dot and wedge products have the simple forms
\begin{align}\label{eq:bivectordot}
  a\cdot \bv{F} &= a_\parallel \cdot \bv{F} = (a_\parallel\cdot b)c - (a_\parallel \cdot c)b, \\
  \label{eq:bivectorwedge}
  a\wedge \bv{F} &= a_\perp \wedge \bv{F} = a_\perp \wedge b \wedge c.
\end{align}
For spacelike $a,b,c$, these expressions have intuitive geometric interpretations.  The contraction $a\cdot \bv{F}$ produces a vector in the plane of $\bv{F}$ that is perpendicular to $a_\parallel$.  Indeed, choosing $b=a_\parallel$ so $\bv{F}=a_\parallel\wedge c$ yields $a\cdot \bv{F} = c$.  The wedge product $a\wedge \bv{F}$ produces a trivector with magnitude equal to the volume of the parallelipiped constructed from dragging the plane segment $\bv{F}$ along a perpendicular vector $a_\perp$.  These constructions are illustrated in Figure~\ref{fig:bivectorproduct}.


\subsubsection{Reversion and inversion}\label{sec:reversion}


Before we continue, we make a brief diversion to define another useful operation on the spacetime algebra known as \emph{reversion} \cite{Hestenes1987}, which reverses the order of all vector products: $(ab)^\sim = ba$.  The reverse distributes across general multivector products recursively $(AB)^{\sim} = \widetilde{B}\widetilde{A}$, and is the algebraic equivalent of the \emph{transpose} operation for matrices (as can be verified using a matrix representation).  The reverse inverts itself $(\widetilde{M})^\sim = M$, which makes it an \emph{involution} on the algebra.  The reverse is also summarized in Table~\ref{tab:operations} for reference.

The reverse of a bivector $\bv{F}$ is its negation $\widetilde{\bv{F}} = -\bv{F}$, which can be seen by splitting each basis element of $\bv{F}$ into orthogonal factors $\gamma_{\mu\nu} = \gamma_{\mu} \gamma_{\nu} = \gamma_{\mu} \wedge \gamma_{\nu}$.  The reversion of the factors then flips the wedge product $\widetilde{\gamma}_{\mu\nu} = \gamma_\nu\gamma_\mu = \gamma_\nu\wedge \gamma_\mu = -\gamma_\mu\wedge \gamma_\nu = -\gamma_{\mu\nu}$, which results in a negation.  Each grade of the general multivector in \eqref{eq:grades} can be reversed in an analogous way according to
\begin{equation}\label{eq:reversegrades}
  \widetilde{M} = \alpha + v - \bv{F} - \mathfrak{T} + \beta I.
\end{equation}
Notably, the pseudoscalar reverses to itself $\widetilde{I} = \gamma_{3210} = \gamma_{0123} = I$.

As a useful application of the reversion, the product of a pure $k$-blade $M = \mean{M}_k$ with its reverse $\widetilde{M}$ produces a scalar.  As an example, for $M = \gamma_{123}$ we have $\widetilde{M} M = \gamma_{321}\gamma_{123} = \gamma_{32}(\gamma_1)^2\gamma_{23} = -\gamma_3(\gamma_2)^2\gamma_3 = \gamma_3^2 = -1$.  The resulting positive magnitude $|M|^2 \equiv |\widetilde{M}M| = 1$ is a product of the magnitudes of the factors of $M$, while the sign $\epsilon_M \equiv \widetilde{M}M/|M|^2 = -1$ is a product of the \emph{signatures} of the factors.  Hence, the reversed-square of a pure $k$-blade 
\begin{align}
  \widetilde{M} M = M \widetilde{M} = \epsilon_{M} |M|^2
\end{align}
produces a notion of \emph{pseudonorm} for $M$ with a net signature $\epsilon_{M}$ and positive magnitude $|M|^2$, in exact analogy with the definition \eqref{eq:contraction} for vectors\footnote{It will be shown in Section~\ref{sec:complexbivectors} that bivectors $\bv{F}$ may not have a well-defined signature of $\pm 1$.  Nevertheless, these reversion equations will still be valid.}.  It follows that if $|M|^2 \neq 0$, then
\begin{equation}\label{eq:bladeinverse}
  M^{-1} \equiv \frac{\widetilde{M}}{\widetilde{M} M}
\end{equation}
is the inverse of a pure $k$-blade $M$ that satisfies $M^{-1}M = M M^{-1} = 1$, in analogy with \eqref{eq:inverse}.  


\subsection{Reciprocal bases, components, and tensors}\label{sec:components}


To connect with the standard tensor-analysis treatments of the electromagnetic field, where it is considered to be an antisymmetric rank-2 tensor $F^{\mu\nu}$, we now briefly consider the various ways of expanding proper multivector quantities in components.  We shall see that we immediately recover and clarify the standard tensor analysis formulas, making the tensor analysis techniques available as a restricted consequence of spacetime algebra.

Given a reference basis $\gamma_\mu$ for $\mathcal{M}_{1,3}$, it is convenient to introduce a \emph{reciprocal basis} of vectors 
\begin{align}
  \gamma^\mu &= (\gamma_\mu)^{-1}, &  
  \gamma^\mu\gamma_\nu &= \gamma^\mu \cdot \gamma_\nu = \delta^\mu_\nu
\end{align}
defined by the algebraic inverse \eqref{eq:inverse}.  This is an equally valid basis of vectors for $\mathcal{M}_{1,3}$ that satisfies the same metric relation $\gamma^\mu \cdot \gamma^\nu = \eta^{\mu\nu} = \eta_{\mu\nu}$.  It also follows from the inverse that $\gamma^0 = \gamma_0$, while $\gamma^j = -\gamma_j$ for $j = 1,2,3$, so the reciprocal basis is the spatial inversion of the reference basis $\gamma_\mu$.

Any multivector $M$ is a geometric object that can be expanded in any basis, including the reciprocal basis.  For example, a vector $v$ has the expansions
\begin{equation}\label{eq:vectorcomponents}
  v = \sum_\mu v^\mu \gamma_\mu = \sum_\mu v_\mu \gamma^\mu
\end{equation}
in terms of the reference basis and the reciprocal basis.  Traditionally, the components $v^\mu = \gamma^\mu \cdot v$ in the reference basis are called the \emph{contravariant} components of $v$, while the components $v_\mu = \gamma_\mu \cdot v$ in the reciprocal basis are the \emph{covariant} components of $v$; we see here that they can be interpreted as different ways of expressing the same geometric object.  Evidently from \eqref{eq:vectorcomponents} we have $v \cdot \gamma_\nu = v^\nu \eta_{\nu \nu} = v_\nu$, which shows how the two component representations are related by the metric (which effectively ``raises'' or ``lowers'' the index).

Since the graded basis elements of the spacetime algebra are constructed with the antisymmetric wedge product from the reference (or reciprocal) vector basis, then one can also expand a $k$-blade into redundant components that will be the same as the components of a rank-$k$ antisymmetric tensor.  For example, a bivector $\bv{F}$ may be expanded into components as 
\begin{equation}\label{eq:tensorcomponents}
  \bv{F} = \frac{1}{2}\sum_{\mu,\nu}F^{\mu\nu}\gamma_\mu\wedge\gamma_\nu = \frac{1}{2}\sum_{\mu,\nu} F_{\mu\nu} \gamma^\mu\wedge\gamma^\nu.
\end{equation}
The components $F^{\mu\nu} = \gamma^\mu \cdot \bv{F} \cdot \gamma^\nu = \gamma^\nu \cdot (\gamma^\mu \cdot \bv{F}) = (\gamma^\nu\wedge\gamma^\mu)\cdot \bv{F} = \gamma^{\nu\mu} \cdot \bv{F} = (\gamma_{\mu\nu})^{-1} \cdot \bv{F}$ can be extracted using the contractions defined in Section~\ref{sec:bivectorvector}, and are usually called the rank (2,0) (contravariant) components of $\bv{F}$, while the components $F_{\mu\nu} = \gamma_\mu \cdot \bv{F} \cdot \gamma_\nu$ are the rank (0,2) (covariant) components of $\bv{F}$.  Similarly, the components $F\indices{^\mu_\nu} = \gamma^\mu \cdot \bv{F} \cdot \gamma_\nu$ and $F\indices{_\mu^\nu} = \gamma_\mu \cdot \bv{F} \cdot \gamma^\nu$ are rank (1,1) (matrix) components for $\bv{F}$.  There are overtly 16 real components in each case, of which only $(16 - 4)/2 = 6$ are nonzero and independent due to the antisymmetry; these six independent components correpond precisely to the components of $\bv{F}$ when expanded more naturally in a bivector basis.  The factor of $1/2$ in \eqref{eq:tensorcomponents} ensures that the redundant components (differing only by a sign) are not double-counted.

The various ranked tensor components corresponding to $\bv{F}$ all refer to the action of the same antisymmetric tensor $\underbar{F}$, which is a multilinear \emph{function}
\begin{align} 
  \underbar{F}(a,b) &\equiv a \cdot \bv{F} \cdot b = b \cdot (a \cdot \bv{F}) = (b\wedge a) \cdot \bv{F} \\
  &= \sum_{\mu\nu} a^\mu F_{\mu\nu} b^\nu = \sum_{\mu\nu} a_\mu F^{\mu\nu} b_\nu = \sum_{\mu\nu} a_\mu F^\mu_\nu b^\nu = \sum_{\mu\nu} a^\mu F_\mu^\nu b_\nu  \nonumber
\end{align}
that takes two vector arguments $a = \sum_\mu a^\mu \gamma_\mu = \sum_\mu a_\mu \gamma^\mu$ and $b = \sum_\nu b^\nu \gamma_\nu = \sum_\nu b_\nu \gamma^\nu$ and produces a scalar through a total contraction with the bivector $\bv{F}$.  We use the underbar notation $\underbar{F}$ for functions to disambiguate them from products of multivectors.  Note that the different ranks of components for the tensor $\underbar{F}$ correspond to different ways of expanding the arguments $a$ and $b$ into different bases.  

Importantly, the electromagnetic field is intrinsically a bivector $\bv{F}$ and \emph{not} its associated antisymmetric tensor $\underbar{F}$, which is the multilinear function that performs contractions with $\bv{F}$ to produce a scalar; the confusion between these two distinct concepts arises because they have the same characteristic components $F^{\mu\nu}$.  Component-based tensor analysis obscures this subtle conceptual distinction by neglecting the $k$-blades themselves in favor of the functions that can be defined by contractions with these $k$-blades, all while emphasizing component descriptions that depend on particular basis expansions.  More distressingly, one cannot construct more general multivectors or the Clifford product using tensor notation.  As a result, we regard component-based tensor analysis as a part of---but not a replacement for---the spacetime algebra used in this report.


\subsection{The pseudoscalar $I$, Hodge duality, and complex structure}\label{sec:complex}


\begin{figure}[t]
  \begin{center}
    \includegraphics[width=0.7\columnwidth]{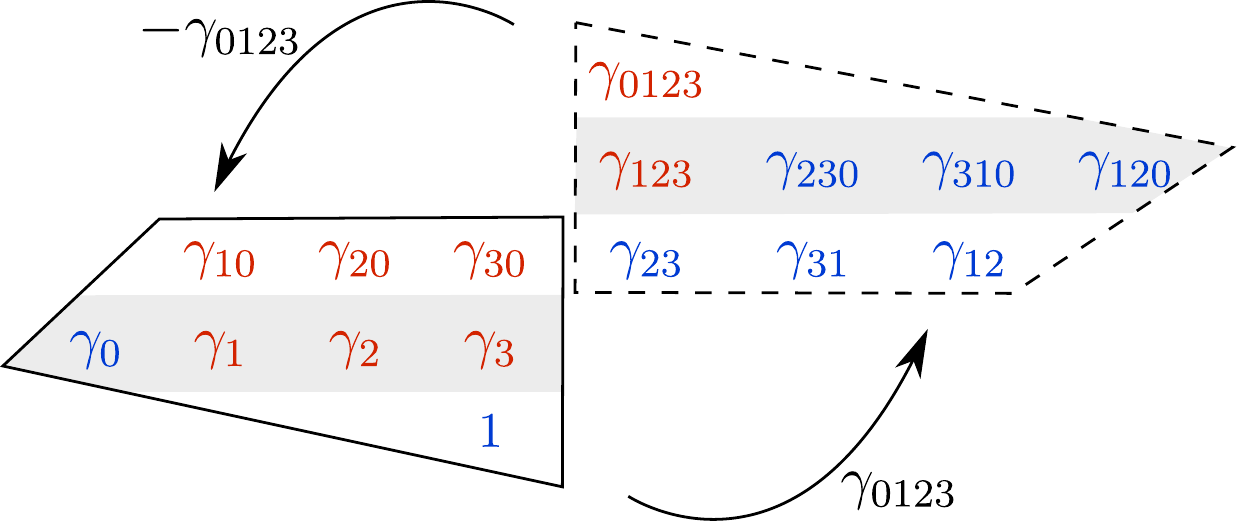}
  \end{center}
  \caption[Hodge duality]{Hodge duality illustrated using the same basis as Figure~\ref{fig:grades}.  Right multiplication by the pseudoscalar $I = \gamma_{0123}$ converts an algebraic element to its dual, which is an orthogonal complement in the geometric sense. For example, the dual of a trivector $\gamma_{123}$ is its unique normal vector $\gamma_{123}(-I) = \gamma_0$.  Dual elements have opposite signatures, indicated here by the flip in color coding.}
  \label{fig:dual1}
\end{figure}

The expansion \eqref{eq:grades} of a multivector in terms of elements of differing grade is similar to the expansion of a complex number into its real and imaginary parts, as we emphasized in Section \ref{sec:multivectors}.  In fact, this similarity is more than an analogy because the pseudoscalar $I$ also satisfies $I^2 = -1$, which makes the subalgebra of the form $\zeta = \alpha + \beta I$ completely equivalent in practice to the complex scalar numbers.  Hence, we do not need to additionally complexify the algebra in order to reap the benefits of complex analysis; the complex algebraic structure automatically appears within the spacetime algebra itself.  

The pseudoscalar $I$ has intrinsic geometric significance that goes well beyond the theory of scalar complex numbers, however, and has several additional interesting and useful properties.  It commutes with elements of even grade (i.e., scalars, pseudoscalars, and bivectors), but \emph{anticommutes} with elements of odd grade (i.e., vectors and trivectors).  More surprisingly, a right product with $I$ is a \emph{duality transformation} from grade $k$ to its orthogonal complement of grade $(4-k)$.  For example, the dual of a vector is its unique oriented orthogonal trivector $\gamma_0 I = \gamma_0\gamma_0\gamma_1\gamma_2\gamma_3 = \gamma_{123}$, while the dual of a trivector is its unique normal vector $\gamma_{123}I = (\gamma_0 I)I = -\gamma_0$.  This duality is illustrated for the entire graded basis in Figure~\ref{fig:dual1}.  We will see in Section~\ref{sec:dualsym} that this duality is intimately connected to the field-exchange \emph{dual symmetry} of the electromagnetic field.

\begin{figure}[t]
  \begin{center}
    \includegraphics[width=\columnwidth]{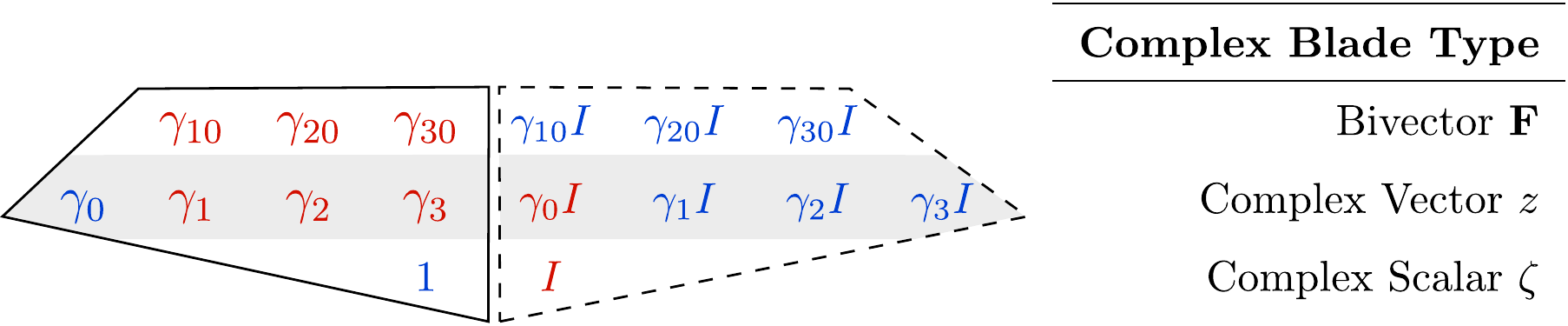}
  \end{center}
  \caption[Complex structure]{Complex structure of the spacetime algebra.  The complementary Hodge-dual halves of the algebra illustrated in Figure~\ref{fig:dual1} become intrinsically complex pairings by factoring out the pseudoscalar $I = \gamma_{0123}$ that satisfies $I^2 = -1$.  The complex scalars and vectors have parts that are independent geometric objects.  However, the bivectors are self-dual and irreducibly complex objects that cannot be further decomposed into frame-independent parts.}
  \label{fig:dual2}
\end{figure}

The duality transformation induced by $I$ is equivalent to the \emph{Hodge-star} transformation in differential forms (though is arguably simpler to work with), and splits the spacetime algebra into two halves that are geometric complements of each other.  Exploiting this duality, we can write any multivector $M$ in an intrinsically complex form (in the sense of $I$) that pairs quantities with their duals
\begin{align}\label{eq:dual}
  M &= \zeta + z + \bv{F}.
\end{align}
We illustrate this decomposition in Figure~\ref{fig:dual2}.  The complex scalar $\zeta$ and complex vector $z$ parts have the form 
\begin{align}
  \zeta &\equiv \alpha + \beta I, & z &\equiv v + w I = v - I w,
\end{align}
where $\alpha,\beta$ are real scalars, and $v,w$ are 4-vectors.  As anticipated, the trivector $\mathfrak{T} = w I = - I w$ has been expressed as the orthogonal complement of a vector (i.e., a \emph{pseudovector}), which is a useful simplification.  

Some care has to be taken with manipulations of complex vectors $z$, since $I$ anti-commutes with vectors.  As a result, complex scalars and vectors only quasi commute 
\begin{align}
  \zeta z = (\alpha + \beta I) z = z (\alpha - \beta I) \neq z \zeta,
\end{align}
while complex scalars and bivectors commute normally: $\zeta \bv{F} = \bv{F} \zeta$.  Thus, the pseudoscalar $I$ is only algebraically equivalent to the usual notion of the scalar imaginary $i$ when restricted to the \emph{even} graded subalgebra of complex scalars and bivectors.


\subsubsection{Bivectors: canonical form}\label{sec:complexbivectors}


A bivector like the electromagnetic field, $\bv{F}$, is an interesting special case for the complex decomposition \eqref{eq:dual} of a general multivector, since the full bivector basis is already \emph{self-dual}.  We can see this by computing the duality for a particular basis
\begin{align}\label{eq:bivectorduality}
  \gamma_{10}I &= -\gamma_{23}, & \gamma_{20}I &= -\gamma_{31}, & \gamma_{30}I &= -\gamma_{12}, \\
  \gamma_{23}I &= \gamma_{10}, & \gamma_{31}I &= \gamma_{20}, & \gamma_{12}I &= \gamma_{30}. \nonumber
\end{align}
Geometrically, these duality relations express the fact that in 4-dimensions the orthogonal complement to any plane is another plane (not a normal vector as in 3-dimensions).  We can express these relations more compactly by collecting cyclic permutations and writing $\gamma_{i0}I = -\epsilon_{ijk}\gamma_{jk}$ and $\gamma_{jk}I = \epsilon_{ijk}\gamma_{i0}$, where $i,j,k = 1,2,3$ and $\epsilon_{ijk}$ is the completely antisymmetric Levi-Civita symbol (no summation implied).  

This self-duality of the bivector basis makes the signature of a general bivector \emph{mixed}.  That is, although each basis bivector has a well-defined signature, $\gamma_{i0}\widetilde{\gamma}_{i0} = (1)(-1) = -1$ or $\gamma_{jk}\widetilde{\gamma}_{jk} = (-1)(-1) = 1$, a general bivector $\bv{F}$ will be a mixture of these two different signatures.  Indeed, consider the simple mixed bivector $\bv{F} = \gamma_{10} + \gamma_{23}$ as an example.  Computing its reversed square produces
\begin{align}
  \widetilde{\bv{F}}\bv{F} &= [\gamma_{01} + \gamma_{32}][\gamma_{10} + \gamma_{23}], \\
  &= \gamma_{01}\gamma_{10} + \gamma_{01}\gamma_{23} + \gamma_{32}\gamma_{10} + \gamma_{32}\gamma_{23}, \nonumber \\
  &= \gamma_1^2\gamma_0^2 + \gamma_{0123} + \gamma_{3210} + \gamma_2^2\gamma_3^2, \nonumber \\
  &= (-1)(1) + I + I + (-1)(-1), \nonumber \\
  &= 2 I. \nonumber
\end{align}
The cross-terms between the basis elements of different signature produce a \emph{complex signature} $\epsilon_{\bv{F}} = I$ attached to its positive magnitude $|\bv{F}|^2 = 2$.  Thus, mixing bases of differing signature is automatically handled by the algebra by making the total signature complex.
  
More generally, the reversed square of any bivector $\bv{F}$ will produce $\bv{F}\widetilde{\bv{F}} = \epsilon_F |\bv{F}|^2$, where $\epsilon_{F} = \alpha + \beta I$ is a complex signature with unit magnitude $\alpha^2 + \beta^2 = 1$.  As such, this signature can be written using a \emph{phase angle} by using the same algebraic techniques as standard complex analysis: $\epsilon_F = \exp(2\varphi I) = \cos2\varphi + I\sin2\varphi$.  It follows that a general bivector $\bv{F}$ must have a canonical (i.e., frame-independent) form with an \emph{intrinsic phase} 
\begin{equation}\label{eq:phase}
  \bv{F} = \bv{f}\, \exp(\varphi I) = \bv{f}\, (\cos\varphi +  I \sin\varphi)
\end{equation}
that mixes the bivector subspaces of different signature.  Here we define $\bv{f}$ to be the \emph{canonical bivector} for $\bv{F}$ that we choose to have a definite negative signature $\bv{f}\widetilde{\bv{f}} = -|\bv{F}|^2$ (for reasons that will become clear in Section \ref{sec:relative}).  Its dual $\bv{f}I$ then has a definite positive signature, while the square produces the positive magnitude $\bv{f}^2 = |\bv{F}|^2$.  It is worth emphasizing that the existence of this canonical complex form of a bivector is not at all apparent when working solely with its real components $F^{\mu\nu}$.

The phase and canonical bivector can be directly extracted from $\bv{F}$ according to
\begin{align}\label{eq:phasecomp}
  \varphi &= \frac{1}{2}\tan^{-1}\frac{\ell_2}{\ell_1}, &
  \bv{f} &= \bv{F}\, \exp(-\varphi I),
\end{align}
where $\ell_1 = \mean{\bv{F}^2}_0 = |\bv{F}|^2\cos2\varphi$ and $\ell_2 = \mean{\bv{F}^2}_4 I^{-1} = |\bv{F}|^2\sin2\varphi$ are the scalar parts of its square $\bv{F}^2 = \ell_1 + \ell_2 I$.  Note that for null bivectors ($\bv{F}^2 = 0$) the decomposition \eqref{eq:phase} is not unique and the phase $\varphi$ becomes degenerate (just as with standard complex analysis).  This potentially degenerate intrinsic phase of a bivector will become very important in our discussion of electromagnetic waves in Section~\ref{sec:waves}.


\subsubsection{Bivectors: products with bivectors}\label{sec:bivectorbivector}


A product between two bivectors $\bv{F}$ and $\bv{G}$ can be split into \emph{three} distinct grades (scalar, bivector, and pseudoscalar), in contrast to the two grades in the vector-vector and vector-bivector products of \eqref{eq:product} and \eqref{eq:productvecbi},
\begin{equation}
  \bv{F}\bv{G} = \mean{\bv{F}\cdot \bv{G}}_0 + [\bv{F},\bv{G}] + \mean{\bv{F} \cdot \bv{G}}_4.
\end{equation}
We will find it instructive to consider the scalar and pseudoscalar parts together as the same complex scalar $\bv{F}\cdot \bv{G}$.

The symmetric part of the product produces this complex scalar, which we can better understand by making the canonical decompositions $\bv{F} = \bv{f}\exp(\vartheta I)$, $\bv{G} = \bv{g}\exp(\varphi I)$ according to \eqref{eq:phase},  
\begin{equation}
  \bv{F}\cdot \bv{G} \equiv \frac{1}{2}(\bv{FG} + \bv{GF}) = (\bv{f}\cdot \bv{g})\, \exp[(\vartheta + \varphi)I].
\end{equation}
Since $\bv{f}$ and $\bv{g}$ have fixed signature, $\bv{f}\cdot \bv{g} = (\bv{fg} + \bv{gf})/2$ is a real scalar.  In fact, it will become clear after we introduce the relative 3-vectors in Section~\ref{sec:bivectorsplit} that this real scalar is completely equivalent to the usual (Euclidean) dot product from non-relativistic 3-vector analysis, which motivates this choice of notation.

It follows that the total dot product between arbitrary bivectors will have scalar and pseudoscalar parts of the form
\begin{align}
  \mean{\bv{F}\cdot \bv{G}}_0 &= (\bv{f}\cdot \bv{g}) \cos(\vartheta + \varphi), \\
  \mean{\bv{F}\cdot \bv{G}}_4 &= (\bv{f}\cdot \bv{g}) \sin(\vartheta + \varphi) I.
\end{align}
Note that the intrinsic phases of $\bv{F}$ and $\bv{G}$ play a critical role in these products, and produce wave-like \emph{interference} factors.  It is worth emphasizing that these factors arise solely from the structure of spacetime itself, and not from any additional assumptions.

The antisymmetric part of $\bv{FG}$ is written as a \emph{commutator bracket} that produces another bivector
\begin{equation}\label{eq:commutatorbracket}
  [\bv{F},\bv{G}] \equiv \mean{\bv{F}\bv{G}}_2 = \frac{1}{2}(\bv{F G} - \bv{G F}).
\end{equation}
We will explore the deep significance of this commutator bracket in Section~\ref{sec:lorentz}, where it will play the role of a Lie bracket for the Lorentz group (motivating this choice of notation).  For completeness here, we also define the \emph{bivector cross product} as the (Hodge) dual of the commutator bracket
\begin{equation}\label{eq:crossproduct}
  \bv{F}\times \bv{G} \equiv [\bv{F}, \bv{G}]I^{-1}.
\end{equation}
After we introduce the relative 3-vectors in \ref{sec:bivectorsplit} it will become clear that this cross product is completely equivalent to the traditional 3-vector cross product (again, motivating this choice of notation).


\subsubsection{Complex conjugation}\label{sec:conjugation}


At this point we define another useful operation: complex conjugation.  It follows from \eqref{eq:phase} that the proper inverse \eqref{eq:bladeinverse} of a bivector $\bv{F}$ has the canonical form
\begin{equation}
  \bv{F}^{-1} = \frac{\bv{f}}{|\bv{F}|^2}\, \exp(-\varphi I)
\end{equation}
which must additionally invert the intrinsic phase of the complex scalar that appears in its signature.

To more compactly write this inverse, we introduce a complex conjugation of the spacetime algebra as a flip in sign of the pseudoscalar.  For a multivector $M$, as in \eqref{eq:dual}, this produces
\begin{equation}\label{eq:conjugation}
  M^* = M|_{I\mapsto -I} = (\alpha - \beta I) + (v - wI) + \bv{f}\, \exp(-\varphi I).
\end{equation}
Geometrically, the flip in pseudoscalar sign corresponds to changing the orientation (handedness) of spacetime.  Since $(M^*)^* = M$, conjugation is also an involution on the algebra.  

For complex vectors, the conjugation is equal to reversion $\widetilde{z} = (v + wI)^\sim = v + I w = v - wI = z^*$, since $I$ anticommutes with vectors, which is a useful property.  However, this anticommutation also forces us to be careful about applying conjugation to products, since it does \emph{not} generally distribute across products that involve vectors, e.g., $(zM)^* \neq M^*z^*$, $(zM)^* \neq z^* M^*$.  Nevertheless, conjugation does distribute over products involving only complex scalars and bivectors, just as in standard complex analysis.  To aid memory about some of these subtler properties, we have collected a few commonly encountered identities in Table~\ref{tab:identities}. 

\begin{table}
  \centering
  \begin{tabular}{l l}
    \hline 
\noalign{\vskip 2mm} 
    \multicolumn{2}{c}{\textbf{Algebraic Identities}} \\
\noalign{\vskip 2mm} 
    \hline \\
    Dot-wedge conversion: & $v\cdot(wI) = -(vI)\cdot w = (v\wedge w)I$ \\
     & $v\wedge(w I) = - (vI)\wedge w = (v\cdot w) I$ \\
     & $(vI)\cdot\bv{F} = v\wedge(\bv{F}I) = (v\cdot\bv{F})I$ \\
     & $(vI)\wedge\bv{F} = v\cdot(\bv{F}I) = (v\wedge\bv{F})I$ \\
     & $v\cdot\bv{F}\cdot w = w\cdot(v\cdot \bv{F}) = (w\wedge v)\cdot\bv{F}$ \\
     \\
    Trace cyclicity: & $\mean{ABC}_0 = \mean{CAB}_0$ \\
    & $\mean{ABC}_4 = \mean{ABCI^{-1}}_0I$ \\
    \\
    Bivector-pseudoscalar commutation: & $\zeta\bv{F} = \bv{F}\zeta$ \\
    & $(\zeta \bv{F})^* = \zeta^* \bv{F}^* = \bv{F}^* \zeta^*$ \\
     \\
    Vector-pseudoscalar anti-commutation: & $(v + wI) = (v - Iw)$ \\
     & $z^* = \widetilde{z}$ \\
     & $\zeta z = z \zeta^*$ \\
     & $(z M)^* \neq z^* M^*$ \\
     \\
    Relative frame splits: 
    & $I = \gamma_{0123} = \rv{\sigma}_1\rv{\sigma}_2\rv{\sigma}_3$ \\
    & $v = (v_0 + \rv{v})\gamma_0 = \gamma_0(v_0 - \rv{v})$ \\
    & $\nabla = \gamma_0(\partial_0 + \rv{\nabla}) = (\partial_0 - \rv{\nabla})\gamma_0$ \\
    & $\bv{F} = \rv{E} + \rv{B}I$ \\
    \\
    3-vector products:
    & $\rv{E}\rv{B} = \rv{E}\cdot\rv{B} + \rv{E}\times\rv{B}I$ \\
    & $\mean{\rv{A}\rv{B}\rv{C}}_4 = \rv{A}\cdot(\rv{B}\times\rv{C})I$ \\
    & $\begin{aligned}\mean{\rv{A}\rv{B}\rv{C}}_2 &= (\rv{A}\cdot\rv{B})\rv{C} - (\rv{A}\times\rv{B})\times\rv{C} \\
      &= (\rv{A}\cdot\rv{B})\rv{C} + \rv{A}(\rv{B}\cdot\rv{C}) - (\rv{A}\cdot\rv{C})\rv{B} \end{aligned}$ \\
      \\
    Relative reversion:
    & $\zeta^\dagger = \zeta^*$ \\
    & $\bv{F}^\dagger = \rv{E} - \rv{B}I$ \\
    & $\gamma_0\widetilde{\bv{F}} = \bv{F}^\dagger\gamma_0$ \\
    & $(\rv{\sigma}_j\rv{\sigma}_k)^\dagger = \rv{\sigma}_k\rv{\sigma}_j$ \\ \\
    \hline 
  \end{tabular}
  \caption[Algebraic identities]{Commonly encountered algebraic identities.  All of these may derived in straightforward ways from the fundamental Clifford product, but we include them here for reference.}
  \label{tab:identities}
\end{table}

With this operation we can write the proper complex generalization of the blade inverse \eqref{eq:bladeinverse}
\begin{align}\label{eq:complexinverses}
  \zeta^{-1} &= \frac{\zeta^*}{\zeta^*\zeta}, & z^{-1} &= (z^*z)^{-1}z^*, & \bv{F}^{-1} &= \frac{\bv{F}^*}{\bv{F}^*\bv{F}},
\end{align}
since the quantities $\zeta^*\zeta = \alpha^2 + \beta^2$, $z^* z = \widetilde{z}z = v^2 + w^2 + 2 (v\cdot w) I$, and $\bv{F}^* \bv{F} = \bv{f}^2 = |\bv{F}|^2$ are all invertible scalars\footnote{Note that $z^* z$ produces a complex scalar, so $(z^* z)^{-1}$ recursively follows the complex scalar rule in \eqref{eq:complexinverses}.  Moreover, the inverse $z^{-1}$ written here is both a left and right inverse, even though $(z^* z)^{-1}$ and $z^*$ do not commute in general.  That is, the explicit right inverse is $z^{-1} = z^* (zz^*)^{-1} = [(zz^*)^{-1}]^* z^* = (z^*z)^{-1} z^*$, which is the same expression.}.


\subsection{Relative frames and paravectors}\label{sec:relative}


\begin{figure}[t]
  \begin{center}
    \subfloat[Spacetime split]{
      \includegraphics[width=0.3\columnwidth]{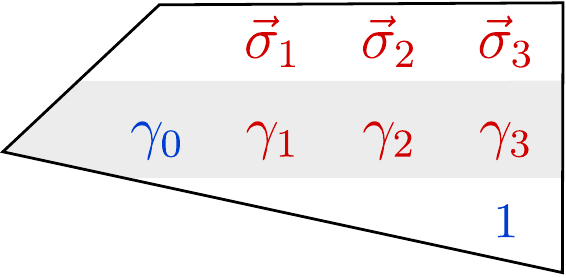}
    }
    \quad
    \subfloat[Relative frame duality]{
      \includegraphics[width=0.5\columnwidth]{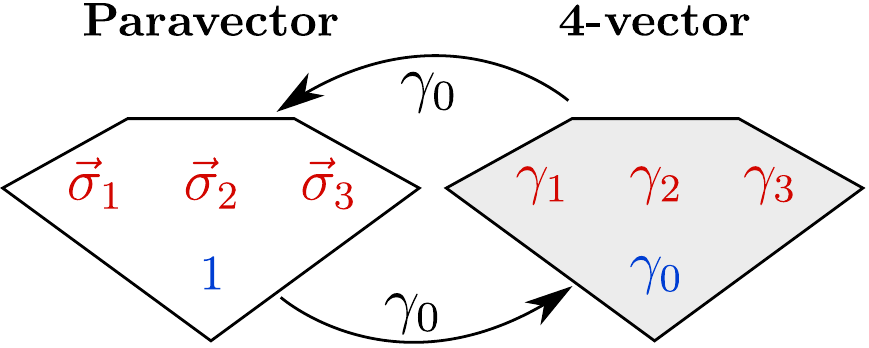}
    }
  \end{center}
  \caption[Spacetime split and paravectors]{Spacetime split and paravectors.  (a) Performing a spacetime split of Figure~\ref{fig:dual2} selects a timelike vector $\gamma_0$ that corresponds to a particular inertial reference frame.  This procedure partitions the proper bivectors into \emph{relative} parts spanned by a 3-vector basis $\rv{\sigma}_i = \gamma_{i0}$ and its dual $\rv{\sigma}_i I = -\epsilon_{ijk}\gamma_{jk}$.  (b) Each half of the split algebra then satisfies a secondary duality relation under right multiplication by $\gamma_0$.  This duality maps proper 4-vectors to relative paravectors, and vice versa.  As a result, the entire spacetime algebra may be expanded in terms of relative paravectors and duality factors, which provides an explicit embedding of traditional nonrelativistic 3-vector analysis inside spacetime algebra.}
  \label{fig:paravectors}
\end{figure}

To connect the spacetime algebra $\mathcal{C}_{1,3}$ with standard 3-vector analysis, we express the duality of the spacetime algebra in another way that selects a particular \emph{inertial reference frame}.  To do this, we decompose the bivector basis into the three elements that involve a specific timelike vector $\gamma_0$,
\begin{equation}\label{eq:relativebasis}
  \rv{\sigma}_i \equiv \gamma_{i0} = \gamma_i\wedge\gamma_0, 
\end{equation}
with $i=1,2,3$, and their duals $\rv{\sigma}_i I = -\epsilon_{ijk}\gamma_{jk}$.  This \emph{spacetime split} of the bivector basis into an inertial frame is illustrated in Figure \ref{fig:paravectors}(a).  The elements $\rv{\sigma}_i$ directly correspond to the three spatial directions $\gamma_i$ orthogonal to $\gamma_0$.  Importantly, they will act as a basis for the 3-vector quantities that are used in standard vector analysis treatments of electromagnetism.  As such, we notate these basis elements using standard vector arrows and refer to them as \emph{relative 3-vectors}.  Geometrically, one can conceptualize the relative 3-vectors that are experienced by an inertial observer as plane segments in spacetime that are obtained by dragging the spatial frame $\gamma_i$ along the timelike direction $\gamma_0$ of the observer's worldline.  As we will make clear in Section~\ref{sec:paulidirac}, the 3-vectors $\rv{\sigma}_i$ also have a deep connection to the standard Pauli spin matrices, which motivates our choice of notation.


\subsubsection{Bivectors: spacetime split and cross product}\label{sec:bivectorsplit}


In terms of this relative 3-vector basis, a bivector takes the simple form 
\begin{align}\label{eq:spacetimesplit}
  \bv{F} &= E_1 \gamma_{10} + E_2 \gamma_{20} + E_3 \gamma_{30} + B_1 \gamma_{32} + B_2 \gamma_{13} + B_3 \gamma_{21} = \rv{E} + \rv{B}I
\end{align}
involving a relative 3-vector and the dual of a relative 3-vector 
\begin{align}
  \rv{E} &= E_1 \rv{\sigma}_1 + E_2 \rv{\sigma}_2 + E_3 \rv{\sigma}_3, &
  \rv{B}I &= \left[B_1 \rv{\sigma}_1 + B_2 \rv{\sigma}_2 + B_3 \rv{\sigma}_3\right]I. 
\end{align}
Unlike the canonical decomposition \eqref{eq:phase}, splitting the bivector into dual relative pieces in this manner requires the specification of a particular timelike vector $\gamma_0$ in the basis \eqref{eq:relativebasis}, which spoils the frame-independence of each relative piece.  Only the total bivector $\bv{F}$ produced in the sum \eqref{eq:spacetimesplit} is a proper geometric object.  We choose the suggestive labels $\rv{E}$ and $\rv{B}$ here to anticipate their correspondence to the electromagnetic field vectors in Section~\ref{sec:maxwell}, but we can already see at this stage the inevitable equivalence in form of $\bv{F}$ to the Riemann-Silberstein complex vector \eqref{eq:RiemannSilberstein}.  As with the complex canonical form of a bivector, the existence of this complex spacetime split is not at all apparent when working solely with its real components $F^{\mu\nu}$.

Briefly, we note that the product between a vector and a bivector that we defined in Section~\ref{sec:bivectorvector} can be used to compute the spacetime split \eqref{eq:spacetimesplit} according to (noting $\gamma_0^{-1} = \gamma_0$)
\begin{equation}\label{eq:bivectorsplit}
  \bv{F} = \bv{F}\gamma_0\gamma_0^{-1} = (\bv{F}\cdot\gamma_0)\gamma_0 + (\bv{F}\wedge\gamma_0)\gamma_0.
\end{equation}
The first term contracts (i.e., lowers the grade of) $\bv{F}$ to a vector perpendicular to $\gamma_0$ and then multiplies by $\gamma_0$ to produce the relative 3-vector basis $\rv{\sigma}_i = \gamma_i\gamma_0$. The second term inflates (i.e., raises the grade of) $\bv{F}$ to a trivector including $\gamma_0$ and then contracts out the $\gamma_0$ part, leaving behind the dual basis $\rv{\sigma}_i I = -\epsilon_{ijk}\gamma_j\gamma_k$.  

Similarly, we can now revisit the cross product between two bivectors, defined in Section~\ref{sec:bivectorbivector} as the dual of the commutator bracket: $\bv{F}\times\bv{G} = [\bv{F},\bv{G}]I^{-1}$.  If we compute this cross product for the relative basis elements $\rv{\sigma}_i$ we find
\begin{align}\label{eq:crossbrackets}
  \rv{\sigma}_1 \times \rv{\sigma}_2 &= \rv{\sigma}_3, & \rv{\sigma}_2 \times \rv{\sigma}_3 &= \rv{\sigma}_1, & \rv{\sigma}_3 \times \rv{\sigma}_1 &= \rv{\sigma}_2.
\end{align}
These are precisely the defining relations for the familiar relative 3-vector cross product, which motivates its standard notation.  Thus, the bivector cross product forms a closed nonassociative (Gibbs) subalgebra for relative 3-vectors that is entirely embedded within the spacetime algebra in a natural way.  Hence, all the standard manipulations with the vector cross product still apply to the relative 3-vectors in the spacetime algebra.  Moreover, the conceptual meaning of the cross product is enriched, since it is induced by the more general associative product of spacetime algebra.


\subsubsection{Paravectors}


Using the spacetime split \eqref{eq:spacetimesplit} for the bivectors, we can rewrite a general multivector \eqref{eq:dual} in the following way
\begin{equation}\label{eq:biquaternion}
  M = \left[(\alpha + \rv{E}) + v\right] + \left[(\beta + \rv{B}) + w\right]I,
\end{equation}
where $\alpha,\beta$ are scalars, $v,w$ are proper 4-vectors, and $\rv{E},\rv{B}$ are relative 3-vectors.  

We have further partitioned each half of the complex expansion \eqref{eq:biquaternion} into even $(\alpha + \rv{E})$ and odd $(v)$ grades.  These partitions have four components each and are also dual to each other upon right multiplication by the chosen timelike unit vector $\gamma_0$.  We illustrate this duality in Figure~\ref{fig:paravectors}(b).  As an important example, the coordinate 4-vector $x$ becomes 
\begin{align}
  x\gamma_0 &= [\cc t \gamma_0 + x_1 \gamma_1 + x_2 \gamma_2 + x_3 \gamma_3]\gamma_0, \\
  &= \cc t + x_1 \gamma_{10} + x_2 \gamma_{20} + x_3 \gamma_{30}, \nonumber \\
  &= \cc t + \rv{x}, \nonumber
\end{align}
which is the sum of a relative scalar time and a 3-vector of the relative spatial coordinates.  This combination of a relative scalar and a relative 3-vector is commonly called a \emph{paravector} \cite{Hestenes1999,Baylis2002} and is often written as a tuple $(\cc t,\rv{x})$ instead of a sum to indicate that the scalar and 3-vector parts of a 4-vector should be considered part of the same unified whole.  Interestingly, the paravector that appears here as a sum of a scalar and a 3-vector is closely related to the original use of quaternions by Maxwell \cite{Maxwell1881} to describe the essential unification of temporal and spatial parts of a 4-vector.  We will comment more on the close relation between the 3-vectors that appear here and quaternions in Section~\ref{sec:lorentz}.

We see that right-multiplication of a 4-vector by $\gamma_0$ isolates the \emph{relative} quantities that correspond to the inertial frame of $\gamma_0$.  Left multiplication flips the sign of the 3-vector part due to the wedge product in the 3-vector basis: $\gamma_0 x = \cc t - \rv{x}$.  Thus, we can consistently recover the invariant coordinate interval in a direct way by using the paravector duality 
\begin{align}
  x^2 = x\gamma_0\gamma_0x = [\cc t + \rv{x}][\cc t - \rv{x}] = (\cc t)^2 - |\rv{x}|^2.
\end{align}
Observe that this style of derivation has completely bypassed the need to expand the expression in explicit components to obtain the final result.  This is a general and convenient property of working with spacetime algebra: transitioning between proper and relative expressions is a component-free and straightforward exercise.  

Using the vector-paravector duality, we can further expand the multivector \eqref{eq:biquaternion} into a fully frame-dependent form relative to $\gamma_0$
\begin{equation}\label{eq:relative}
  M = \left[(\alpha + \rv{E}) + (\delta + \rv{p})\gamma_0\right] + \left[(\beta + \rv{B}) + (\omega + \rv{a})\gamma_0\right]I,
\end{equation}
that involves only paravectors (with scalar parts $\alpha,\delta,\beta,\omega$ and vector parts $\rv{E},\rv{p},\rv{B},\rv{a}$ discussed below) and the appropriate duality factors $\gamma_0$ and $I$.  These paravectors are the quantities usually considered in frame-dependent treatments of relativistic systems that use traditional vector analysis.  However, the geometric origins of the quantities involved become obscured when they are lifted outside the spacetime algebra.  

\begin{table}
  \centering
  \begin{tabular}{l l  c c c c }
    \hline
\noalign{\vskip 2mm} 
    \multicolumn{2}{c}{\textbf{Multivector Part}} & \textbf{Parity} & \textbf{Scalar} & \textbf{4-vector} & \textbf{Bivector} \\
\noalign{\vskip 2mm} 
    \hline 
\noalign{\vskip 2mm} 
    $\alpha$ & proper scalar & $\times$ & $\checkmark$ & $\times$ & $\times$ \\
\noalign{\vskip 1mm} 
    $\beta I$  & proper pseudoscalar & $\checkmark$ & $\checkmark$ & $\times$ & $\times$ \\
\noalign{\vskip 1mm} 
    $\delta$ & relative scalar & $\times$ & $\times$ & $\checkmark$ & $\times$ \\
\noalign{\vskip 1mm} 
    $\omega I$ & relative pseudoscalar & $\checkmark$ & $\times$ & $\checkmark$ & $\times$ \\
\noalign{\vskip 1mm} 
    $\rv{p}$ & polar 3-vector & $\checkmark$ & $\times$ & $\checkmark$ & $\times$ \\
\noalign{\vskip 1mm} 
    $\rv{E}$ & polar 3-vector & $\checkmark$ & $\times$ & $\times$ & $\checkmark$ \\
\noalign{\vskip 1mm} 
    $\rv{a}$ & axial 3-vector & $\times$ & $\times$ & $\checkmark$ & $\times$ \\
\noalign{\vskip 1mm} 
    $\rv{B}$ & axial 3-vector & $\times$ & $\times$ & $\times$ & $\checkmark$ \\
\noalign{\vskip 2mm} 
    \hline 
\noalign{\vskip 3mm} 
    \multicolumn{6}{c}{$M = \left[(\alpha + \rv{E}) + (\delta + \rv{p})\gamma_0\right] + \left[(\beta + \rv{B}) + (\omega + \rv{a})\gamma_0\right]I$} \\
\noalign{\vskip 2mm} 
\hline
 \end{tabular}
 \caption[Transformation properties]{Transformation properties of the relative parts of a multivector $M$ in a particular frame.  Each checkmark indicates that the associated quantity changes sign under a spatial parity inversion, or is a proper scalar quantity that does not depend upon a specific Lorentz frame, or has components that change covariantly with the frame like parts of a proper 4-vector, or has components that change like parts of a proper bivector.  These transformation properties are traditionally used to distinguish the distinct relative parts of a spacetime multivector.}
 \label{tab:transformations}
\end{table}

We can recover some distinguishing information about the scalars by examining their behavior under frame transformations and spatial reflections, which we summarize in Table~\ref{tab:transformations}.  The scalar $\alpha$ is invariant under both frame transformations and spatial reflections, so is a proper scalar.  The scalar $\beta$ flips sign under spatial reflections but does not change under a frame transformation, so is a proper pseudoscalar.  The scalar $\delta$ is invariant under spatial reflections but changes under a frame transformation, so is a relative scalar (i.e., temporal) part of a 4-vector.  The scalar $\omega$ flips sign under a spatial reflection and changes under a frame transformation, so is a relative pseudoscalar part of a 4-pseudovector.  

Similarly, we can distinguish the 3-vectors by looking at spatial reflections and frame transformation properties.  The vectors $\rv{E},\rv{p}$ flip sign under a spatial reflection, so are \emph{polar} relative 3-vectors.  The vectors $\rv{B},\rv{a}$ do not flip sign under a spatial reflection, so are \emph{axial} relative 3-vectors.  However, while the combination of components $(\delta,\rv{p})$ transforms as a proper 4-vector and $(\omega,\rv{a})$ transforms as a proper 4-pseudovector, the combination of components in $(\rv{E},\rv{B})$ transforms as what is usually considered as an antisymmetric rank-2 tensor (i.e., a bivector), as we indicated in Section~\ref{sec:spacetime} and further elucidated in Section~\ref{sec:components}.  

\begin{table}
  \centering
  \begin{tabular}{l l}
    \hline
\noalign{\vskip 2mm} 
    \multicolumn{2}{c}{\textbf{Multivectors}} \\
\noalign{\vskip 2mm} 
    \hline \\
    \textbf{Graded Form} & $M = \alpha + v + \bv{F} + w I + \beta I$ \\
    \\
     & $\alpha$ : scalar \\
     & $v$ : vector \\
     & $\bv{F}$ : bivector \\
     & $wI$ : pseudovector \\
     & $\beta I$ : pseudoscalar \\
     \\
     \hline
    \\
    \textbf{Complex Form} & $M = \zeta + z + \bv{F}$ \\
    \\
     & $\zeta = \alpha + \beta I$ : complex scalar \\
     & $z = v + w I$ : complex vector \\
     & $\bv{F} = \bv{f}\,\exp(\varphi I)$ : (irreducibly) complex bivector \\
     \\
     \hline
    \\
    \textbf{Relative Form} & $M = \left[(\alpha + \rv{E}) + (\delta + \rv{p})\gamma_0\right] + \left[(\beta + \rv{B}) + (\omega + \rv{a})\gamma_0\right]I$ \\
    \\
    & $\alpha$ : proper scalar \\
    & $\rv{E}$ : relative polar 3-vector part of $\bv{F}$ \\
    & $\delta$ : relative scalar part of $v$ \\
    & $\rv{p}$ : relative polar 3-vector part of $v$ \\
    & $\beta I$  : proper pseudoscalar \\
    & $\rv{B}$ : relative axial 3-vector part of $\bv{F}$ \\
    & $\omega I$ : relative pseudoscalar part of $w I$ \\
    & $\rv{a}$ : relative axial 3-vector part of $w I$ \\ \\
    \hline 
 \end{tabular}
 \caption[Multivector expansions]{Multivector expansions into distinct geometric grades, intrinsically complex parts, and relative parts that depend upon a particular choice of Lorentz frame.}
 \label{tab:multivectors}
\end{table}

For convenience, we also summarize the various expansions of a general multivector in Table \ref{tab:multivectors}.  Comparing the proper expression of \eqref{eq:grades} to the frame-dependent expression of \eqref{eq:relative} makes the geometric origins of all these transformation properties clear.  The proper combinations of relative quantities correspond to distinct geometric grades of a spacetime multivector.  This geometric structure is left implicit in standard vector or tensor analysis methods, even though the transformation properties remain.  The spacetime algebra makes this structure explicit, intuitive, and easily manipulated in calculations.


\subsubsection{Relative reversion}


To isolate the distinct quantities that are relevant for a particular frame, it is useful to define a \emph{relative reversion} involution
\begin{align}\label{eq:relativereversion}
  M^\dagger &= \gamma_0 \widetilde{M} \gamma_0 \\
  &= \left[(\alpha + \rv{E}) + (\delta - \rv{p})\gamma_0\right] + \left[-(\beta + \rv{B}) + (\omega - \rv{a})\gamma_0\right]I. \nonumber
\end{align}
Comparing with \eqref{eq:relative}, note that this $\gamma_0$-dependent involution has the effect of the proper complex conjugation for the complex scalars 
\begin{align}
  \zeta^\dagger &= (\alpha + \beta I)^\dagger = \alpha - \beta I = \zeta^*,
\end{align}
but has the effect of conjugation \emph{with respect to the spacetime split} for the complex bivectors
\begin{align}\label{eq:relativeconjugation}
 \bv{F}^\dagger &= (\rv{E} + \rv{B}I)^\dagger = \rv{E} - \rv{B}I \neq \bv{F}^*.
\end{align}
Importantly, this type of conjugation is not frame-independent, but depends on the relative frame $\gamma_0$ due to the spacetime split.

In contrast, the relative reversion of a 4-vector only flips the sign of the relative 3-vector part 
\begin{align}
  v^\dagger &= [(\delta + \rv{p})\gamma_0]^\dagger = (\delta - \rv{p})\gamma_0, \\
  (wI)^\dagger &= [(\omega + \rv{a})\gamma_0I]^\dagger = (\omega - \rv{a})\gamma_0I, \nonumber
\end{align}
which is intrinsically different from complex conjugation.

Using the appropriate relative reversions, we can then extract the relative parts of each geometric object in a systematic way
\begin{align}
  \alpha &= \frac{\zeta + \zeta^\dagger}{2}, & \beta &= \frac{\zeta - \zeta^\dagger}{2 I}, \\ \relax
  \rv{E} &= \frac{\bv{F} + \bv{F}^\dagger}{2}, & \rv{B} &= \frac{\bv{F} - \bv{F}^\dagger}{2 I}, \nonumber \\ \relax
  \delta &= \frac{v + v^\dagger}{2} \gamma_0^{-1}, & \rv{p} &= \frac{v - v^\dagger}{2}\gamma_0^{-1}, \nonumber \\ \relax
  \omega &= \frac{wI + (wI)^\dagger}{2}I^{-1}\gamma_0^{-1}, & \rv{a} &= \frac{wI - (wI)^\dagger}{2}I^{-1}\gamma_0^{-1}. \nonumber 
\end{align}
Note that we have to be careful about the order of the inversions in the case when elements do not commute.


\subsection{Bivectors: commutator bracket and the Lorentz group}\label{sec:lorentz}


We now revisit the commutator bracket that we briefly defined for bivectors in Section~\ref{sec:bivectorbivector}.  Surprisingly, this commutator bracket makes the bivectors a closed nonassociative \emph{Lie algebra} \cite{Hestenes1987}.  Indeed, this feature has motivated the notation of the bracket.  To see which Lie algebra the bivectors belong to, we compute the $15$ following bracket relations between all six bivector basis elements:
\begin{align}\label{eq:bivectorbrackets}
  [\gamma_{12},\gamma_{23}] &= \gamma_{31}, & [\gamma_{23},\gamma_{31}] &= \gamma_{12}, & [\gamma_{31},\gamma_{12}] &= \gamma_{23}, \\ 
  [\gamma_{10},\gamma_{20}] &= - \gamma_{12}, & [\gamma_{20},\gamma_{30}] &= -\gamma_{23}, & [\gamma_{30},\gamma_{10}] &= -\gamma_{31}, \nonumber \\ 
  [\gamma_{31},\gamma_{10}] &= \gamma_{30}, & [\gamma_{12},\gamma_{20}] &= \gamma_{10}, & [\gamma_{23},\gamma_{30}] &= \gamma_{20}, \nonumber \\ 
  [\gamma_{12},\gamma_{10}] &= \gamma_{20}, & [\gamma_{23},\gamma_{20}] &= \gamma_{30}, & [\gamma_{31},\gamma_{30}] &= \gamma_{10}, \nonumber \\
  [\gamma_{23},\gamma_{10}] &= 0, & [\gamma_{31},\gamma_{20}] &= 0, & [\gamma_{12},\gamma_{30}] &= 0. \nonumber
\end{align}

We can simplify these bracket relations to a more familiar form by introducing the temporary (but standard) notations $\bv{S}_1 = \gamma_{23}$, $\bv{S}_2 = \gamma_{31}$, $\bv{S}_3 = \gamma_{12}$ and $\bv{K}_1 = \gamma_{10}$, $\bv{K}_2 = \gamma_{20}$, $\bv{K}_3 = \gamma_{30}$, and combining cyclic permutations
\begin{align}\label{eq:lorentzbrackets}
  [\bv{S}_i,\bv{S}_j] &= \epsilon_{ijk} \bv{S}_k, & [\bv{K}_i,\bv{K}_j] &= -\epsilon_{ijk} \bv{S}_k, & [\bv{S}_i,\bv{K}_j] &= \epsilon_{ijk} \bv{K}_k.
\end{align}
These are precisely the Lie bracket relations that define the generators for the \emph{Lorentz group}, where $\bv{S}_i$ generate spatial rotations, and $\bv{K}_i$ generate boosts.  That is, the closed bivector subalgebra under the commutator bracket is precisely the Lie algebra of the Lorentz group.  This connection between bivectors and the Lorentz group is extremely useful in practice for manipulating objects in spacetime.

As a consequence of this correspondence, exponentiating the commutator bracket of a spacetime bivector produces a restricted (proper, orthochronous) Lorentz transformation that is continuously connected to the identity.  For example, to spatially rotate a bivector $\bv{F}$ by an angle $\theta$ in the plane $\bv{S}_3 = \rv{\sigma}_3 I^{-1}$ (i.e., around the relative axis $\rv{\sigma}_3$), then we would use the following exponentiation
\begin{align}\label{eq:lorentzrotation}
  \exp(\theta\, [\bv{S}_3,\cdot])\,\bv{F} &= \bv{F} + \theta[\bv{S}_3,\bv{F}] + \frac{\theta^2}{2}[\bv{S}_3,[\bv{S}_3,\bv{F}]] + \dots, \\
  &= \exp(\theta\, \bv{S}_3 /2)\,\bv{F}\,\exp(-\theta\,\bv{S}_3/2), \nonumber \\
  &= \exp(-I\theta\, \rv{\sigma}_3 /2)\,\bv{F}\,\exp(I\theta\,\rv{\sigma}_3/2). \nonumber
\end{align}
The ability to resum an exponentiated commutator back into a double-sided product is an important identity that is commonly used in quantum-mechanics\footnote{This equivalence and the resulting half-angle is a manifestation of the double-covering of the spatial rotation group SO(3) by the special unitary group SU(2), which is in turn isomorphic to the Spin(3) group generated by the rotation bivectors $\bv{S}_i = \rv{\sigma}_i I^{-1}$.}.  Indeed, the final form of the rotation in \eqref{eq:lorentzrotation} is identical in form to the spatial rotation of a quantum mechanical spin, where the vector $\rv{\sigma}_3$ is usually interpreted as a Pauli matrix.  We will return to this point in the next few sections.  

As another important historical note, this double-sided product also appears as the proper form of a spatial rotation when using \emph{quaternions}, where 
\begin{align}
  \bv{i} &\equiv \bv{S}_1 = \rv{\sigma}_1I^{-1}, &
  \bv{j} &\equiv -\bv{S}_2 = -\rv{\sigma}_2I^{-1}, &
  \bv{k} &\equiv \bv{S}_3 = \rv{\sigma}_3I^{-1} 
\end{align}
are the usual ``imaginary'' quaternion elements\footnote{Note the odd change in sign for $\bv{j}$: Hamilton accidentally defined a left-handed set, which has caused significant confusion about the proper application of quaternions in physics, and has obfuscated their equivalence to the rotation generators made apparent here.} that all square to $-1$ and satisfy Hamilton's defining relation $\bv{i}\bv{j}\bv{k} = -1$ \cite{Hamilton1853,Hamilton1901}.  Note that the quaternion basis differs from the 3-vector basis by a factor of $I^{-1}$ (i.e., the dual-basis), which is the fundamental reason that Maxwell's attempt to use them for describing relative spatial directions \cite{Maxwell1881} was fraught with problems and eventually fell out of favor (despite leaving behind the legacy of using $\hat{\imath}$, $\hat{\jmath}$, and $\hat{k}$ to denote the spatial unit vectors).  Quaternions are fundamentally related to spatial rotations, not spatial directions.

Another important and enlightening example of the Lorentz group is the boost from a frame with timelike vector $\gamma_0$ to a different frame moving at a relative velocity $\rv{v} = |v| \rv{\sigma}_3$ with respect to $\gamma_0$.  Recall that a relative vector such as $\rv{v} = |v| \gamma_3\wedge\gamma_0$ is really a bivector, so it can directly act as a generator for the appropriate Lorentz boost.  Indeed, note that the relative unit vector $\rv{\sigma}_3 = \gamma_{30}$ is also equal to a boost generator $\rv{\sigma}_3 = \bv{K}_3$.  The proper hyperbolic angle (i.e., \emph{rapidity}) for the boost rotation is determined by the magnitude of the velocity $\alpha = \tanh^{-1}\!|v|/\cc$, while the proper plane of rotation is simply the relative unit vector itself $\rv{\sigma}_3$.  Hence, the active Lorentz boost that rotates a bivector $\bv{F}$ from the $\rv{v}$ frame back into the reference $\gamma_0$ frame has the form
\begin{align}\label{eq:lorentzboost}
  \bv{F}' = \exp(-\alpha\,[\rv{\sigma}_3,\cdot])\,\bv{F} = \exp(-\alpha\,\rv{\sigma}_3/2)\,\bv{F}\,\exp(\alpha\,\rv{\sigma}_3/2).
\end{align}
Again, the conversion of the exponentiated commutator to a double-sided product makes the expression simpler.  Note that the only difference in form between the boost \eqref{eq:lorentzboost} and the spatial rotation \eqref{eq:lorentzrotation} is the factor of $I$ in the exponent that changes the signature of the spacetime rotation.

To simplify \eqref{eq:lorentzboost}, we split the bivector $\bv{F} = \bv{F}_\parallel + \bv{F}_\perp$ into pieces parallel $\bv{F}_\parallel$ and perpendicular $\bv{F}_\perp$ to $\rv{v}$, implying that $\rv{v}$ commutes with $\bv{F}_\parallel$ but anti-commutes with $\bv{F}_\perp$.  The Lorentz transformation then becomes
\begin{align}
  \bv{F}' &= \exp(-\alpha\, \rv{\sigma}_3/2)\,\bv{F}\,\exp(\alpha\,\rv{\sigma}_3/2), \\
  &= \bv{F}_\parallel + \exp(-\alpha\,\rv{\sigma}_3)\,\bv{F}_\perp, \nonumber \\
  &= \bv{F}_\parallel + \exp(-\tanh^{-1}\!|v|/\cc\,\rv{\sigma}_3)\, \bv{F}_\perp, \nonumber \\
  &= \bv{F}_\parallel + [\cosh(\tanh^{-1}\!|v|/\cc) - \sinh(\tanh^{-1}\!|v|/\cc)\,\rv{\sigma}_3]\,\bv{F}_\perp, \nonumber \\
  &= \bv{F}_\parallel + \frac{1 - \rv{v}/\cc}{\sqrt{1 - (v/\cc)^2}}\,\bv{F}_\perp, \nonumber
\end{align}
where the boost dilation factor $\gamma = (1 - (v/\cc)^2)^{-1/2}$ appears to scale only the perpendicular part of $\bv{F}$.  If we make a spacetime split $\bv{F}_\perp = \rv{E}_\perp + \rv{B}_\perp I$ of the perpendicular bivector, then the transformed version is
\begin{align}\label{eq:relativelorentzboost}
  \bv{F}'_\perp &= \gamma(1 - \rv{v}/\cc)\bv{F}_\perp = \gamma(1 - \rv{v}/\cc)(\rv{E}_\perp + \rv{B}_\perp I), \\
  &= \gamma[ (\rv{E}_\perp + \frac{\rv{v}}{\cc}\times\rv{B}_\perp) + (\rv{B}_\perp - \frac{\rv{v}}{\cc}\times\rv{E}_\perp)I ], \nonumber
\end{align}
where we have simplified the bivector product $\rv{v}\rv{E}_\perp = \rv{v}\times\rv{E}_\perp I$, since $\rv{v}\cdot\rv{E}_\perp = 0$ (and similarly for $\rv{B}$).  Up to choices of units, this is precisely the transformation that we expect from a boost of the electromagnetic field \cite{Landau1975,Jackson1999}.


\subsubsection{Spinor representation}


Recall that in \eqref{eq:lorentzboost} we converted the exponentiated commutator bracket into a double-sided product.  Such a double-sided product is a more natural algebraic representation of a group transformation that can be extended from bivectors to the entire algebra.  Specifically, if we notate the product factors as $U = \exp(-\alpha\,\rv{\sigma}_3/2)$, then the transformation in \eqref{eq:lorentzboost} has the form
\begin{align}\label{eq:lorentztransform}
  \bv{F}' = U\,\bv{F}\,\widetilde{U},
\end{align}
where we have used $\widetilde{\rv{\sigma}}_3 = -\rv{\sigma}_3$.  The factors satisfy the normalization condition $U\widetilde{U} = \widetilde{U}U = 1$.  Hence, this form of the transform makes it clear how the transformation preserves the product structure of the algebra, 
\begin{align}
  \bv{GH} \mapsto U\bv{G}\widetilde{U}U\bv{H}\widetilde{U} = U(\bv{GH})\widetilde{U},
\end{align}
precisely in the same manner that unitary representations of group transformations preserve product structure in quantum mechanics.

We call the quantity $U$ that appears in \eqref{eq:lorentztransform} a \emph{rotor} to emphasize that a Lorentz transformation has the geometric form of a \emph{rotation} through spacetime.  Specifically, the rotation generators $\bv{S}_i = \rv{\sigma}_i I^{-1}$ square to $-1$ and produce spatial rotations, while the boost generators $\bv{K}_i = \rv{\sigma}_i$ square to $+1$ and produce hyperbolic rotations in spacetime.  Since the rotor representation of a Lorentz transformation produces a geometrically meaningful rotation, the bivector $\bv{F}$ can be replaced in \eqref{eq:lorentztransform} with any geometric object that can be rotated.  Indeed, \emph{any multivector} $M$ can be Lorentz transformed according to $M' = U M \widetilde{U}$.

A rotor $U$ is an example of a \emph{spinor} $\psi$, which is an element of the even-graded subalgebra of the spacetime algebra \cite{Crumeyrolle1990} that may be decomposed as  
\begin{equation}
  \psi = \zeta + \bv{F}, 
\end{equation}
where $\zeta = \alpha + \beta I$ is a complex scalar and $\bv{F}$ is a bivector.  This even-graded subalgebra is closed under the full Clifford product.  As an example of this decomposition, the boost rotor $\exp(-\alpha\, \rv{\sigma}_3/2)$ can be expanded in terms of hyperbolic trigonometric functions as $\cosh(\alpha/2) - \rv{\sigma}_3\,\sinh(\alpha/2)$, which is a sum of a scalar and a bivector.  Similarly, the spatial rotor $\exp(-\theta\,I\rv{\sigma}_3/2)$ can be expanded in terms of circular trigonometric functions as $\cos(\theta/2) - I\rv{\sigma}_3\,\sin(\theta/2)$.

Spinors also produce transformations of any multivector according to a double-sided product that preserves the product structure of the algebra.  For example, if $\psi\widetilde{\psi} = \widetilde{\psi}\psi = 1$, then $M \mapsto \psi M \widetilde{\psi}$ is a \emph{spinor representation} of a group transformation.  Hence, the rotor form of \eqref{eq:lorentztransform} can also be understood as a spinor representation of the restricted Lorentz transformation.  Similarly, if $\psi \psi^* = \psi^* \psi = 1$, then $M \mapsto \psi M \psi^*$ is another kind of group transformation that preserves the product structure of the algebra.  

An example transformation of this latter type that we will find particularly useful in what follows is the \emph{global phase rotation} characterized by the spinor $\psi = \exp(-\theta I / 2)$.  Interestingly, this phase rotation only affects complex vectors since $I$ commutes with complex scalars and bivectors.  That is, for a general multivector $M = \zeta + z + \bv{F}$ in complex form, we have 
\begin{align}\label{eq:globalphase}
  \psi M \psi^* &= \zeta + z\,\exp(\theta I) + \bv{F},
\end{align}
which is a proper group transformation since $\psi\psi^* = \psi^*\psi = 1$.  This phase transformation is responsible for the field-exchange \emph{dual symmetry} of the electromagnetic field, which we will detail in Section \ref{sec:dualsym}.


\subsection{Pauli and Dirac matrices}\label{sec:paulidirac}


A vast notational simplification of the Lorentz bracket relations \eqref{eq:lorentzbrackets} occurs when we use the duality relations to rewrite the bivectors $\bv{K}_i = \rv{\sigma}_i$ and their duals $\bv{S}_i = \rv{\sigma}_i I^{-1}$ in terms of a particular Lorentz frame of relative 3-vectors.  The pseudoscalar $I$ commutes with all bivectors and squares to $-1$, so all $15$ bracket relations reduce to simple variations of the three fundamental relations
\begin{equation}\label{eq:paulibrackets}
  [\rv{\sigma}_i,\rv{\sigma}_j] = \epsilon_{ijk} \rv{\sigma}_k I,
\end{equation}
which are simply another way of expressing the 3-vector cross products \eqref{eq:crossbrackets} in the relative frame.  Observe that the relations \eqref{eq:paulibrackets} are also precisely the bracket relations that define the Pauli spin matrices used in quantum mechanics.  Moreover, note that the relative 3-vector basis satisfies the relation
\begin{equation}
  \rv{\sigma}_i \rv{\sigma}_j + \rv{\sigma}_j \rv{\sigma}_i = 2 \delta_{ij},
\end{equation}
which indicates that the relative 3-vectors $\rv{\sigma}_i$ form a \emph{Euclidean} Clifford subalgebra $\mathcal{C}_{3,0}$ in their own right.

In terms of this frame-dependent split, the proper pseudoscalar $I$ can be written as $I = \rv{\sigma}_1\rv{\sigma}_2\rv{\sigma}_3$.  That is, $I$ is also the appropriate pseudoscalar (apparent unit 3-volume) for any relative frame, which is a nontrivial property.  It also commutes with every element of this relative frame, making it completely equivalent in practice to a scalar imaginary $i$.  Similarly, the relative reversion operation $\psi^\dagger = \gamma_0\widetilde{\psi}\gamma_0$ that we defined in \eqref{eq:relativereversion} is the true reversion of the relative basis $\rv{\sigma}_i$ within this subalgebra: $(\rv{\sigma}_j\rv{\sigma}_k)^\dagger = \rv{\sigma}_k\rv{\sigma}_j$.  As such, it also acts as the appropriate complex conjugation for the frame-dependent subalgebra according to \eqref{eq:relativeconjugation}.  The fact that this relative complex conjugation is distinct from the global complex conjugation is not at all obvious when working solely within a relative frame.  This subalgebra and relative conjugation is precisely what appears in the usual treatments of the Riemann-Silberstein vector (e.g., \cite{Bialynicki-Birula2013}).

Collecting these properties together, it is clear that the familiar Pauli matrices and complex scalar imaginary $i$ can be viewed as a matrix representation of a particular spacetime split of the spinor subalgebra of spacetime.  The matrix product simulates the noncommutative vector product.  Indeed, this direct connection to the Pauli matrices has motivated our notation $\rv{\sigma}_i$ for the relative 3-vector basis \cite{Hestenes2005}.  Within the spacetime algebra, however, $\rv{\sigma}_i$ are not matrices---they are a bivector basis associated with the three spatial directions in a particular frame.  Hence matrix constructions like $\sum_i a^i \rv{\sigma}_i$ that appear in quantum mechanics can be understood to define relative 3-vectors $\rv{a}$ in spacetime algebra (e.g., the reference spatial direction along which a spin can be aligned).  There is nothing intrinsically quantum mechanical about these constructions.  

\emph{Notably, the representation-free nature of spacetime algebra has exposed the deep structural relationship between the relative spatial directions, the cross product of relative 3-space, the Lie bracket relations of the Lorentz group, the quaternion algebra, the commutation relations of the Pauli spin matrices, and even the imaginary unit $i$ that appears in these commutation relations.}  Using standard mathematical methods, these topics do not appear to have such an obvious connection.  Indeed, the natural appearance of all these topics solely from a systematic construction of the geometry of spacetime highlights the deep geometric significance of their relation to one another.

As we briefly indicated at the beginning of this introduction, the embedding of Clifford algebras $\mathcal{C}_{3,0} \subset \mathcal{C}_{1,3}$ motivated our choice of metric signature for spacetime to be $(+,-,-,-)$.  In fact, this embedding is part of a larger sequence of Clifford algebra embeddings $\mathcal{C}_{2,4}\supset\mathcal{C}_{4,1}\supset\mathcal{C}_{1,3}\supset\mathcal{C}_{3,0}\supset\mathcal{C}_{0,2}\supset\mathcal{C}_{0,1}\supset\mathcal{C}_{0,0}$ illustrated in Table~\ref{tab:clifford}, which corresponds to, in order: conformal space (Penrose twistors), relativistic electron (Dirac spinors), spacetime/electromagnetism (Maxwell spinors), relative 3-space (Pauli spinors), quaternions (Hamilton spinors), complex numbers (Schr\"odinger spinors), and the real numbers.  The connection between this sequence and the successive approximations of quantum mechanical particles has been emphasized in \cite{Hiley2012,Hiley2001,Hiley2010a,Hiley2010b}.  The other choice of spacetime signature $(-,+,+,+)$ produces a distinct algebra $\mathcal{C}_{3,1}$ that does not belong to this sequence of algebras.

\begin{table}
  \centering
  \begin{tabular}{c  c  c }
    \hline
\noalign{\vskip 2mm} 
    \textbf{Clifford Algebras} & \textbf{Physics Connections} & \textbf{Spinor Types} \\
\noalign{\vskip 2mm} 
    \hline 
\noalign{\vskip 2mm} 
    $\mathcal{C}_{2,4}$ & Conformal space & Penrose twistors \\
\noalign{\vskip 1mm} 
    $\mathcal{C}_{4,1}$ & Dirac electron & Dirac spinors \\
\noalign{\vskip 1mm} 
    \hline
\noalign{\vskip 1mm} 
    $\bm{\mathcal{C}_{1,3}}$ & \textbf{Spacetime / EM} & \textbf{Maxwell spinors} \\
\noalign{\vskip 1mm} 
    \hline
\noalign{\vskip 1mm} 
    $\mathcal{C}_{3,0}$ & Relative 3-space & Pauli spinors \\
\noalign{\vskip 1mm} 
    $\mathcal{C}_{0,2}$ & Quaternion rotations & Hamilton spinors \\
\noalign{\vskip 1mm} 
    $\mathcal{C}_{0,1}$ & Complex (wave-like) phases & Schr\"odinger spinors \\
\noalign{\vskip 1mm} 
    $\mathcal{C}_{0,0}$ & Real numbers & Scalars \\
\noalign{\vskip 2mm} 
    \hline 
 \end{tabular}
 \caption[Clifford algebra embeddings]{The Clifford algebras in the nesting sequence of closed subalgebras $\mathcal{C}_{2,4}\supset\mathcal{C}_{4,1}\supset\mathcal{C}_{1,3}\supset\mathcal{C}_{3,0}\supset\mathcal{C}_{0,2}\supset\mathcal{C}_{0,1}\supset\mathcal{C}_{0,0}$, and their connection to common physics concepts, as well as the distinct types of spinors in quantum mechanics.  Note how the $(+,-,-,-)$ signature of the spacetime algebra $\mathcal{C}_{1,3}$ appears naturally in the middle of this sequence.}
 \label{tab:clifford}
\end{table}

At this point we revisit the dot-product relation \eqref{eq:dot} for the 4-vectors $\gamma_\mu$ of the spacetime algebra $\mathcal{C}_{1,3}$ 
\begin{equation}
  \gamma_\mu \gamma_\nu + \gamma_\nu \gamma_\mu = 2\eta_{\mu\nu}.
\end{equation}
This relation was historically used to define the Dirac $\gamma$-matrices that appear in the Dirac equation for the relativistic electron.  We can thus view the Dirac matrices as a representation of the noncommutative spacetime algebra of 4-vectors $\gamma_\mu$, where the matrix product simulates the noncommutative product.  Indeed, our notation $\gamma_\mu$ is motivated by this connection \cite{Hestenes1966}.  As with the Pauli matrices, however, there is no need to construct such a matrix representation for the spacetime algebra to be well-defined.  Matrix constructions like $\slashed{p} = \sum_\mu p^\mu \gamma_\mu$ that appear in the Dirac theory can be understood to define proper 4-vectors $p$ in the spacetime algebra.  Again, there is nothing intrinsically quantum mechanical about these constructions.


\section{Spacetime calculus}\label{sec:calculus}

\epigraph{4.5in}{
Modern physical developments have required a mathematics that continually shifts its foundation and gets more abstract. 
Non-euclidean geometry and noncommutative algebra, which were at one time were considered to be purely fictions of the mind and pastimes of logical thinkers, have now been found to be very necessary for the description of general facts of the physical world. 
It seems likely that this process of increasing abstraction will continue in the future. 
}{Paul A. M. Dirac \cite{Dirac1931}}


With the spacetime algebra $\mathcal{C}_{1,3}$ now in hand, we consider how to construct \emph{fields} on which we can perform calculus.  This is the minimal extension of the algebra that is required to describe electromagnetism.  Formally speaking, we replicate the spacetime algebra $\mathcal{C}_{1,3}$ at every point $x$ on a flat spacetime manifold so that it acts as a \emph{tangent algebra} in which multivectors $M(x)$ can be locally defined.  Collecting all the spacetime points $x$ and their tangent algebras $\mathcal{C}_{1,3}(x)$ produces a \emph{Clifford fiber bundle}.  This bundle is the appropriate mathematical space to define \emph{multivector fields} $M(x)$, such as the electromagnetic field, and to perform calculus on those fields.  

The astonishing feature of the spacetime calculus that we will briefly construct over this bundle is that it permits the \emph{directed integration} of multivector quantities by treating the points $x$ themselves as vectors.  This generalization recovers the usual scalar integration permitted by differential geometry as a special case, but also recovers the theorems of complex analysis and their generalizations.  Moreover, it permits a coordinate-free definition of the vector derivative $\nabla$ and a simple statement of the fully general fundamental (Stokes) theorem of calculus.  We give a brief review here to give the essential background, but emphasize that more extensive and rigorous treatments may be found in \cite{Hestenes1987,Hestenes2005,Doran2007,Hestenes1968,Hestenes1968b,Sobczyk1989}.


\subsection{Directed integration}


Each vector point $x = \sum_\mu x^\mu \gamma_\mu$ on the oriented spacetime manifold corresponds to an oriented 4-volume $\textrm{d}^4\!x = \textrm{d}x_0\wedge\textrm{d}x_1\wedge\textrm{d}x_2\wedge\textrm{d}x_3$ that can be decomposed into infinitesimal vector segments $\textrm{d}x_\mu = |\textrm{d}x|\gamma_\mu$ (i.e., differentials) that point along coordinate directions\footnote{Formally speaking, the manifold can be constructed such that each oriented $k$-blade $\textrm{d}^k\!x$ has an oriented boundary $\partial \textrm{d}^k\!x = \sum_j (\textrm{d}^{k-1}\!x^+_j - \textrm{d}^{k-1}\!x^-_j)$ of $(k-1)$-blades centered at pairs of boundary points $x^\pm_j$.  For example, a vector segment $\textrm{d}x$ has two ordered boundary points separated by the scalar interval $|\textrm{d}x|$, a plane segment $\textrm{d}^2\!x$ has two ordered pairs of boundary line segments each separated by $|\textrm{d}x|$, and a small cube of volume $\textrm{d}^3\!x$ has three pairs of boundary sides separated by $|\textrm{d}x|$.  These nested boundary relationships are formally constructed as an oriented chain complex via a limiting refinement procedure $\Delta^k\! x \to \textrm{d}^k\! x$ that shrinks finite oriented volumes down to infinitesimal size \cite{Sobczyk1989}.  The orientation information for each $\Delta^k\! x$ is encoded as a $k$-blade in the spacetime algebra at the corresponding point $x$.}.

One can then define a \emph{directed integration} on the manifold as a Riemann summation of multivector products over a particular region $S_k$ of dimension $k$.  To do this, one encloses each point $x$ of $S_k$ in a finite oriented $k$-volume $\Delta^k\!x = |\Delta x|^k\,I_k$, where $I_k = \bigwedge_j \gamma_j$ is the unit $k$-volume element that encodes the orientation of the region $S_k$ at the point $x$.  One then computes a sum over all these points in the limit that the bounding intervals $|\Delta x|$ vanish and the number of points $x$ increases to infinity in the standard way \cite{Hestenes1968}
\begin{equation}
  \int_{S_k} M_1(x) \textrm{d}^k\!x M_2(x) = \lim_{|\Delta x|\to 0} \sum_{x\in S_k} M_1(x) \Delta^k\!x M_2(x).
\end{equation}
The noncommutativity of the multivectors $M_1(x)$, $M_2(x)$ with the directed volume segment $\textrm{d}^k\!x$ permits both right and left products in the integrand, which contrasts sharply with the usual definition of a scalar integral.

To understand the difference between scalar integration and directed integration, consider a $k$-volume $S_k$ that has no boundary.  The directed integral of this closed region must vanish $\int_{S_k} \textrm{d}^k\! x = 0$.  For example, if one sums small vectors around a curve, the net vector sum will be the vector connecting the start and end point of the curve; for a closed loop with no boundary points, this sum is simply zero. The volume of $S_k$, on the other hand, is given by the standard scalar integral of the undirected volume elements, which can be obtained from the directed integral by inverting the directed measure
\begin{equation}\label{eq:volume}
  \text{Vol}(S_k) = \int_{S_k} |\textrm{d}^k\!x| = \int_{S_k} \textrm{d}^k\!x \, I^{-1}_k
\end{equation}
using the unit $k$-volume element $I_k$ for the measure $\textrm{d}^k\!x = |\textrm{d}^k\!x|I_k$.  For a closed loop, this scalar integral will evaluate to the net length of the loop.  As a side note, the implicit wedge products in $\textrm{d}^k\!x$ that are used to compute the scalar measure $|\textrm{d}^k\!x|$ in \eqref{eq:volume} are what produces the Jacobian determinant factor in the scalar integral when the coordinates are changed.


\subsection{Vector derivative: the Dirac operator}


If the spacetime points $x$ are expanded into coordinates $x = \sum_\mu x^\mu \gamma_\mu$ using a standard basis $\gamma_\mu$, then a proper \emph{vector derivative} $\nabla$ can be defined using the reciprocal basis $\gamma^\mu$
\begin{equation}\label{eq:nabla}
  \nabla = \sum_\mu \gamma^\mu \stackrel{\leftrightarrow}{\partial}_\mu,
\end{equation}
where $\stackrel{\leftrightarrow}{\partial}_\mu \,=\, \stackrel{\leftarrow}{\partial}_\mu \!+\! \stackrel{\rightarrow}{\partial}_\mu$ is the usual partial derivative $\partial_\mu = \partial/\partial x^\mu$ along the coordinate $x^\mu = \gamma^\mu \cdot x$ that generally acts both to the left and to the right (one can think of this bidirectionality as a special case of the product rule for derivatives).  The form of \eqref{eq:nabla} is precisely the same as the \emph{Dirac operator} that is usually constructed with gamma matrices and used in the Dirac equation.  However, we see here that this operator has the natural interpretation as the proper vector derivative for a spacetime manifold \emph{with no matrix representation required}.    


\subsubsection{Coordinate-free definition}


For completeness, we emphasize that the directed integration permits a coordinate-free definition of the vector derivative in \eqref{eq:nabla} as a generalized difference quotient \cite{Hestenes1968}
\begin{equation}\label{eq:invariantnabla}
  M_1(x)\nabla M_2(x) = \lim_{\text{Vol}(S_4) \to 0} \frac{\int_{\partial S_4} M_1(x') [\textrm{d}^3\!x' I^{-1}(x)] M_2(x')}{\text{Vol}(S_4)}.
\end{equation}
Here the numerator is the \emph{directed} integral of the fields $M_1(x)$ and $M_2(x)$ over the \emph{boundary} $\partial S_4$ of the spacetime region $S_4$, while the denominator is the scalar integral of $S_4$ that produces the volume $\text{Vol}(S_4)$ in \eqref{eq:volume}.  The local duality factor of the unit 4-volume $I^{-1}(x)$ in the numerator produces the normal vectors dual to the (trivector) boundary surfaces of $\partial S_4$.  

The resulting derivative $\nabla$ is thus a proper 4-vector algebraically that acts as a total derivative on both fields.  When expanding it in local coordinates, one precisely obtains the form of the Dirac operator \eqref{eq:nabla}.  In practice, one often needs to consider only the one-sided action of $\nabla$ on a field, usually from the left, which amounts to setting $M_1(x) = 1$ in \eqref{eq:invariantnabla}.  Note that the algebraic and differential properties of $\nabla$ can be treated independently according to \eqref{eq:nabla}.  When it is not clear from context, it is conventional to use overdot notation to indicate which quantities are being differentiated in an expression, with the position of $\nabla$ indicating its algebraic role as a vector (e.g., $\dot{\nabla}(a\cdot\dot{b})$ only differentiates $b$, but constructs a proper 4-vector) \cite{Doran2007}.


\subsubsection{The relative gradient and d'Alembertian}


The derivative $\nabla$ can be restricted along any $k$-blade $M = \mean{M}_k$.  Specifically, the \emph{tangential derivative} along $M$ is defined as 
\begin{align}
  \nabla_{M} = M^{-1} (M \cdot \nabla),
\end{align}
which computes only the derivative along the surface $M$.  When $M$ is a unit vector $a$, then $a = a^{-1}$, so the tangential derivative recovers the usual directional derivative from vector calculus.  Similarly, the complementary derivative is $\nabla_{M I^{-1}} = M^{-1} (M \wedge \nabla)$, which computes the derivative along the orthogonal surface $M I^{-1}$ dual to $M$.  The total derivative decomposes into a sum of these complementary derivatives $\nabla = \nabla_{M} + \nabla_{M I^{-1}}$ since $M^{-1} (M \cdot \nabla) + M^{-1} (M \wedge \nabla) = (M^{-1} M) \nabla = \nabla$.

Most usefully for our purposes here, we can decompose the derivative in terms of a frame relative to $\gamma_0$ by using tangential derivatives, 
\begin{equation}\label{eq:threederivative}
  \nabla = \gamma_0^2 \nabla = \gamma_0(\partial_0 + \rv{\nabla}) = (\partial_0 - \rv{\nabla})\gamma_0,
\end{equation}
where $\nabla_{\gamma_0} = \gamma_0(\gamma_0\cdot\nabla) = \gamma_0\partial_0$ is the tangential derivative along $\gamma_0$ and $\nabla_{\gamma_0 I^{-1}} = \gamma_0(\gamma_0 \wedge \nabla) = \gamma_0 \rv{\nabla}$ is its dual.  The frame-dependent 3-vector derivative 
\begin{equation}\label{eq:threegradient}
  \rv{\nabla} = \gamma_0 \wedge \nabla = \sum_i \rv{\sigma}_i \partial_i
\end{equation}
that appears in the decomposition \eqref{eq:threederivative} is precisely the relative 3-\emph{gradient} operator used in standard vector analysis.  

This decomposition of $\nabla$ into a relative frame also makes it clear that $\nabla^2$ is the d'Alembertian operator
\begin{equation}
  \nabla^2 = (\partial_0 - \rv{\nabla})\gamma_0^2(\partial_0 + \rv{\nabla}) = \partial_0^2 - \rv{\nabla}^2 = \Box.
\end{equation}
In this sense, $\nabla$ is the natural ``square-root'' of the d'Alembertian operator in spacetime, which was originally suggested by Dirac \cite{Dirac1928}.


\subsubsection{Example: field derivatives}


The derivative of a scalar field $\alpha(x)$ produces the total gradient vector field 
\begin{equation}\label{eq:scalarfield}
  \nabla\alpha(x) = \gamma_0[\partial_0\alpha(x) + \rv{\nabla}\alpha(x)],
\end{equation}
that includes the proper derivative along the timelike coordinate.  Taking the dual of this equation via right multiplication by $I$ immediately produces the derivative of a pseudoscalar field $\alpha(x)I$.  It follows that the derivative of a complex scalar field $\zeta(x) = \alpha(x) + \beta(x)I$ is a complex vector field $z(x) = \nabla\zeta(x) = \nabla\alpha(x) + [\nabla\beta(x)]I$.

The derivative of a vector field $v(x) = [v_0(x) + \rv{v}(x)]\gamma_0$ produces a scalar field and a bivector field (i.e., a spinor field)
\begin{align}\label{eq:vectorfield}
  \nabla v(x) &= \nabla \cdot v(x) + \nabla \wedge v(x), \\ 
  &= (\partial_0 - \rv{\nabla})\gamma_0^2(v_0 - \rv{v}), \nonumber \\
  &= [\partial_0 v_0 + \rv{\nabla}\cdot\rv{v}] + [-(\partial_0\rv{v} + \rv{\nabla}v_0) + \rv{\nabla}\times\rv{v}I]. \nonumber
\end{align}
The scalar part $\nabla \cdot v$ is the total \emph{divergence} of $v(x)$, while the bivector part $\nabla\wedge v$ is the proper generalization of the \emph{curl} of $v(x)$ to spacetime.  Taking the dual of this equation produces the derivative of a trivector field $v(x)I$, which produces a pseudoscalar field and a bivector field (i.e., a spinor field).  It follows that taking the derivative of a complex vector field $z = a + g I$ generally produces a spinor field $\nabla z(x) = \psi(x) = \zeta(x) + \bv{F}(x)$ with a complex scalar part $\zeta = \nabla \cdot a + (\nabla \cdot g) I$ and a bivector part $\bv{F} = \nabla \wedge a + (\nabla \wedge g)I$.

The derivative of a bivector field $\bv{F}(x) = \rv{E}(x) + \rv{B}(x)I$ produces a vector field and a trivector field
\begin{align}\label{eq:bivectorfield}
  \nabla \bv{F}(x) &= \nabla \cdot \bv{F}(x) + \nabla \wedge \bv{F}(x), \\
  &= \gamma_0(\partial_0 + \rv{\nabla})(\rv{E} + \rv{B} I), \nonumber \\
  &= \gamma_0\left[\rv{\nabla}\cdot\rv{E} + \partial_0 \rv{E} - \rv{\nabla}\times\rv{B}\right] + \gamma_0\left[\rv{\nabla}\cdot\rv{B} + \partial_0 \rv{B} + \rv{\nabla}\times\rv{E}\right]I, \nonumber
\end{align}
which can be understood as a single complex vector field.  Since derivatives of both complex scalars $\zeta$ and bivectors $\bv{F}$ produce complex vector fields $z$, it follows that spinors $\psi = \zeta + \bv{F}$ and complex vectors $z$ are duals under the action of the derivative $\nabla$.

Note that we can also write \eqref{eq:bivectorfield} in terms of the intrinsic phase of $\bv{F}(x)$ defined in Eq.~\eqref{eq:phase}
\begin{align}
  \nabla \bv{F}(x) &=  \nabla[\bv{f}(x) \exp([\varphi(x) + \varphi_0] I)], \\
  &= [\nabla \bv{f}(x)] \exp([\varphi(x) + \varphi_0] I) + [\nabla \varphi(x)] \bv{F}(x) I \nonumber
\end{align}
to show that the intrinsic bivector derivative $\nabla \bv{f}$ and the gradient of the \emph{local} phase $\nabla \varphi$ contribute to the complex vector field.  However any \emph{global} intrinsic phase $\exp(\varphi_0 I)$ of $\bv{F}$ is a constant overall factor.  This distinction between local and global phases holds for complex scalar and vector fields as well.


\subsection{Fundamental theorem}


The fundamental theorem of calculus readily follows from the integral definition of the derivative \eqref{eq:invariantnabla} and takes a particularly elegant form in terms of the tangential derivative along some surface $S_k$ of dimension $k$ \cite{Hestenes1968}
\begin{equation}\label{eq:fundamentaltheorem}
  \int_{S_k}\!\! M_1(x) \textrm{d}^k\!x \nabla_{S_k} M_2(x) = \int_{\partial S_k}\!\! M_1(x) \textrm{d}^{k-1}\!x M_2(x).
\end{equation}
This is the generalized \emph{Stokes theorem} of calculus that states that the boundary integral is equal to the volume integral of the derivative.  Computing this for various dimensional surfaces and $k$-blade fields reproduces all the standard variations of Stokes theorem, including the divergence theorem, the various Green's theorems, and the usual fundamental theorem of calculus in one dimension \cite{Hestenes1968b}.  
 

\subsubsection{Example: Cauchy integral theorems}


More surprisingly, if we consider the integration of appropriate Green's functions that invert the tangential derivative $\nabla_{S_k}$, then we obtain and generalize the celebrated Cauchy integral theorems of complex analysis \cite{Hestenes1968b,Doran2007}.  
First, for any field $M(x)$ that satisfies $\nabla_{S_2} M(x) = 0$ everywhere in a closed planar region $S_2$, the following boundary integral vanishes according to \eqref{eq:fundamentaltheorem}
\begin{equation}
  \oint_{\partial S_2}\!\! \textrm{d}x\, M(x) = 0.
\end{equation}
Second, the Green's function that inverts the planar derivative is $\nabla_{S_2}[2\pi(x-x')]^{-1} = \delta(x-x')$.  Hence, if $N(x) = \alpha_{-1}x^{-1} + M(x)$ is a modification of $M(x)$ that adds a pole, then we obtain the residue theorem
\begin{equation}
  \oint_{\partial S_2}\!\! \textrm{d}x\, N(x) = \alpha_{-1} \int_{S_2}\!\! |\textrm{d}^2x|\bv{I}_2 \nabla_{S_2}x^{-1} = 2\pi \bv{I}_2 \alpha_{-1},
\end{equation}
where $\bv{I}_2$ is the unit bivector for $S_2$ that squares to $-1$.  Third, multiplying $M(x)$ with the Green's function and then integrating produces the remaining Cauchy integral theorem
\begin{equation}
  \oint_{\partial S_2} \!\! \textrm{d}x\, (x-x')^{-1}M(x) = \int_{S_2}\!\! \bv{I}_2|\textrm{d}^2\!x|\nabla(x-x')^{-1}M(x) = 2\pi \bv{I}_2 M(x').
\end{equation}

To rewrite these formulas in the standard form from complex analysis, we can expand the 2-vector $x = \alpha r + \beta y = \zeta r$, where $r$ is a unit 2-vector that points along the ``real axis,'' $y$ is a unit 2-vector that points along the ``imaginary axis,'' and $\zeta = \alpha + \beta \bv{I}_2$ is the ``complex scalar'' corresponding to $x$, with $\bv{I}_2 = yr$ being the unit planar area.  Importantly, however, by considering different dimensional surfaces and using the fundamental theorem \eqref{eq:fundamentaltheorem} with the correct Green's functions for derivatives along different $k$-volumes $\nabla_{S_k}$, it is possible to generalize the Cauchy theorems above to the whole spacetime algebra \cite{Hestenes1968b,Doran2007}.


\section{Maxwell's equation in vacuum}\label{sec:maxwell}

\epigraph{4.5in}{One scientific epoch ended and another began with James Clerk Maxwell.}{Albert Einstein \cite{Arianrhod2006}} 


\epigraph{4.5in}{The ultimate importance of the Maxwell theory is far greater than its immediate achievement in explaining and unifying the phenomena of electricity and magnetism. Its ultimate importance is to be the prototype for all the great triumphs of twentieth-century physics.}{Freeman J. Dyson \cite{Dyson1999}}


With the spacetime algebra developed in the previous two sections now in hand, we can explore the physical structure of electromagnetism as expressed with this proper geometric language.  Wherever possible, we shall remind the reader of the previous mathematical results as they become relevant to the physics in what follows.  As we shall see, all of electromagnetism essentially emerges without any additional effort directly from the geometric structure of spacetime algebra that we have already developed.

To proceed, we briefly examine the derivative of a bivector field that we calculated in Eq.~\eqref{eq:bivectorfield} to make the following observation:  the simplest differential equation that one can write for $\bv{F}$ 
\begin{equation}\label{eq:maxwell}
  \nabla \bv{F} = 0
\end{equation}
is equivalent to all four of Maxwell's equations in vacuum, up to appropriate choices of constants that define units for the bivector components.  This almost automatic appearance of Maxwell's vacuum equation is remarkable, since the deceptively simple differential equation \eqref{eq:maxwell} is trivially postulated solely from our careful development of the algebraic and differential \emph{structure} of a spacetime manifold, with no reference to additional physical assumptions.  We will return to this intriguing point later in Section~\ref{sec:lagrangian}.


\subsection{Relative frame form}


We can prove the equivalence of \eqref{eq:maxwell} to the usual form of Maxwell's equations in a relative frame with timelike vector $\gamma_0$ by writing $\bv{F}$ in terms of relative polar ($\rv{E}$) and axial ($\rv{B}$) parts obtained via the contractions in Eq.~\eqref{eq:bivectorsplit}
\begin{align}\label{eq:maxwellsplit}
  \bv{F} = \bv{F}\gamma_0\gamma_0^{-1} = (\bv{F}\cdot\gamma_0)\gamma_0 + (\bv{F}\wedge\gamma_0)\gamma_0 = \rv{E} + \rv{B}I
\end{align}
Applying the identity \eqref{eq:bivectorfield} to \eqref{eq:maxwell} then yields 
\begin{align}
  \nabla \bv{F}(x) &= \gamma_0\left[\rv{\nabla}\cdot\rv{E} + \partial_0 \rv{E} - \rv{\nabla}\times\rv{B}\right] + \gamma_0\left[\rv{\nabla}\cdot\rv{B} + \partial_0 \rv{B} + \rv{\nabla}\times\rv{E}\right]I, 
\end{align}
The independent vector and trivector parts of \eqref{eq:maxwell} can be separated, after which the duality factors $\gamma_0$ and $I$ cancel, yielding the two paravector equations for the $\gamma_0$ frame\footnote{These two paravector equations are in fact equivalent to Maxwell's preferred quaternion formulation of electromagnetism \cite{Maxwell1881}.  However, as mentioned in Section~\ref{sec:lorentz}, in place of the basis 3-vector directions $\rv{\sigma}_i$ he used the (left-handed set of) basis quaternions $(\bv{i},\bv{j},\bv{k}) = (\rv{\sigma}_1 I^{-1}, -\rv{\sigma}_2 I^{-1},\rv{\sigma}_3 I^{-1})$, which geometrically correspond to rotation planes and thus caused numerous subtle problems.}
\begin{align}
  \rv{\nabla}\cdot\rv{E} + (\partial_0 \rv{E} - \rv{\nabla}\times\rv{B})&= 0, \\
  \rv{\nabla}\cdot\rv{B} + (\partial_0 \rv{B} + \rv{\nabla}\times\rv{E})&= 0. \nonumber
\end{align}
Finally, the relative scalar and 3-vector parts of each paravector can be separated to produce the standard Gibbs 3-vector form for the four vacuum Maxwell equations \cite{Jackson1999,Landau1975}
\begin{align}\label{eq:maxwellfour}
  \rv{\nabla}\cdot\rv{E} &= 0 , ~~~~~~~\rv{\nabla}\times\rv{B} = \partial_0 \rv{E}, \\
  \rv{\nabla}\cdot\rv{B} &= 0 , ~~~~~~~ \rv{\nabla}\times\rv{E} = - \partial_0 \rv{B}. \nonumber
\end{align}

Rescaling these vectors with constant factors that set the (SI) units $\rv{E} \mapsto \rv{E}\sqrt{\epsilon_0}$, $\rv{B} \mapsto \rv{B}/\sqrt{\mu_0}$, and expanding $\partial_0 \mapsto \cc^{-1}\partial_t$ makes the equivalence apparent.  Furthermore, with these replacements and the identity $\cc = 1/\sqrt{\epsilon_0\mu_0}$, it is clear that the proper bivector field 
\begin{equation}\label{eq:emfield}
  \bv{F} = \sqrt{\epsilon_0} \rv{E} + \rv{B} I/\sqrt{\mu_0} = \sqrt{\epsilon_0}(\rv{E} + \cc \rv{B}I) = (\rv{E}/\cc + \rv{B}I)/\sqrt{\mu_0}
\end{equation}
is precisely equivalent to the Riemann-Silberstein complex electromagnetic field vector \eqref{eq:RiemannSilberstein}.  Importantly, however, the pseudoscalar factor of $I$ that appears here has rich physical meaning that is not apparent when using the scalar imaginary $i$ appearing in other treatments (e.g., \cite{Bialynicki-Birula2013}).  For brevity, we will omit the constants in the discussion to follow, since they may be easily restored when needed.  The various formulations of the electromagnetic field and Maxwell's vacuum equation are compared in Tables~\ref{tab:emfield} and \ref{tab:vacuummaxwell} for reference, along with a reminder of the proper unit scalings for both SI and CGS conventions.

\begin{table}
  \centering
  \begin{tabular}{l l}
    \hline
\noalign{\vskip 2mm} 
    \multicolumn{2}{c}{\textbf{Electromagnetic Field}} \\
\noalign{\vskip 2mm} 
    \hline \\
    Spacetime Algebra: & {$\begin{aligned}\Aboxed{\bv{F} &= \rv{E} + \rv{B}I} \\
      \Aboxed{\bv{G} &= \bv{F}I^{-1} = \rv{B} - \rv{E}I}\end{aligned}$} \\
    \\
    Differential Forms: & {$\begin{aligned}\bv{F} &= \frac{1}{2}\sum_{\mu\nu}F_{\mu\nu}dx^\mu\wedge dx^\nu \\
      \bv{G} &= \star \bv{F}\end{aligned}$} \\
    \\
    Tensor Components: & {$\begin{aligned}G_{\mu\nu} &= -\frac{1}{2}\sum_{\alpha\beta}\epsilon_{\mu\nu\alpha\beta}F^{\alpha\beta} \\
      E_i &= F_{i0}, \qquad B_i = G_{i0}\end{aligned}$} \\
      \\
      Gibbs 3-Vectors: & {$\begin{aligned}&(\rv{E}, \rv{B}) \\
        &(\rv{B}, -\rv{E})\end{aligned}$} \\ 
    \\
    \hline
    \\
    SI Units: & {$\begin{aligned}\rv{E} &\mapsto \sqrt{\epsilon_0}\,\rv{E}, &\quad \rv{B} &\mapsto \rv{B}/\sqrt{\mu_0}, &\quad \cc &\equiv 1/\sqrt{\epsilon_0\mu_0}\end{aligned}$} \\
    \\
    CGS Units: & {$\begin{aligned}\rv{E} &\mapsto \rv{E}/\sqrt{4\pi}, &\quad \rv{B} &\mapsto \rv{B}/\sqrt{4\pi}\end{aligned}$} \\
    \\
    \hline 
 \end{tabular}
 \caption[Electromagnetic field descriptions]{The electromagnetic field $\bv{F}$ and its dual $\bv{G}$, as expressed in various formalisms.  In spacetime algebra $\bv{F}$ is a bivector, and its dual $\bv{G}$ is obtained by right-multiplication with the pseudoscalar $I^{-1}$.  In differential forms $\bv{F}$ is a 2-form (antisymmetric rank-2 tensor) with components $F_{\mu\nu}$, and its dual $\bv{G}$ is obtained with the Hodge-star transformation $\star$.  In component notation, this Hodge-star is the total contraction of $F_{\mu\nu}$ with the fully anti-symmetric Levi-Civita symbol $\epsilon_{\mu\nu\alpha\beta}$.  The standard Gibbs 3-vector notation is reference-frame-dependent, but the 3-vectors of $\rv{E}$ and $\rv{B}$ can still be appropriately paired.}
 \label{tab:emfield}
\end{table}

\begin{table}
  \centering
  \begin{tabular}{l l}
    \hline
\noalign{\vskip 2mm} 
    \multicolumn{2}{c}{\textbf{Maxwell's Equation in Vacuum}} \\
\noalign{\vskip 2mm} 
    \hline \\
    Spacetime Algebra: & {$\boxed{ \nabla \bv{F} = 0 } \qquad \boxed{ \bv{F} = \rv{E} + \rv{B}I }$} \\
    \\
    & $\qquad\qquad\quad \Downarrow$ \\
    \\
    & {$\begin{aligned}\nabla\cdot\bv{F} &= 0, &\quad \nabla \wedge \bv{F} &= 0\end{aligned}$} \\
    \\
    & $\qquad\qquad\quad \Downarrow$ \\
    \\
    & {$\begin{aligned}\rv{\nabla}\cdot\rv{E} + (\partial_0 \rv{E} - \rv{\nabla}\times\rv{B}) &= 0 \\
      \rv{\nabla}\cdot\rv{B} + (\partial_0 \rv{B} + \rv{\nabla}\times\rv{E}) &= 0\end{aligned}$} \\
    \\
    & $\qquad\qquad\quad \Downarrow$ \\
    \\
    & {$\begin{aligned}\rv{\nabla}\cdot\rv{E} &= 0 , &\quad \rv{\nabla}\times\rv{B} &= \partial_0 \rv{E} \\
      \rv{\nabla}\cdot\rv{B} &= 0, &\quad \rv{\nabla}\times\rv{E} &= - \partial_0 \rv{B}\end{aligned}$} \\
    \\
    Differential Forms: & {$\begin{aligned}\textrm{d}\bv{G} &= \textrm{d}{\star }\bv{F} = 0, &\quad \textrm{d}\bv{F} &= \textrm{d}{\star}\bv{G} = 0\end{aligned}$} \\
    \\
    Tensor Components: & {$\begin{aligned}\partial_\nu F^{\mu\nu} &= 0, &\quad \partial_\nu G^{\mu\nu} &= 0\end{aligned}$} \\
      \\
    Gibbs 3-Vectors: & {$\begin{aligned}\rv{\nabla}\cdot\rv{E} &= 0 , &\quad \rv{\nabla}\times\rv{B} &= \partial_0 \rv{E} \\
      \rv{\nabla}\cdot\rv{B} &= 0 , &\quad \rv{\nabla}\times\rv{E} &= - \partial_0 \rv{B}\end{aligned}$} \\
    \\
    \hline
    \\
    SI Units: & {$\begin{aligned}\rv{E} &\mapsto \sqrt{\epsilon_0}\,\rv{E}, &\quad \rv{B} &\mapsto \rv{B}/\sqrt{\mu_0}, &\quad \cc \equiv 1/\sqrt{\epsilon_0\mu_0}\end{aligned}$} \\
    \\
    CGS Units: & {$\begin{aligned}\rv{E} &\mapsto \rv{E}/\sqrt{4\pi}, &\quad \rv{B} &\mapsto \rv{B}/\sqrt{4\pi}\end{aligned}$} \\
    \\
    \hline 
 \end{tabular}
 \caption[Maxwell's vacuum equation]{Maxwell's vacuum equation, as expressed in various formalisms.  Notably, only spacetime algebra permits the formulation as the single boxed equation for the electromagnetic bivector field $\bv{F}$, which can be algebraically expanded into various other forms as needed.  Shown are successive expansions that are equivalent to the formalisms of differential forms, paravectors / quaternions, and Gibbs 3-vectors, respectively, to show how they are all naturally contained within spacetime algebra.}
 \label{tab:vacuummaxwell}
\end{table}

Equation \eqref{eq:maxwell} is manifestly proper, but the four equations \eqref{eq:maxwellfour} are not so obviously invariant under changes of the inertial reference frame.  We can see the effect of choosing a different frame by splitting $\bv{F} = \rv{E}' + \rv{B}'I$ along a different timelike vector $\gamma_0'$ according to \eqref{eq:bivectorsplit}.  This produces different relative 3-vector fields 
\begin{align}\label{eq:passivelorentzboost}
  \rv{E}' &= (\bv{F}\cdot\gamma_0')\gamma_0' = (\rv{E}\cdot\gamma_0')\gamma_0' + (\rv{B}I\cdot\gamma_0')\gamma_0', \\
  \rv{B}'I &= (\bv{F}\wedge\gamma_0')\gamma_0' = (\rv{E}\wedge\gamma_0')\gamma_0' + (\rv{B}I\wedge\gamma_0')\gamma_0', \nonumber
\end{align}
which must also satisfy \eqref{eq:maxwellfour} (proved by the same derivation).  These different relative fields accordingly mix the components of $\rv{E}$ and $\rv{B}$, even though they produce the \emph{same} proper bivector field $\bv{F} = \rv{E} + \rv{B}I = \rv{E}' + \rv{B}'I$.  The subtle and important role of the pseudoscalar $I$ in managing the implicit relative bases $\rv{\sigma}_i = \gamma_i\gamma_0$ and $\rv{\sigma}_i' = \gamma_i'\gamma_0'$ becomes apparent when comparing these different frame expansions: using a scalar imaginary $i$ would not preserve the frame invariance in the same way.

The mixing of components from such a change of frame will have precisely the same form as a Lorentz boost, such as the one derived in \eqref{eq:relativelorentzboost}.  To prove this, we note that we can always write the new frame vector $\gamma_0'$ as a Lorentz boost of the old frame $\gamma_0$ with some relative velocity $\rv{v}$.  This boost has the spinor representation $\gamma_0' = \psi \gamma_0 \widetilde{\psi}$ with the rotation spinor $\psi = \exp(-\alpha \hat{v}/2)$, where the velocity unit 3-vector $\hat{v} = \rv{v}/|v|$ also acts as the plane of rotation for the boost, and $\alpha = \tanh^{-1}|v|/\cc$ is the appropriate rapidity (rotation angle).  This Lorentz transformation produces the relation
\begin{align}
  \gamma_0' = \psi\gamma_0\widetilde{\psi} = \exp(-\alpha \hat{v})\gamma_0 = \frac{1 - \rv{v}/\cc}{\sqrt{1-(|v|/\cc)^2}}\,\gamma_0.
\end{align}
After defining $\gamma = [1 - (|v|/\cc)^2]^{-1/2}$ to be the usual time-dilation factor for brevity, we can then directly expand $\bv{F}$ into the relative fields of the $\gamma_0'$ frame using \eqref{eq:maxwellsplit} 
\begin{align}\label{eq:relativelorentzboost2}
  \bv{F} &= \bv{F}\gamma_0'\gamma_0'^{-1} = \gamma^2 \bv{F}\gamma_0\gamma_0(1-\rv{v}/\cc)(1+\rv{v}/\cc) \\
  &= \gamma^2(\rv{E} + \rv{B}I)(1-\rv{v}/\cc)(1+\rv{v}/\cc), \nonumber \\
  &= \gamma^2[\rv{E} + \rv{B}I - \rv{E}\cdot(\rv{v}/\cc) - \rv{B}\cdot(\rv{v}/\cc) I + \rv{B}\times(\rv{v}/\cc) - \rv{E}\times(\rv{v}/\cc) I](1+\rv{v}/\cc), \nonumber \\
  &= [\rv{E}_\parallel + \gamma(\rv{E}_\perp + \rv{v}\times\rv{B}_\perp/\cc)] + [\rv{B}_\parallel + \gamma(\rv{B}_\perp - \rv{v}\times\rv{E}_\perp/\cc)] I, \nonumber
\end{align}
which is the same form as the direct Lorentz boost of $\bv{F}$ derived in \eqref{eq:relativelorentzboost}.


\subsection{Global phase degeneracy and dual symmetry}\label{sec:dualsym}


Notably, the Maxwell's vacuum equation Eq.~\eqref{eq:maxwell} is invariant under \emph{global} phase transformations, such as those considered in \eqref{eq:globalphase}.  To understand the implications of this symmetry of Maxwell's equation, we will briefly revisit how this transformation acts on the electric field itself.

Specifically, we apply a phase rotation using the spinor representation $\psi = \exp(-\theta I / 2)$ to Eq.~\eqref{eq:maxwell} to find
\begin{align}\label{eq:dualsymmetry}
  \psi(\nabla \bv{F})\psi^* = e^{-\theta I /2}(\nabla \bv{F})e^{\theta I/2} = \nabla(e^{\theta I/2}\bv{F}e^{\theta I/2}) = \nabla \bv{F}\,\exp(\theta I) = 0,
\end{align}
which shows that the phase rotation cancels.  Note that this phase transformation is equivalent to modifying the \emph{intrinsic} phase of $\bv{F} \mapsto \bv{F} \exp(\theta I)$.  This added phase mixes $\rv{E}$ and $\rv{B}$ at every point $x$ in the same way\footnote{Note that we could consider converting this global gauge transformation of $\bv{F}$ into a local gauge transformation in the usual way by defining a covariant derivative using an auxiliary scalar field (see Sections~\ref{sec:gauge} and \ref{sec:symmetry}), but we shall not do so here.  However, see \cite{Vasconcellos2014} for an interesting connection between performing this procedure here and the electroweak theory of particle physics.} while leaving the inertial frame the same
\begin{equation}
  \bv{F} \exp(\theta I) = (\cos\theta \rv{E} - \sin\theta \rv{B}) + (\sin\theta \rv{E} + \cos\theta \rv{B})I.
\end{equation}

The interesting special case of the angle $\theta = -\pi/2$ exchanges the relative fields $\rv{E}$ and $\rv{B}$ up to a sign at every point $x$.  The resulting bivector is dual to $\bv{F}$ and is typically given a special notation 
\begin{align}\label{eq:dualfield}
  \bv{G} = \bv{F}I^{-1} = \rv{B} - \rv{E}I,
\end{align}
which can also be written in tensor component notation as 
\begin{align}
  G^{\mu\nu} = \star F^{\mu\nu} = -\frac{1}{2}\sum_{\alpha,\beta}\epsilon^{\mu\nu\alpha\beta}F_{\alpha\beta}
\end{align}
using the Hodge-star operation from differential forms (or the completely antisymmetric Levi-Civita symbol $\epsilon^{\mu\nu\alpha\beta}$) as shown in Table~\ref{tab:emfield} \cite{Bliokh2013}.  The dual bivector $\bv{G}$ has been particularly useful for studies in magnetic monopoles \cite{Dirac1931,Dirac1948,Cabibbo1962,Schwinger1966b,Rohrlich1966,Zwanziger1968,Zwanziger1971,Han1971,Mignani1975,Deser1976,
Deser1982,Gambini1979,Gambini1980,Schwartz1994,Pasti1995,Singleton1995,Singleton1996,Khoudeir1996,Kato2002,Shnir2005} and optical helicity \cite{Afanasiev1996,Trueba1996,Cameron2012,Cameron2012b,Bliokh2013,Philbin2013,Fernandez2012,Fernandez2013}, and will play an interesting role in the Lagrangian treatment described in Section \ref{sec:lagrangian}.  

This exchange freedom of the vacuum Maxwell equation \cite{Heaviside1892,Larmor1897} has historically been called the \emph{dual symmetry} of the vacuum Maxwell fields \cite{Calkin1965,Zwanziger1968,Deser1976,Cameron2012b,Bliokh2013,Cameron2012,Fernandez2012,Fernandez2013}, and has prompted considerable study into similar dualities in field and string theories beyond standard electromagnetism \cite{Born1934,Schrodinger1935,Ferrara1977,Gaillard1981,Sen1993,Schwarz1994,Pasti1995b,Gibbons1995,Hull1995,Deser1998,Cremmer1998,Figueroa-O'Farrill,Aschieri2008,Aschieri2014}.  We see here that it appears as a structural feature of a bivector field on spacetime that exploits a global phase symmetry of the equation of motion.  Note that the bivector magnitude $|\bv{F}|^2 \equiv \bv{F}^* \bv{F} = \bv{G}^*\bv{G}$ is always manifestly invariant under dual symmetry.


\subsection{Canonical form}\label{sec:canonical}


We now revisit the expansion of the electromagnetic field into its canonical form, introduced in Section~\ref{sec:complexbivectors}.  These expansions do not require Maxwell's equation, but will be useful in what follows.  

Computing the square of $\bv{F} = \rv{E} + \rv{B}I$ produces two proper scalar fields as a single complex scalar
\begin{equation}\label{eq:lorentzscalars}
  \bv{F}^2 = |\bv{F}|^2 \exp(2\varphi I) = (|\rv{E}|^2 - |\rv{B}|^2) + 2(\rv{E}\cdot\rv{B})I.
\end{equation}
These scalars are precisely those used to construct Lagrangian densities for the electromagnetic field \cite{Battesti2013}.  The first scalar is precisely the usual electromagnetic Lagrangian, while the second term is the ``axion'' contribution that has been extensively discussed in particle physics \cite{Weinberg1978}.  We will return to the issue of electromagnetic Lagrangians in Section~\ref{sec:lagrangian}.

Denoting the two proper scalars as 
\begin{align}
  \ell_1 &= \mean{\bv{F}^2}_0 = |\rv{E}|^2 - |\rv{B}|^2, \\
  \ell_2 &= \mean{\bv{F}^2}_4\, I^{-1} = 2(\rv{E}\cdot\rv{B}),
\end{align}
they determine the intrinsic local phase of $\bv{F}$ according to \eqref{eq:phasecomp}
\begin{equation}
  \varphi = \frac{1}{2}\tan^{-1}\frac{\ell_2}{\ell_1},
\end{equation}
which demonstrates that this phase is also a proper scalar field.

The canonical bivector corresponding to $\bv{F}$ can then be computed according to \eqref{eq:phasecomp} 
\begin{align}
  \bv{f} &= \bv{F} \exp(-\varphi I) = (\cos\varphi \rv{E} + \sin\varphi \rv{B}) + (-\sin\varphi \rv{E} + \cos\varphi \rv{B})I,
\end{align}
where we can expand the trigonometric functions in terms of $\ell_1$ and $\ell_2$ as
\begin{align}
  \cos\varphi &= \sqrt{\frac{1}{2}\left(1 + \frac{\ell_1}{\sqrt{\ell_1^2 + \ell_2^2}}\right)}, &
  \sin\varphi &= \sqrt{\frac{1}{2}\left(1 - \frac{\ell_1}{\sqrt{\ell_1^2 + \ell_2^2}}\right)}.
\end{align}
This transformation performs a rotation of the relative fields in the complex plane of the spacetime split at each local point $x$, while leaving the chosen frame of $\gamma_0$ invariant.  

The proper complex conjugate of the field can be computed in a similar way according to \eqref{eq:conjugation}, 
\begin{align}
  \bv{F}^* &= \bv{f}\exp(-\varphi I) = \bv{F} \exp(-2\varphi I) = \frac{(\ell_1 \rv{E} + \ell_2 \rv{B}) + (-\ell_2 \rv{E} + \ell_1 \rv{B})I }{\sqrt{\ell_1^2 + \ell_2^2}}.
\end{align}
It follows that the dual-symmetric field magnitude has the explicit and intuitive form
\begin{align}\label{eq:magnitudelorentzscalars}
  |\bv{F}|^2 &= \bv{F}^* \bv{F} = \sqrt{\ell_1^2 + \ell_2^2}.
\end{align}

Note that when $\ell_2 \propto \rv{E}\cdot\rv{B} = 0$ the proper phase vanishes $\varphi = 0$ and the bivector field is purely canonical $\bv{F} = \bv{F}^* = \bv{f} = \rv{E} + \rv{B}I$.  Indeed the pseudoscalar part in the pseudonorm \eqref{eq:lorentzscalars} vanishes in this case.  The magnitude $|\bv{F}|^2 = |\ell_1|$ becomes the pseudonorm involving only the traditional Lagrangian density $\ell_1 = |\rv{E}|^2 - |\rv{B}|^2$, albeit in a manifestly dual-symmetric form (due to the absolute value).

Similarly, note that when $\ell_1 = |\rv{E}|^2 - |\rv{B}|^2 = 0$ the proper phase is $\varphi = \pi/4$ and the canonical bivector field has the symmetrized form 
\begin{align}
  \bv{f} = \frac{1}{\sqrt{2}}[(\rv{E} + \rv{B}) + (-\rv{E} + \rv{B})I].
\end{align}
The conjugate becomes the dual $\bv{F}^* = \bv{F} I^{-1} = \rv{B} - \rv{E}I$ up to the sign of $\ell_2$, and the magnitude becomes the other proper scalar $|\bv{F}|^2 = |\ell_2| = 2|\rv{E}\cdot\rv{B}|$ in a manifestly dual-symmetric form.

When $\ell_1 = \ell_2 = 0$ then $|\bv{F}|^2 = 0$, making $\bv{F}$ a null bivector.  The phase $\varphi(x)$ correspondingly becomes degenerate.  Similarly, the phase of $\bv{F}^*$ will also be degenerate, but will always be the negative of the degenerate phase of $\bv{F}$.  For such a null field, the conditions $|\rv{E}| = |\rv{B}|$ and $\rv{E}\cdot\rv{B} = 0$ hold according to \eqref{eq:lorentzscalars}.  


\subsection{Electromagnetic waves}\label{sec:waves}


Although a null vacuum field $\bv{F}$ has a globally degenerate phase, it must still satisfy Maxwell's equation \eqref{eq:maxwell}.  The constraining derivative therefore breaks the \emph{local} phase degeneracy by connecting the local values of $\varphi(x)$ at nearby points $x$.  Only the global phase remains arbitrary in the form of the dual symmetry of the solution.




To see how Maxwell's equation \eqref{eq:maxwell} breaks the phase degeneracy of a null field for a specific and important solution, consider a field $\bv{F} = \bv{f} \exp[\varphi(x) I]$ with a \emph{constant} canonical bivector $\bv{f}$.  Maxwell's equation \eqref{eq:maxwell} then takes the simpler form
\begin{align}\label{eq:maxwellnull}
  \nabla \bv{F} = [\nabla\varphi]\,\bv{f} = 0.
\end{align}
If $\nabla\varphi \neq 0$ and $\bv{f}\neq 0$ then both factors must be null factors that cannot be inverted, making $\bv{F}$ a null bivector.  Since $\bv{f}$ is constant for all $x$, then $[\nabla\varphi]$ must not vary with $x$ in order to satisfy \eqref{eq:maxwellnull}; hence the phase must be linear $\varphi = \varphi_0 \pm x\cdot k$ in terms of a null wavevector $k = \nabla\varphi$ with units inverse to $x$.  Furthermore, the constraint $k\, \bv{f} = 0$ implies that the constant bivector $\bv{f}$ can be written in terms of this null wavevector as 
\begin{align}
  \bv{f} = sk = s\wedge k = -k\wedge s
\end{align}
for some constant spacelike vector $s$ orthogonal to $k$: $s\cdot k = 0$.  Since the global phase is arbitrary, we set $\varphi_0 = 0$, and thus obtain the (monochromatic) \emph{plane wave solutions}
\begin{align}\label{eq:planewave}
  \bv{F}(x) &= (sk)\, \exp[\pm(k\cdot x) I].
\end{align}

Importantly, this complex form of the plane wave solution arises with no \emph{ad hoc} introduction of the complex numbers \cite{Hestenes1971}, unlike the usual treatment of electromagnetic plane waves.  The factor of $I$ is the intrinsic pseudoscalar for spacetime, and both the ``real'' and ``imaginary'' parts of the expression produce meaningful and necessary parts of the same proper bivector.  Note that the six components of $\bv{F}$ are now entirely contained in the specification of the null vector $k$ and the spacelike orientation vector $s$ orthogonal to $k$.

If we perform a spacetime split of $k = \gamma_0(\omega - \rv{k})$ to put \eqref{eq:planewave} in more standard notation (setting $\cc=1$ for convenience), we find that the null condition $k^2 = (\omega + \rv{k})(\omega - \rv{k}) = \omega^2 - |\rv{k}|^2 = 0$ implies the usual \emph{dispersion relation} $|\rv{k}| = |\omega|$ in any frame\footnote{Note that choosing a different frame will boost the frequency $\omega$ (i.e., red- or blue-shift it from the Doppler effect), but will leave the null vector $k$ and the dispersion relation invariant.}.  Hence, we can factor out a frame-dependent frequency as a magnitude $k = \omega\, k_0$ to isolate a purely directional null factor $k_0 = \gamma_0(1 - \rv{\kappa})$ that contains the relative unit vector $\rv{\kappa} \equiv \rv{k}/|\rv{k}|$ for the chosen frame.  

It follows from this decomposition that Maxwell's equation \eqref{eq:maxwellnull} can also be written as the eigenvalue constraint $\rv{\kappa}\bv{f} = \bv{f}$, which makes $\bv{f}$ an eigenbivector of the relative unit vector $\rv{\kappa}$ (in any frame).  Performing a spacetime split $\bv{f} = sk = \rv{E} + \rv{B}I$ then reduces this constraint to two copies of the equation $\rv{\kappa}\rv{E} = \rv{B}I$, in accordance with the null condition $|\rv{E}| = |\rv{B}|$, which implies $\rv{\kappa}\times\rv{E} = \rv{B}$ since there is no scalar term.  Therefore, $\rv{\kappa}\cdot\rv{E} = \rv{\kappa}\cdot\rv{B} = 0$ and the three relative vectors $\rv{\kappa}\rv{E}\rv{B} = +I\abs{\rv{E}}\abs{\rv{B}}$ form a \emph{right-handed} oriented set of basis vectors in any relative frame.  Consequently, the appropriately scaled directional factor $\omega s$ for the null bivector $\bv{f} = sk = \omega sk_0$ is simply the purely spatial part of its spacetime split, $\omega s = \rv{E}\gamma_0$.  

The canonical bivector for the plane wave \eqref{eq:planewave} is therefore 
\begin{align}
  \bv{f} = \omega sk_0 = (\rv{E}\gamma_0)k_0 = \rv{E}(1-\rv{\kappa}) = (1+\rv{\kappa})\rv{E}.
\end{align}
Hence, in a relative frame $\rv{E}\gamma_0$ determines a reference orientation and notion of amplitude $|\rv{E}|$ for $\bv{f}$, while the purely directional null factor $k_0 = \gamma_0(1-\rv{\kappa})$ provides the proper wave orientation.  The full relative-frame version of \eqref{eq:planewave} thus has the familiar form
\begin{align}\label{eq:planewaverelative}
  \bv{F} &= \rv{E}\,(1 - \rv{\kappa})\,\exp[\pm(\omega t - \rv{k}\cdot\rv{x})I], \\
  &= (\rv{E} + \rv{\kappa}\times\rv{E}I)\,\exp[\mp(\rv{k}\cdot\rv{x} - \omega t)I], \nonumber \\
  &= [\rv{E}\cos(\rv{k}\cdot\rv{x} - \omega t) \mp \rv{\kappa}\times\rv{E}\sin(\rv{k}\cdot\rv{x} - \omega t)] \nonumber \\
  &\quad + [\rv{E}\sin(\rv{k}\cdot\rv{x} - \omega t) \pm \rv{\kappa}\times\rv{E}\cos(\rv{k}\cdot\rv{x} - \omega t)]I. \nonumber
\end{align}
We emphasize that an overall frequency scaling factor $\omega$ has been absorbed into the amplitude of $\rv{E}$ in the frame $\gamma_0$ by convention, so this relative amplitude will also change with the reference frame due to the Doppler effect.  Indeed, we will see later in Section \ref{sec:symemstress} that the energy-density of the field (which contains the square of this relative amplitude) is a frame-dependent quantity.

The relative form \eqref{eq:planewaverelative} makes it clear that the pure plane wave solution \eqref{eq:planewave} with $-I$ is \emph{right-hand circularly polarized} in the traditional sense.  In each relative spatial plane orthogonal to the relative wavevector direction $\rv{\kappa}$ (i.e., $\rv{\kappa}\cdot\rv{x}=0$), the relative vectors $\rv{E}$ and $\rv{B}= \rv{\kappa}\times\rv{E}$ rotate around the axis $\rv{\kappa}$ at a frequency $\omega$ with increasing $t$.  This rotation shows that a circularly polarized wave has a definite \emph{helicity} \cite{Nye1974,Allen1992,Allen1999,Andrews2008,Torres2011,Andrews2013,O'Neil2002,Garces2003,Curtis2003, Zhao2007,Adachi2007,Roy2014,Bliokh2014,Arlt2003,Leach2006,Yeganeh2013,Terborg2013,Dennis2008, Berry2009,Bekshaev2011,Kocsis2011,Dennis2013,Bliokh2013,Bliokh2013a,Barnett2013, Huard1978,Huard1979,Bliokh2012,Bliokh2013b,Bekshaev2014, Tang2010,Hendry2010,Bliokh2011,Tang2011,Hendry2012,Schaferling2012,Canaguier2013,Cameron2012c,Tkachenko2014,Bliokh2013c}, which is a manifestation of the spin-1 nature of the electromagnetic field.  Choosing the opposite sign of the pseudoscalar $+I$ in \eqref{eq:planewave} correspondingly produces a \emph{left-hand circularly polarized} plane wave by flipping the sign of the phase in \eqref{eq:planewaverelative}, effectively flipping both the propagation direction $\rv{\kappa}$ and the direction of rotation with increasing $t$.  It is now also clear that the dual-symmetric global phase freedom of the solution \eqref{eq:planewave} corresponds to the arbitrary phase offset of the polarization rotation \eqref{eq:planewaverelative}; that is, the relative vectors $\rv{E}$ and $\rv{B}$ may be arbitrarily rotated in unison around the propagation axis $\rv{\kappa}$ without changing the solution (in any frame).

To go beyond the monochromatic plane wave solution, we observe that in addition to flipping the sign of $I$ in \eqref{eq:planewave}, we can also scale the magnitude $\omega$ of $k$ arbitrarily.  We can thus exploit the fact that Maxwell's equation is linear to construct new solutions as superpositions of all possible values of $\omega$.  It follows that for any given directional null vector $k$ there exists an infinite number of corresponding (polychromatic) \emph{wave-packet} solutions determined by scalar spectral weight functions $\alpha(\omega)$ 
\begin{align}\label{eq:wavepacket}
  \bv{F} &= sk\, \int_{-\infty}^\infty \! \textrm{d}\omega\, \alpha(\omega)\,\exp[-\omega\,(k\cdot x) I], \\
  &= sk\, \int_0^\infty \! \textrm{d}\omega\,[\alpha_+(\omega)\,\exp[-\omega\,(k\cdot x) I] + \alpha_-(\omega)\,\exp[\omega\,(k\cdot x) I]], \nonumber
\end{align}
where $\alpha_\pm(\omega) = \alpha(\pm|\omega|)$.  The function $\alpha(\omega)$ can absorb the magnitude of $s$ for each choice of $\omega$, making $s^2 = -1$ a spacelike reference unit vector by convention.  This total field $\bv{F}$ is a linear superposition of the right- and left-hand circularly polarized plane waves at each positive scalar frequency $|\omega|$, and still satisfies Maxwell's vacuum equation \eqref{eq:maxwell} by construction.  Note that when $\alpha_+ = \alpha_-^*$ the integral becomes purely real and the resulting polarization is linear. In general, coefficients $\alpha_+$ and $\alpha_-$ determine the Jones vector of the wave polarization in the basis of circular poalrizations, which are attached to the coordinate frame spacified by the vector $s$ orthogonal to $k$.

The two signed solutions with $\pm |\omega|$ are traditionally called the \emph{positive-frequency} and \emph{negative-frequency} plane wave solutions of Maxwell's vacuum equation, respectively.  Here we see that the positive factor $|\omega|$ arises from the degeneracy of the null vector $k$ with respect to scaling.  However, the sign of $\pm 1$ directly corresponds to a choice of handedness for $I$ that determines a particular circular polarization (i.e., \emph{helicity}).  These features are invariant in \eqref{eq:wavepacket} because null vectors and the handedness of $I$ do not depend on any particular reference frame.  

The solution \eqref{eq:wavepacket} can also be understood as a Fourier transform of the spectral function $\alpha(\omega)$, which can be simply evaluated to obtain
\begin{align}\label{eq:fourierpacket}
  \bv{F} &= sk\, \widetilde{\alpha}(-k_0\cdot x) = (\rv{E}_0 + \rv{\kappa}\times\rv{E}_0I)\,\widetilde{\alpha}(\rv{\kappa}\cdot \rv{x} - ct),
\end{align}
in terms of the relative unit-vector direction $\rv{E}_0$.  The (complex scalar) amplitude $\widetilde{\alpha}$ is the Fourier-transform of $\alpha(\omega)$ that completely determines the packet shape of the traveling wave.  The bivector prefactor provides an overall linear polarization for a traveling wave packet by default, with electric and magnetic parts phase-offset by an angle of $\pi/2$.  This polarization can be made elliptic or circular, however, when $\widetilde{\alpha}$ has additional complex structure (as demonstrated in \eqref{eq:planewaverelative} for the monochromatic circularly polarized case).  

We can also go beyond a simple plane wave solution altogether to construct a more general (not necessarily null) solution from \eqref{eq:planewave} as a superposition of plane waves in all null wavevector directions $k$.  To do this we can rewrite the null factor as $k = \gamma_0(|\rv{k}| - \rv{k})$ in a particular reference frame $\gamma_0$, after which the components of $\rv{k}$ can be integrated as a 3-vector to obtain
\begin{align}\label{eq:superposition}
  \bv{F} &= \iiint_{-\infty}^\infty \! \textrm{d}^3\rv{k}\,\,s(\rv{k})\,k_0(\rv{k})\,[\alpha_+(\rv{k})\,e^{-(k\cdot x) I} + \alpha_-(\rv{k})\,e^{(k\cdot x) I}],
\end{align}
where $k_0 = k/|\rv{k}| = \gamma_0(1 - \rv{\kappa})$ is a purely directional factor in the frame $\gamma_0$, and where the reference directions $s(\rv{k}) = \rv{E}_0(\rv{k})\gamma_0$ are orthogonal to each $\rv{\kappa}$: $s(\rv{k})\cdot k(\rv{k}) \propto \rv{E}_0(\rv{k})\cdot\rv{\kappa} = 0$.  This integral \eqref{eq:superposition} can be understood as the construction of $\bv{F}$ through a spatial Fourier transform.

The functions $\alpha_\pm(\rv{k})$ encode the spectral decomposition and magnitude of the field.  The unit vectors $s(\rv{k})$ encode the reference polarizations for each traveling mode of the field.  Notably, the functions $\alpha_{\pm}(\rv{k})$ that appear here become the foundation for the canonical (``second'') quantization of the field in quantum field theory (after normalization), where they become the raising and lowering operators for specific field modes up to appropriate scaling factors \cite{Cohen1997}.  We will not further explore this quantization procedure here, but the interested reader can find excellent discussions of how to quantize the field using the complex Riemann-Silberstein vector in Refs.~\cite{Smith2007,Bialynicki-Birula2013}.


\section{Potential representations}\label{sec:potentials}


\epigraph{4.5in}{By performing particular operations on symbols, we acquire the possibility of expressing the same thing in many different forms.}{James Clerk Maxwell \cite{Maxwell1870}}


For more general solutions of Maxwell's equation in vacuum \eqref{eq:maxwell}, it can be useful to construct the bivector field $\bv{F}$ from auxiliary \emph{potential} fields.  There are three types of potential that are particularly convenient to use: 1) a vector potential, 2) a bivector potential, and 3) a scalar Hertz potential.  We will consider each in turn, and show their explicit relationship.


\subsection{Complex vector potential}


The standard vector potential construction follows from the ansatz
\begin{align}\label{eq:potential}
  \bv{F} &= \nabla z,
\end{align}
where $z$ is some initially unspecified multivector field.  This ansatz is compared with other formalisms in Table~\ref{tab:potentials}.  Due to the fundamental theorem \eqref{eq:fundamentaltheorem}, this ansatz can only describe \emph{conservative} fields that vanish when integrated around the boundary of any closed surface, such as a loop.  

\begin{table}
  \centering
  \begin{tabular}{l l}
    \hline
\noalign{\vskip 2mm} 
    \multicolumn{2}{c}{\textbf{Vector Potentials}} \\
\noalign{\vskip 2mm} 
    \hline \\
    Spacetime Algebra: & {$\boxed{\bv{F} = \nabla z}  \qquad \boxed{z = a_e + a_m I} $} \\
    \\
    & $\qquad\qquad\quad \Downarrow$ \\
    \\
    & $\bv{F} = \bv{F}_e + \bv{F}_mI$ \\
    & $\bv{F}_e = \nabla\wedge a_e, \quad \bv{F}_m = \nabla\wedge a_m$ \\
    & $\nabla \cdot a_e = \nabla \cdot a_m = 0$\\
    \\
    & $\qquad\qquad\quad \Downarrow$ \\
    \\
    & {$\begin{aligned}\bv{F} &= \rv{E} + \rv{B} I, & a_e &= (\phi_e + \rv{A})\gamma_0, & a_m &= (\phi_m + \rv{C})\gamma_0 \end{aligned}$} \\
    & $\rv{E} = - \rv{\nabla} \phi_e -\partial_0\rv{A} - \rv{\nabla}\times\rv{C}$ \\
    & $\rv{B} = - \rv{\nabla} \phi_m -\partial_0\rv{C} + \rv{\nabla}\times\rv{A}$ \\
    \\
    Differential Forms: 
    & $\bv{F} = \bv{F}_e + i\,{\star}\,\bv{F}_m$ \\
    & $\bv{F}_e = \textrm{d}\bv{a}_e, \quad \bv{F}_m = \textrm{d}\bv{a}_m$ \\
    \\
    Tensor Components: 
     & {$\begin{aligned} (F_e)_{\mu\nu} &= \frac{1}{2}(\partial_\mu (a_e)_\nu - \partial_\nu (a_e)_\mu) \\
       (F_m)_{\mu\nu} &= \frac{1}{2}(\partial_\mu (a_m)_\nu - \partial_\nu (a_m)_\mu) \end{aligned}$} \\
      \\
      Gibbs 3-Vectors: & {$\begin{aligned} \rv{E} &= -\rv{\nabla} \phi_e - \partial_0\rv{A} - \rv{\nabla}\times\rv{C} \\ 
        \rv{B} &= -\rv{\nabla} \phi_m - \partial_0\rv{C} + \rv{\nabla}\times\rv{A}\end{aligned}$} \\ \\
    \hline 
 \end{tabular}
 \caption[Vector potentials]{Vector potential representations of the vacuum electromagnetic field, as expressed in various formalisms.  Notably, spacetime algebra most naturally motivates an intrinsically complex vector potential $z$.  Nevertheless, one can also manually simulate this structure using differential forms with a scalar imaginary $i$ and the Hodge star $\star$ (e.g., \cite{Bliokh2013,Vasconcellos2014}), after interpreting the independent electric ($a_e$) and magnetic ($a_m$) vector potentials as one-forms $\bv{a}_{e,m}$ with the same components.  As we discuss in Section~\ref{sec:constituentconstraint}, imposing the additional constraint $\bv{F}_e = \bv{F}_m I = \bv{F}/\sqrt{2}$ can optionally symmetrize the contributions of the two vector potentials so that they describe the same (vacuum) field $\bv{F}$ \cite{Bliokh2013,Vasconcellos2014}.}
 \label{tab:potentials}
\end{table}

Since $\bv{F}$ is a pure bivector field, $\nabla z$ must satisfy the following conditions.  First, $z$ must be a vector field to produce a bivector with a derivative.  Second,  
the \emph{dual symmetry} of $\bv{F}$ forces $z$ to be correspondingly \emph{complex}.  Specifically, a phase-rotation of $\bv{F}$ induces the transformation 
\begin{align}
  \bv{F} \mapsto \bv{F}\exp(\theta I) = (\nabla z)\exp(\theta I) = \nabla (ze^{\theta I}) = \nabla (e^{-\theta I/2}ze^{\theta I/2}),
\end{align}
which is a phase rotation $\psi z \psi^*$ of $z$ with a spinor representation $\psi = \exp(-\theta I /2)$, exactly as used in \eqref{eq:dualsymmetry} and \eqref{eq:globalphase}.

It follows that $z$ must be a complex vector field of the form
\begin{align}\label{eq:complexpotential}
  z = a_e + a_m I
\end{align}
with separate (``electric'' $a_e$ and ``magnetic'' $a_m$) parts\footnote{Note that we will use the subscripts $e$ and $m$ throughout this report to distinguish the quantities that will become naturally associated with electric and magnetic \emph{charges} in Section~\ref{sec:source} (not to be confused with the relative electric and magnetic fields, which can both be produced by either type of charge). To recover traditional electromagnetism, all quantities with magnetic subscripts can be neglected, but we keep the discussion general.}, that satisfies the equation
\begin{align}\label{eq:potentialexpanded}
  \nabla z &= \nabla\wedge a_e + \nabla \cdot(a_m I) = \nabla\wedge a_e + (\nabla\wedge a_m)I = \bv{F},
\end{align}
along with the additional \emph{Lorenz-FitzGerald conditions}\footnote{This condition is also known as the \emph{Lorenz gauge}, which only partially fixes the gauge freedom to be more natural for manifestly relativistic treatments of the field.}
\begin{align}
  \label{eq:lorenzconditions}
  \nabla \cdot a_e &= \nabla \cdot a_m = 0.
\end{align}
As a result, the potential fields must be manifestly \emph{transverse} $\nabla a_e = \nabla \wedge a_e$ and $\nabla a_m = \nabla \wedge a_m$.  

This transversality implies that there is an \emph{additional gauge freedom} in the definition \eqref{eq:potential} of $z$ due to the invariance under the replacement $z' = z + \nabla\zeta$.  Provided that\footnote{Note that if $\zeta = \alpha + \beta I$ such that $\nabla^2\zeta \neq 0$, then we obtain $\nabla z' = \bv{F} + \nabla^2\zeta$ with $\bv{F} = \nabla\wedge(a + \nabla\alpha) + [\nabla\wedge(g + \nabla\beta)] I = \mean{\nabla z'}_2$.  Adding this projection in the field definition provides a more general (but less natural) gauge freedom that does not preserve the Lorenz-FitzGerald conditions.} $\nabla^2\zeta = 0$, each new $z'$ will produce the same field according to $\nabla z' = \nabla z + \nabla^2 \zeta = \nabla z = \bv{F}$.  Moreover, each new $z'$ will also satisfy the Lorenz-FitzGerald conditions since $\nabla\cdot\nabla\zeta = \nabla^2\zeta = 0$.  


\subsubsection{Plane wave vector potential}\label{sec:planewavepotential}


If we revisit the null plane wave solution of \eqref{eq:planewave}, and use the relation $\bv{F} = \nabla z$, we can infer that the potential $z$ for a polarized plane wave must have the simple form (being careful to remember that $I$ anticommutes with 4-vectors)
\begin{align}\label{eq:planewavepotential}
  z &= \exp[\mp(k\cdot x)I](Is) = (Is)\exp[\pm(k\cdot x)I],
\end{align}
where the spacelike vector $s$ determines a reference polarization direction and amplitude.  To verify that this is the appropriate potential, we take its derivative 
\begin{align}
  \nabla z &= (\nabla e^{\mp(k\cdot x)I})Is = (kI)e^{\mp(k\cdot x)I}I s = -kse^{\pm(k\cdot e)I} = (sk)e^{\pm(k\cdot e)I},
\end{align}
which reproduces \eqref{eq:planewave}.  A second derivative will annihilate the null factor $k$ to produce Maxwell's equation in the form $\nabla^2 z = 0$.  

The complexity of the vector potential $z$ is critical for obtaining \eqref{eq:planewavepotential}.  There are intrinsic phases in the traveling wave factor $\exp(\pm(k\cdot x)I)$.  Furthermore, this is the vector potential for a null solution $\bv{F}^2 = (\nabla z)^2 = 0$, so the global intrinsic phase of the solution must be degenerate as a general structural property of null bivectors.

Moreover, we know from the derivation of the plane wave \eqref{eq:planewaverelative} that the directional vector $s = \rv{E}\gamma_0$ should be orthogonal to $k$ in any frame.  This constraint is another way of expressing the Lorenz-FitzGerald condition $\nabla \cdot z = 0$ for the gauge freedom since 
\begin{align}
  \nabla\cdot z = \pm I (s\cdot\nabla)\exp[\pm(k\cdot x) I] = \pm(s\cdot k)\exp[\pm(k\cdot x) I] = 0.
\end{align}
We also know from the derivation of \eqref{eq:planewaverelative} that the vector $s$ arises as a factor of a null bivector that can always be written in a relative frame as purely spatial vector $s = \rv{E}_0\gamma_0$.  It then follows that $\rv{E}_0\cdot\rv{k} = 0$, which makes the plane wave potential manifestly \emph{transverse} in any relative frame: $\rv{\nabla}\cdot(z\gamma_0) = \pm(\rv{E}_0\cdot\rv{k})\exp[\mp(\rv{k}\cdot\rv{x} - \omega t)I] = 0$.  This property will become important in Section \ref{sec:bivectorpotential}.

It follows from \eqref{eq:planewavepotential} that the vector potential for the null wave packet in \eqref{eq:wavepacket} has the form
\begin{align}\label{eq:wavepacketpotential}
  z &= Is\, \int_{-\infty}^\infty \! \frac{\textrm{d}\omega}{\omega}\,\alpha(\omega)\,e^{-\omega\,(k\cdot x) I}, \\
  &= Is\, \int_0^\infty \! \frac{\textrm{d}\omega}{\omega}\,\left[\alpha_+(\omega)\,e^{-\omega\,(k\cdot x) I} - \alpha_-(\omega)\,e^{\omega\,(k\cdot x) I}\right], \nonumber
\end{align}
where the derivative produces a sign flip between the spectral functions $\alpha_\pm$.  The potential for the superposition of plane waves in \eqref{eq:superposition} thus has the corresponding form
\begin{align}\label{eq:superpositionpotential}
  z &= I\iiint_{-\infty}^\infty \! \frac{\textrm{d}^3\rv{k}}{|\rv{k}|}\,s(\rv{k})\,[\alpha_+(\rv{k})\,e^{-(k\cdot x) I} - \alpha_-(\rv{k})\,e^{(k\cdot x) I}],
\end{align}
where the remaining factor of $|\rv{k}|$ can be absorbed into $\alpha_\pm$ as units and convention dictate.  These potentials are also manifestly transverse in any frame since $s(\rv{k})\cdot k = 0$ by construction for each null plane wave propagating in direction $\rv{k}$.


\subsubsection{Relative vector potentials}


We perform a spacetime split of the vector potentials $a_e = \gamma_0(\phi_e - \rv{A})$ and $a_m = \gamma_0(\phi_m - \rv{C})$ into \emph{relative scalar potentials} $\phi_e,\phi_m$, and \emph{relative vector potentials} $\rv{A},\rv{C}$.  Note that for these relative quantities, we try to conform to existing notation conventions.  The notation $\rv{C}$ for the magnetic vector potential has become standard in studies of dual symmetry in vacuum \cite{Calkin1965,Zwanziger1968,Deser1976,Cameron2012b,Bliokh2013,Cameron2012,Fernandez2013}.  With these relative potentials, we can expand the field into a relative form
\begin{align}
  \bv{F} &= \nabla z = (\partial_0 - \rv{\nabla})[(\phi_e - \rv{A}) + (\phi_m - \rv{C})I], \\
  &= [-\partial_0\rv{A} - \rv{\nabla} \phi_e - \rv{\nabla}\times\rv{C}] + [-\partial_0\rv{C} - \rv{\nabla}\phi_m + \rv{\nabla}\times\rv{A}]I, \nonumber
\end{align}
where we have used the Lorenz-FitzGerald conditions \eqref{eq:lorenzconditions} written in the relative frame, 
\begin{align}
  \partial_0 \phi_e + \nabla\cdot\rv{A} = \partial_0 \phi_m + \nabla\cdot\rv{C} = 0.
\end{align}
This split implies that the relative fields may be written as 
\begin{align}\label{eq:relativepotentials}
  \rv{E} &= -\rv{\nabla} \phi_e - \partial_0\rv{A} - \rv{\nabla}\times\rv{C}, &
  \rv{B} &= -\rv{\nabla} \phi_m - \partial_0\rv{C} + \rv{\nabla}\times\rv{A}, 
\end{align}
in terms of the relative potentials.

Under a gauge transform $z' = z + \nabla\zeta$, with $\zeta = \zeta_a + \zeta_g I$, these relative vector potentials become
\begin{align}\label{eq:gaugetransform}
  z'\gamma_0 &= (\phi_e + \rv{A}) + (\phi_m - \rv{C})I + (\partial_0 - \rv{\nabla})(\zeta_a - \zeta_g I), \\
  &= (\phi_e + \partial_0 \zeta_a) + (\rv{A} - \rv{\nabla}\zeta_a) + (\phi_m - \partial_0 \zeta_g)I - (\rv{C} - \rv{\nabla}\zeta_g)I. \nonumber
\end{align}
Thus, the relative scalar and vector potentials can become mixed both by Lorentz transformations and gauge transformations in general.  One must fix both the inertial reference frame and the gauge in order to completely fix the form of the potential.


\subsubsection{Constituent fields}\label{sec:constituent}


The decomposition of the field \eqref{eq:potentialexpanded} can be interpreted as constructing the total field $\bv{F} = \bv{F}_e + \bv{F}_m I$ from \emph{two constituent bivector fields} that are derived from each vector potential separately.  The ``electric'' part, $\bv{F}_e = \nabla \wedge a_e$, will become naturally associated with electric charge in Section~\ref{sec:source}, while the ``magnetic'' part, $\bv{F}_m = \nabla \wedge a_m$, will be correspondingly associated with magnetic charge.  However, the dual symmetry of vacuum makes these constituent fields interchangable.  For example, we can exchange their roles by taking the dual of $\bv{F}$ as in \eqref{eq:dualfield}
\begin{align}
  \bv{G} = \bv{F}I^{-1} = (\nabla\wedge a_m) - (\nabla\wedge a_e)I = \bv{F}_m - \bv{F}_e I = \rv{B} - \rv{E}I.
\end{align}

If we perform a spacetime split of each piece of the total field, $\bv{F}_e = \rv{E}_e + \rv{B}_e I$ and $\bv{F}_m I = \rv{E}_m + \rv{B}_m I$ (so $\bv{F}_m = \rv{B}_m - \rv{E}_m I$), then we find that the relative fields can be written as
\begin{align}
  \rv{E}_e &= -\rv{\nabla} \phi_e - \partial_0\rv{A}, & \rv{B}_e &= \rv{\nabla}\times\rv{A}, \\
  \rv{E}_m &= -\rv{\nabla}\times\rv{C},            & \rv{B}_m &= -\rv{\nabla} \phi_m - \partial_0\rv{C}. \nonumber
\end{align}
The total relative fields include contributions from both potentials, $\rv{E} = \rv{E}_e + \rv{E}_m$, $\rv{B} = \rv{B}_e + \rv{B}_m$.

Note that we will use the subscripts of $e$ and $m$ throughout this report to signify the eventual associations of physical quantities with either electric or magnetic \emph{charges}.  Traditionally, the electromagnetic field couples only to \emph{electric} charges, so is described by $\bv{F}_e$ and its associated electric vector potential $a_e$.  Therefore, to recover traditional electromagnetism, we can simply neglect all ``magnetic'' quantities with an $m$ subscript in what follows.  However, in the absence of charges (as in vacuum) there is no such constraint, so we keep the formulas dual-symmetric.


\subsection{Bivector potential}\label{sec:bivectorpotential}


We can make an important simplification of the gauge freedom of the complex vector potential $z$ if we introduce the \emph{locally transverse} $\rv{A}_\perp,\rv{C}_\perp$ and \emph{longitudinal} $\rv{A}_\parallel,\rv{C}_\parallel$ parts of the relative vector potentials $\rv{A} = \rv{A}_\perp + \rv{A}_\parallel$ and $\rv{C} = \rv{C}_\perp + \rv{C}_\parallel$ in a particular frame $\gamma_0$.  These pieces of the vector potentials are defined to satisfy 
\begin{align}
  \rv{\nabla}\cdot\rv{A}_\perp &= \rv{\nabla}\cdot\rv{C}_\perp = 0, \\
  \rv{\nabla}\times\rv{A}_\parallel &= \rv{\nabla}\times\rv{C}_\parallel = 0.
\end{align}
As a result, the gauge transform \eqref{eq:gaugetransform} produces
\begin{align}\label{eq:gaugetransformperp}
  z'_{\perp}\gamma_0 &= \rv{A}_\perp - \rv{C}_\perp I = z_\perp \gamma_0, \\
  \label{eq:gaugetransformpar}
  z'_{\parallel}\gamma_0 &= (\phi_e + \partial_0 \zeta_a) + (\rv{A}_\parallel - \rv{\nabla}\zeta_a) + (\phi_m - \partial_0 \zeta_g)I - (\rv{C}_\parallel - \rv{\nabla}\zeta_g)I.
\end{align}
That is, in any local frame the \emph{transverse} part of the vector potential is \emph{gauge-invariant}.  

This invariance leads to an interesting observation.  After separating the vector potential $z$ into a transverse and longitudinal piece in a particular frame, we can define a corresponding transverse \emph{bivector potential} (e.g., as in \cite{Bliokh2013})
\begin{align}
  \bv{Z} &= z_\perp \gamma_0 = \rv{A}_\perp - \rv{C}_\perp I,
\end{align}
and longitudinal \emph{spinor potential}
\begin{align}
  \psi_z &= z_\parallel \gamma_0 = (\phi_e - \phi_m I) + (\rv{A}_\parallel - \rv{C}_\parallel I).
\end{align}
Both of these potentials can now be easily manipulated as proper geometric objects, even though their definitions involve a particular frame $\gamma_0$.  Moreover, only the spinor potential $\psi_z$ depends on the choice of gauge according to \eqref{eq:gaugetransformperp}, which makes the bivector $\bv{Z}$ a \emph{gauge-invariant} potential (restricted to the $\gamma_0$ frame).  This separation of the total potential into a gauge-invariant part $\bv{Z}$ and a gauge-dependent part $\psi_z$ in a particular frame is known as the \emph{Chen decomposition} \cite{Chen2008,Leader2014}, and is important for the meaningful separation of the orbital and spin parts of the angular momentum (as we shall see in Section~\ref{sec:noether}).

According to Maxwell's vacuum equation \eqref{eq:maxwell}, we can then write
\begin{align}
  \nabla \bv{F} = \nabla^2 z = \nabla^2(\bv{Z} + \psi_z)\gamma_0 = 0.
\end{align}
In the absence of sources, the frame-dependent factor of $\gamma_0$ simply cancels to further simplify the analysis.  Moreover, since $\psi_z$ can be changed arbitrarily by the choice of gauge, according to \eqref{eq:gaugetransformpar}, we can always fix the gauge to set the longitudinal potential $\psi_z$ to zero in the local frame (i.e., choosing the Coulomb, or radiation gauge).  After fixing the gauge in this manner, we can write Maxwell's vacuum equation \eqref{eq:maxwell} solely in terms of the transverse bivector potential 
\begin{align}\label{eq:bivectormaxwell}
  \nabla^2 \bv{Z} &= \nabla^2(\rv{A}_\perp - \rv{C}_\perp I) = 0.
\end{align}
This bivector potential is summarized in Table~\ref{tab:bivectorpot} along with several of its derivatives (from the next section) for reference.

\begin{table}
  \centering
  \begin{tabular}{l l}
    \hline
\noalign{\vskip 2mm} 
    \multicolumn{2}{c}{\textbf{Bivector Potentials for Vacuum Fields}} \\
\noalign{\vskip 2mm} 
    \hline \\
    Transverse Bivector Potential: & $\bv{Z} = z_\perp \gamma_0 = \rv{A}_\perp - \rv{C}_\perp I$ \\
    Field Relation: & $\bv{F} = (\nabla \bv{Z})\gamma_0 \qquad (z_\parallel = 0)$ \\
    Vacuum Equation: & $\nabla^2 \bv{Z} = 0$ \\
    \\
    \hline
    \\
    Hertz Bivector Potential: & $\bv{Z} = -[\rv{\nabla},\bv{\Pi}] = -\rv{\nabla}\times\bv{\Pi}I$ \\
    Field Relation: & $\bv{F} = \nabla\gamma_0[\rv{\nabla},\bv{\Pi}] = (I\partial_0 + \rv{\nabla}\times)(\rv{\nabla}\times\bv{\Pi})$ \\
    Vacuum Equation: & $\nabla^2(\rv{\nabla}\times\rv{\Pi}) = (\partial^2_0 - \rv{\nabla}^2)(\rv{\nabla}\times\bv{\Pi}) = 0$ \\
    \\
    \hline
    \\
    Hertz Scalar Potential: & $\bv{\Pi} = \rv{\Pi}\Phi$ \\
    Field Relation: & $\bv{F} = \nabla\gamma_0[\rv{\nabla},\rv{\Pi}]\Phi = (I\partial_0 + \rv{\nabla}\times)(\rv{\nabla}\times\rv{\Pi})\Phi$ \\
    Vacuum Equation: & $\nabla^2\Phi = (\partial_0^2 - \rv{\nabla}^2)\Phi = 0$ \\
    \\
    \hline 
 \end{tabular}
 \caption[Bivector potentials]{Various bivector potential representations of the vacuum electromagnetic field.  All such representations fix a particular reference frame $\gamma_0$, and fix the gauge to make the longitudinal part of the vector potentials vanish $z_\parallel = 0$ (i.e., fixing the Coloumb gauge).  With this simplification, the bivector potential $\bv{Z}$ contains the same information as the vacuum field $\bv{F}$ without added freedom.  As such, expanding this fundamental bivector potential into various forms permits alternative strategies for solving Maxwell's vacuum equation. }
 \label{tab:bivectorpot}
\end{table}


\subsubsection{Plane wave bivector potential}


As a notable special case, we showed that the plane wave potential \eqref{eq:planewavepotential} must be manifestly transverse in any relative frame in Section \ref{sec:planewavepotential}.  Thus we can immediately write the corresponding gauge-invariant bivector potential for a plane wave
\begin{align}
  \bv{Z} = z_\perp\gamma_0 = (Is\gamma_0) \exp[\mp(k\cdot x)I] = I\exp[\mp(k\cdot x) I]\rv{E}_0.
\end{align}
Note that this bivector potential is already essentially equivalent to the total field bivector $\bv{F}$.  There is no remaining arbitrariness in the potential, which emphasizes that the bivector potential is a gauge-invariant representation in the $\gamma_0$ frame.

The corresponding bivector potentials for the plane wave superpositions follow in a similar way from \eqref{eq:wavepacketpotential} and \eqref{eq:superpositionpotential}
\begin{align}\label{eq:wavepacketbipotential}
  \bv{Z} &= I\rv{E}_0\, \int_{-\infty}^\infty \! \frac{\textrm{d}\omega}{\omega}\,\alpha(\omega)\,e^{\omega\,(k\cdot x) I}, \\
  &= I\rv{E}_0\, \int_0^\infty \! \frac{\textrm{d}\omega}{\omega}\,[\alpha_+(\omega)\,e^{\omega\,(k\cdot x) I} - \alpha_-(\omega)\,e^{-\omega\,(k\cdot x) I}], \nonumber \\
  \bv{Z} &= I\iiint_{-\infty}^\infty \! \frac{\textrm{d}^3\rv{k}}{|\rv{k}|}\,\rv{E}_0(\rv{k})\,[\alpha_+(\rv{k})\,e^{(k\cdot x) I} - \alpha_-(\rv{k})\,e^{-(k\cdot x) I}].
\end{align}
The only appreciable difference in structure from the vector potentials is the flip in sign of the phase factors associated with each $\alpha_\pm$, which occurs since $\gamma_0$ and $I$ anticommute.  Moreover, since the same functions $\alpha_\pm$ occur here as in the vector potentials in Section \ref{sec:planewavepotential} and the plane wave fields in Section \ref{sec:waves}, the same procedure can be used to canonically quantize the bivector potential directly, if desired.


\subsubsection{Field correspondence}


The relative fields associated with $\bv{Z}$ still depend upon the defining frame $\gamma_0$
\begin{align}\label{eq:bivectorcorrespondence}
  \bv{F} = (\nabla \bv{Z})\gamma_0 = \nabla z_\perp.
\end{align}
Thus, both the transversality properties of $\bv{Z}$ and its relation to the field $\bv{F}$ depend on the frame $\gamma_0$.  Nevertheless, for any frame $\gamma_0$ it is possible to construct the transverse $\bv{Z}$ for that frame.

Using the correspondence \eqref{eq:bivectorcorrespondence}, the relative fields have the form \cite{Candlin1965,Afanasiev1996,Trueba1996,Cameron2012b}
\begin{align}\label{eq:coulombpotentials}
  \rv{E} &= -\partial_0\rv{A}_\perp - \rv{\nabla}\times\rv{C}_\perp, & \rv{B} &= -\partial_0\rv{C}_\perp + \rv{\nabla}\times\rv{A}_\perp,
\end{align}
which can be understood as the appropriate simplification of \eqref{eq:relativepotentials}.  Interestingly, the equations \eqref{eq:coulombpotentials} plus the transversality conditions $\rv{\nabla}\cdot\rv{A}_\perp = \rv{\nabla}\cdot\rv{C}_\perp = 0$ resemble the four Maxwell's equations in vacuum \eqref{eq:maxwellfour} that we started with \cite{Cameron2012b,VanEnk2013}.  This resemblance stems from the fact that the derivative of the bivector potential $\bv{Z}$ has the same structural form \eqref{eq:bivectorfield} that also produced Maxwell's equations \eqref{eq:maxwellfour}.


\subsection{Hertz potentials}


Solving equations with vector and bivector potentials can be simpler than solving the main bivector equation \eqref{eq:maxwell}, but it would be simpler still to solve equations using a scalar potential.  Hence, we are motivated to make another ansatz to further reduce the vector potential $z$ or the bivector potential $\bv{Z}$ by using another derivative.  

A naive attempt at performing such a reduction is to assume that $z = \nabla \phi$ for some complex scalar potential $\phi$. However, this attempt fails since the resulting field $\bv{F} = \nabla^2\phi$ would be a scalar.  Hence, we instead consider simplifying the bivector potential $\bv{Z}$ further.


\subsubsection{Hertz bivector potential}


A suitable reduction was found by Hertz \cite{Hertz1889}, who (after translating to the language of spacetime algebra) exploited the fact that the commutator bracket in Eq.~\eqref{eq:commutatorbracket} for bivectors produces another bivector.  That is, if we write $z = \bv{Z}\gamma_0 = -\gamma_0\bv{Z}^\dagger$, with a bivector potential $\bv{Z}^\dagger = \rv{A}_\perp + \rv{C}_\perp I$, then we can make the ansatz
\begin{align}\label{eq:hertzpotential}
  \bv{Z}^\dagger = -[\rv{\nabla},\bv{\Pi}] = -\rv{\nabla}\times\bv{\Pi} I,
\end{align}
making $\bv{Z}^\dagger$ the dual of the relative curl of a second-order\footnote{Curiously, we can perform this trick as many times as we like by making each relative potential the curl of another generating potential.  This ability to keep increasing the number of derivatives was also noticed in \cite{Cameron2012b,VanEnk2013}.  We could also use the form $\bv{Z}^\dagger = \nabla\bv{\Pi}\gamma_0$ to mirror the field definition \eqref{eq:bivectorcorrespondence}, which would add an extra time derivative to the curl.} bivector potential $\bv{\Pi}$.  Note the restriction to a particular relative frame $\gamma_0$ implicit in the relative gradient $\rv{\nabla} = \gamma_0 \wedge \nabla$; this is the same fixed frame used to define $\bv{Z}$. 

The second-order bivector potential $\bv{\Pi} = \rv{\Pi}_e + \rv{\Pi}_m I$ is known as a \emph{Hertz potential} for the field, and can be decomposed into electric $\rv{\Pi}_e$ and magnetic $\rv{\Pi}_m$ relative vector parts.  The resulting field equation has the form \cite{Bialynicki-Birula2013}
\begin{align}
  \bv{F} &= \nabla z = \nabla \bv{Z}\gamma_0 = - \nabla \gamma_0 \bv{Z}^\dagger, \\
  &= \nabla \gamma_0 [\rv{\nabla},\bv{\Pi}] = (\partial_0 - \rv{\nabla})(\rv{\nabla}\times\bv{\Pi}) I = (I\partial_0 + \rv{\nabla}\times)(\rv{\nabla}\times\bv{\Pi}), \nonumber
\end{align}
where the last equality uses the fact that the relative divergence of a curl vanishes.

Maxwell's equation in vacuum then mandates
\begin{align}
  \nabla \bv{F} = \nabla^2 z = -\gamma_0\nabla^2 \bv{Z}^\dagger = -\gamma_0\nabla^2(\rv{\nabla}\times\bv{\Pi})I = 0.
\end{align}
That is, the relative curl of the Hertz potential must satisfy the d'Alembert wave equation
\begin{align}
  \nabla^2(\rv{\nabla}\times\bv{\Pi}) = (\partial_0^2 - \rv{\nabla}^2)(\rv{\nabla}\times\bv{\Pi}) = 0.
\end{align}
This general construction by Hertz is quite useful for applications, such as describing dipole radiation fields or rectangular waveguide modes \cite{Essex1977}.  However, the construction still involves a bivector potential, so we will simplify it further.


\subsubsection{Hertz complex scalar potential}


We can avoid the complication of the Hertz bivector potential $\bv{\Pi} = \rv{\Pi}\Phi$ by factoring it into a \emph{complex scalar potential} $\Phi$ and a fixed reference unit-3-vector $\rv{\Pi}$ in the relative frame.  This reference direction is usually chosen to have physical significance, such as the propagation axis for a paraxial beam or the dipole-radiation axis  \cite{Bialynicki-Birula2013,Berry2011b}. The Maxwell constraint on the scalar function $\Phi$ can then be written
\begin{align}
  \nabla^2(\rv{\nabla}\times\rv{\Pi})\Phi = (\rv{\nabla}\times\rv{\Pi})\nabla^2\Phi = 0,
\end{align}
making it sufficient for the potential $\Phi$ to be a complex scalar solution of the d'Alembert wave equation
\begin{align}\label{eq:dalembert}
  \nabla^2\Phi = (\partial_0^2 - \rv{\nabla}^2) \Phi = 0.
\end{align}
Since this relativistic scalar wave equation is well-understood, and may be systematically solved by separation of variables in a myriad of coordinate systems (e.g., \cite{Kalnins1978}), constructing solutions of Maxwell's equation from solutions of this scalar wave equation is a powerful method for generating nontrivial solutions to Maxwell's equation in vacuum \cite{Bialynicki-Birula2013}.

Given such a wave solution $\Phi$, an associated electromagnetic field is 
\begin{align}
  \bv{F} &= \nabla\gamma_0[\rv{\nabla},\rv{\Pi}]\Phi = (I\partial_0 + \rv{\nabla}\times)(\rv{\nabla}\times\rv{\Pi})\Phi.
\end{align}
As we mentioned earlier, the frame-specific differential factor $\gamma_0[\rv{\nabla},\rv{\Pi}]$ is a consequence of fixing the frame to define the transverse potential $\bv{Z}$.  The resulting bivector field $\bv{F}$ will always be a proper (frame-independent) geometric object.  This Hertz potential technique has been used to great effect by Hertz \cite{Hertz1889}, Whittaker \cite{Whittaker1904}, Debye \cite{Debye1909}, and others \cite{Nisbet1955,Synge1965,Bialynicki-Birula2013}.  

The use of dual-symmetric fields and spacetime algebra makes the origin of the Hertz potentials particularly transparent.  In particular, their construction relies upon further decomposing a \emph{complex} 4-vector potential $z$ into a transverse bivector potential $\bv{Z}$, after which that potential can be decomposed into a second-order bivector potential $\bv{\Pi}$, which in turn can be decomposed into a complex scalar potential $\Phi$.  Notably, the starting vector potential $z$ is only complex when the dual symmetry of the vacuum Maxwell equation is preserved.


\section{Maxwell's equation with sources}\label{sec:source}

\epigraph{4.5in}{The theory I propose may therefore be called a theory of the \emph{Electromagnetic Field} because it has to do with the space in the neighbourhood of the electric or magnetic bodies, and it may be called a \emph{Dynamical Theory}, because it assumes that in the space there is matter in motion, by which the observed electromagnetic phenomena are produced.}{James Clerk Maxwell \cite{Maxwell1865}}


The simplest modification to Maxwell's equation in vacuum \eqref{eq:maxwell} that we can make is to add a source field.  Since the derivative of a bivector field can only produce a complex vector field, this modification produces 
\begin{align}\label{eq:maxwellsources}
  \nabla \bv{F} &= j,
\end{align}
where the complex current $j = j_e + j_m I$ acts as the appropriate source field, and admits both vector (electric) and trivector (magnetic) parts.  Note that this equation is no longer intrinsically invariant under the simple phase-rotation $\bv{F} \mapsto \bv{F}\exp(\theta I)$ due to the presence of the source current $j$; we will clarify this important point in Section~\ref{sec:dualsymmetrysource}.

Taking a second derivative decouples \eqref{eq:maxwellsources} into the set of equations
\begin{align}
  \nabla^2 \bv{F} &= (\nabla \wedge j_e) + (\nabla \wedge j_m)I, \\
  \label{eq:currentcontinuity}
  \nabla \cdot j_e &= \nabla \cdot j_m = 0. 
\end{align}
The first equation (for the bivector part) indicates that $\bv{F}$ satisfies a \emph{wave equation} with a complex source equal to the curl of $j$.  The second equation (for the scalar part) indicates the source satisfies a (charge-current) \emph{continuity equation}.

\begin{table}
  \centering
  \begin{tabular}{l l}
    \hline
\noalign{\vskip 2mm} 
    \multicolumn{2}{c}{\textbf{Maxwell's Equation with Sources}} \\
\noalign{\vskip 2mm} 
    \hline \\
    Spacetime Algebra: & {$\boxed{\nabla \bv{F} = \nabla^2 z = j} \qquad \boxed{j = j_e + j_m I}$} \\
    \\
    & $\qquad\qquad\qquad\qquad \Downarrow$ \\
    \\
    & {$\begin{aligned}\nabla\cdot\bv{F} &= \nabla^2 a_e = j_e, &\quad \nabla \wedge \bv{F} &= \nabla^2 a_m I = j_m I\end{aligned}$} \\
    \\
    & $\qquad\qquad\qquad\qquad \Downarrow$ \\
    \\
    & {$\begin{aligned}\bv{F} &= \rv{E} + \rv{B}I, &\quad j_e &= (\cc\rho_e + \rv{J}_e)\gamma_0, &\quad j_m &= (\cc\rho_m + \rv{J}_m)\gamma_0 \\
      & &\quad a_e &= (\phi_e + \rv{A})\gamma_0, &\quad a_m &= (\phi_m + \rv{C})\gamma_0 \end{aligned}$} \\
    \\
    & {$\begin{aligned}\rv{\nabla}\cdot\rv{E} + (\partial_0 \rv{E} - \rv{\nabla}\times\rv{B}) &= \nabla^2(\phi_e - \rv{A}) = \cc\rho_e - \rv{J}_e \\
      \rv{\nabla}\cdot\rv{B} + (\partial_0 \rv{B} + \rv{\nabla}\times\rv{E}) &= \nabla^2(\phi_m - \rv{C}) = \cc\rho_m - \rv{J}_m \end{aligned}$} \\
    \\
    & $\qquad\qquad\qquad\qquad \Downarrow$ \\
    \\
    & {$\begin{aligned}\rv{\nabla}\cdot\rv{E} &= \nabla^2 \phi_e = \cc\rho_e , &\quad \partial_0 \rv{E} - \rv{\nabla}\times\rv{B} &= -\nabla^2\rv{A} = -\rv{J}_e \\
      \rv{\nabla}\cdot\rv{B} &= \nabla^2\phi_m = \cc\rho_m , &\quad \partial_0 \rv{B} + \rv{\nabla}\times\rv{E} &= -\nabla^2\rv{C} = -\rv{J}_m \end{aligned}$} \\
    \\
\noalign{\vskip 2mm} 
    Differential Forms: & {$\begin{aligned} \bv{F} &= \bv{F}_e + i\,{\star}\,\bv{F}_m, & &{\star}\,\textrm{d}\,{\star }\,\bv{F}_e = \star\,\textrm{d}\,{\star }\,\textrm{d}\bv{a}_e = -\bv{j}_e \\ & & &{\star}\,\textrm{d}\,{\star }\,\bv{F}_m = \star\,\textrm{d}\,{\star }\,\textrm{d}\bv{a}_m = -\bv{j}_m\end{aligned}$} \\
    \\
\noalign{\vskip 2mm} 
    Tensor Components: & {$\begin{aligned}\partial_\nu F_e^{\mu\nu} &= j^\mu_e, & \partial_\nu F_m^{\mu\nu} &= j^\mu_m, & F^{\mu\nu} = F_e^{\mu\nu} + i{\star}F_m^{\mu\nu} \end{aligned}$} \\
    \\
\noalign{\vskip 2mm} 
    Gibbs 3-Vectors: & {$\begin{aligned}\rv{\nabla}\cdot\rv{E} &= \cc\rho_e , &\quad \partial_0 \rv{E} - \rv{\nabla}\times\rv{B} &= -\rv{J}_e \\
      \rv{\nabla}\cdot\rv{B} &= \cc\rho_m , &\quad \partial_0 \rv{B} + \rv{\nabla}\times\rv{E} &= -\rv{J}_m \end{aligned}$} \\
    \\
    \hline
    \\
    {$\begin{aligned} & \text{SI Units :} \\ & (\cc \equiv 1/\sqrt{\epsilon_0\mu_0})\end{aligned}$} & {$\begin{aligned}(\rv{E},\rv{C},\phi_e) &\mapsto \sqrt{\epsilon_0}\,(\rv{E},\rv{C},\phi_e), &  (\rv{B},\rv{A},\phi_m) &\mapsto (\rv{B},\rv{A},\phi_m)/\sqrt{\mu_0} \\
      (\rv{J}_e,\rho_e) &\mapsto \sqrt{\mu_0}\,(\rv{J}_e,\rho_e), & (\rv{J}_m,\rho_m) &\mapsto \sqrt{\mu_0}\,(\rv{J}_m,\rho_m)/\cc  \end{aligned}$} \\
    \\
    CGS Units: & {$\begin{aligned}\rv{E} &\mapsto \rv{E}/\sqrt{4\pi}, &\quad \rv{B} &\mapsto \rv{B}/\sqrt{4\pi}, &\quad j &\mapsto \sqrt{4\pi}\,j/\cc \end{aligned}$} \\
    \\
    \hline 
 \end{tabular}
 \caption[Maxwell's equation with sources]{Maxwell's equation with both electric and magnetic sources, as expressed in various formalisms.  As with the vacuum case in Tables~\ref{tab:vacuummaxwell} and \ref{tab:potentials}, only spacetime algebra permits the formulation as the single boxed equation involving either the bivector field $\bv{F}$ or the complex vector potential field $z = a_e + a_m I$, while treating both electric sources $j_e$ and magnetic sources $j_m$ on equal footing as a single complex source $j = j_e + j_m I$.  In contrast, the differential-form approach becomes strained with a nonzero magnetic source (one-form) $\bv{j}_m$ \cite{Cabibbo1962}, but simplifies to a form similar to the vacuum case in Table~\ref{tab:vacuummaxwell} for purely electric sources $\bv{j}_e$. Note that for the SI unit scalings, we use the Ampere-meter convention for magnetic charges. }
 \label{tab:sourcemaxwell}
\end{table}


\subsection{Relative frame form}


After breaking the source $j = j_e + j_m I$ into vector and trivector parts, the equation \eqref{eq:maxwellsources} implies the two independent equations
\begin{align}
  \nabla\cdot \bv{F} &= j_e, & \nabla\wedge \bv{F} &= j_m I,
\end{align}
which, under the spacetime splits $\bv{F} = \rv{E} + \rv{B}I$, $j_e = \gamma_0(\cc\rho_e - \rv{J}_e)$, and $j_m = \gamma_0(\cc\rho_m - \rv{J}_m)$, expand to the four equations
\begin{align}
  \rv{\nabla}\cdot\rv{E} &= \cc\rho_e, & \partial_0 \rv{E} - \rv{\nabla}\times\rv{B} &= - \rv{J}_e, \\
  \rv{\nabla}\cdot\rv{B} &= \cc\rho_m, & \partial_0 \rv{B} + \rv{\nabla}\times\rv{E} &= - \rv{J}_m, \nonumber
\end{align}
according to \eqref{eq:bivectorfield}.  These are the familiar Maxwell's equations that include both electric and magnetic charges \cite{Dirac1931,Dirac1948,Cabibbo1962,Schwinger1966b,Rohrlich1966,Zwanziger1968,Zwanziger1971,Han1971,Mignani1975,Deser1976,Deser1982,Gambini1979,
Gambini1980,Schwartz1994,Pasti1995,Singleton1995,Singleton1996,Khoudeir1996,Kato2002,Shnir2005}.  Similarly, the continuity equation for the sources \eqref{eq:currentcontinuity} produces the two equations
\begin{align}
  \cc\partial_0 \rho_e + \rv{\nabla}\cdot\rv{J}_e &= 0, & \cc\partial_0 \rho_m + \rv{\nabla}\cdot\rv{J}_m &= 0,
\end{align}
which each have the familiar form \cite{Jackson1999}.


\subsection{Vector potentials revisited}


If $\bv{F}$ is generated by a vector potential, $\bv{F} = \nabla z = \nabla(a_e + a_m I)$, then Maxwell's equation \eqref{eq:maxwellsources} is equivalent to a complex vector wave equation with a source
\begin{equation}
  \nabla^2 z = j.
\end{equation}
In this form it becomes clear that the electric potential $a_e$ couples directly to the electric source $j_e$, while the magnetic potential $a_m$ couples directly to the magnetic source $j_m$.  Hence the total complex field $\bv{F} = (\nabla\wedge a_e) + (\nabla\wedge a_m)I = \bv{F}_e + \bv{F}_m I$ can be equivalently constructed by solving the two independent equations
\begin{align}
  \nabla \bv{F}_e &= \nabla^2 a_e = j_e, & \nabla \bv{F}_m = \nabla^2 a_m &= j_m
\end{align}
that each have the same form as \eqref{eq:maxwellsources}, but involve only non-complex vector fields and bivectors of fixed signature.  Evidently the constituent fields $\bv{F}_e$ and $\bv{F}_m$ are \emph{noninteracting, completely independent, and distinguished only by how they couple to distinct types of electromagnetic charge}.  In particular, note how the $e$ and $m$ subscripts for all quantities are consistently matched; for traditional electromagnetism that considers only electric charges, the ``magnetic'' terms (with subscripts $m$) can thus be neglected.

In terms of a relative frame with $a_e = (\phi_e + \rv{A})\gamma_0$ and $j_e = (\cc\rho_e + \rv{J}_e)\gamma_0$, the wave equation $\nabla^2 a_e = j_e$ expands to
\begin{align}\label{eq:potentialwavesraw}
  (\partial_0^2 - \rv{\nabla}^2)(\phi_e + \rv{A})\gamma_0 = (\cc\rho_e + \rv{J}_e)\gamma_0,
\end{align}
which further splits into the two wave equations \cite{Jackson1999}
\begin{align}\label{eq:potentialwavesdecoupled}
  (\partial_0^2 - \rv{\nabla}^2) \phi_e &= \cc\rho_e, & (\partial_0^2 - \rv{\nabla}^2)\rv{A} = \rv{J}_e. 
\end{align}
These two equations are implicitly coupled by the Lorenz-FitzGerald condition $\nabla\cdot a_e = \partial_0 \phi_e + \rv{\nabla}\cdot\rv{A} = 0$, which can be applied to the first equation in order to produce the modified form 
\begin{align}\label{eq:potentialwaves}
  \rv{\nabla}^2 \phi_e + \partial_0(\rv{\nabla}\cdot\rv{A}) &= -\cc\rho_e, & (\partial_0^2 - \rv{\nabla}^2)\rv{A} = \rv{J}_e. 
\end{align}
A similar split also applies to the equation $\nabla^2 a_m = j_m$ involving the magnetic potential.


\subsection{Lorentz force}


Similar to the appearance of Maxwell's equation \eqref{eq:maxwellsources}, the corresponding Lorentz force law arises quite naturally from the structure of spacetime itself.  To see this, consider what happens when the field $\bv{F} = \rv{E} + \rv{B}I$ is contracted with a 4-vector $w = (w_0 + \rv{w})\gamma_0$.  According to \eqref{eq:dotwedgebivector}, we have the general structure
\begin{align}\label{eq:fieldvectorcontraction}
  \bv{F}\cdot w &= \frac{1}{2}[\bv{F}w - w\bv{F}], \\
  &= \frac{1}{2}[(\rv{E} + \rv{B}I)(w_0 + \rv{w}) + (w_0 + \rv{w})(\rv{E} - \rv{B}I)]\gamma_0, \nonumber \\
  &= (\rv{E}\cdot\rv{w})\gamma_0 + [w_0\rv{E} + \rv{w}\times\rv{B}]\gamma_0. \nonumber 
\end{align}

Now consider what happens if $w$ is a proper velocity vector $w = dx/d\tau$, which implies $w^2 = \cc^2$.  Inserting the spacetime split of $x$ yields $dx/d\tau = d/d\tau[\cc t + \rv{x}]\gamma_0 = [d(\cc t)/d\tau + d\rv{x}/d\tau]\gamma_0$.  Denoting $dt/d\tau = \gamma$ as the time dilation factor, we see that $d\rv{x}/d\tau = \gamma d\rv{x}/dt = \gamma \rv{v}$ is the scaled relative velocity $\rv{v}$.  From $w^2 = \cc^2 = w_0^2 - |\rv{w}|^2 = \gamma^2(\cc^2-|\rv{v}|^2)$ we find the standard formula $\gamma = (1-|\rv{v}/\cc|^2)^{-1/2}$.  Hence, we have the relative components
\begin{align}\label{eq:fourvelocity}
  w_0 &= \gamma \cc = \frac{\cc}{\sqrt{1-|\rv{v}/\cc|^2}}, & \rv{w} &= \gamma \rv{v},
\end{align}
as well as the elegant and useful relations \cite{Doran2007}
\begin{align}
  \gamma &= \frac{dt}{d\tau} = \frac{w}{\cc}\cdot\gamma_0, & \frac{d\rv{x}}{d\tau} &= w\wedge\gamma_0, & \frac{\rv{v}}{\cc} &= \frac{w\wedge\gamma_0}{w\cdot\gamma_0}.
\end{align}

Inserting \eqref{eq:fourvelocity} into \eqref{eq:fieldvectorcontraction} produces the suggestive relation
\begin{align}
  \bv{F}\cdot w &= \gamma \cc\left[\rv{E}\cdot\rv{v}\right]\gamma_0 + \gamma\left[\cc\rv{E} + \rv{v}\times\rv{B}\right]\gamma_0.
\end{align}
Inverting factors and multiplying by a scalar charge $q_0$ then produces
\begin{align}\label{eq:lorentzforcerelative}
  [\bv{F}\cdot (q_0w)]\frac{d\tau}{dt}\gamma_0 &= q_0\cc\,\rv{E}\cdot\rv{v} + q_0\left[\cc\rv{E} + \rv{v}\times\rv{B}\right].
\end{align}
The vector part on the right hand side is precisely\footnote{The choice of units for $\rv{E}$, $\rv{B}$, and $q_0$ will induce appropriate scaling factors (see Table~\ref{tab:lorentz}).} the \emph{Lorentz force} $d\rv{p}/dt$ in a relative frame with $\rv{p} = m\rv{v}$, while the scalar part on the right hand side is precisely the corresponding rate of work $d\mathcal{E}/d(\cc t)$ with $\mathcal{E} = \gamma m \cc^2$. It follows that the simple contraction of $\bv{F}$ and a 4-vector current $q_0w$ produces the proper Lorentz force
\begin{align}\label{eq:lorentzforce}
  \frac{dp}{d\tau} &= \bv{F}\cdot (q_0w),
\end{align}
where $p = (\mathcal{E}/\cc + \rv{p})\gamma_0$ is the proper 4-vector energy-momentum.

\begin{table}
  \centering
  \begin{tabular}{l l}
    \hline
\noalign{\vskip 2mm} 
    \multicolumn{2}{c}{\textbf{Lorentz Force Law}} \\
\noalign{\vskip 2mm} 
    \hline \\
    Proper Force Law: & $\boxed{\frac{dp}{d\tau} = \mean{\bv{F} j}_1} = \frac{1}{2}\left[\bv{F} j + (\bv{F} j)^\sim\right] = q_e\,\bv{F}\cdot w + q_m\,(\bv{F}\wedge w)I$ \\
    \\
    & $\displaystyle p \equiv mw, \quad j \equiv qw, \quad q \equiv q_0 e^{\theta I} = q_e + q_m I$ \\
    \\
    Proper Velocity: & $\displaystyle w \equiv \frac{dx}{d\tau} = \gamma(\cc + \rv{v})\gamma_0, \quad w^2 = \cc^2$ \\
    \\
    & $\displaystyle \gamma\, \cc \equiv \frac{dt}{d\tau} = w \cdot \gamma_0 = \cc(1 - |\rv{v}/\cc|^2)^{-1/2}$ \\
    \\
    & $\displaystyle \gamma\, \rv{v} \equiv \frac{d\rv{x}}{d\tau} = w \wedge \gamma_0$ \\
    \\
    \hline
    \\
    Relative Force Law: & $\displaystyle\frac{dp}{d\tau}\frac{d\tau}{dt}\gamma_0 = \frac{d\mathcal{E}}{d(\cc t)} + \frac{d\rv{p}}{dt}$ \\
    \\
    & $\displaystyle\mathcal{E} \equiv \gamma m \cc^2, \quad \rv{p} \equiv m\rv{v}, \quad q_e = q_0\,\cos\theta, \quad q_m = q_0\,\sin\theta$ \\
    \\
    & $\displaystyle \frac{d\mathcal{E}}{d(\cc t)} = q_e \cc\,\rv{E}\cdot\rv{v} + q_m \cc\,\rv{B}\cdot \rv{v} $ \\
    \\
    & $\displaystyle \frac{d\rv{p}}{dt} = q_e \cc\,\rv{E} + q_m \cc\,\rv{B} + q_e \,\rv{v}\times\rv{B} - q_m\, \rv{v}\times\rv{E}$ \\
    \\
    Relative Velocity: & $\displaystyle \frac{\rv{v}}{\cc} \equiv \frac{w \wedge \gamma_0}{w \cdot \gamma_0}, \quad \gamma \equiv \frac{w}{\cc} \cdot \gamma_0$ \\ 
    \\
    \hline
    \\
    {$\begin{aligned} & \text{SI Units:} \\ & (\cc\equiv 1/\sqrt{\epsilon_0\mu_0}) \end{aligned}$} & {$\begin{aligned}\rv{E} &\mapsto \sqrt{\epsilon_0}\,\rv{E}, & \rv{B} &\mapsto \rv{B}/\sqrt{\mu_0}, & q_e &\mapsto \sqrt{\mu_0}\,q_e, & q_m &\mapsto \sqrt{\mu_0}\,q_m/\cc \end{aligned}$} \\
    \\
    CGS Units: & {$\begin{aligned}\rv{E} &\mapsto \rv{E}/\sqrt{4\pi}, &\quad \rv{B} &\mapsto \rv{B}/\sqrt{4\pi}, &\quad q_0 &\mapsto \sqrt{4\pi}\,q_0/\cc \end{aligned}$} \\
    \\
    \hline 
 \end{tabular}
 \caption[Lorentz Force Law]{The Lorentz Force Law, as expressed in spacetime algebra.  We show the proper (frame-independent) formulation that supports both electric and magnetic charges, combined as a single complex charge $q = q_e + q_m I$.  We also show the expansion of this force law in terms of a particular reference frame $\gamma_0$.  Notably, the proper and relative velocities are related by elegant contraction formulas in spacetime algebra.}
 \label{tab:lorentz}
\end{table}

We can expand upon this result by considering what happens if the vector current $q_0w$ is allowed to be a \emph{complex} vector current $j = q_0w\exp(\theta I)$ similar to \eqref{eq:maxwellsources}.  Since $\bv{F}\cdot(w I) = (\bv{F}\wedge w)I$ is the proper contraction that produces a vector, we must also consider the wedge of the bivector field with the velocity $w$
\begin{align}\label{eq:fieldvectorwedge}
  \bv{F}\wedge w &= \frac{1}{2}[\bv{F}w + w\bv{F}], \\
  &= \frac{1}{2}[(\rv{E} + \rv{B}I)(w_0 + \rv{w}) - (w_0 + \rv{w})(\rv{E} - \rv{B}I)]\gamma_0, \nonumber \\
  &= [\rv{B}\cdot\rv{w}]\gamma_0I^{-1} + [w_0\rv{B} - \rv{w}\times\rv{E}]\gamma_0I^{-1}. \nonumber 
\end{align}
The total contraction of $\bv{F}$ with $j = q_0 w \exp(\theta I) = j_e + j_mI$ then has the form
\begin{align}\label{eq:lorentzforcecomplex}
  \frac{dp}{d\tau} &=  \mean{\bv{F} j}_1 = \bv{F}\cdot j_e + (\bv{F}\wedge j_m)I = \frac{1}{2}\left[\bv{F}j + (\bv{F}j)^\sim\right],
\end{align}
where the grade-1-projection makes the expression particularly elegant, while the last equality follows from the reversion properties $\widetilde{\bv{F}} = -\bv{F}$ and $\widetilde{j} = j^*$ (recall Sections~\ref{sec:multivectors}, \ref{sec:reversion} and \ref{sec:conjugation}).  We summarize this proper Lorentz force that admits a complex source $j$ in Table~\ref{tab:lorentz} for reference.

We must be careful to be fully consistent, however.  That is, the physical momentum $p$ is a vector quantity that can be decomposed into a mass-scaled velocity $p = m w$ (at least for classical point particles).  Thus, if we make the current complex $q_0w\mapsto q_0w\exp(\theta I) = j$, as in \eqref{eq:lorentzforcecomplex}, then the \emph{charge} must become complex
\begin{align}
  q = q_0 \exp(\theta I) = q_e + q_m I,
\end{align}
leading to distinct electric and magnetic parts of the charge 
\begin{align}
  q_e &= q_0\cos\theta, & q_m &= q_0\sin\theta,
\end{align}
such that the observed charges have a well-defined ratio $q_m/q_e = \tan\theta$.  (We will clarify the intimate connection between this ratio and the dual symmetry of the field in the next section.)

Inserting the relation $j = w[q_0\exp(\theta I)]$ into \eqref{eq:lorentzforcecomplex} with $w = dx/d\tau$ as before yields the relative expression
\begin{align}
  \frac{dp}{d\tau}\frac{d\tau}{dt}\gamma_0 &= \frac{d\mathcal{E}}{d(\cc t)} + \frac{d\rv{p}}{dt},
\end{align}
where $\mathcal{E} = \gamma m \cc^2$, $\rv{p} = m\rv{v}$, and where
\begin{subequations}\label{eq:lorentzforcemonopole}
\begin{align}
  \frac{d\mathcal{E}}{d(\cc t)} &= q_e \cc\,\rv{E}\cdot\rv{v} + q_m \cc\,\rv{B}\cdot\rv{v}, \\
  \frac{d\rv{p}}{dt} &= q_e \cc\,\rv{E} + q_m \cc\,\rv{B} + q_e\, \rv{v}\times\rv{B} - q_m \, \rv{v}\times\rv{E}. 
\end{align}
\end{subequations}
After restoring units, these equations are precisely the correction to the Lorentz force law that properly accounts for the addition of magnetic monopoles \cite{Dirac1931,Dirac1948,Cabibbo1962,Schwinger1966b,Rohrlich1966,Zwanziger1968,Zwanziger1971,Han1971,Mignani1975,Deser1976,Deser1982,Gambini1979,Gambini1980,Schwartz1994,Pasti1995,Singleton1995,Singleton1996,Khoudeir1996,Kato2002,Shnir2005}.  This monopole description arises directly from the complexity of the current $j$ and asserting that the total field $\bv{F}$ should couple to this complex current.  We summarize these relative expressions in Table~\ref{tab:lorentz} for reference, along with both the SI and CGS unit conventions.


\subsection{Dual symmetry revisited}\label{sec:dualsymmetrysource}


We are now in a position to consider the meaning of dual symmetry in the presence of an electromagnetic source current $j$.  The two equations that we must examine are Maxwell's equation \eqref{eq:maxwellsources} and the Lorentz force law \eqref{eq:lorentzforcecomplex}, which we reproduce here for convenience in terms of both the field $\bv{F}$ and the vector potential $z$
\begin{align}
  \nabla\bv{F} &= \nabla^2 z = j, & \frac{dp}{d\tau} &= \mean{\bv{F}j}_1 = \mean{(\nabla z) j}_1.
\end{align}
Keep in mind that $j = qw$ and $p = mw$ for classical point particles, where $w = dx/d\tau$ is the proper velocity. 

It is now apparent that both these equations are invariant under a simultaneous phase rotation of both the vector potential $z$ and the current $j$, using the spinor $\psi = \exp(-\theta I / 2)$ from \eqref{eq:dualsymmetry}
\begin{align}
  z &\mapsto \psi z\psi^*, & j &\mapsto \psi j \psi^*.
\end{align}
Specifically, note that since $\psi z \psi^* = z\exp(\theta I)$ and $\psi j \psi^* = j \exp(\theta I)$, Maxwell's equation becomes
\begin{align}
  (\nabla^2 z)\exp(\theta I) = j\exp(\theta I),
\end{align}
so the phase factors cancel.  Similarly, the Lorentz force law is also invariant
\begin{align}
  \mean{(\nabla ze^{\theta I}) je^{\theta I}}_1 = \mean{(\nabla z)je^{-\theta I}e^{\theta I}}_1 = \mean{(\nabla z) j}_1.
\end{align}

Thus, while it may initially appear that the dual symmetry of Maxwell's equation \eqref{eq:maxwellsources} is broken in the presence of a source current $j$, this is only strictly true if the description of the current is left unchanged.  An equivalent physical picture is obtained if the current is also changed (via the same phase rotation of the coupling charge $q$).  As such, choosing electric charges to be the primary type of charge is an arbitrary convention. \emph{Regardless of how one describes the fundamental charges, the same forces (and thus the same physics) will still be described, as long as the fields are correspondingly transformed} \cite{Shnir2005}.

In addition to this equivalence of representation, there is another important subtlety that appears when we consider a field solution of Maxwell's equation $\nabla \bv{F} = j$.  Specifically, any solution to this equation will be the combination of a homogeneous solution $\bv{F}_0$ of the vacuum equation $\nabla \bv{F}_0 = 0$ and a particular solution that includes the source $j$.  We can consider these two independent solutions to correspond to distinct vector potentials $z = z_0 + z_s$ that satisfy distinct equations of motion
\begin{align}
  \nabla^2 z_0 &= 0, & \nabla^2 z_s &= j.
\end{align}
However, as noted above, we can always rotate the source $j$ to become purely \emph{electric} $j_e$ using some phase rotation $\psi = \exp(-\theta I/2)$ 
\begin{align}
  \nabla^2 (\psi z_s \psi^*) &= \psi j \psi^* = j_e.
\end{align}
After this rotation, there is no magnetic source, so the trivector part of the rotated potential $\mean{\psi z_s \psi^*}_3$ must satisfy the vacuum Maxwell equation, and therefore should be absorbed into the homogeneous equation as part of the vacuum potential $z_0$.  That is, the rotated source potential is a \emph{pure (electric) vector} potential $a_{s,e}$, while the homogeneous vacuum potential will still be generally \emph{complex}
\begin{align}
  \psi z_s \psi^* &= a_{s,e}, & \psi z_0 \psi^* &= a_{0,e} + a_{0,m} I.
\end{align}
Thus, there is a profound difference between the independent source and vacuum parts of a field solution.  The source piece must have a description as a pure vector potential matched with an electric source; this pair can then be jointly rotated into equivalent descriptions with both electric and magnetic sources as desired.  In contrast, the vacuum piece is always manifestly dual-symmetric and thus admits an irreducibly complex vector potential description (such that no phase rotation exists that will make the vacuum potential a pure vector).  Notably, this decomposition of the potential into distinct vacuum and matter parts is important in treatments of the quantized electromagnetic field \cite{Cohen1997}.  We will return to this decomposition in Sections~\ref{sec:constituentconstraint}, \ref{sec:symmetry} and Figure~\ref{fig:symmetrybreaking}.


\subsection{Fundamental conservation laws}\label{sec:symtensor}


Before turning our attention to macroscopic fields, we first provide a brief but enlightening derivation of the symmetric energy-momentum stress tensor\footnote{This symmetric tensor is also known as the Belinfante--Rosenfeld stress-energy tensor, or the kinetic stress-energy tensor \cite{Soper1976,Thide2014,Bliokh2013,Leader2014}.}, as well as its associated angular momentum tensor \cite{Thide2014,Soper1976}.  These derivations will impose general constraints on the form of any local modifications to these tensors, which will become important in their Lagrangian derivation via Noether's theorem in Section~\ref{sec:lagrangian}.


\subsubsection{Energy-momentum stress tensor}\label{sec:symemstress}


Conceptually, the energy-momentum tensor is a function $\underbar{T}(b)$ that maps a constant vector direction $b$ into the \emph{flux} of energy-momentum that is passing through the hypersurface orthogonal to $b$.  This flux of energy-momentum should satisfy a continuity relation 
\begin{align}\label{eq:energymomentumcontinuity}
  b\cdot \frac{dp}{d\tau} + \nabla\cdot\underbar{T}(b) = 0,
\end{align}
stating that the rate of change of the energy-momentum (the Lorentz force \eqref{eq:lorentzforcecomplex}) projected along the direction of $b$ should correspond to the divergence of the energy-momentum flux through the surface orthogonal to $b$.  (This statement is analogous to how the current continuity equation $\cc\partial_0\rho + \rv{\nabla}\cdot\rv{J} = 0$ from \eqref{eq:currentcontinuity} relates the rate of change of the charge density to the divergence of the charge flux, or current.)

We can simplify \eqref{eq:energymomentumcontinuity} using the \emph{adjoint} tensor\footnote{The underbar and overbar notation for a tensor and its adjoint is a useful notational device \cite{Doran2007}, so we adopt it here.} $\overline{T}$ that corresponds to the tensor $\underbar{T}$ according to the vector dot product relation $d \cdot \underbar{T}(b) = \overline{T}(d) \cdot b$.  Applying this adjoint relation to \eqref{eq:energymomentumcontinuity} yields \cite{Doran2007}
\begin{align}
  b\cdot\left[\frac{dp}{d\tau} + \overline{T}(\nabla)\right] = 0,
\end{align}
keeping in mind that $\nabla$ still differentiates the adjoint tensor $\overline{T}$.  Since $b$ is arbitrary, we can dispense with it to find a direct constraint equation for the adjoint, in which we can substitute the Lorentz force law \eqref{eq:lorentzforcecomplex} to find
\begin{align}
  \overline{T}(\nabla) = -\frac{dp}{d\tau} = -\mean{\bv{F}j}_1.
\end{align}
After substituting Maxwell's equation \eqref{eq:maxwellsources} for the source $j$, we obtain 
\begin{align}
  \overline{T}(\nabla) = -\frac{1}{2}\left[\bv{F}(\nabla\bv{F}) + (\widetilde{\bv{F}}\nabla)\widetilde{\bv{F}}\right] = -\frac{1}{2}\bv{F}\nabla\bv{F}.
\end{align}
Therefore, we can infer that the adjoint of the energy-momentum stress tensor must have the elegant quadratic form
\begin{align}\label{eq:symemstress}
  \overline{T}_{\text{sym}}(b) = -\frac{1}{2}\bv{F}b\bv{F} = \frac{1}{2}\bv{F}b\widetilde{\bv{F}} = \frac{1}{2}\bv{f}b\widetilde{\bv{f}},
\end{align}
which is the simplest dual-symmetric form that involves the field bivector $\bv{F} = \bv{f}\exp(\varphi I)$.  This form is also \emph{symmetric} (in the sense of the adjoint $\overline{T}_{\text{sym}} = \underbar{T}_{\text{sym}}$) since 
\begin{align}
  \overline{T}_{\text{sym}}(d)\cdot b = \mean{\overline{T}_{\text{sym}}(d)b}_0 = -\mean{\bv{F}d\bv{F}b}_0 = -\mean{d\bv{F}b\bv{F}}_0 = d\cdot\overline{T}_{\text{sym}}(b)
\end{align}
by the cyclic property of the scalar projection \eqref{eq:cyclicscalar}.  Any proposed modification to this fundamental symmetric tensor must have a vanishing divergence, according to the general continuity equation \eqref{eq:energymomentumcontinuity}.  This symmetric tensor is summarized for reference in Table~\ref{tab:symtensors}.

To verify that this symmetric energy-momentum stress tensor gives us the results we expect, we compute the energy-momentum flux through the surface perpendicular to $\gamma_0$ (restoring the SI unit scaling factors from \eqref{eq:emfield}),
\begin{align}\label{eq:symemflux}
  \overline{T}_{\text{sym}}(\gamma_0) &= \frac{1}{2}\bv{F}\gamma_0\widetilde{\bv{F}} = \frac{1}{2}\bv{F}\bv{F}^\dagger\gamma_0, \\
  &= \frac{1}{2}\left[\rv{E} + \rv{B}I\right]\left[\rv{E} - \rv{B}I\right]\gamma_0, \nonumber \\
  &= \left[\frac{1}{2}\left(|\rv{E}|^2 + |\rv{B}|^2\right) + \rv{E}\times\rv{B}\right]\gamma_0 \equiv (\varepsilon + \rv{P})\gamma_0. \nonumber 
\end{align}
The first term is the energy density $\varepsilon$ of the field, which is relative to a particular frame $\gamma_0$, as we anticipated in Section \ref{sec:waves}.  The second term is the Poynting vector $\rv{P}$ that gives the corresponding relative field momentum \cite{Doran2007}.  The two together construct the proper field energy-momentum $\underbar{T}(\gamma_0) = (\varepsilon + \rv{P})\gamma_0$ that will be measured by an observer traveling along a worldline pointing in the $\gamma_0$ direction.  

As another example, we can compute the energy-momentum flux through a surface perpendicular to $\gamma_i$ ($i=1,2,3$) using $\gamma_i = \gamma_i\gamma_0^2 = \rv{\sigma}_i\gamma_0$,
\begin{align}
  \overline{T}_{\text{sym}}(\gamma_i) &= \frac{1}{2}\bv{F}\gamma_i\widetilde{\bv{F}} = \frac{1}{2}\bv{F}\rv{\sigma}_i\bv{F}^\dagger\gamma_0, \\
  &= \frac{1}{2}\left[\rv{E} + \rv{B}I\right]\rv{\sigma}_i\left[\rv{E} - \rv{B}I\right]\gamma_0, \nonumber \\
  &= \left[\frac{1}{2}\left(\rv{E}\rv{\sigma}_i\rv{E} + \rv{B}\rv{\sigma}_i\rv{B}\right) + \frac{1}{2}\left(\rv{E}\rv{\sigma}_i\rv{B} - \rv{B}\rv{\sigma}_i\rv{E}\right)I^{-1}\right]\gamma_0. \nonumber
\end{align}
This form emphasizes the bilinear nature of the tensor with respect to the fields.  The vector products can be expanded into a more familiar form by using the identity
\begin{align}
  \rv{E}\rv{\sigma}_i\rv{B} &= \rv{B}(\rv{E}\cdot\rv{\sigma}_i) + \rv{E}(\rv{B}\cdot\rv{\sigma}_i) - (\rv{E}\cdot\rv{B})\rv{\sigma}_i - (\rv{E}\times\rv{B})\cdot\rv{\sigma}_i I,
\end{align}
to produce
\begin{align}\label{eq:symemstressmaxwell}
  \overline{T}_{\text{sym}}(\gamma_i) &= -\left[\rv{E}\times\rv{B}\cdot\rv{\sigma}_i\right]\gamma_0 \\
  &\quad + \left[\rv{E}(\rv{E}\cdot\rv{\sigma}_i) + \rv{B}(\rv{B}\cdot\rv{\sigma}_i) - \frac{1}{2}\left(|\rv{E}|^2 + |\rv{B}|^2\right)\rv{\sigma}_i\right]\gamma_0. \nonumber
\end{align}
The first term is the projection of the Poynting vector $\rv{P}$ onto the direction $\rv{\sigma}_i$.  The second term is the contraction of the Maxwell stress tensor along the direction $\rv{\sigma}_i$.

\begin{table}
  \centering
  \begin{tabular}{l l}
    \hline
\noalign{\vskip 2mm} 
    \multicolumn{2}{c}{\textbf{Symmetric Energy-Momentum Tensor}} \\
\noalign{\vskip 2mm} 
    \hline \\
    Spacetime Algebra: & $\boxed{\overline{T}_{\text{sym}}(b) = \underbar{T}_{\text{sym}}(b) = \frac{1}{2}\bv{F}b\widetilde{\bv{F}}}$ \\
    \\
    & $\qquad\qquad\quad \Downarrow$ \\
    \\
    & $\displaystyle \overline{T}_{\text{sym}}(\gamma_0) = \frac{1}{2}\bv{F}\gamma_0\widetilde{\bv{F}} = \frac{1}{2}\bv{F}\bv{F}^\dagger\gamma_0 = (\varepsilon + \rv{P})\gamma_0$ \\
    \\
    & $\displaystyle \overline{T}_{\text{sym}}(\gamma_i) = \frac{1}{2}\bv{F}\rv{\sigma}_i\bv{F}^\dagger\gamma_0 = [\rv{P}\cdot\rv{\sigma}_i + \rv{E}(\rv{E}\cdot\rv{\sigma}_i) + \rv{B}(\rv{B}\cdot\rv{\sigma}_i) - \varepsilon\,\rv{\sigma}_i]\gamma_0$ \\
    \\
    & $\displaystyle \varepsilon = \frac{1}{2}\left(|\rv{E}|^2 + |\rv{B}|^2\right), \quad \rv{P} = \rv{E}\times\rv{B}$ \\
    \\
    \\
    Tensor Components: & $\displaystyle T^{\mu\nu}_{\text{sym}} = T^{\nu\mu}_{\text{sym}} = \sum_\alpha F^{\mu\alpha}F\indices{_\alpha^\nu} - \frac{1}{4}\eta^{\mu\nu}\sum_{\alpha,\beta}F^{\alpha\beta}F_{\beta\alpha} $ \\
    \\
    \hline
\noalign{\vskip 2mm} 
    \multicolumn{2}{c}{\textbf{Associated Angular Momentum Tensor}} \\
\noalign{\vskip 2mm} 
    \hline \\
    \\
    Spacetime Algebra: & $\boxed{\overline{M}_{\text{sym}}(b) = x \wedge \overline{T}_{\text{sym}}(b)} \qquad \boxed{\underbar{M}_{\text{sym}}(\bv{R}) = \underbar{T}_{\text{sym}}(\bv{R}\cdot x)}$ \\
    \\
    & $\qquad\qquad\quad \Downarrow$ \\
    \\
    & $\displaystyle \overline{M}_{\text{sym}}(\gamma_0) = -\rv{N} + \rv{J}I^{-1}$, \\
    \\
    & $\displaystyle \rv{N} = (\cc t)\rv{P} - \varepsilon\,\rv{x}, \quad \rv{J} = \rv{x}\times\rv{P}$ \\
    \\
    \\
    Tensor Components: & $\displaystyle M_{\text{sym}}^{\alpha\beta\gamma} = x^{\alpha}\, T_{\text{sym}}^{\beta\gamma} - x^{\beta}\,T_{\text{sym}}^{\alpha\gamma}$ \\
    \\
    \hline 
 \end{tabular}
 \caption[Symmetric tensors]{The symmetric energy-momentum tensor and its associated angular momentum tensor, shown in both spacetime algebra and tensor component notation.  Notably, spacetime algebra emphasizes that the energy-momentum tensor $\overline{T}_{\text{sym}}$ is a simple quadratic form of the field $\bv{F}$, which maps a vector direction $b$ into the energy-momentum flux passing through a surface orthogonal to $b$, exactly as the energy-momentum current for the Dirac equation of the electron \cite{Berestetskii1982,Doran2007,Hiley2012}.  It is a symmetric tensor, meaning it equals its adjoint $\underbar{T}_{\text{sym}}$, and is easily expanded algebraically into the usual expressions containing the energy-density of the field $\varepsilon$, the Poynting vector $\rv{P}$, and the Maxwell stress tensor.  Similarly, the associated angular momentum tensor $\overline{M}_{\text{sym}}$ is a function that maps a vector direction $b$ into the angular momentum flux through a surface orthogonal to $b$, and expands to the usual boost-momentum $\rv{N}$ and rotational-angular-momentum $\rv{J}$ vectors for the field.  The adjoint of this tensor $\underbar{M}_{\text{sym}}$ is a function that maps a plane of rotation (bivector) $\bv{R}$ into an energy-momentum flux.}
 \label{tab:symtensors}
\end{table}


\subsubsection{Angular momentum tensor}\label{sec:symamstress}


Conceptually, angular momentum is a bivector that indicates a plane of rotation.  The rotation in question is that of a coordinate vector as it rotates to track movement associated with a linear momentum.  For a classical point particle, the plane segment containing this rotation is produced explicitly by a wedge product $\bv{M} = x\wedge p$ between a coordinate 4-vector $x$ and a 4-momentum $p$ \cite{Landau1975}.  Indeed, after introducing the spacetime splits $p = (\mathcal{E}/\cc + \rv{p})\gamma_0$, $x = (\cc t + \rv{x})\gamma_0$ we have
\begin{align}\label{eq:angularmomentum}
  x\wedge p = \frac{xp - px}{2} = [\mathcal{E}\rv{x}/\cc - (\cc t) \rv{p}] + \rv{x}\times\rv{p} I^{-1} \equiv -\rv{N} + \rv{L}I^{-1}.
\end{align}
The second term $\rv{L}I^{-1}$ of \eqref{eq:angularmomentum} involves $I$, and is part of the angular momentum that agrees with the usual notion of the relative angular momentum as the cross product $\rv{L}=\rv{x}\times\rv{p}$.  Within spacetime algebra, however, we see that this cross product is the Hodge-dual of the proper angular momentum bivector $\rv{L}I^{-1}$ that generates \emph{spatial rotations} as a Lorentz transformation, just as discussed in \eqref{eq:lorentzrotation} of Section \ref{sec:lorentz}.  In contrast, the first term $-\rv{N}$ of \eqref{eq:angularmomentum} is a bivector of opposite signature that generates \emph{boost rotations}, as discussed in \eqref{eq:lorentzboost} of Section \ref{sec:lorentz}.  The total invariant notion of angular momentum in spacetime includes both spatial rotation and boost components, since the 4-dimensional rotations in spacetime are Lorentz transformations that can be of either type.

With this intuition, the angular momentum tensor\footnote{Specifically, this is the Belinfante-Rosenfeld angular momentum tensor \cite{Soper1976,Bliokh2013,Leader2014,Thide2014}.} for the electromagnetic field is straightforward to obtain directly from the adjoint symmetric energy-momentum tensor $\overline{T}_{\text{sym}}(b)$.  As seen in the last section, this latter tensor produces an energy-momentum density associated with a particular direction $b$.  Hence, taking the wedge product of that energy-momentum density will produce its associated angular momentum density
\begin{align}
  \overline{M}_{\text{sym}}(b) &= x\wedge \overline{T}_{\text{sym}}(b).
\end{align}
As with the energy-momentum tensor, the angular momentum tensor $\overline{M}_{\text{sym}}$ is a function that maps a constant vector direction $b$ into the angular momentum flux passing through the hypersurface orthogonal to $b$.  This tensor is also summarized for reference in Table~\ref{tab:symtensors}.

Applying the angular momentum tensor to the timelike direction $\gamma_0$ produces the intuitive result
\begin{align}\label{eq:symamflux}
  \overline{M}_{\text{sym}}(\gamma_0) &= x\wedge \overline{T}_{\text{sym}}(\gamma_0) = [\varepsilon \rv{x} - (\cc t)\rv{P}] + \rv{x}\times\rv{P} I^{-1} \equiv -\rv{N} + \rv{J}I^{-1},
\end{align}
with the energy-density and Poynting vector 
as in \eqref{eq:angularmomentum}, and where $\rv{N}$ is the \emph{boost angular momentum} and $\rv{J}$ is the \emph{rotational angular momentum} of the electromagnetic field. 
According to commonly accepted notations, we denote the rotational angular momentum of the field as $\vec{J}$ (not to be confused with the charge current!), because it includes both the orbital ($\vec{L}$) and spin ($\vec{S}$) parts  \cite{Allen1999,Andrews2008,Torres2011,Andrews2013} (see Section~\ref{sec:canonicalAM} below).

Just as the energy-momentum tensor $\overline{T}_{\text{sym}}$ has an adjoint tensor $\underbar{T}_{\text{sym}}$, the angular momentum tensor $\overline{M}_{\text{sym}}$ also has an adjoint.  We can compute this adjoint by contracting the tensor with any pure bivector $\bv{R}$, which should be understood as the plane segment for a rotation (i.e., generating a Lorentz transformation as in Section~\ref{sec:lorentz})
\begin{align}
  \bv{R}\cdot \overline{M}_{\text{sym}}(b) = \mean{\bv{R}x\overline{T}_{\text{sym}}(b)}_0 = \mean{\underbar{T}_{\text{sym}}(\mean{\bv{R}x}_1)b}_0 = \underbar{T}_{\text{sym}}(\bv{R}\cdot x)\cdot b,
\end{align}
yielding the adjoint tensor
\begin{align}\label{eq:symamfluxadjoint}
  \underbar{M}_{\text{sym}}(\bv{R}) &= \underbar{T}_{\text{sym}}(\bv{R}\cdot x).
\end{align}
This adjoint tensor is a function that takes the plane of rotation $\bv{R}$, contracts it with the coordinate direction $x$ to find the orthogonal direction $b$ such that $\bv{R} = b\wedge x_\parallel = b x_\parallel$ (see the discussion in Section~\ref{sec:bivectorvector}), and then returns the energy-momentum flux $\underbar{T}_{\text{sym}}(b)$ through the hypersurface perpendicular to that direction.

As an example, applying $\underbar{M}_{\text{sym}}$ to a bivector $\bv{R}=\gamma_i\gamma_0 = \rv{\sigma}_i$ that generates boost rotations produces
\begin{align}
\underbar{M}_{\text{sym}}(\rv{\sigma}_i) = \underbar{T}_{\text{sym}}(\rv{\sigma}_i\cdot x) &= (ct)\,\underbar{T}_{\text{sym}}(\gamma_i) - x_i\,\underbar{T}_{\text{sym}}(\gamma_0), 
\end{align}
which is the correct energy-momentum flux through the hyperplane perpendicular to the part of the coordinate vector $x$ that lies in the boost rotation plane $\bv{R}=\rv{\sigma}_i$.


\subsection{Macroscopic fields}


The preceding discussion has assumed the free propagation of a (\emph{micro}scopic) field $\bv{F}$ through vacuum.  In the presence of a medium, however, it is useful to partition the total field into a \emph{macro}scopic distinction between a \emph{free} part $\bv{F}_f$ that is coupled to freely moving charges in the conduction bands of the medium, and a \emph{bound} part $\bv{F}_b$ that is coupled to the bound charges held rigid by the lattice structure of the medium \cite{Jackson1999,Landau1984}.  Such a partitioning of the fields and sources must have the general form
\begin{align}\label{eq:macroscopicsplit}
  \bv{F} &= \bv{F}_f + \bv{F}_b, & j &= j_f + j_b.
\end{align}

To see how the macroscopic fields in \eqref{eq:macroscopicsplit} are expressed in a relative frame, we consider the spacetime splits (using conventional (SI) units) 
\begin{align}
  \bv{F}_f &= \rv{D}/\sqrt{\epsilon_0} + \sqrt{\mu_0}\rv{H}I, & \bv{F}_b &= -\rv{\mathcal{P}}/\sqrt{\epsilon_0} + \sqrt{\mu_0}\rv{\mathcal{M}}I.
\end{align}
Here $\rv{D}$ is the usual electric \emph{displacement}, $\rv{H}$ is the macroscopic magnetic field, $\rv{\mathcal{P}}$ is the \emph{electric polarization} field of the medium, and $\rv{\mathcal{M}}$ is the \emph{magnetization} field of the medium\footnote{Note that we use calligraphic script here for the polarization and magnetization to disambiguate them from momentum and angular momentum in other sections.  They will only be used within this section.}.  From the definition \eqref{eq:macroscopicsplit} we see that the total electromagnetic field $\bv{F} = \sqrt{\epsilon_0}\,\rv{E} + \rv{B}I/\sqrt{\mu_0}$ is related to the macroscopic fields in the following way
\begin{align}\label{eq:macrofields}
  \epsilon_0 \rv{E} &= \rv{D} - \rv{\mathcal{P}},  & \frac{\rv{B}}{\mu_0} &= \rv{H} + \rv{\mathcal{M}},
\end{align}
which agrees with the standard definition of the electric polarization and magnetization vectors \cite{Jackson1999,Landau1984}.

Imposing Maxwell's equation \eqref{eq:maxwellsources} on the partitions in \eqref{eq:macroscopicsplit} then couples the fields and sources in a nontrivial way
\begin{align}\label{eq:macroscopicsplit2}
  \nabla \bv{F}_f &= j_f - (\nabla \bv{F}_b - j_b).
\end{align}
We can then \emph{define} the split \eqref{eq:macroscopicsplit} to force the final bracketed term of \eqref{eq:macroscopicsplit2} to vanish, which purposefully couples the free macroscopic field with only the \emph{free} charge-currents.  This definition forces the separation of the total field into two equations
\begin{align}\label{eq:maxwellsplitsources}
  \nabla \bv{F}_f &= j_f, &  \nabla \bv{F}_b &= j_b.
\end{align}
The first of these split equations is Maxwell's \emph{macroscopic} equation for free charges.  The second is a constraint equation that defines the \emph{electromagnetic polarization field} that is bound to the medium \cite{Jackson1999,Landau1984}.  Note that the bound charges $j_b$ will generally respond to the total field $\bv{F}$, so will shift around while staying confined to the lattice structure, creating an indirect coupling between the two split equations in \eqref{eq:maxwellsplitsources}.  We will return to this important point later.

In terms of a relative frame, the free fields $\rv{D}$ and $\rv{H}$ satisfy the macroscopic Maxwell's equations in \eqref{eq:maxwellsplitsources}
\begin{align}
  \rv{\nabla}\cdot\rv{D} &= \cc\rho_{e,f}, & \partial_0 \rv{D} - \rv{\nabla}\times\rv{H} &= - \rv{J}_{e,f}, \\
  \rv{\nabla}\cdot\rv{H} &= \cc\rho_{m,f}, & \partial_0 \rv{H} + \rv{\nabla}\times\rv{D} &= - \rv{J}_{m,f}, \nonumber
\end{align}
that involve only the \emph{free} charge-currents.  Similarly, the bound fields $\rv{\mathcal{P}}$ and $\rv{\mathcal{M}}$ satisfy the constraint equations from \eqref{eq:maxwellsplitsources}
\begin{align}\label{eq:maxwellsplitsources2}
  -\rv{\nabla}\cdot\rv{\mathcal{P}} &= \cc\rho_{e,b}, & -\partial_0 \rv{\mathcal{P}} - \rv{\nabla}\times\rv{\mathcal{M}} &= - \rv{J}_{e,b}, \\
  \rv{\nabla}\cdot\rv{\mathcal{M}} &= \cc\rho_{m,b}, & \partial_0 \rv{\mathcal{M}} - \rv{\nabla}\times\rv{\mathcal{P}} &= - \rv{J}_{m,b}, \nonumber
\end{align}
that involve only the \emph{bound} charge-currents.  

\begin{table}
  \centering
  \begin{tabular}{l l}
    \hline
\noalign{\vskip 2mm} 
    \multicolumn{2}{c}{\textbf{Macroscopic Maxwell's Equations}} \\
\noalign{\vskip 2mm} 
    \hline \\
    \textbf{Micro}scopic: 
    & {$\boxed{\nabla \bv{F} = j}  \qquad \boxed{\bv{F} = \rv{E} + \rv{B}I} $} \\ \\
    & {$\begin{aligned} \rv{\nabla}\cdot\rv{E} &= c\rho_e, & \quad \partial_0 \rv{E} - \rv{\nabla}\times\rv{B} &= -\rv{J}_e \\
      \rv{\nabla}\cdot\rv{B} &= c\rho_m, & \quad \partial_0 \rv{B} + \rv{\nabla}\times\rv{E} &= -\rv{J}_m 
      \end{aligned}$} \\
    \\
    \textbf{Macro}scopic: 
    & {$\boxed{\bv{F} = \bv{F}_f + \bv{F}_b}  \quad \boxed{\bv{F}_f = \rv{D} + \rv{H}I}  \quad \boxed{\bv{F}_b = -\rv{\mathcal{P}} + \rv{\mathcal{M}} I}$} \\
      \\
      & {$\begin{aligned} \rv{E} &= \rv{D} - \rv{\mathcal{P}} &\quad \rv{B} &= \rv{H} + \rv{\mathcal{M}} &\quad j &= j_f + j_b \end{aligned}$} \\
    \\
    Free fields: & $\boxed{\nabla\bv{F}_f = j_f}$ \\
    \\
    & {$\begin{aligned} \rv{\nabla}\cdot\rv{D} &= \cc\rho_{e,f}, & \quad \partial_0 \rv{D} - \rv{\nabla}\times\rv{H} &= -\rv{J}_{e,f} \\
      \rv{\nabla}\cdot\rv{H} &= \cc\rho_{m,f}, & \quad \partial_0 \rv{H} + \rv{\nabla}\times\rv{D} &= -\rv{J}_{m,f} \end{aligned}$} \\
    \\
    Bound fields: & $\boxed{\nabla \bv{F}_b = j_b}$ \\
    \\
    & {$\begin{aligned} -\rv{\nabla}\cdot\rv{\mathcal{P}} &= \cc\rho_{e,b}, &\quad -\partial_0 \rv{\mathcal{P}} - \rv{\nabla}\times\rv{\mathcal{M}} &= -\rv{J}_{e,b} \\
      \rv{\nabla}\cdot\rv{\mathcal{M}} &= \cc\rho_{m,b}, & \quad \partial_0 \rv{\mathcal{M}} - \rv{\nabla}\times\rv{\mathcal{P}} &= -\rv{J}_{m,b} \end{aligned}$} \\
    \\
    Reduced set: & {$\boxed{\nabla\cdot \bv{F}_f = j_f} \quad \boxed{\nabla \wedge \bv{F} = j_m I} \quad \boxed{\nabla\cdot \bv{F}_b = j_b}$} \\
    \\
    & {$\begin{aligned}\rv{\nabla}\cdot\rv{D} &= \cc\rho_{e,f}, &\quad \partial_0 \rv{D} - \rv{\nabla}\times\rv{H} &= -\rv{J}_{e,f} \\
      \rv{\nabla}\cdot\rv{B} &= \cc\rho_m, &\quad \partial_0 \rv{B} + \rv{\nabla}\times\rv{E} &= -\rv{J}_m \\
      -\rv{\nabla}\cdot\rv{\mathcal{P}} &= \cc\rho_{e,b}, &\quad -\partial_0 \rv{\mathcal{P}} - \rv{\nabla}\times\rv{\mathcal{M}} &= -\rv{J}_{e,b} \end{aligned}$} \\
    \\
    \hline
    \\
    {$\begin{aligned} & \text{SI Units :} \\ & (\cc \equiv 1/\sqrt{\epsilon_0\mu_0})\end{aligned}$} & {$\begin{aligned}\rv{E} &\mapsto \sqrt{\epsilon_0}\,\rv{E}, & \rv{B} &\mapsto \rv{B}/\sqrt{\mu_0}, & j_e &\mapsto \sqrt{\mu_0}\,j_e, & j_m &\mapsto \sqrt{\mu_0}\,j_m/\cc \\
      \rv{D} &\mapsto \rv{D}/\sqrt{\epsilon_0}, & \rv{H} &\mapsto \sqrt{\mu_0}\rv{H}, & \rv{\mathcal{P}} &\mapsto \rv{\mathcal{P}}/\sqrt{\epsilon_0}, & \rv{\mathcal{M}} &\mapsto \sqrt{\mu_0}\,\rv{\mathcal{M}} \end{aligned}$} \\
    \\
    CGS Units: & {$\begin{aligned}\rv{E} &\mapsto \rv{E}/\sqrt{4\pi}, & \rv{B} &\mapsto \rv{B}/\sqrt{4\pi}, & j &\mapsto \sqrt{4\pi}\,j/\cc & & \\
      \rv{D} &\mapsto \rv{D}/\sqrt{4\pi}, & \rv{H} &\mapsto \rv{H}/\sqrt{4\pi}, & \rv{\mathcal{P}} &\mapsto \sqrt{4\pi}\,\rv{\mathcal{P}}, & \rv{\mathcal{M}} &\mapsto \sqrt{4\pi}\,\rv{\mathcal{M}} \end{aligned}$} \\
    \\
    \hline 
 \end{tabular}
 \caption[Maxwell's macroscopic equations]{Macroscopic Maxwell's equations. The total microscopic field $\bv{F}$ is partitioned into a free part $\bv{F}_f$ and a bound part $\bv{F}_b$, each coupled to free $j_f$ and bound $j_b$ charge-currents in a medium, respectively.  The bound field contains the electric polarization $\rv{\mathcal{P}}$ and magnetization $\rv{\mathcal{M}}$ densities of the medium, while the free field contains the electric displacement $\rv{D}$ and magnetic field strength $\rv{H}$.  Out of the total set of equations implied by Maxwell's microscopic equation, a reduced set of the indicated three equations is typically considered, augmented by phenomenological constitutive relations between $\bv{F}$ and $\bv{F}_f$. }
 \label{tab:macroscopicmaxwell}
\end{table}

The subtlety that arises from this separation of the total source into free and bound parts is that the partition could implicitly contain complementary \emph{induced} charge-currents $j_i$ that emerge from complicated interactions with the medium.  That is, one can generally arrange for the relation
\begin{align}\label{eq:inducedchargecurrent}
  j_f &= j_{f,0} + j_i, & 
  j_b &= j_{b,0} - j_i,
\end{align}
such that the total source is $j = j_f + j_b = j_{f,0} + j_{b,0}$, so any induced part cancels and is only visible to the separation of the macroscopic fields.  Interestingly, the induced charge-current $j_i$ could change the apparent nature of the separated charge-current.  For example, even if one considers a purely electric (vector) total charge-current, $j = j_e$, an exotic material (e.g., \cite{Castelnovo2008,Giblin2011,Braun2014,Paulsen2014,Ray2014}) could induce an effective \emph{magnetic} (trivector) part, $j_i = j_{m,i}I$, to the total free charge-current, $j_f = j_{e,f} + j_{m,i}I$.  The free macroscopic field $\bv{F}_f$ will then be sensitive to this induced magnetic source, which will appear as apparent magnetic monopoles embedded in the medium \cite{Shnir2005}. 

To see this behavior more clearly in the relative frame, consider the case when the total current $j = j_e = (\cc\rho_e + \rv{J}_e)\gamma_0$ is purely electric (as is traditional and expected), but a magnetic part $j_iI = (\cc\rho_{m,i} + \rv{J}_{m,i})\gamma_0 I$ is then induced by an exotic medium (e.g., \cite{Castelnovo2008,Giblin2011,Braun2014,Paulsen2014,Ray2014}).  In this case, according to \eqref{eq:maxwellsources} and \eqref{eq:macroscopicsplit}, the following macroscopic and microscopic equations will be satisfied 
\begin{align}\label{eq:macroscopicrelativefull}
  \rv{\nabla}\cdot\rv{D} &= \cc\rho_{e,f}, & \partial_0 \rv{D} - \rv{\nabla}\times\rv{H} &= - \rv{J}_{e,f}, \\
  \rv{\nabla}\cdot\rv{H} &= \cc\rho_{m,i}, & \partial_0 \rv{H} + \rv{\nabla}\times\rv{D} &= - \rv{J}_{m,i}, \nonumber \\
  \epsilon_0 \rv{\nabla}\cdot\rv{E} &= \cc\rho_e, & \epsilon_0\partial_0 \rv{E} - \rv{\nabla}\times\rv{B}/\mu_0 &= - \rv{J}_e, \nonumber \\
  \rv{\nabla}\cdot\rv{B} &= 0, & \partial_0 \rv{B} + \rv{\nabla}\times\rv{E} &= 0. \nonumber
\end{align}
Note that the divergence equation for $\rv{H}$ and the curl equation for $\rv{D}$ acquire induced \emph{magnetic source} terms, even though the corresponding equations for the total fields $\rv{B}$ and $\rv{E}$ show no evidence of such a magnetic source.  

The standard practice \cite{Jackson1999,Landau1984} to simplify the large set of equations \eqref{eq:macroscopicrelativefull} is to drop the middle four equations and consider only the remaining four equations that are hybridized between the macro- and microscopic fields and involve only the free electric charge
\begin{align}\label{eq:macroscopicrelative}
  \rv{\nabla}\cdot\rv{D} &= \cc\rho_{e,f}, & \partial_0 \rv{D} - \rv{\nabla}\times\rv{H} &= - \rv{J}_{e,f}, \\
  \rv{\nabla}\cdot\rv{B} &= 0, & \partial_0 \rv{B} + \rv{\nabla}\times\rv{E} &= 0. \nonumber
\end{align}
These equations are then supplemented by the first constraint equation in \eqref{eq:maxwellsplitsources2} that involves the bound electric charge
\begin{align}\label{eq:macroscopicrelative2}
  -\rv{\nabla}\cdot\rv{\mathcal{P}} &= \cc\rho_{e,b}, & -\partial_0 \rv{\mathcal{P}} - \rv{\nabla}\times\rv{\mathcal{M}} &= - \rv{J}_{e,b}.
\end{align}
All six of this reduced set of equations then involve only the \emph{electric} part of the charge, regardless of any magnetic charge that may have been induced by the medium.  This simplification avoids any consideration of the induced magnetic source in favor of a purely electric description.  

Note that the six standard equations \eqref{eq:macroscopicrelative} and \eqref{eq:macroscopicrelative2} for macroscopic fields can be equivalently written in a frame-independent way as the triplet of equations \cite{Doran2007}
\begin{align}\label{eq:macroscopicproper}
  \nabla\cdot \bv{F}_f &= j_f, & \nabla\wedge \bv{F} &= j_m I, & \nabla\cdot \bv{F}_b &= j_b.
\end{align}
The first equation involves the free electric charge. The second becomes the Bianchi identity for the total field when the total magnetic source $j_m$ is zero.  The third is the constraint for the bound electric charge.  It is important to remember, however, that these three equations arise as only part of the fully separated equations in \eqref{eq:maxwellsplitsources}, which are parts of Maxwell's single equation for the total (microscopic) field \eqref{eq:maxwellsources}.  All these relationships are summarized for reference in Table~\ref{tab:macroscopicmaxwell}.

The information that is lost by neglected the middle equations in \eqref{eq:macroscopicrelativefull} can be phenomenologically restored by determining a \emph{constitutive relation} between the macroscopic fields $\rv{D}$ and $\rv{H}$ and the total microscopic fields $\rv{E}$ and $\rv{B}$.  As mentioned previously, such a relation should exist because the configuration of the bound charge-current $j_b$ will generally depend upon the total field $\bv{F}$ in the medium, which in turn alters the constraint for the electromagnetic polarization in \eqref{eq:macroscopicsplit} that determines the macroscopic fields.  Determining the effective functional dependence of this polarization field is a nontrivial and material-specific phenomenological problem \cite{Jackson1999,Landau1984}.  In the simplest case there is no medium (and thus no electromagnetic polarization), so $\rv{D} = \epsilon_0 \rv{E}$ and $\rv{H} = \rv{B}/\mu_0$ according to \eqref{eq:macrofields}.  In the next simplest case the relation is linear and one can define new constants $\epsilon$ and $\mu$ such that $\rv{D} = \epsilon \rv{E}$ and $\rv{H} = \rv{B}/\mu$.  More generally, the appropriate constitutive relations depend on the total field $\bv{F}$, and are frequency-dependent (i.e., dispersive) \cite{Jackson1999,Landau1984,Milonni2004}.

It is also worth remarking that there is a considerable interest to artificial media and materials that restore the dual symmetry of vacuum on the level of macroscopic medium parameters (e.g., $\epsilon$=$\mu$) \cite{Bialynicki-Birula1996,Savchenko1994,Fernandez2012,Fernandez2013,Zambrana2013,Philbin2013,Tischler2014}. The conservation of the optical helicity associated with such symmetry (see Section~\ref{sec:helicity} below) provides a new insight into the light propagation and scattering dynamics.


\section{Lagrangian formalism}\label{sec:lagrangian}

\epigraph{4.5in}{It is impossible to study this remarkable theory without experiencing at times the strange feeling that the equations and formulas somehow have a proper life, that they are smarter than we, smarter than the author himself, and that we somehow obtain from them more than was originally put into them.}{Heinrich Hertz \cite{Hertz1889bis}}

\epigraph{4.5in}{You can recognize truth by its beauty and simplicity. When you get it right, it is obvious that it is right---at least if you have any experience---because usually what happens is that more comes out than goes in. 
}{Richard P. Feynman \cite{Cole1985}}


It is worth emphasizing that up to this point we have introduced \emph{no physical postulates} other than the need for spacetime itself (and thus its associated algebra).  We made the casual observation that a proper spacetime bivector $\bv{F}$ naturally splits into polar and axial 3-vector parts $\bv{F} = \rv{E} + \rv{B}I$ with respect to a particular reference frame in Eq.~\eqref{eq:spacetimesplit}, and noted that these parts transform under a Lorentz boost exactly as we expect the electromagnetic field to transform in Eqs.~\eqref{eq:relativelorentzboost} and \eqref{eq:relativelorentzboost2}.  We then found that the structure of the simplest bivector differential equation $\nabla \bv{F} = 0$ in Eq~\eqref{eq:maxwell} is precisely Maxwell's equation in vacuum, while the next simplest modification $\nabla \bv{F} = j$ in Eq.~\eqref{eq:maxwellsources} is Maxwell's equation that properly includes both electric and magnetic sources, albeit in a compact form as a single complex vector current $j$.  We also made the casual observation that contracting the bivector $\bv{F}$ with such a complex current $j$ gives the correct Lorentz force equation $dp/d\tau = \mean{\bv{F}j}_1$ that supports both electric and magnetic charges in Eq.~\eqref{eq:lorentzforcecomplex}.  Combining the Lorentz force law and Maxwell's equation using a general conservation constraint in Eq.~\eqref{eq:energymomentumcontinuity} then produces the correct symmetric energy-momentum stress tensor $\underbar{T}(\gamma_0) = \bv{F}\gamma_0\widetilde{\bv{F}}/2$ in Eq.~\eqref{eq:symemstress} as a simple quadratic form of the field (which is the same functional form as the energy-momentum current of the Dirac electron \cite{Berestetskii1982,Doran2007,Hiley2012}).  

Despite the richness of these emergent characteristics of the spacetime geometry, all suggestive notation, terminology, and constant unit factors have been window-dressing for features that are essentially inevitable when considering the simplest dynamics one can ascribe to any bivector field in spacetime, bereft of physical interpretation \emph{a priori}.  The fact that these features exactly correspond to the electromagnetic theory of both electric and magnetic charges is remarkable unto itself.  More importantly, however, this fact prompts a careful reassessment of the physical foundations and postulates of classical electromagnetic field theory. 

The more complete physical foundation of electromagnetism comes from a Lagrangian field-theory perspective.  Fundamentally, the electromagnetic field $\bv{F}$ emerges as a necessary consequence of preserving a local U(1) gauge symmetry (i.e., a local phase rotation) in the Lagrangian for a complex (charged) matter field \cite{Weinberg1995}, as we will detail in Section \ref{sec:gauge}.  However, with such a Lagrangian approach it becomes clear that the \emph{vector potential} $z$ acts as the dynamical gauge field that preserves the charge symmetry, with the usual electromagnetic field $\bv{F} = \nabla z$ arising only as a derivative quantity.  Thus, the vector potential $z$ will be our new starting point, from which we will systematically derive the corresponding conserved physical quantities of the field.


\subsection{Traditional vacuum-field Lagrangian}


As we have shown in Sections~\ref{sec:dualsym} and \ref{sec:potentials}, the dual symmetry of the vacuum Maxwell's equation $\nabla^2 z = 0$ implies that the vector potential should be invariant under the global phase symmetry $z \mapsto z \exp(\theta I)$ in vacuum, and thus must be intrinsically \emph{complex}.  This symmetry poses a problem for the traditional electromagnetic Lagrangian, which can be written in various equivalent forms:
\begin{align}\label{eq:traditional}
  \mathcal{L}_{\text{trad}}(x) &= \frac{\mean{(\nabla z)^2}_0}{2} = \frac{\mean{\bv{F}^2}_0}{2} = \frac{\ell_1}{2} = \frac{|\rv{E}|^2 - |\rv{B}|^2}{2} = \frac{1}{4}\sum_{\mu\nu}F_{\mu\nu}F^{\nu\mu}.
\end{align}
We will motivate why this Lagrangian appears from a gauge theory perspective in Section~\ref{sec:gauge}. Traditionally $z = a_e$ is considered to be a pure (electric) vector potential that generates a single associated constituent field $\bv{F}_e = \nabla\wedge a_e$ \cite{Soper1976,Jackson1999,Thide2011}.  Indeed, recall that the subscripts $e$ and $m$ throughout this report refer to the electric and magnetic types of \emph{charge}, so in the traditional electromagnetism (with only electric charges) all ``magnetic'' (subscript $m$) quantities can be neglected.  However, we keep $z = a_e + a_m I$ complex here, since no electric or magnetic charge exists in \emph{vacuum} to select either $a_e$ or $a_m$ as a preferred vector potential.

The traditional Lagrangian \eqref{eq:traditional}  is \emph{not dual-symmetric}, and, hence, is problematic for the description of dual-symmetric vacuum fields that require such a complex potential.  To see this explicitly, we expand the complex vector potential $z$ in \eqref{eq:traditional} into its constituent parts, impose the Lorenz-FitzGerald conditions $\nabla \cdot a_e = \nabla \cdot a_m = 0$ discussed in Section~\ref{sec:potentials}, and define the two constituent fields $\bv{F}_e = \nabla\wedge a_e = \rv{E}_e + \rv{B}_e I$ and $\bv{F}_m = \nabla\wedge a_m = \rv{B}_m - \rv{E}_m I$ such that $\bv{F} = \nabla z = \bv{F}_e + \bv{F}_m I$, as in Section \ref{sec:constituent}, then the traditional Lagrangian expands into the form
\begin{align}\label{eq:traditionalconstituent}
  \mathcal{L}_{\text{trad}}(x) &= \frac{\mean{\bv{F}_e^2}_0 - \mean{\bv{F}_m^2}_0}{2} + \mean{\bv{F}_e\cdot\bv{F}_m I}_0, \\
  &= \frac{1}{2}[|\rv{E}_e|^2 - |\rv{B}_e|^2 + |\rv{E}_m|^2 - |\rv{B}_m|^2] - [\rv{E}_e\cdot\rv{B}_m + \rv{B}_e\cdot\rv{E}_m], \nonumber
\end{align}
which shows an intrinsic interplay between the constituent fields\footnote{Again, the subscripts of $e$ and $m$ on the constituent relative fields indicate that they would normally be associated with either electric or magnetic \emph{charges}, which do not exist in vacuum.} that alters the original structure of \eqref{eq:traditional}.

This additional structure appears because the traditional Lagrangian \eqref{eq:traditional} is not invariant under the dual symmetry phase rotation $z\mapsto z\exp(\theta I)$.  Instead, it transforms as 
\begin{align}
  \mathcal{L}_{\text{trad}}(x) \mapsto \frac{\ell_1}{2}\cos2\theta - \frac{\ell_2}{2}\sin2\theta,
\end{align}
and oscillates between the two proper scalars $\ell_1 = |\rv{E}|^2 - |\rv{B}|^2$ and $\ell_2 = 2\rv{E}\cdot\rv{B}$ of the total field as the phase is changed \cite{Bliokh2013}.  (As we noted in Section~\ref{sec:canonical}, the second term that appears here is the ``axion'' term proposed in \cite{Weinberg1978} using precisely this dual-symmetric phase-rotation.)  The lack of global phase invariance of the Lagrangian \eqref{eq:traditional} is problematic for maintaining the dual-symmetric complexity of the vector potential $z$.  However, one can always avoid this problem by explicitly \emph{breaking} the dual-symmetry to keep $z = a_e$ a pure (electric) vector potential, as in the traditional approach.  In this case the constituent fields associated with the magnetic vector potential $a_m$ vanish ($\rv{E}_m = \rv{B}_m = 0$) in \eqref{eq:traditionalconstituent}.

We are thus motivated to consider a simple correction to the traditional Lagrangian that addresses these shortcomings for the description of a vacuum field.  We wish to modify the Lagrangian to make it manifestly dual-symmetric under a global phase rotation of $z$.


\subsection{Dual-symmetric Lagrangian}


We can correct the traditional Lagrangian in a simple way to make it properly dual-symmetric by adding a complex conjugate to the traditional expression
\begin{align}\label{eq:maxwelllagrangian}
  \mathcal{L}_{\text{dual}}(x) &= \frac{1}{2}\mean{(\nabla z)(\nabla z^*)}_0.
\end{align}
Notably, this modification makes the Lagrangian a simple quadratic form for the complex vector field $z$, which is the expected form for the kinetic energy of such a field.  This form is thus a logical choice for a properly dual-symmetric Lagrangian.  We must now verify that this modification produces sensible results.

In terms of constituent fields, \eqref{eq:maxwelllagrangian} expands into the following equivalent forms:
\begin{align}\label{eq:maxwelllagrangianpotentials}
  \mathcal{L}_{\text{dual}}(x) &= \frac{\mean{(\nabla a_e)^2}_0}{2} + \frac{\mean{(\nabla a_m)^2}_0}{2} = \frac{\mean{\bv{F}_e^2}_0}{2} + \frac{\mean{\bv{F}_m^2}_0}{2}, \\
  &= \frac{1}{2}(|\rv{E}_e|^2 - |\rv{B}_e|^2) + \frac{1}{2}(|\rv{B}_m|^2 - |\rv{E}_m|^2), \nonumber \\
  &= \frac{1}{4}\sum_{\mu\nu}[ (F_e)_{\mu\nu}(F_e)^{\nu\mu} + (F_m)_{\mu\nu}(F_m)^{\nu\mu}]. \nonumber
\end{align}
Perhaps surprisingly, the dual-symmetric form of \eqref{eq:maxwelllagrangian} is simply the sum of \emph{two copies of the traditional Lagrangian} \eqref{eq:traditionalconstituent}, one for each independent constituent field $a_e$ and $a_m$ in the complex vector potential $z = a_e + a_m I$.  Notably, a dual-symmetric Lagrangian of precisely this form was previously considered in \cite{Singleton1995,Singleton1996,Kato2002,Vasconcellos2014} from a different starting point.  The traditional and dual-symmetric Lagrangians are compared for reference in Table~\ref{tab:lagrangians}.

\begin{table}
  \centering
  \begin{tabular}{l l}
    \hline
\noalign{\vskip 2mm} 
    \multicolumn{2}{c}{\textbf{Electromagnetic Lagrangian Densities}} \\
\noalign{\vskip 2mm} 
    \hline \\
    \textbf{Traditional:} & {$\begin{aligned}z &= a_e \equiv a & \bv{F} &= \nabla a_e = \bv{F}_e\end{aligned}$} \\ \\
      & {$\begin{aligned}\Aboxed{\mathcal{L}_{\text{trad}}(x) &= \frac{\mean{(\nabla a)^2}_0}{2}} = \frac{\mean{\bv{F}^2}_0}{2}
      = \frac{1}{4}\sum_{\mu\nu}F_{\mu\nu}F^{\nu\mu}\end{aligned}$} \\
    \\
    \hline
    \\
    \textbf{Dual-symmetric:} & {$\begin{aligned}z &= a_e + a_m I & \bv{F} &= \nabla z = \bv{F}_e + \bv{F}_m I\end{aligned}$} \\ \\
      & {$\begin{aligned}\Aboxed{\mathcal{L}_{\text{dual}}(x) &= \frac{\mean{(\nabla z)(\nabla z^*)}_0}{2}} 
      = \frac{\mean{\bv{F}_e^2}_0}{2} + \frac{\mean{\bv{F}_m^2}_0}{2} \\
      &= \frac{1}{4}\sum_{\mu\nu}[(F_e)_{\mu\nu}(F_e)^{\nu\mu} + (F_m)_{\mu\nu}(F_m)^{\nu\mu}]\end{aligned}$} \\
    \\
    \hline
    \\
    \textbf{Constrained:} & {$\boxed{\bv{F}_e \to \bv{F}_m I \to \bv{F}_{\text{meas}}/\sqrt{2} = \bv{G}_{\text{meas}}I/\sqrt{2}}$} \\ \\
    & {$\begin{aligned}\mathcal{L}_{\text{dual}}(x) &\to \frac{\mean{\bv{F}_{\text{meas}}^2}_0}{4} + \frac{\mean{\bv{G}_{\text{meas}}^2}_0}{4} 
      = \frac{1}{8}\sum_{\mu\nu}[F_{\mu\nu}F^{\nu\mu} + G_{\mu\nu}G^{\nu\mu}]\end{aligned}$} \\
    \\
    \hline 
 \end{tabular}
 \caption[Electromagnetic Lagrangians]{Electromagnetic Lagrangians in the traditional and dual-symmetric field theories.  The traditional Lagrangian is a kinetic term for only the constituent vector potential $a_e$ associated with \emph{electric} charge.  The dual-symmetric Lagrangian is a kinetic term for the full complex vector potential $z$, and preserves the dual symmetry of vacuum.  The electric $a_e$ and magnetic $a_m$ vector potentials contribute to this Lagrangian as \emph{independent} fields that each satisfy a copy of the traditional Lagrangian \cite{Singleton1995,Singleton1996,Kato2002,Vasconcellos2014}. In the constrained formalism, the two vector potentials of the dual-symmetric Lagrangian describe the \emph{same} (measurable) vacuum field $\bv{F}_{\text{meas}}$ in a symmetrized way \cite{Bliokh2013,Cameron2012}. Note that with this constraint the dual-symmetric Lagrangian formally vanishes; however, the constraint can be implemented using a Lagrange multiplier, or as a final step.}
 \label{tab:lagrangians}
\end{table}


\subsection{Constrained dual-symmetric Lagrangian}\label{sec:constituentconstraint}


The fact that the dual-symmetrized Lagrangian \eqref{eq:maxwelllagrangian} explicitly involves the second constituent field $\bv{F}_m$ as an independent but necessary field prompts the question whether this second field is really independent in vacuum, or should be further constrained.  In particular, if one believes the \emph{total} dual-symmetric field $\bv{F} = \bv{F}_e + \bv{F}_m I$ should be the true electromagnetic field, then one can postulate that the different constituent fields $\bv{F}_e = \nabla \wedge a_e$ and $\bv{F}_m = \nabla \wedge a_m$ actually describe the \emph{same} total field.  We can thus consider imposing an additional nontrivial \emph{constraint} to force these two constituent fields to give equal contributions to the measurable vacuum field
\begin{align}\label{eq:fieldconstraint}
  \bv{F}_e = \bv{F}_m I = \frac{\bv{F}_{\text{meas}}}{\sqrt{2}},
\end{align}
so $\rv{E}_e = \rv{E}_m = \rv{E}_{\text{meas}}/\sqrt{2}$ and $\rv{B}_e = \rv{B}_m = \rv{B}_{\text{meas}}/\sqrt{2}$, and the total field magnitude is preserved:
\begin{align}
  |\bv{F}|^2 = \bv{F}^*\bv{F} = \bv{F}_e^*\bv{F}_e + \bv{F}_m^*\bv{F}_m \to \bv{F}^*_{\text{meas}}\bv{F}_{\text{meas}} = |\bv{F}_{\text{meas}}|^2.
\end{align}
With this constraint we find that $\bv{F}_m = \bv{F}_{\text{meas}} I^{-1}/\sqrt{2} = \bv{G}_{\text{meas}}/\sqrt{2}$, so the two constituent fields become precisely proportional to the measurable field $\bv{F}_{\text{meas}}$ and its Hodge-dual $\bv{G}_{\text{meas}} = \bv{F}_{\text{meas}} I^{-1}$.  

Inserting the constraint \eqref{eq:fieldconstraint} into \eqref{eq:maxwelllagrangianpotentials} produces 
\begin{align}\label{eq:lagrangianvanishing}
  \mathcal{L}_{\text{dual}}(x) &\to \frac{\mean{\bv{F}_{\text{meas}}^2}_0 + \mean{\bv{G}_{\text{meas}}^2}_0}{4} = \frac{1}{8}\sum_{\mu\nu}(F_{\mu\nu}F^{\nu\mu} + G_{\mu\nu}G^{\nu\mu}).
\end{align}
The functional form of this constrained Lagrangian density is then precisely equal to the dual-symmetric form recently postulated in \cite{Bliokh2013,Cameron2012}.  Curiously, this constrained form identically \emph{vanishes} since $\bv{G}_{\text{meas}}^2 = (\bv{F}_{\text{meas}}I^{-1})^2 = -\bv{F}_{\text{meas}}^2$. Thus, the constraint \eqref{eq:fieldconstraint} should be imposed either as a Lagrange multiplier constraint in the dual-symmetric Lagrangian \eqref{eq:maxwelllagrangian}, or as a final step in derivations that retain both constituent fields $\bv{F}_e$ and $\bv{F}_m$ \cite{Bliokh2013,Cameron2012}.  

Notably, the constraint \eqref{eq:fieldconstraint} also forces the vector potential $z$ to be \emph{irreducibly complex}.  Indeed, if one could rotate $z$ to a pure vector potential $a_0$ with some phase rotation $z = a_0\exp(\theta I)$, then the constraint \eqref{eq:fieldconstraint} would force the measurable field $\bv{F}_{\text{meas}}$ to identically vanish.  To see this, note that $z = a_e + a_m I = a_0\exp(\theta I)$, so $a_e = a_0\cos\theta$ and $a_m = a_0\sin\theta$.  Therefore, $\bv{F}_e = \cos\theta\,\bv{F}_0$ and $\bv{F}_m = \sin\theta\,\bv{F}_0$ where $\bv{F}_0 = \nabla\wedge a_0$.  The constraint \eqref{eq:fieldconstraint} then implies: $\cos\theta\,\bv{F}_0 = \sin\theta\,\bv{F}_0 I = \bv{F}_{\text{meas}}/\sqrt{2}$.  Due to the factor of $I$, this relation is only satisfiable if $\bv{F}_0 = 0$ for any $\theta$, making $\bv{F}_{\text{meas}} = 0$.  Therefore, to obtain nontrivial vacuum fields that satisfy the constraint \eqref{eq:fieldconstraint}, the vector potential $z$ must remain complex under \emph{all} phase rotations.  

However, the irreducible complexity of $z$ forced by the constraint \eqref{eq:fieldconstraint} presents a significant problem for \emph{coupling to matter}.  The equation $\nabla z^* = \nabla(a_e - a_m I) = \bv{F}_e - \bv{F}_m I \to 0$ vanishes with this constraint, which is inconsistent with Maxwell's source equation \eqref{eq:maxwellsources}, written in its conjugated form $\nabla^2 z^* = j^*$.  Specifically, $\nabla^2 z^* = \nabla(\nabla z^*) \to 0$ with the constraint, so any source $j$ must vanish.  Thus, the constraint \eqref{eq:fieldconstraint} can only describe \emph{vacuum} fields that are not coupled to matter, which corroborates our general discussion about the fundamental difference between the irreducibly complex vacuum and reducibly complex source parts of the complex vector potential $z$ in Section~\ref{sec:dualsymmetrysource}

Since fields must couple to matter (charges) to be physically measured, the constraint \eqref{eq:fieldconstraint} in its current form is problematic. Either we must modify the constraint \eqref{eq:fieldconstraint} to properly accommodate sources (see, e.g., \cite{Cameron2014} for ideas regarding how this modification could be accomplished), or the constrained Lagrangian becomes inconsistent with the presence of sources. 
At the same time, the independent ``electric'' and ``magnetic'' parts of the unconstrained dual-symmetric Lagrangian \eqref{eq:maxwelllagrangianpotentials} can be coupled to the corresponding charges. In this case one can speculate that the presence of ``electric matter'' explicitly breaks the dual-symmetry by coupling to only one of the constituent fields $\bv{F}_e = \nabla \wedge a_e$.  In the absence of other fields in the model, the remaining constituent field $\bv{F}_m$ in the dual-symmetric Lagrangian \eqref{eq:maxwelllagrangian} would then be undetectable and only contribute to gravity as an invisible and noninteracting energy density (e.g., a \emph{dark light} contribution to dark matter \cite{Vasconcellos2014}).  We will revisit this issue of the second constituent field more carefully in Section \ref{sec:symmetry}.


\subsection{Maxwell's equation of motion}


We can obtain the equation of motion for $z$ implied by all three candidate Lagrangians (Table~\ref{tab:lagrangians}) by treating the constituent vector potentials $a_e$ and $a_m$ independently, and applying the appropriate Euler--Lagrange equation \cite{Doran2007, Soper1976} to each one.  For example,
\begin{align}\label{eq:eulerlagrange}
  \sum_\mu\frac{\partial}{\partial x^\mu}\left(\frac{\partial\mathcal{L}}{\partial(\partial_\mu a_e)}\right) - \frac{\partial \mathcal{L}}{\partial a_e} = \nabla^2 a_e = 0,
\end{align}
is the proper equation of motion for $a_e$, where $x = \sum_\mu x^\mu \gamma_\mu$.  The equation for $a_m$ can be obtained similarly, yielding $\nabla^2 a_m = 0$.  It follows that the equation of motion for $z = a_e + a_m I$ is the Maxwell's vacuum equation \eqref{eq:maxwell} in a manifestly dual-symmetric form,
\begin{align}\label{eq:maxwellvector}
  \nabla^2 z &= \nabla \bv{F} = 0,
\end{align}
for any of the three candidate Lagrangians. This means, that the above three Lagrangian theories are indistinguishable from each other on the level of vacuum equations of motion. However, they do differ from each other in other field-theory aspects \cite{Bliokh2013}.


\subsection{Field invariants: Noether currents}\label{sec:noether}


According to Noether's theorem \cite{Noether1918}, any continuous symmetry of a Lagrangian leads to a conservation law.  We can use this theorem to systematically find the \emph{canonical} conserved currents for the vacuum electromagnetic field, using each of the three Lagrangian densities \eqref{eq:traditional}, \eqref{eq:maxwelllagrangian}, and \eqref{eq:lagrangianvanishing} in Table~\ref{tab:lagrangians}.

Recall the general procedure for obtaining the conserved currents \cite{Soper1976,Doran2007}:
\begin{enumerate}
  \item Suppose one transforms each field $a$ (and possibly the coordinates $x$) in a Lagrangian density $\mathcal{L}[a(x),\nabla a(x)]$ according to a group transformation $\underbar{U}_{\alpha,g}$ that depends on a continuous scalar parameter $\alpha$, as well as a generating element $g$
    \begin{subequations}\label{eq:transformations}
    \begin{align}
      a'(x') &= \underbar{U}_{\alpha,g}[a(x')], \\
      x' &= \underbar{U}^{-1}_{\alpha,g}(x) = \overline{U}_{\alpha,g}(x), \\
      \nabla' &= \underbar{U}_{\alpha,g}(\nabla).
    \end{align}
    \end{subequations}
    If $\mathcal{L}[a'(x'), \nabla' a'(x')] = \mathcal{L}[a(x),\nabla a(x)]$ then $\underbar{U}_{\alpha,g}$ is a symmetry of the Lagrangian density\footnote{Note that if both the coordinates $x$ and the fields $a$ are transformed, then one should typically transform $\nabla$ in the same way as $a$ so that the kinetic terms $\nabla a$ transform similarly to $a$.  Since $\nabla$ and $x$ are inversely related, this implies that $x$ and $a$ should transform oppositely with the group transformation.}.

  \item The infinitesimal field and coordinate increments are then
    \begin{subequations}\label{eq:increments}
    \begin{align}
      \delta a(g) &\equiv \partial_\alpha a'(x')|_{\alpha=0} = \partial_\alpha a'(x)|_{\alpha=0} + (\delta x(g) \cdot \nabla)a(x),\\
      \delta x(g) &\equiv \partial_\alpha x'|_{\alpha=0},
    \end{align}
    \end{subequations}
    and depend on the chosen generator $g$.

  \item The change in the transformed Lagrangian $\mathcal{L}'$ is therefore
    \begin{align}\label{eq:noethertransform}
      \partial_\alpha\mathcal{L}'(x')|_{\alpha = 0} &= \smean{\frac{\partial\mathcal{L}}{\partial a}\,\delta a(g)}_0 + \sum_\mu\smean{\frac{\partial\mathcal{L}}{\partial(\partial_\mu a)}\partial_\mu\delta a(g)}_0 + (\delta x(g) \cdot \nabla)\mathcal{L},
    \end{align}
    which must vanish from the symmetry.  In the presence of multiple independent fields (such as $a_e$ and $a_m$), the right hand side contains corresponding terms for each independent field.

  \item After applying the Euler--Lagrange equation \eqref{eq:eulerlagrange} for each $a$ in the first term (i.e., essentially using Maxwell's equation), \eqref{eq:noethertransform} simplifies to a divergence
    \begin{align}\label{eq:noether}
      \partial_\alpha\mathcal{L}'|_{\alpha = 0} &= \nabla\cdot \underbar{J}_N(g) = 0,
    \end{align}
    of a \emph{conserved Noether current} (tensor) that depends on the generator $g$ 
    \begin{align}\label{eq:noethercurrent}
      \underbar{J}_N(g) &\equiv \sum_\mu \gamma_\mu \smean{\frac{\partial\mathcal{L}}{\partial(\partial_\mu a)}\delta a(g)}_0 + \mathcal{L}\,\delta x(g).
    \end{align}

  \item In any reference frame $\underbar{J}_N(g) = [\xi_N(g) + \rv{J}_N(g)]\gamma_0$, it follows from this conservation law that $\partial_0 \xi_N(g) - \rv{\nabla}\cdot\rv{J}_N(g) = 0$.  Hence, we can integrate over the spatial coordinates (assuming $\rv{J}_N(g)$ vanishes at infinity) to find a conserved \emph{Noether charge} 
    \begin{align}\label{eq:noethercharge}
       \underbar{Q}_N(g) &= \int d^3x\,\,\xi_N(g), ~~~~~\partial_0\, \underbar{Q}_N(g) = 0.
     \end{align}
\end{enumerate}

We now use this procedure to find the Noether currents for the various electromagnetic Lagrangians.  Note that it will be sufficient to compute everything using the dual-symmetric Lagrangian \eqref{eq:maxwelllagrangian}, and then constrain the final results appropriately.  For the traditional Lagrangian \eqref{eq:traditional}, we will simply neglect all terms related to the constituent field $\bv{F}_m$ associated with magnetic charges.  For the constrained Lagrangian, we will enforce the additional constraint $\bv{F}_e = \bv{F}_m = \bv{F}/\sqrt{2}$, Eq.~\eqref{eq:fieldconstraint}, which makes both vector potentials $a_e$ and $a_m$ correspond to the same electromagnetic field.  We will distinguish the expressions for the constrained Lagrangian using an arrow symbol ``$\to$'' for clarity in what follows.


\subsection{Canonical energy-momentum stress tensor}


We first consider translations of spacetime, which will produce the conservation of the energy-momentum of the field. Suppose we translate the position vectors $x$ in the Lagrangian density by a small amount $\alpha$ in an arbitrary direction $b$, so $x' = x - \alpha b = \exp[-\alpha(b\cdot\nabla)]x$.  This translation leaves the Lagrangian \eqref{eq:maxwelllagrangian} invariant since $\nabla' = \nabla$.  It follows from \eqref{eq:increments} that the increment for $x$ is simply $b$, while the increment for $a_e$ is the corresponding directional derivative
\begin{subequations}
\begin{align}
  \delta x[b] &= -b, \\
  \delta a_e[b] &= -(b\cdot\nabla)a_e,
\end{align}
\end{subequations}
and similarly for $a_m$.  Computing the change from this translation according to \eqref{eq:noethertransform} thus produces the conserved Noether current 
\begin{align}\label{eq:canonicalemstress}
  \underbar{T}(b) & = \sum_\mu \gamma_\mu\smean{\frac{\partial\mathcal{L}}{\partial(\partial_\mu a_e)}(-b\cdot\nabla)a_e + \frac{\partial\mathcal{L}}{\partial(\partial_\mu a_m)}(-b\cdot\nabla)a_m}_0 - b\mathcal{L}, \\
  &= \sum_\mu \gamma_\mu\smean{(\partial_\mu a_e)(-b\cdot\nabla)a_e + (\partial_\mu a_m)(-b\cdot\nabla)a_m}_0 - b\mathcal{L}, \nonumber \\
  &= \nabla a_e\cdot [(-b\cdot\nabla)a_e]  + \nabla a_m\cdot[(-b\cdot\nabla)a_m] - b\mathcal{L}, \nonumber \\
  &= (b\cdot\dot{\nabla})\mean{\dot{a}_e\bv{F}_e + \dot{a}_m\bv{F}_m}_1 - \frac{b}{2}(\mean{\bv{F}_e^2}_0 + \mean{\bv{F}_m^2}_0), \nonumber \\
  &\to \frac{1}{2}(b\cdot\dot{\nabla})\mean{\dot{a}_e\bv{F} + \dot{a}_m\bv{G}}_1, \nonumber
\end{align}
Recall that the overdots indicate that only the potentials are being differentiated. This conserved tensor $\underbar{T}(b)$ is the \emph{canonical energy-momentum stress tensor} \cite{Soper1976,Bliokh2013}, which was first derived while preserving dual symmetry by Gaillard and Zumino \cite{Gaillard1981}.   

The canonical energy-momentum tensor is \emph{not} symmetric with respect to its adjoint: $\underbar{T} \neq \overline{T}$, which can be obtained as
\begin{align}\label{eq:canonicalemadjoint}
  \overline{T}(b) = \partial_d\mean{\underbar{T}(d)b}_0 &= \dot{\nabla}\mean{(\dot{a}_e\bv{F}_e + \dot{a}_m\bv{F}_m)b}_0 - \frac{b}{2}(\mean{\bv{F}_e^2}_0 + \mean{\bv{F}_m^2}_0), \\
  &\to \frac{1}{2}\dot{\nabla}\mean{(\dot{a}_e\bv{F} + \dot{a}_m\bv{G})b}_0 . \nonumber 
\end{align}
This adjoint form involves the scalar (real) parts of total bivector contractions. 

Importantly, in contrast to the symmetric energy-momentum tensor in \eqref{eq:symemstress}, the canonical tensor in \eqref{eq:canonicalemstress} and \eqref{eq:canonicalemadjoint} explicitly involves the vector-potentials $a_e$ and $a_m$, making it \emph{gauge-dependent}. Note also that the first summand (involving $\bv{F}_e$ and $a_e$) in the non-constrained tensor is equivalent to the conserved canonical tensor for the traditional Lagrangian \eqref{eq:traditional}. Hence, the two summands in the unconstrained tensor are \emph{separately conserved} energy-momentum tensors for the two constituent fields $\bv{F}_e$ and $\bv{F}_m$.  We detail the canonical energy-momentum tensor in Table~\ref{tab:canemtensor} for reference.

For comparison, we also compute the \emph{symmetric} energy-momentum tensor from \eqref{eq:symemstress} explicitly in terms of the constituent fields to find
\begin{align}\label{eq:symemconstituent}
  \underbar{T}_{\text{sym}}(b) = \overline{T}_{\text{sym}}(b) &= \frac{1}{2}\bv{F}b\widetilde{\bv{F}} = -\frac{1}{2}(\bv{F}_e + \bv{F}_m I)b(\bv{F}_e + \bv{F}_m I), \\
  &= \frac{1}{2}\bv{F}_eb\widetilde{\bv{F}}_e + \frac{1}{2}\bv{F}_mb\widetilde{\bv{F}}_m. \nonumber
\end{align}
The cross-term involves a vector minus its reverse, so it vanishes to leave the sum of the symmetric tensors for each constituent field.  This form as a sum is consistent with the canonical tensors derived from Noether's theorem.  In particular the identity \eqref{eq:symemconstituent} implies that the energy density $\varepsilon = \varepsilon_e + \varepsilon_m$ and the Poynting vector $\rv{P} = \rv{P}_e + \rv{P}_m$ decompose into sums of the corresponding quantities for the constituent fields. 

It is instructive to find the difference between the symmetric (Section~\ref{sec:symemstress}) and canonical energy-momentum tensors:
\begin{align}
  \underbar{K} \equiv \underbar{T}_{\rm sym}-\underbar{T}, ~~~~~ \overline{K} \equiv \overline{T}_{\rm sym}-\overline{T}.
\end{align}
Using the identity $(b\cdot\nabla)a = b\cdot(\nabla a) + \nabla(b\cdot a)$, as well as the identity $2[b \cdot (\nabla a)]\cdot(\nabla a) - b\mean{(\nabla a)^2}_0 = -(\nabla a)b(\nabla a)$, we find the explicit forms
\begin{align}
  \label{eq:emcorrection}
  \underbar{K}(b) &= (\bv{F}_e\cdot\nabla)(a_e\cdot b) + (\bv{F}_m\cdot\nabla)(a_m\cdot b), \\
  &\to \frac{1}{2}[(\bv{F}\cdot\nabla)(a_e\cdot b) + (\bv{G}\cdot\nabla)(a_m\cdot b)], \nonumber \\
  \label{eq:emcorrectionadjoint}
  \overline{K}(b) &= \mean{b\bv{F}_e\dot{\nabla}}_0\dot{a}_e + \mean{b\bv{F}_m\dot{\nabla}}_0\dot{a}_m, \\
  &\to \frac{1}{2}\left[\mean{b\bv{F}\dot{\nabla}}_0\dot{a}_e + \mean{b\bv{G}\dot{\nabla}}_0\dot{a}_m\right]. \nonumber 
\end{align}
Since both the symmetric and canonical energy-momentum tensors satisfy the continuity equations, the difference term $\underbar{K}(b)$ must vanish under a divergence, i.e., it does not contribute to the energy-momentum transport. Using the vacuum Maxwell's equation \eqref{eq:maxwell}, we can make this property apparent by writing \eqref{eq:emcorrection} as a total derivative:
 \begin{align}
  \underbar{K}(b) &= -\nabla[(b\cdot a_e)\bv{F}_e + (b\cdot a_m)\bv{F}_m], \\
  &\to -\frac{1}{2}\nabla[(b\cdot a_e)\bv{F} + (b\cdot a_m)\bv{G}]. \nonumber
\end{align}

Historically, the canonical energy-momentum tensor $\underbar{T}$ that is produced by Noether's theorem was considered problematic due to its non-symmetric and gauge-dependent form (as well as the enigmatic appearance of the spin in the angular momentum tensor, as we shall see in the next section).  Then, in 1940, Belinfante and Rosenfeld \cite{Belinfante1939,Belinfante1940,Rosenfeld1940} suggested a general symmetrization procedure by adding a ``virtual'' contribution $\underbar{K}$ (i.e., a total divergence) to the canonical tensor, which resulted in the familiar symmetric tensor $\underbar{T}_{\rm sym}$. Usually, the symmetric Belinfante--Rosenfeld energy-momentum tensor is considered to be the physically meaningful tensor, because it is gauge-invariant and is naturally coupled to gravity \cite{Soper1976}. However, recent studies of the \emph{local} momentum and angular momentum densities of optical fields have shown that it is actually the \emph{canonical} momentum density from $\underbar{T}$ (in the Coulomb gauge) that determines the optical pressure on small dipole particles or atoms that is observed in experiments \cite{O'Neil2002,Garces2003,Curtis2003,Huard1978,Huard1979,Kocsis2011,Arlt2003,Leach2006,Yeganeh2013,Bliokh2013,Bliokh2013a,Barnett2013,Bliokh2013b,Bliokh2013c, Bekshaev2014}. Furthermore, as we will see in the next section, the observable separation of the spin and orbital angular momenta of light \cite{Allen1992,Allen1999,Andrews2008,Torres2011,Andrews2013,O'Neil2002,Garces2003,Curtis2003, Zhao2007,Adachi2007} is also described by the \emph{canonical} tensors rather than the symmetrized Belinfante--Rosenfeld tensors \cite{Bliokh2013,Bliokh2014}. Notably, this need for the canonical energy-momentum and corresponding angular momentum tensors has also been actively discussed in relation to the separation of the spin and orbital angular momenta of gluon fields in quantum chromodynamics (QCD) \cite{Leader2014}.

\begin{table}
  \centering
  \begin{tabular}{l l}
    \hline
\noalign{\vskip 2mm} 
    \multicolumn{2}{c}{\textbf{Canonical Energy-Momentum Tensor}} \\
\noalign{\vskip 2mm} 
    \hline \\
    \multicolumn{2}{c}{$\begin{aligned}\underbar{T}(\gamma_0) &= (\varepsilon + \rv{P})\gamma_0, & \overline{T}(\gamma_0) &= (\varepsilon + \boxed{\rv{P}_o})\gamma_0, & \rv{P} &= \rv{P}_o + \boxed{\rv{P}_s} \end{aligned}$} \\  
    \\
    \hline 
    \\
    \textbf{Traditional:} & {$\begin{aligned}\Aboxed{\underbar{T}(b) &= (b\cdot\dot{\nabla})\mean{\dot{a}\bv{F}}_1 - \frac{b}{2}\mean{\bv{F}^2}_0} &
      \Aboxed{\overline{T}(b) &= \dot{\nabla}\mean{\dot{a}\bv{F}b}_0 - \frac{b}{2}\mean{\bv{F}^2}_0} \end{aligned}$} \\
    \\
    $z = a_e \equiv a$ & {$\begin{aligned}\rv{P}_o & = \dot{\rv{\nabla}}(\rv{E}\cdot\dot{\rv{A}}), & \rv{P}_s & = -(\rv{E}\cdot\rv{\nabla})\rv{A} \end{aligned}$} \\
    \\
    $\bv{F} = \nabla a_e = \bv{F}_e$ & {$\begin{aligned}T^{\mu\nu} &= \sum_\alpha F^{\alpha\mu}\partial^\nu a_\alpha - \frac{\eta^{\mu\nu}}{4}\sum_{\alpha\beta}F_{\alpha\beta}F^{\beta\alpha}\end{aligned} $} \\
    \\
    \hline
    \\
    \textbf{Dual-symmetric:} & {$\begin{aligned}\Aboxed{\underbar{T}(b) &= (b\cdot\dot{\nabla})\mean{\dot{a}_e\bv{F}_e + \dot{a}_m\bv{F}_m}_1 - \frac{b}{2}\left[\mean{\bv{F}_e^2}_0 + \mean{\bv{F}_m^2}_0\right]} \\
    \Aboxed{\overline{T}(b) &= \dot{\nabla}\mean{(\dot{a}_e\bv{F}_e + \dot{a}_m\bv{F}_m)b}_0 - \frac{b}{2}\left[\mean{\bv{F}_e^2}_0 + \mean{\bv{F}_m^2}_0\right]}\end{aligned}$} \\
    \\
    $z = a_e + a_m I$ & {$\begin{aligned} \rv{P}_o &= \dot{\rv{\nabla}}\left[\rv{E}_e\cdot\dot{\rv{A}} + \rv{B}_m\cdot\dot{\rv{C}}\right] &  \rv{P}_s &= -[(\rv{E}_e\cdot\rv{\nabla})\rv{A} + (\rv{B}_m\cdot\rv{\nabla})\rv{C}] \end{aligned}$} \\
    \\
    {$\begin{aligned}\bv{F} &= \nabla z \\ &= \bv{F}_e + \bv{F}_m I \end{aligned}$} & {$\begin{aligned}T^{\mu\nu} &= \sum_\alpha\left( (F_e)^{\alpha\mu}\partial^\nu (a_e)_\alpha  + (F_m)^{\alpha\mu}\partial^\nu (a_m)_\alpha  \right) \\
      &- \frac{\eta^{\mu\nu}}{4}\sum_{\alpha\beta}\left[(F_e)_{\alpha\beta}(F_e)^{\beta\alpha} + (F_m)_{\alpha\beta}(F_m)^{\beta\alpha}\right]\end{aligned} $} \\
    \\
    \hline
    \\
    \textbf{Constrained:} & {$\begin{aligned}\Aboxed{\underbar{T}(b) &\to \frac{(b\cdot\dot{\nabla})}{2}\mean{\dot{a}_e\bv{F} + \dot{a}_m\bv{G}}_1} & \Aboxed{\overline{T}(b) &\to \frac{\dot{\nabla}}{2}\mean{(\dot{a}_e\bv{F} + \dot{a}_m\bv{G})b}_0} \end{aligned}$} \\
    \\
    $\displaystyle \bv{F}_e = \bv{F}_m I \to \frac{\bv{F}}{\sqrt{2}}$ & {$\begin{aligned} \rv{P}_o &\to \frac{1}{2}\dot{\rv{\nabla}}\left[\rv{E}\cdot\dot{\rv{A}} + \rv{B}\cdot\dot{\rv{C}}\right] &  \rv{P}_s &\to -\frac{1}{2}[(\rv{E}\cdot\rv{\nabla})\rv{A} + (\rv{B}\cdot\rv{\nabla})\rv{C}] \end{aligned}$} \\
    \\
    $\bv{F}=\bv{G}I$ & {$\begin{aligned}T^{\mu\nu} &\to \frac{1}{2}\sum_\alpha\left( F^{\alpha\mu}\partial^\nu (a_e)_\alpha  + G^{\alpha\mu}\partial^\nu (a_m)_\alpha  \right) \end{aligned} $} \\
    \\
    \hline 
 \end{tabular}
 \caption[Canonical energy-momentum tensor]{Canonical energy-momentum stress tensors, conserved by translation symmetry.  Unlike the symmetric tensor in Table~\ref{tab:symtensors}, the canonical tensor naturally separates the Poynting vector $\rv{P}$ into an \emph{orbital} part $\rv{P}_o$ and a separate \emph{spin} part $\rv{P}_s$ that does not contribute to energy-momentum transport (for all three Lagrangians) \cite{Berry2009,Ohanian1986,Mita2000,Bliokh2013}.  The traditional Lagrangian produces field-asymmetric parts that are \emph{electric}-biased \cite{Soper1976,Bliokh2013}. The dual-symmetric Lagrangian in Table~\ref{tab:lagrangians} adds the complementary constituent field that is \emph{magnetic}-biased. Constraining this Lagrangian makes both contributions describe the same total field \cite{Bliokh2013}. }
 \label{tab:canemtensor}
\end{table}

To obtain more familiar expressions for the canonical energy-momentum tensor, we expand Eqs.~\eqref{eq:canonicalemstress} and \eqref{eq:canonicalemadjoint} in terms of a relative frame, using the spacetime splits $\bv{F}_{e,m} = \rv{E}_{e,m} + \rv{B}_{e,m} I$ (that become constrained to $\bv{F} = \rv{E} + \rv{B}I$, with $\bv{G} = \bv{F}I^{-1} = \rv{B} - \rv{E} I$), as well as the (transverse) Coulomb gauge $a_e = \rv{A}\gamma_0$ and $a_m = \rv{C}\gamma_0$ for the potentials\footnote{Recall from Section~\ref{sec:bivectorpotential} that this choice of gauge is equivalent to using the gauge-invariant transverse bivector potential $\bv{Z} = z_\perp\gamma_0$ in the frame $\gamma_0$, and then fixing the gauge to make the gauge-dependent longitudinal spinor potential $\psi_z = z_\parallel\gamma_0$ vanish, since it is arbitrary and thus unmeasurable.}:
\begin{align}\label{eq:emtensorrel}
  \underbar{T}(\gamma_0)\gamma_0 &= \varepsilon + \rv{P}, \\
 \underbar{T}(\gamma_i)\gamma_0 &= -(\rv{\sigma}_i\cdot\dot{\rv{\nabla}})\left[\rv{E}_e\cdot\dot{\rv{A}} + \dot{\rv{A}}\times\rv{B}_e + \rv{B}_m\cdot\dot{\rv{C}} - \dot{\rv{C}}\times\rv{E}_m\right], \nonumber \\
 &\to -\frac{1}{2}(\rv{\sigma}_i\cdot\dot{\rv{\nabla}})\left[\rv{E}\cdot\dot{\rv{A}} + \rv{B}\cdot\dot{\rv{C}} + \dot{\rv{A}}\times\rv{B} - \dot{\rv{C}}\times\rv{E}\right], \nonumber 
\end{align}
\begin{align}\label{eq:emadjointrel}
  \overline{T}(\gamma_0)\gamma_0 &= \varepsilon + \dot{\rv{\nabla}}\left[\rv{E}_e\cdot\dot{\rv{A}} + \rv{B}_m\cdot\dot{\rv{C}}\right], \\
  &\to \varepsilon + \frac{1}{2}\dot{\rv{\nabla}}\left[\rv{E}\cdot\dot{\rv{A}} + \rv{B}\cdot\dot{\rv{C}}\right], \nonumber\\
  \overline{T}(\gamma_i)\gamma_0 &= -\rv{P}\cdot\rv{\sigma}_i - \dot{\rv{\nabla}}\left[(\dot{\rv{A}}\times\rv{B}_e - \dot{\rv{C}}\times\rv{E}_m)\cdot\rv{\sigma}_i\right], \nonumber \\
  &\to -\rv{P}\cdot\rv{\sigma}_i - \frac{1}{2}\dot{\rv{\nabla}}\left[(\dot{\rv{A}}\times\rv{B} - \dot{\rv{C}}\times\rv{E})\cdot\rv{\sigma}_i\right]. \nonumber
\end{align}
Here $\varepsilon = \frac{1}{2}(|\rv{E}|^2 + |\rv{B}|^2)$ and $\rv{P} = \rv{E}\times\rv{B}$ are the energy density and Poynting vector derived in Section \ref{sec:symemstress}, and we have used the fact that the energy densities and Poynting vectors for the constituent fields, $\varepsilon_e,\varepsilon_m,\rv{P}_e,\rv{P}_m$, sum to the total field quantities according to \eqref{eq:symemconstituent}: $\varepsilon = \varepsilon_e + \varepsilon_m$, $\rv{P} = \rv{P}_e + \rv{P}_m$. 
The vector part of $\overline{T}(\gamma_0)\gamma_0$ represents the canonical (orbital) momentum density $\vec{P}_o$, i.e., $\overline{T}(\gamma_0)\gamma_0 = \varepsilon + \vec{P}_o$ (we describe this quantity below).

The associated Noether charges from \eqref{eq:noethercharge} are the integrals of the energy and momentum densities over all space, i.e., the total energy and momentum of the field:
\begin{align}\label{eq:integralEM}
\int \varepsilon \; d^3x = {\rm const},~~~~~~ \int \vec{P} \; d^3x = \int \vec{P}_o \; d^3x = {\rm const}.
\end{align}
These 4 conserved charges are derived from \eqref{eq:noethercharge} and \eqref{eq:canonicalemstress} by choosing the generator $b$ to be the 4 orthogonal vector directions $\gamma_\mu$.  Note that these equations are valid only for sufficiently localized square-integrable fields. Importantly, the second equation shows that the difference between the canonical and Poynting momentum densities (i.e., the difference between the canoncial and symmetric energy-momentum tensors) does not affect the integral momentum value.

We also expand the correction Belinfante tensor \eqref{eq:emcorrection} and \eqref{eq:emcorrectionadjoint} in a similar way:
\begin{align}
  \label{eq:emcorrectionrel}
  \underbar{K}(\gamma_0)\gamma_0 &= 0, \\
\underbar{K}(\gamma_i)\gamma_0 &= -\left[(\rv{E}_e\cdot\rv{\nabla})(\rv{A}\cdot \rv{\sigma}_i) + (\rv{B}_m\cdot\rv{\nabla})(\rv{C}\cdot \rv{\sigma}_i)\right] \nonumber \\
  &\quad - \frac{1}{2}\left[\rv{E}_e(\rv{E}_e\cdot \rv{\sigma}_i) + \rv{B}_e(\rv{B}_e\cdot \rv{\sigma}_i) + \rv{E}_m(\rv{E}_m\cdot \rv{\sigma}_i) + \rv{B}_m(\rv{B}_m\cdot \rv{\sigma}_i)\right] \nonumber \\
  &\quad + \left[ \rv{B}_e\times\rv{\nabla}(\rv{A}\cdot \rv{\sigma}_i) - \rv{E}_m\times\rv{\nabla}(\rv{C}\cdot \rv{\sigma}_i)\right], \nonumber \\  
  &\to -\frac{1}{2}\left[(\rv{E}\cdot\rv{\nabla})(\rv{A}\cdot \rv{\sigma}_i) + (\rv{B}\cdot\rv{\nabla})(\rv{C}\cdot \rv{\sigma}_i)\right] \nonumber \\
  &\quad - \frac{1}{2}\left[\rv{E}(\rv{E}\cdot \rv{\sigma}_i) + \rv{B}(\rv{B}\cdot \rv{\sigma}_i) - \rv{B}\times\rv{\nabla}(\rv{A}\cdot \rv{\sigma}_i) + \rv{E}\times\rv{\nabla}(\rv{C}\cdot \rv{\sigma}_i)\right], \nonumber
\end{align}
\begin{align}
  \label{eq:emcorrectionadjointrel}
  \overline{K}(\gamma_0)\gamma_0 &= -\left[(\rv{E}_e\cdot\rv{\nabla})\rv{A} + (\rv{B}_m\cdot\rv{\nabla})\rv{C}\right], \\
  &\to -\frac{1}{2}\left[(\rv{E}\cdot\rv{\nabla})\rv{A} + (\rv{B}\cdot\rv{\nabla})\rv{C}\right], \nonumber \\
\overline{K}(\gamma_i)\gamma_0 &= \frac{1}{2}\left[(\rv{\sigma}_i\cdot\rv{E}_e)\rv{E}_e + (\rv{\sigma}_i\cdot\rv{B}_e)\rv{B}_e + (\rv{\sigma}_i\cdot\rv{E}_m)\rv{E}_m + (\rv{\sigma}_i\cdot\rv{B}_m)\rv{B}_m\right] \nonumber \\
&\quad - \left[\rv{\sigma}_i\cdot(\rv{B}_e\times\rv{\nabla})\rv{A} - \rv{\sigma}_i\cdot(\rv{E}_m\times\rv{\nabla})\rv{C}\right], \nonumber \\
&\to \frac{1}{2}\left[(\rv{\sigma}_i\cdot\rv{E})\rv{E} + (\rv{\sigma}_i\cdot\rv{B})\rv{B}\right] - \frac{1}{2}\left[\rv{\sigma}_i\cdot(\rv{B}\times\rv{\nabla})\rv{A} - \rv{\sigma}_i\cdot(\rv{E}\times\rv{\nabla})\rv{C}\right]. \nonumber
\end{align}

For all the above tensors, the terms involving only the ``electric'' parts (i.e., $\bv{F}_e$, $a_e$, $\rv{E}_e$, $\rv{B}_e$, and $\rv{A}$) are equivalent to the corresponding tensors for traditional electromagnetism \cite{Soper1976,Bliokh2013}. The dual-symmetric Lagrangian adds complementary ``magnetic'' terms to these traditional tensors.  In turn, the constrained dual-symmetric Lagrangian makes both these halves correspond to the same field, producing dual-symmetrized canonical tensors that agree with those obtained very recently in \cite{Bliokh2013}.


\subsubsection{Orbital and spin momentum densities}


In the above expressions for the energy-momentum tensors, it is important to analyze the energy-momentum densities, which are given by the vector terms in $\underbar{T}_{\text{sym}}(\gamma_0)\gamma_0$, $\underbar{K}_{\text{sym}}(\gamma_0)\gamma_0$, and their adjoints. As we shall see in the next section, these densities will be responsible for the generation of the angular momentum of the field in a way similar to \eqref{eq:symamflux}, but with a clear separation of the spin and orbital contributions.

First, observe in \eqref{eq:emcorrectionrel} that the correction to the timelike component vanishes: $\underbar{K}(\gamma_0) = 0$, so the canonical tensor $\underbar{T}(\gamma_0)$ yields precisely the same energy-momentum density as the symmetric tensor $\underbar{T}_{\text{sym}}(\gamma_0)$.  That is, both the correct energy density $\varepsilon = (|\rv{E}|^2 + |\rv{B}|^2)/2$ and Poynting vector $\rv{P} = \rv{E}\times\rv{B}$ are obtained in \eqref{eq:emtensorrel}.  These quantities satisfy a continuity equation, i.e., the Poynting theorem, or energy-transport equation \cite{Jackson1999}.

In contrast, the adjoint correction in \eqref{eq:emcorrectionadjointrel} is nonzero. Remarkably, the vector terms in $\overline{T}(\gamma_0)\gamma_0$ and $\overline{K}(\gamma_0)\gamma_0$ are the so-called \emph{canonical (orbital) and spin momentum densities} \cite{Berry2009,Bekshaev2011,Bliokh2012,Bliokh2013,Bliokh2013b}:
\begin{align}\label{eq:orbitalmomentum}
  \rv{P}_o &= \overline{T}(\gamma_0)\gamma_0 - \varepsilon, \\
  &= \dot{\rv{\nabla}}[\rv{E}_e\cdot\dot{\rv{A}} + \rv{B}_m\cdot\dot{\rv{C}}], \nonumber\\
&\to \frac{1}{2}\dot{\rv{\nabla}}[\rv{E}\cdot\dot{\rv{A}} + \rv{B}\cdot\dot{\rv{C}}], \nonumber \\ \nonumber\\
\label{eq:spinmomentum}
  \rv{P}_s &= \overline{K}(\gamma_0)\gamma_0, \\
  &= -[(\rv{E}_e\cdot\rv{\nabla})\rv{A} + (\rv{B}_m\cdot\rv{\nabla})\rv{C}], \nonumber\\
&\to -\frac{1}{2}[(\rv{E}\cdot\rv{\nabla})\rv{A} + (\rv{B}\cdot\rv{\nabla})\rv{C}]. \nonumber
\end{align}
Together, the orbital and spin momentum densities form the Poynting vector in the corresponding sector of the symmetric tensor $\overline{T}_{\rm sym}(\gamma_0)\gamma_0$: $\rv{P}_o + \rv{P}_s = \rv{P}$. Furthermore, the wedge products of these momentum densities generate, respectively, the orbital and spin angular momenta of the field \cite{Bekshaev2011,Bliokh2013,Bliokh2013b,Ohanian1986,Mita2000,Soper1976}, as we shall see in the next section. 
It follows from Eq.~\eqref{eq:integralEM} that the spin momentum density does not contribute to the inetgral momentum of the localized field:
\begin{align}
\int \vec{P}_s \; d^3x = 0.
\end{align}

Importantly, although it is the total Poynting vector that is gauge-invariant and coupled to gravity \cite{Soper1976}, the spin and orbital degrees of freedom manifest themselves drastically differently in local interactions with matter. Optical and quantum measurements of the momentum density of light reveal the canonical (orbital) momentum (in the Coulomb gauge) \eqref{eq:orbitalmomentum}, while the spin momentum \eqref{eq:spinmomentum} remains largely unobservable and virtual \cite{Bliokh2013,Bliokh2013a,Bliokh2013b,Bliokh2013c}. 
The appearance of the canonical momentum density in optics can be explained by the fact that it has a clear physical interpretation. Namely, for monochromatic optical fields, it is given by the local expectation value of the quantum momentum operator $\hat{\rv{p}} \propto -i\rv{\nabla}$, which makes it proportional to the local gradient of the phase, i.e., the \emph{local wavevector} of the field \cite{Berry2009,Bliokh2013a}. In contrast, the Poynting vector does not have such intuitively-clear wave interpretation.


\subsection{Canonical angular momentum tensor}\label{sec:canonicalAM}


We are now in a position to consider rotational symmetries of spacetime, which produces the conservation of the relativistic angular momentum of the field. Suppose that we perform a general restricted Lorentz transformation (i.e., a spatial rotation, boost, or combination thereof).  As we discussed in Section \ref{sec:lorentz}, the bivectors themselves are the generators of this group, so the appropriate transformation is $\underbar{U}(a,\alpha) = U_\alpha a \widetilde{U}_\alpha$, where $U_\alpha = \exp(\alpha \bv{R}/2)$ and $\bv{R}$ is a unit bivector corresponding to the plane of rotation.  A symmetry of the Lagrangian follows if we first transform the fields $a' = \underbar{U}(a,\alpha) = U_\alpha a \tilde{U}_\alpha$ then counter-transform the coordinates $x' = \overline{U}(x,\alpha) = \tilde{U}_\alpha x U_\alpha$, as discussed in \eqref{eq:transformations}.

From this symmetry transformation, the corresponding increments \eqref{eq:increments} have the form
\begin{subequations}
\begin{align}
  \delta x(\bv{R}) &= -\bv{R}\cdot x, \\
  \delta a(\bv{R}) &= \bv{R}\cdot a + [(-\bv{R}\cdot x)\cdot\nabla]a. 
\end{align}
\end{subequations}
The change \eqref{eq:noethertransform} of the Lagrangian thus produces the conserved Noether current
\begin{align}\label{eq:angmomtensgen}
  \underbar{M}(\bv{R}) &= \sum_\mu\gamma_\mu\smean{(\partial_\mu a_e)\delta a_e(\bv{R}) + (\partial_\mu a_m)\delta a_m(\bv{R})}_0 - \delta x(\bv{R})\mathcal{L}, \\
  &= \left(a_e\cdot \bv{R} + [(\bv{R}\cdot x)\cdot\nabla]a_e\right)\cdot\bv{F}_e + \left(a_m\cdot \bv{R} + [(\bv{R}\cdot x)\cdot\nabla]a_m\right)\cdot\bv{F}_m  - (\bv{R}\cdot x)\cdot\nabla\mathcal{L}, \nonumber \\
  &\to \frac{1}{2}\left[(a_e\cdot \bv{R} + [(\bv{R}\cdot x)\cdot\nabla]a_e)\cdot\bv{F} + (a_m\cdot \bv{R} + [(\bv{R}\cdot x)\cdot\nabla]a_m)\cdot \bv{G}\right]. \nonumber
\end{align}
This conserved current is the \emph{canonical angular momentum tensor}, which transforms bivectors (i.e., Lorentz transformation generators) into vectors (linear momenta), as we discussed in Section \ref{sec:symamstress}.

After comparing \eqref{eq:angmomtensgen} to the form of $\underbar{T}$ in \eqref{eq:canonicalemstress}, we can make the simplification
\begin{align}\label{eq:angmomtens}
  \underbar{M}(\bv{R}) = \underbar{L}(\bv{R}) + \underbar{S}(\bv{R}), ~~~~~ 
\underbar{L}(\bv{R}) = \underbar{T}(\bv{R}\cdot x),
\end{align}
where
\begin{align}\label{eq:spinmom}
  \underbar{S}(\bv{R}) &\equiv (a_e\cdot \bv{R})\cdot\bv{F}_e + (a_m\cdot \bv{R})\cdot\bv{F}_m, \\
  &\to \frac{1}{2}[(a_e\cdot \bv{R})\cdot\bv{F} + (a_m\cdot \bv{R})\cdot\bv{G}]. \nonumber
\end{align}
Here, $\underbar{L}(\bv{R})$ is the \emph{orbital angular momentum} of the field (including the boost momentum), which is related to the canonical energy-momentum tensor in the same way as in Eq.~\eqref{eq:symamfluxadjoint} for the symmetric (Belinfante--Rosenfeld) tensors. The additional term $\underbar{S}(\bv{R})$ that appears is the \emph{spin angular momentum} tensor.  Importantly, this spin tensor describes \emph{intrinsic} (i.e., independent of the radius-vector $x$) angular momentum, and it \emph{cannot} be obtained from the energy-momentum tensor. 

The corresponding adjoint angular momentum tensors transform unit vectors $b$ (linear momentum directions) into bivectors (associated angular momenta), so can be easier to interpret physically.  The symmetric and canonical adjoint angular momentum tensors have the intuitive forms
\begin{align}\label{eq:angmomtensadjoint}
  \overline{M}_{\text{sym}}(b) &= x\wedge\overline{T}_{\text{sym}}(b), & \overline{M}(b) &= x\wedge \overline{T}(b) + \overline{S}(b) \equiv \overline{L}(b)+\overline{S}(b), 
\end{align}
where the adjoint spin angular momentum tensor is
\begin{align}\label{eq:spinmomadjoint}
  \overline{S}(b) &= a_e\wedge(b \cdot \bv{F}_e)+a_m\wedge(b\cdot \bv{F}_m),  \\
  &\to \frac{1}{2}[a_e\wedge (b\cdot \bv{F})+ a_m\wedge(b\cdot \bv{G}))]. \nonumber
\end{align}
We detail these angular momentum tensors in Table~\ref{tab:canamtensor} for reference.

\begin{table}
  \centering
  \begin{tabular}{l l}
    \hline
\noalign{\vskip 2mm} 
    \multicolumn{2}{c}{\textbf{Canonical Angular Momentum Tensor}} \\
\noalign{\vskip 2mm} 
    \hline \\
    \multicolumn{2}{c}{$\begin{aligned} \Aboxed{\underbar{M}(\bv{R}) &= \underbar{L}(\bv{R}) + \underbar{S}(\bv{R})} &
      \Aboxed{\overline{M}(b) &= \overline{L}(b) + \overline{S}(b)} \\
      \underbar{L}(\bv{R}) &= \underbar{T}(\bv{R}\cdot x) & \overline{L}(b) &= x\wedge \overline{T}(b) \\
      \\
      \overline{S}(\gamma_0) &= \boxed{\rv{S}}I^{-1} & \overline{L}(\gamma_0) &= [\varepsilon \rv{x} - (\cc t)\boxed{\rv{P}_o}] + \rv{x}\times\boxed{\rv{P}_o} I^{-1} \\
      M^{\alpha\beta\gamma} &= L^{\alpha\beta\gamma} + S^{\alpha\beta\gamma} & L^{\alpha\beta\gamma} & = x^{\alpha}T^{\beta\gamma}\, - x^{\beta}T^{\alpha\gamma} \end{aligned}$} \\
    \\
    \hline
    \\
    \textbf{Traditional:} & {$\begin{aligned} \Aboxed{\underbar{S}(\bv{R}) &= (a\cdot \bv{R})\cdot\bv{F}} & \Aboxed{\overline{S}(b) &= a\wedge(b\cdot \bv{F})} \end{aligned}$} \\
    \\
    $z = a_e \equiv a$ & $\rv{S} = \rv{E}\times\rv{A} $ \\
    \\
    $\bv{F} = \nabla a_e = \bv{F}_e$ & {$\begin{aligned} S^{\alpha\beta\gamma} &= F^{\gamma\alpha}a^\beta - F^{\gamma\beta}a^{\alpha} \end{aligned} $} \\
    \\
    \hline
    \\
    \textbf{Dual-symmetric:} &  {$\begin{aligned}\Aboxed{\underbar{S}(\bv{R}) &= (a_e\cdot\bv{R})\cdot\bv{F}_e + (a_m\cdot\bv{R})\cdot\bv{F}_m} \\
      \Aboxed{\overline{S}(b) &= a_e\wedge(b\cdot \bv{F}_e) + a_m\wedge(b\cdot \bv{F}_m)} \end{aligned}$} \\ 
    \\
    $z = a_e + a_m I$ & $\rv{S} = \rv{E}_e\times\rv{A} + \rv{B}_m\times\rv{C}$ \\
    \\
    $\bv{F} = \nabla z = \bv{F}_e + \bv{F}_m I$ & {$\begin{aligned}S^{\alpha\beta\gamma} &= F_e^{\gamma\alpha}a_e^\beta - F_e^{\gamma\beta}a_e^{\alpha} + F_m^{\gamma\alpha}a_m^\beta - F_m^{\gamma\beta}a_m^{\alpha}\end{aligned} $} \\
    \\
    \hline
    \\
    \textbf{Constrained:} & {$\begin{aligned}\Aboxed{\underbar{S}(\bv{R}) &= \frac{1}{2}[(a_e\cdot\bv{R})\cdot\bv{F} + (a_m\cdot\bv{R})\cdot\bv{G}]} \\
      \Aboxed{\overline{S}(b) &= \frac{1}{2}[a_e\wedge(b\cdot \bv{F}) + a_m\wedge(b\cdot \bv{G})]} \end{aligned}$} \\ 
    \\
    $\displaystyle \bv{F}_e = \bv{F}_m \to \frac{\bv{F}}{\sqrt{2}}$ & $\displaystyle \rv{S} \to \frac{1}{2}[\rv{E}\times\rv{A} + \rv{B}\times\rv{C}]$ \\
    \\
    $\bv{F} = \bv{G}I$ & {$\begin{aligned}S^{\alpha\beta\gamma} &\to \frac{1}{2}\left[F^{\gamma\alpha}a_e^\beta - F^{\gamma\beta}a_e^{\alpha} + G^{\gamma\alpha}a_m^\beta - G^{\gamma\beta}a_m^{\alpha}\right]\end{aligned} $} \\
    \\
    \hline 
 \end{tabular}
 \caption[Canonical angular momentum tensor]{Canonical angular momentum tensors, conserved by Lorentz transformations.  Unlike the symmetric tensor in Table~\ref{tab:symtensors}, the canonical tensor naturally separates into an extrinsic \emph{orbital} part $\underline{L}$ that contains the wedge product of the orbital part $\rv{P}_o$ of the Poynting vector $\rv{P}$ in Table~\ref{tab:canemtensor}, as well as a separate \emph{intrinsic spin} part $\underline{S}$ containing the spin vector $\rv{S}$.  As with the energy-momentum tensor, the traditional Lagrangian produces electric-biased quantities, while the dual-symmetric Lagrangian adds complementary magnetic-biased terms. Constraining the Lagrangian symmetrizes the field contributions. }
 \label{tab:canamtensor}
\end{table}
The appearance of the spin angular momentum in \eqref{eq:spinmom} and \eqref{eq:spinmomadjoint} indicates the principal difference between the canonical and symmetrized pictures of the angular momentum in field theory. Namely, the Belinfante--Rosenfeld angular momentum tensor $\overline{M}_{\rm sym}$ in \eqref{eq:angmomtensadjoint} (and Section \ref{sec:symamstress}) is simply the wedge product of the symmetric energy-momentum tensor $\overline{T}_{\rm sym}$ with a coordinate radius $x$, so does not contain any new information about local properties of the field. In contrast, the canonical angular momentum tensor $\overline{M}$ has both an extrinsic orbital part $\overline{L}$ and an independent intrinsic spin part $\overline{S}$, which signifies additional degrees of freedom for the field. Critically, the \emph{spin and orbital degrees of freedom are separated only in the canonical picture}. As we mentioned before, these degrees of freedom have been historically considered to have no separate meaning in both orthodox quantum electrodynamics and classical electromagnetic field theory, but nonetheless they clearly correspond to separate observable effects in local light-matter interactions in optical experiments \cite{Allen1999,Andrews2008,Torres2011,Andrews2013,O'Neil2002,Garces2003,Curtis2003, Zhao2007,Adachi2007}. Furthermore, the integral values  (Noether charges) of the spin and orbital angular momenta are \emph{separately conserved} quantities \cite{Enk1994}, and the corresponding Noether currents $\overline{L}$ and $\overline{S}$ can also be brought to meaningful separately conserved forms \cite{Bliokh2014}.
Thus, akin to the local momentum densities, the optical angular momentum properties are described by the \emph{canonical} Noether currents rather than the symmetrized Belinfante--Rosenfeld currents \cite{Bliokh2013,Bliokh2013a,Bliokh2013b,Bliokh2014} (see also \cite{Leader2014} for a discussion of this same issue in QCD).

For simplicity, we only compute the timelike components of these tensors in a relative frame (in the Coulomb gauge):
\begin{align}\label{eq:angmomrel}
  \overline{M}_{\text{sym}}(\gamma_0) &= [\varepsilon \rv{x} - (\cc t)\rv{P}] + \rv{x}\times\rv{P} I^{-1} \equiv \vec{N}+\vec{J} I^{-1}, 
\end{align}
\begin{align}
  \label{eq:orbangmomrel}
  \overline{L}(\gamma_0) &= [\varepsilon \rv{x} - (\cc t)\rv{P}_o] + \rv{x}\times\rv{P}_o I^{-1}\equiv \vec{N}_o+\vec{L} I^{-1},
\end{align}
\begin{align}
  \label{eq:spinangmomrel}
  \overline{S}(\gamma_0) & = \left[\rv{E}_e\times\rv{A} + \rv{B}_m\times\rv{C}\right] I^{-1} \equiv \vec{S} I^{-1}, \\
  &\to \frac{1}{2}[\rv{E}\times\rv{A} + \rv{B}\times\rv{C}]I^{-1}. \nonumber
\end{align}
For the orbital angular momentum $\overline{L}$, the associated orbital linear momentum $\rv{P}_o$ is given by \eqref{eq:orbitalmomentum}, while the energy density $\varepsilon$ and full Poynting vector $\rv{P}$ are the usual expressions.  
Recall from our discussion in Section \ref{sec:symamstress} that each bivector term in the angular momentum that involves $I$ generates spatial rotations, while the remaining bivectors generate boosts. One can then clearly see that the intrinsic spin angular momentum \eqref{eq:spinangmomrel} is a purely spatial rotation (around the axis $\rv{S}$) that appears explicitly only in the canonical angular momentum tensor \cite{Bliokh2013}. 

The Noether charges (integral conserved quantities) associated with the symmetric and canonical angular-momentum tensors are as follows:
\begin{align}\label{eq:integralAM}
\int \vec{J} \; d^3x = \int (\vec{L}+\vec{S}) \; d^3x = {\rm const},~~~~~~ \int \vec{N} \; d^3x = \int \vec{N}_o \; d^3x = {\rm const}.
\end{align}
These 6 conserved charges represent the integral angular momentum and boost momentum of the field, and are derived from \eqref{eq:noethercharge} and \eqref{eq:angmomtens} by choosing the generator $\bv{R}$ to be the 6 orthogonal bivector directions $\gamma_{\mu\nu}$.
Together with the 4 energy-momentum integrals \eqref{eq:integralEM} they form 10 Poincar\'{e} invariants of the field associated with the 10-parameter Poincar\'{e} group of spacetime symmetries \cite{Bialynicki-Birula1996,Thide2014,Soper1976,Bliokh2013}.

The seeming absence of the intrinsic spin contribution in the Belinfante--Rosenfeld angular momentum \eqref{eq:angmomrel} causes the so-called ``spin-of-a-plane-wave paradox'' \cite{Soper1976,Ohanian1986,Mita2000,Zambrini2005,Allen2002,Bliokh2013b}. Namely, the local density of the total angular momentum $\overline{M}_{\text{sym}}(\gamma_0)$ vanishes in a circularly polarized plane electromagnetic wave, while the spin density $\overline{S}(\gamma_0)$ is non-zero, as it should be.  The explanation of this paradox lies in the fact that, in contrast to the extrinsic orbital angular momentum \eqref{eq:orbangmomrel}, the relation between the spin angular momentum density \eqref{eq:spinangmomrel} and the spin part of the momentum density \eqref{eq:spinmomentum} is essentially \emph{nonlocal}:
\begin{align}  
  \int{\rv{S}}\, d^3 x&= \int{ \rv{x}\times\rv{P}_s}\,d^3 x,
\end{align}
so applies only for sufficiently localized fields, such as wave packets. 

We can see that the concepts of the canonical (orbital) momentum and independent spin density is more physically meaningful and intuitive than the concept of the Poynting vector and its corresponding Belinfante--Rosenfeld angular momentum. Akin to the canonical momentum density, the spin angular momentum density has a clear physical meaning in the case of monochromatic optical fields. Namely, it is proportional to the ellipticity of the local polarization of the field and is directed along its normal axis. In other words, the rotation of the electric and magnetic field vectors in optical fields generates a well-defined spin angular momentum density.

Most importantly for our study, the separation of the spin and orbital degrees of freedom is closely related to the \emph{dual symmetry} between the electric and magnetic contributions. The constrained forms of the spin momentum density in \eqref{eq:spinangmomrel}, as well as the orbital momentum density in \eqref{eq:orbitalmomentum}, are dual-symmetric in the fields (as recently emphasized in \cite{Bliokh2013}), while the traditional Lagrangian produces dual-asymmetric spin and orbital characteristics, containing only the electric part (e.g., $\bv{F}_e$ and $a_e$) \cite{Soper1976,Bliokh2013}. This asymmetry is closely related to the fact that the spin and orbital degrees of freedom of electromagnetic waves are coupled to dual-asymmetric matter via the \emph{electric}-dipole coupling. Nonetheless, for vacuum fields, the dual-symmetric canonical Noether currents provide a more natural and self-consistent picture \cite{Bliokh2013}.


\subsection{Helicity pseudocurrent}\label{sec:helicity}


In addition to the Poincar\'{e} symmetries of spacetime (translations and rotations) considered above, the Lagrangian $\mathcal{L}_{\text{dual}}$ in \eqref{eq:maxwelllagrangian} exhibits one more continuous symmetry, namely, the \emph{dual symmetry}. Hence, we can determine the proper conserved quantity that should be associated with the Lagrangian invariance with respect to the dual ``electric-magnetic'' rotations $\psi = \exp(\theta I/2)$. Importantly, this symmetry does \emph{not} exist in the traditional Lagrangian \eqref{eq:traditional} since only the \emph{electric} vector potential $a_e$ is usually considered, which has no complex phase.

The proper field increment for the transformations $a' = \psi a \psi^*$ is 
\begin{align}
  \delta a(I) = a I^{-1}.
\end{align}
Notably, this field increment is simply the Hodge-dual of $a$. Noether's theorem \eqref{eq:noethercurrent} thus produces a conserved \emph{helicity pseudovector current} 
\begin{align}
  \underbar{X}(I) &= \sum_\mu\gamma_\mu\smean{(\partial_\mu a_e)a_eI^{-1} + (\partial_\mu a_m)a_mI^{-1}}_4, \\
  &= \sum_\mu\gamma_\mu[(\partial_\mu a_e)\cdot a_e + (\partial_\mu a_m)\cdot a_m]I^{-1}, \nonumber \\
  &= a_e\wedge(\nabla a_e) + a_m\wedge(\nabla a_m), \nonumber \\
  &= a_e\wedge\bv{F}_e + a_m\wedge\bv{F}_m, \nonumber \\
  &= [a_e\cdot(\bv{F}_eI^{-1}) + a_m \wedge(\bv{F}_m I^{-1})]I, \nonumber \\
  &\to \frac{1}{2}[a_e\cdot \bv{G} + a_m\cdot \bv{F}]I. \nonumber
\end{align}
Interestingly, after factoring out $I$ from the pseudovector, each constituent potential becomes coupled to the \emph{Hodge-dual} of its associated field.  If the additional constraint in the last line is imposed, then the vector potentials effectively cross-couple through the total field $\bv{F} = \bv{G} I$.  We detail this pseudovector current in Table~\ref{tab:canhelicity} for reference.  It is also worth noting that the Weinberg--Witten theorem \cite{Weinberg1980,Stepanovsky1998,Smith2007} that forbids conserved vector currents for massless spin-1 electromagnetic fields is not violated here, for two reasons: first, the conserved helicity pseudovector current here is not a grade-1 vector, but is rather a grade-3 pseudovector; second, as noted by Gaillard and Zumino \cite{Gaillard1981}, this conserved current is not gauge-invariant.

\begin{table}
  \centering
  \begin{tabular}{l l}
    \hline
\noalign{\vskip 2mm} 
    \multicolumn{2}{c}{\textbf{Helicity Pseudocurrent}} \\
\noalign{\vskip 2mm} 
    \hline \\
    \multicolumn{2}{c}{$\underbar{X}(I) = [\chi I + \rv{S} I^{-1} ]\gamma_0$} \\
    \\
    \hline \\
    \\
    \textbf{Traditional:} & $z = a_e \equiv a, \quad \bv{F} = \nabla a_e = \bv{F}_e$ \\
    \\
    & {\bf No dual symmetry of the Lagrangian.} \\
    \\
    \hline
    \\
    \textbf{Dual-symmetric:} & $z = a_e + a_m I, \quad \bv{F} = \nabla z = \bv{F}_e + \bv{F}_m I$ \\
    \\
    & {$\begin{aligned}\Aboxed{\underbar{X}(I) &= [a_e\cdot(\bv{F}_e I^{-1}) + a_m\cdot(\bv{F}_mI^{-1})]I} \\
    \\
    \chi &= \rv{A}\cdot\rv{B}_e - \rv{C}\cdot\rv{E}_m \end{aligned}$} \\
    \\
    \hline
    \\
    \textbf{Constrained:} & $\bv{F}_e \to \bv{F}_m \to \bv{F}/\sqrt{2} = \bv{G}I/\sqrt{2}$ \\
    \\
    & {$\begin{aligned}\Aboxed{\underbar{X}(I) &\to \frac{1}{2}[a_e\cdot\bv{G} + a_m\cdot\bv{F}]I} \\
    \\
    \chi &\to \frac{1}{2}(\rv{A}\cdot\rv{B} - \rv{C}\cdot\rv{E}) \end{aligned}$} \\
    \\
    \hline 
 \end{tabular}
 \caption[Canonical helicity pseudocurrent]{Canonical helicity pseudocurrent, conserved by the continuous \emph{dual symmetry} of the vacuum Lagrangian.  This symmetry does not exist for the traditional Lagrangian, for which the conserved \emph{integral} helicity can only be obtained from the symmetry of the \emph{action} \cite{Calkin1965,Deser1976,Deser1982}.  The pseudoscalar part of the helicity pseudocurrent is the helicity density $\chi$, while the relative pseudovector part is precisely the spin density $\rv{S}$ from Table~\ref{tab:canamtensor}.  Curiously, only the \emph{constrained} Lagrangian produces the field-symmetrized expression for the helicity density that is known in the literature \cite{Calkin1965,Deser1976,Afanasiev1996,Trueba1996,Cameron2012,Cameron2012b,Bliokh2013,Philbin2013}; this Lagrangian also produces the corresponding dual-symmetric spin density $\rv{S}$, which differs from the dual-asymmetric spin produced by the traditional Lagrangian \cite{Bliokh2013} (see Table~\ref{tab:canamtensor}).}
 \label{tab:canhelicity}
\end{table}

Writing this result in a relative frame using the Coulomb gauge yields
\begin{align}\label{eq:helicity}
  \underbar{X}(I) = (\chi I + \rv{S} I^{-1})\gamma_0,~~~~~
  \chi &= \rv{A}\cdot\rv{B}_e - \rv{C}\cdot\rv{E}_m, \\
  &\to \frac{1}{2}(\rv{A}\cdot\rv{B} - \rv{C}\cdot\rv{E}). \nonumber
\end{align}
The pseudoscalar term is the \emph{helicity density} $\chi$ of the electromagnetic field. When integrated over the spatial coordinates, it produces the conserved integral helicity, i.e., the corresponding Noether charge \eqref{eq:noethercharge}:
\begin{align}\label{eq:integralH}
 \int\chi\;d^3 x = {\rm const}. 
\end{align}
This integral helicity is equal to the difference between the numbers of right-hand and left-hand circularly polarized photons \cite{Calkin1965,Candlin1965,Oconnell1965,Afanasiev1996,Trueba1996,Bliokh2011,Cameron2012b}. 

The pseudovector term in the helicity pseudocurrent \eqref{eq:helicity} is the helicity flux density  $\overline{S}(\gamma_0) = \rv{S} I^{-1}$, which is precisely equal to the \emph{spin density} \eqref{eq:spinangmomrel} \cite{Cameron2012b,Bliokh2013,Cameron2012}. Intriguingly, the constrained form of the helicity $\chi$ matches the known expression for the conserved optical helicity in vacuum \cite{Calkin1965,Afanasiev1996,Trueba1996,Bliokh2011,Cameron2012b,Bliokh2013,Cameron2012}.  Indeed, the field-asymmetric electric part (or magnetic part) is not independently conserved since the corresponding symmetry in the Lagrangian does not exist unless the dual-symmetric form \eqref{eq:maxwelllagrangian} is used.  This fact gives strong indirect support for the requirement of the additional field $\bv{F}_m$ in the Lagrangian: without its complementary contribution, the conserved optical helicity $\chi$ is not obtainable as a Noether current from the Lagrangian\footnote{We note, however, that by carefully considering the symmetries of the integrated \emph{action} it is still possible to derive the conserved integral helicity \eqref{eq:integralH} using the traditional Lagrangian \cite{Calkin1965,Deser1976,Deser1982}.  Curiously, this alternative method reproduces the \emph{constrained} form of the dual-symmetric helicity density that we derive here, even though the form of the spin vector $\rv{S}$ for the traditional Lagrangian is not the same as for the constrained dual-symmetric Lagrangian \cite{Bliokh2013}.}. 


\subsection{Sources and symmetry-breaking}\label{sec:lagrangiansource}


Thus far we have focused on what can be said by constructing the general dual-symmetric Lagrangian in \eqref{eq:maxwelllagrangian} for a pure vacuum field that corresponds to an intrinsically complex vector potential $z$ with the appropriate gauge freedom.  We found that we could reproduce (and generalize) all the traditional results that stem from the Lagrangian \eqref{eq:traditional} using this ansatz.  In particular, we derived dual-symmetrized formulas for the various canonical Noether currents (i.e., the energy-momentum, angular-momentum, spin, and helicity currents). However, an isolated vacuum field is an incomplete picture, since such a field can only be measured through coupling to charged probes.

We now detail how the dual symmetry of the vacuum field is broken by the gauge mechanism that couples it to charge.  We will find that a charge essentially introduces a preferred angle of rotation for the dual symmetry, and thus will couple only to a (conventional) electric part of the field that is governed by a pure electric vector potential $a_e$.  The magnetic potential $a_m$ can remain after breaking the symmetry, but it is uncoupled to the electric charge and is thus unmeasurable (except by its effect on gravity, in principle \cite{Vasconcellos2014}).  However, one can speculate that this extra potential is still required for the complete description of the vacuum field, and may be measurable by its coupling to other fields not considered in isolated electromagnetism (e.g., gravity).


\subsubsection{Gauge mechanism}\label{sec:gauge}


Physically, a pure vector potential $a_e$ acts as a \emph{connection field} for the proper parallel transport of a coupled charged field across spacetime, which provides spacetime with an effective \emph{curvature} that corresponds to the electromagnetic field $\bv{F} = \nabla\wedge a_e$ \cite{Yang1954,Yang1980,Yang2014,Yang2014b,Chern1974,THooft1972,Witten1988,Weinberg1995,Nakahara2003,Wu2006}.  This curvature alters the geodesics for the charged field from its typically straight lines, which can be understood as the effect of the electromagnetic force on the charged field.  The gauge freedom of $a_e$ does not affect this induced curvature, so is not directly observable in terms of this physical force.  However, the influence of $a_e$ as a dynamical field can be indirectly observed in other topologically-induced field dynamics, such as the Aharonov--Bohm effect \cite{Aharonov1959,Konopinski1978,Tonomura1986}.

From this point of view, a gauge field like $a_e$ can be understood as a dynamical mechanism for effectively \emph{warping the spacetime metric} by using a simpler flat spacetime metric as an static embedding space.  The curvature of the effective metric then depends on the charge configuration, in an entirely analogous manner to how the curvature of the actual spacetime metric depends on the configuration of the energy-momentum density in the gravitational theory of General Relativity \cite{Soper1976,Landau1975}.  Indeed, it has been shown that all currently measured predictions of General Relativity may be reproduced by a simpler dynamical gauge field reformulation over a flat and static spacetime metric \cite{Lasenby1998,Hestenes2005,Doran2007}; this gauge theory of gravity only appreciably differs in its (as-yet-unmeasured) predictions for spin-fields and torsion, and in its inability to handle global changes to the topology of spacetime (which may be unmeasurable or unphysical anyway).

To make the gauge origins of electromagnetism evident, we consider a complex scalar field $\phi(x)$ as the simplest prototype for a charged matter spinor field \cite{Weinberg1995}.  The complexity of the field is the origin of its charge; indeed, the two possible signs of the charge correspond to the two possible orientations of the pseudoscalar $I$ of the spacetime 4-volume.  

The field Lagrangian density for this complex field $\phi$ must be a proper scalar field so it must involve only quadratic terms like
\begin{align}\label{eq:barelagrangian}
  \mathcal{L}(x) &= \nabla\phi^* \cdot \nabla\phi - m^2 \phi^*\phi.
\end{align}
According to the Euler-Lagrange equations $\nabla \cdot (\delta\mathcal{L}/\delta\nabla\phi^*) = \delta\mathcal{L}/\delta\phi^*$, this bare Lagrangian produces the relativistic Klein-Gordon equation for the scalar field: $(\nabla^2 + m^2)\phi = 0$, as well as its complex conjugate.  This Lagrangian is invariant under the \emph{global} phase transformation $\phi(x) \mapsto \phi(x) \exp(\theta I)$ (which is in fact identical to the dual-symmetry phase transformation for the electromagnetic field from Section~\ref{sec:dualsym}).

The idea for a gauge field then arises from the observation that the mass term involving $\phi^*\phi = |\phi|^2$ is also invariant under a \emph{local} phase rotation of the complex field $\phi \mapsto \exp[q_e\theta(x)I]\phi$, but the kinetic term that involves derivatives $\nabla\phi$ is not.  Here $q_e = \pm|q_e|$ is an appropriate coupling strength (i.e., charge) that also indicates the orientation for the pseudoscalar $I$.  Performing the phase rotation with the derivative yields $\nabla\phi \mapsto \exp(q_e\theta I)(\nabla\phi + q_e[\nabla\theta]\phi I)$.  The extra term involving the gradient of the local phase spoils the invariance of the Lagrangian.

To fix this problem, the phase gradient can be absorbed into the definition of an auxiliary vector field $a_e$ that must be the same type of blade as the phase gradient $\nabla \theta$ (i.e., a pure vector field).  Specifically, we can modify the derivative in the Lagrangian to a \emph{covariant derivative} that includes this field 
\begin{align}
  D &\equiv \nabla - q_e a_e I.
\end{align}
The extra term from the phase rotation of $\phi$ can then be canceled by a corresponding modification of the vector field 
\begin{align}\label{eq:vecpotgauge}
  a_e \mapsto a_e + \nabla\theta.
\end{align}
Since this shift of the vector field must not change anything fundamental, the field $a_e$ must be defined only up to the addition of an arbitrary gradient term, which is the origin of the gauge freedom of $a_e$.  After the introduction of this gauge field, the phase rotation commutes with the covariant derivative
\begin{align}
  D\phi \mapsto (D - q_e[\nabla\theta]I)(\exp[q_e\theta I]\phi) = \exp(q_e\theta I)D\phi. 
\end{align}

We then rewrite the Lagrangian \eqref{eq:barelagrangian} in a properly invariant form by replacing all bare derivatives with covariant derivatives
\begin{align}\label{eq:covariantlagrangian}
  \mathcal{L}(x) &= D\phi^*\cdot D\phi - m^2 \phi^*\phi \\
  &= \nabla\phi^*\cdot \nabla\phi+q_e w\cdot a_eI + (|q_e a_e|^2-m^2)|\phi|^2, \nonumber \\
  w &\equiv (\nabla\phi^*)\phi -(\nabla\phi)\phi^* = 2[\phi_1(\nabla\phi_2) - \phi_2 (\nabla\phi_1)].
\end{align}
Note that the addition of the gauge field results in a linear coupling between $a_e$ and the proper 4-current $q_e\,w$, which is the charge $q_e$ times the proper 4-velocity $w$ of the complex field $\phi = \phi_1 + \phi_2 I$.  The gauge field also renormalizes the mass term; however, this term is an artifact of the quadratic kinetic term for the scalar field. Spinor fields with linear kinetic terms (such as the Dirac electron field) do not acquire such a term.

The linear coupling in \eqref{eq:covariantlagrangian} implies that $a_e$ must \emph{itself} be a dynamical field with its own kinetic term.  However, the form of this kinetic term is initially unspecified.  Adding the simplest quadratic contraction of $\nabla a_e$ as a kinetic term produces the gauge-invariant \emph{scalar electrodynamics} Lagrangian
\begin{align}\label{eq:lagrangian}
  \mathcal{L}(x) &= D\phi^*\cdot D\phi + \frac{1}{2}\mean{(\nabla a_e)^2}_0 - m^2 \phi^*\phi \\
  &= \nabla\phi^*\cdot \nabla\phi + \frac{1}{2}\mean{(\nabla a_e)^2}_0 + q_e w\cdot a_e I + (|q_e a_e|^2 - m^2)|\phi|^2. \nonumber
\end{align}
If we define the bivector field $\bv{F}_e = \nabla a_e = \nabla\wedge a_e$ as before, then we find that the postulated kinetic term for $a_e$ in \eqref{eq:lagrangian} is precisely the traditional Lagrangian \eqref{eq:traditional}.  

Applying the Euler-Lagrange equation $\nabla\cdot(\delta\mathcal{L}/\delta\nabla a_e) = \delta\mathcal{L}/\delta a_e$ to the gauge-invariant Lagrangian \eqref{eq:lagrangian} produces Maxwell's equation \eqref{eq:maxwellsources} as the equation of motion for the dynamical gauge field $a_e$
\begin{align}
  \nabla^2 a_e &= \nabla \bv{F}_e = j_e,
\end{align}
where the pure vector current (for scalar electrodynamics) has the explicit form
\begin{align}
  j_e &= q_e w + j_a, & j_a &= [q_e\,|\phi|^2] [q_e\,a_e].
\end{align}
The charge-current of the field $q_e w$ is modified in this case by an extra nonlinear term $j_a$ that couples the charge density $q_e|\phi|^2$ back to the gauge field itself.  Again, this term does not appear for fields with linear kinetic terms (like the Dirac electron).  

Importantly, the extra term that is being canceled by the gauge approach is the (vector) phase gradient $\nabla \theta$ arising from the transformation $\exp(q_e\theta I)$. As such, only a pure vector potential $a_e$ can be modified as in \eqref{eq:vecpotgauge} to cancel this extra term, in accordance with the standard treatment of electromagnetism.  An additional trivector potential $a_m I$ is thus not motivated by the usual gauge transformation $\exp(q_e\theta I)$.


\subsubsection{Breaking dual symmetry}\label{sec:symmetry}


Notably, there is no overt restriction from the gauge mechanism itself that forbids adding a second kinetic term to the Lagrangian that corresponds to the second potential $a_m$ (as in \eqref{eq:maxwelllagrangian}) and then treating $z = a_e + a_m I$ as an intrinsically complex vector field.  The gauge mechanism only motivates the necessary introduction of $a_e$, but it does not specify whether $a_e$ is the \emph{complete} gauge field.  The gauge mechanism only implies that if additional symmetry for $a_e$ exists in vacuum prior to coupling, then it must be \emph{broken} by the coupling to the charged field.

In fact, we already know that the electromagnetic theory does in fact arise from precisely such a symmetry-breaking process.  In the Glashow-Weinberg-Salam electroweak theory \cite{Glashow1961,Weinberg1961,Salam1968} the vector potential $a_e$ is itself a hybridization of two massless vector boson fields $w_0$ and $b_0$ that each obey Lagrangians $\mean{(\nabla w_0)^2}_0/2$ and $\mean{(\nabla b_0)^2}_0/2$ that are identical in form to the traditional electromagnetic Lagrangian \eqref{eq:traditional}.  After constructing a complex boson field for this electroweak doublet $z = b_0 + w_0 I$, it then behaves identically to the (unconstrained) dual-symmetric formalism explored in this report.  In particular, the sum of Lagrangians for $b_0$ and $w_0$ can be rewritten as the single complex field Lagrangian $\mean{(\nabla z)(\nabla z^*)}_0/2$, which is precisely our dual-symmetric form \eqref{eq:maxwelllagrangian} motivated by the intrinsic complex structure of spacetime.  Furthermore, there is evidence that performing a gauge-procedure (as in the previous section) to convert the global dual symmetry of this doublet to a local symmetry in its own right recovers additional structure of the electroweak theory \cite{Vasconcellos2014}.

\begin{figure}[t]
  \begin{center}
    \includegraphics[width=0.4\columnwidth]{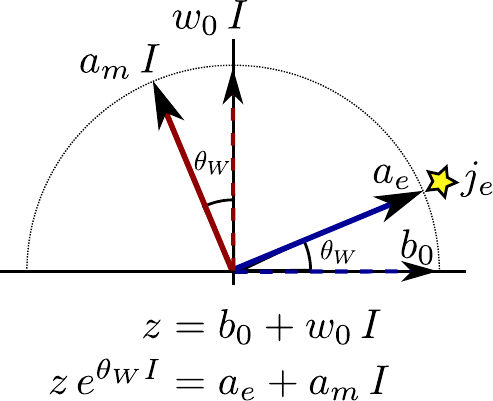}
  \end{center}
  \caption[Breaking dual symmetry]{Breaking dual symmetry. The neutral complex vector potential $z = b_0 + w_0 I$ that includes the neutral $b_0$ and $w_0$ bosons in the electroweak theory is dual-symmetric in vacuum, meaning that it is invariant under any global phase rotation $z \mapsto z\,\exp(\theta I)$.  The Brout-Englert-Higgs-Guralnik-Hagan-Kibble (BEHGHK, or Higgs) mechanism fixes a particular (Weinberg) angle $\theta_W$ that couples the field to a purely electric source, breaking this dual symmetry.  The vector part $a_e = \mean{z\exp(\theta_W I)}_1$ of the doublet becomes the electric vector potential for the photon ($\gamma$) that couples to vector electric source $j_e$.  The pseudovector part $a_m I = \mean{z\exp(\theta_W I)}_3$ acquires mass from the Higgs mechanism to become the neutral ($Z_0$) boson of the weak force, which is the analogy to the magnetic vector potential that appears in the dual-symmetric formalism.}
  \label{fig:symmetrybreaking}
\end{figure}

For the electroweak theory, the Brout-Englert-Higgs-Guralnik-Hagan-Kibble (BEHGHK, or Higgs) mechanism \cite{Englert1964,Higgs1964,Guralnik1964} is responsible for breaking the dual symmetry of this doublet, as schematically shown in Fig.~\ref{fig:symmetrybreaking}.  In effect, the gauge coupling to the scalar BEHGHK boson (Higgs) field fixes a specific dual symmetry (Weinberg) rotation angle $\theta_W$ that fixes $a_e = \mean{z\exp(\theta_W I)}_1$ and $a_m = \mean{z\exp(\theta_W I)}_3I^{-1}$ as the appropriate matter-coupled vector fields.  The field $a_e$ becomes the massless electromagnetic field vector potential for the photon ($\gamma$), while the field $a_m$ acquires mass and corresponds to the neutral ($Z_0$) boson of the weak force.  The details of this emergent coupling are much more involved than the simple electromagnetic model considered here, but the essential character of the broken dual symmetry is identical.

We can therefore understand the dual-symmetric formulation of electromagnetism as a simpler model of the neutral vacuum vector boson fields before the symmetry is broken by coupling to matter.  In this dual-symmetric model, the gauge field $a_m$ is not coupled to matter at all, but is still left over as a massless part of the total vacuum field.  It is needed to preserve the symmetry of the vacuum, which \emph{a priori} has no preferred reference axis for the dual-symmetric rotation $\exp(\theta I)$.  The introduction of charged matter breaks this symmetry to force a particular reference axis for the dual symmetry, exactly as discussed in Section \ref{sec:dualsymmetrysource}.  It follows that the only quantities that can be measured (according to this model) from all the calculated conserved quantities will be the electric quantities (involving the potential $a_e$, its field $\bv{F}_e$, and the source $j_e$) just as with the standard formulation of electromagnetism.  This does not, however, prevent the magnetic parts ($a_m$,$\bv{F}_m$,$j_m$) from being measurable in principle using other interactions that are not considered in this simple model, such as the additional isospin interaction that is acquired by the neutral ($Z_0$) boson that mediates the weak force.

The more remarkable note for the purposes of this report is that this dual-symmetric structure appears almost entirely from the assumption of spacetime.  It is the geometry of spacetime itself that produced the intrinsically dual-symmetric bivector field $\bv{F}$ and associated complex vector potential $z$, as well as the especially simple forms of Maxwell's equation $\nabla \bv{F} = \nabla^2 z = j$ and the Lorentz force law $dp/d\tau = \mean{\bv{F}j}_1$.  The use of spacetime algebra makes the geometric origin of all these quantities especially apparent.


\clearpage

\section{Conclusion}\label{sec:conclusion}

\epigraph{4.5in}{It requires a much higher degree of imagination to understand the electromagnetic field than to understand invisible angels. \ldots~I speak of the $\rv{E}$ and $\rv{B}$ fields and wave my arms and you may imagine that I can see them \ldots~[but] I cannot really make a picture that is even nearly like the true waves.}{Richard P. Feynman \cite{Feynman1964}}


In this report we have presented a detailed introduction to the natural (Clifford) algebra of the Minkowski spacetime geometry, and have demonstrated its practicality and power as a tool for studying electromagnetism.  Spacetime algebra not only preserves the existing electromagnetic formalisms, but also generalizes and unifies them into one comprehensive whole.  This unification produces tremendous insight about the electromagnetic theory in a comparatively simple way.  Below, we briefly review a few of the most important features that have emerged from the spacetime algebra approach, noting that we have also summarized a more complete list in the introductory Section~\ref{sec:insights}, which we encourage the reader to review.

In particular, we emphasize that spacetime has an \emph{intrinsic complex structure}, where the \emph{ad hoc} scalar imaginary $i$ is replaced with the physically meaningful unit 4-volume (pseudoscalar) $I$.  Within this algebra, the electromagnetic field is an intrinsically complex bivector field $\bv{F} = \rv{E} + \rv{B}I$ on spacetime, which is a reference-frame-independent generalization of the Riemann-Silberstein complex field vector.  Similarly, the vector potential for this field becomes complex, with associated complex bivector and scalar (Hertz) potentials.  In all these representations, Maxwell's equations reduce to a single, and inexorable, equation.  Similarly, the proper Lorentz force reduces to the simplest contraction of a current with the field, while the symmetric energy-momentum tensor is the simplest quadratic function of the field (identical in form to the energy-momentum current of the relativistic Dirac electron) and produces the associated angular momentum tensor as a wedge product.

The global phase of each complex field representation is arbitrary in vacuum (just like the global phase of a quantum wavefunction), and is responsible for the continuous dual (electric-magnetic field-exchange) symmetry of the vacuum electromagnetic field.  Adding a source breaks this dual symmetry of the vacuum; however, making the charge of the source similarly complex (with separate electric and magnetic parts) preserves a similar global phase symmetry of the total system.  Choosing the source to be purely electric fixes this arbitrary global phase by convention, but the same physics will be described for any choice of global phase, which will produce equivalent descriptions that use both electric and magnetic charges.  The dual symmetry of the complex field can be broken, however, by coupling to matter or other fields (as is the case in the electroweak theory).

In the Lagrangian treatment, the dual symmetry of the vacuum motivates the complex vector potential to be the primary dynamical field.  Its constituent ``electric'' and ``magnetic'' parts effectively satisfy two separate traditional electromagnetic Lagrangians. As such, the traditionally conserved Noether currents (i.e., the canonical energy-momentum and angular-momentum tensors) are reproduced, but in an appropriately field-symmetrized form.  The additional dual symmetry also immediately produces a conserved helicity current that cannot be easily derived in the traditional electromagnetic theory. Notably, the canonical Noether currents explictly separate the orbital and spin parts of the local linear and angular momentum densities of the field, which are currently under active discussion in relation to both theoretical and experimental aspects of optics, QED, and QCD.

We also emphasize that the representation-free nature of spacetime algebra exposes deep structural relationships between mathematical topics that are seemingly disconnected in standard approaches. A detailed list of the remarkable and beneficial features of the spacetime algebra approach is provided in Section~\ref{sec:insights}. Some of the most poignant examples of this unification include: 
\begin{itemize}
  \item The Hodge-star duality operation from differential forms is equivalent to multiplication by $I^{-1}$.  
  \item The 4 Dirac matrices $\gamma_\mu$ from the relativistic electron theory are representations of unit 4-vectors in spacetime algebra (which require no such matrix representation).  
  \item The 3 Pauli matrices $\rv{\sigma}_i=\gamma_i\gamma_0$ are representations of relative spatial directions, which are unit plane segments (bivectors) that drag the spatial coordinates $\gamma_i$ along the proper-time direction $\gamma_0$. 
  \item Their duals $\rv{\sigma}_i I^{-1}$ are spacelike planes that are equivalent to the quaternions. (There is thus nothing intrinsically quantum mechanical about any of these objects.)  
  \item The commutator brackets of these 6 bivectors are precisely the Lie algebra of the Lorentz group, and reduce to the Pauli spin commutation relations in a relative frame.
  \item The spin commutation relations are simply the Hodge-duals of the usual 3-vector cross product relations.  
  \item Exponentiating the bivector for a spatial plane produces a rotation spinor, while exponentiating a timelike plane produces a Lorentz boost spinor.  
  \item These spinors can Lorentz-transform any geometric object with a double-sided product, in precisely the same way as the unitary transformations familiar from quantum mechanics, or the quaternionic formulation of spatial rotations.
\end{itemize}

It is our hope that this review will prompt continued research into the applications of spacetime algebra for both electromagnetic theory and other relativistic field theories.  We feel that the language of spacetime algebra is practical, mature, and provides tremendous physical insight that can be more difficult to obtain when using other mathematical formulations of spacetime.  Moreover, we feel that it is a natural extension of the existing electromagnetic formalisms in use, which gives it the distinct advantage of preserving (and augmenting) prior intuition and experience gained from working with those formalisms. 

\section*{Acknowledgments}
  JD is indebted to Alexander Korotkov for the opportunity to complete this research. We thank Sophia Lloyd, Mauro Cirio, and Alexei Deriglazov for valuable commentary. We also thank Edoardo D'Anna and Laura Dalang for their assistance with figures. This work was partially supported by the ARO, RIKEN iTHES Project, MURI Center for Dynamic Magneto-Optics, JSPS-RFBR contract no. 12-02-92100, Grant-in-Aid for Scientific Research (S). It was also partially funded by the ARO MURI Grant No. W911NF-11-1-0268, as well as the ODNI-IARPA-ARO Grant No. W911NF-10-1-0334. All statements of fact, opinion, or conclusions contained herein are those of the authors and should not be construed as representing the official views or policies of IARPA, the ODNI, or the U.S. Government. 


\bibliographystyle{elsarticle-num}

\end{document}